\PassOptionsToPackage{square,comma,numbers,sort&compress}{natbib}
\documentclass[final,11pt]{elsarticleMod} 
\usepackage[T1]{fontenc}
\usepackage{natbib}
\usepackage{dsfont}               %
\usepackage{mathrsfs}             %
\usepackage{slashed}              %
\usepackage{amsmath,empheq}
\usepackage{amssymb}
\usepackage{amsbsy}
\usepackage{amsfonts}
\usepackage{multicol}
\usepackage{relsize}
\usepackage{nicefrac}
\usepackage{tabularx}
\usepackage{arydshln}  %
\numberwithin{equation}{section}
\numberwithin{table}{section}
\numberwithin{figure}{section}
\let\oldcite\cite
\renewcommand{\cite}[1]{\mbox{\oldcite{#1}}}
\usepackage{titlesec} 
\usepackage{sectsty}
\titleformat{\section}{\normalfont\Large\bfseries}{\thesection}{1em}{}
\titleformat{\subsection}{\normalfont\large\bfseries}{\thesubsection}{1em}{}
\titleformat{\subsubsection}{\normalfont\normalsize\bfseries}{\thesubsubsection}{1em}{}
\usepackage{amsmath,amssymb}
\usepackage{scalerel} %
\usepackage{graphicx}
\usepackage{lineno}
\usepackage{MnSymbol}
\usepackage{multirow}
\usepackage{subfigure,slashed}
\usepackage[dvipsnames]{xcolor}
\usepackage{setspace}
\usepackage{array}
\usepackage{verbatim}
\usepackage{cancel}
\usepackage{paralist}
\usepackage[normalem]{ulem}
\usepackage{hyperref}
\usepackage{caption}
\usepackage{pdfpages}
\hypersetup{
    pdftitle={Ruiz, et al, Target mass corrections in lepton-nucleus DIS --     Prog.Part.Nucl.Phys. 136 (2024) 104096 [arXiv:2301.07715]},  %
    pdfauthor={Ruiz, et al},
    colorlinks=true,
}
\usepackage{calligra}
\DeclareMathAlphabet{\mathcalligra}{T1}{calligra}{m}{n}
\DeclareFontShape{T1}{calligra}{m}{n}{<->s*[2.2]callig15}{}
\newcommand{\scripty}[1]{\ensuremath{\mathcalligra{#1}}}
\def\rr{\scripty{r}}   %

\graphicspath{{}}

\newcommand{\pMass}{\ensuremath{M_p}} %

\newcommand{\GeV}{{\rm ~GeV}}

\newcommand{\TMC}{{\rm TMC}}

\newcommand{\stkout}[1]{\ifmmode\text{\color{red}\sout{\ensuremath{#1}}}\else{\color{red}\sout{#1}}\fi}

\newcommand{\etaW}{\ensuremath{\eta_{{}_W}}}
\newcommand{\fhead}[1]{\textbf{#1}}
\newcommand{\orcid}[1]{\,\href{https://orcid.org/#1}{\includegraphics[width=9pt]{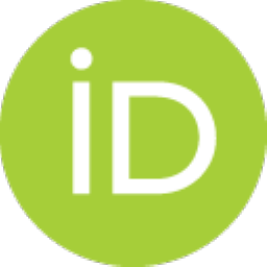}}}
\newcommand{\orcidTJ}{0000-0002-1334-7607} %
\newcommand{\orcidAA}{0000-0002-2077-6557} %
\newcommand{\orcidPD}{0000-0001-7960-7953} %
\newcommand{\orcidTH}{0000-0002-2729-0015} %
\newcommand{\orcidTK}{0000-0002-7516-8292} %
\newcommand{\orcidMK}{0000-0002-4665-3088} %
\newcommand{\orcidKK}{0000-0003-1412-447X} %
\newcommand{\orcidAK}{0000-0002-4090-0084} %
\newcommand{\orcidJM}{0000-0001-9343-9351} %
\newcommand{\orcidFM}{0000-0002-3888-1697} %
\newcommand{\orcidFO}{0000-0001-6799-2436} %
\newcommand{\orcidJO}{0000-0002-7351-0218} %
\newcommand{\orcidIS}{0000-0003-0373-474X} %
\newcommand{\orcidJY}{0000-0001-8366-0968} %
\newcommand{\orcidRR}{0000-0002-3316-2175} %
\newcommand{\orcidPR}{0000-0002-8570-5506} %
\newcommand{\orcidCL}{0000-0001-7509-5655} %

\def\smu{{Department of Physics, Southern Methodist University,
    Dallas, TX 75275-0175, U.S.A.}}
\def\jlab{{Jefferson Lab, Newport News, VA 23606, U.S.A.}}

\def\krakow{{Institute of Nuclear Physics Polish Academy of Sciences, PL-31342 Krakow, Poland}}
\def\muenster{{Institut f{\"u}r Theoretische Physik, Westf{\"a}lische Wilhelms-Universit{\"a}t M{\"u}nster,
  \\Wilhelm-Klemm-Stra{\ss}e 9, D-48149 M{\"u}nster, Germany}}
\def\lpsc{{Laboratoire de Physique Subatomique et de Cosmologie, Université Grenoble-Alpes, 
    \\CNRS/IN2P3, 53 avenue des Martyrs, 38026 Grenoble, France}}
\def\hampton{{Hampton University, Hampton, VA 23668, USA}}
\def\fnal{{Fermi National Accelerator Laboratory, Batavia, Illinois 60510, USA}}

\def\fsu{{Department of Physics, Florida State University, Tallahassee, Florida 32306-4350, USA}}

\def\wales{{School of Physics, The University of New South Wales, Sydney NSW 2052, Australia}}
\def\jyv{{University of Jyväskylä, Department of Physics, P.O.\ Box 35, FI-40014, Finland}}
\def\helsinki{{Helsinki Institute of Physics, P.O.\ Box 64, FI-00014 University of Helsinki, Finland}}
\def\argonne{{High Energy Physics Division, Argonne National Laboratory, Argonne, Illinois 60439, USA}}
\def\fzj{{Institut für Energie- und Klimaforschung, Forschungszentrum Jülich GmbH, 52425 Jülich, Germany}}
 \modulolinenumbers[5]

\journal{IFJPAN-IV-2022-18, SMU-HEP-22-12, MS-TP-22-49, ANL-180568, FNAL-PUB-23-142-ND} %

\begin{document}
\begin{frontmatter}

\title{
Target mass corrections in lepton--nucleus DIS:\\
{\Large theory and applications to nuclear PDFs}
} 

\author[a]{R.~Ruiz\orcid{\orcidRR}\corref{corauthor}}\ead{rruiz@ifj.edu.pl}
\author[b,c]{K.F.~Muzakka\orcid{\orcidFM}}\ead{k.muzakka@fz-juelich.de}
\author[d]{C.~L\'eger\orcid{\orcidCL}}\ead{leger@lpsc.in2p3.fr}
\author[b]{P.~Risse\orcid{\orcidPR}}\ead{risse.p@uni-muenster.de}
\author[e,f]{A.~Accardi\orcid{\orcidAA}}\ead{accardi@jlab.org}
\author[b,g,h]{P.~Duwent\"aster\orcid{\orcidPD}}\ead{pitduwen@jyu.fi}
\author[i]{T.J.~Hobbs\orcid{\orcidTH}}\ead{tim@anl.gov}
\author[b]{T.~Je\v{z}o\orcid{\orcidTJ}}\ead{tomas.jezo@uni-muenster.de}
\author[e]{C.~Keppel\orcid{\orcidTK}}\ead{keppel@jlab.org}
\author[b,j]{M.~Klasen\orcid{\orcidMK}}\ead{michael.klasen@uni-muenster.de}
\author[b]{K.~Kova\v{r}\'{i}k\orcid{\orcidKK}}\ead{karol.kovarik@uni-muenster.de}
\author[a]{A.~Kusina\orcid{\orcidAK}}\ead{Aleksander.Kusina@ifj.edu.pl}
\author[k]{J.G.~Morf\'{i}n\orcid{\orcidJM}}\ead{morfin@fnal.gov}
\author[l]{F.I.~Olness\orcid{\orcidFO}\corref{corauthor}}\ead{olness@smu.edu}
\author[m]{J.F.~Owens,\orcid{\orcidJO}}\ead{owens@hep.fsu.edu}
\author[d]{I.~Schienbein\orcid{\orcidIS}\corref{corauthor}}\ead{schien@lpsc.in2p3.fr}
\author[d]{J.Y.~Yu.\orcid{\orcidJY}}\ead{yu@physics.smu.edu}
\address[a]{\krakow}
\address[b]{\muenster}
\address[c]{\fzj}
\address[d]{\lpsc}
\address[e]{\jlab}
\address[f]{\hampton}
\address[g]{\jyv}
\address[h]{\helsinki} 
\address[i]{\argonne}
\address[j]{\wales}
\address[k]{\fnal}
\address[l]{\smu}
\address[m]{\fsu}

\cortext[corauthor]{Corresponding authors}

\begin{abstract}
Motivated by the wide range of kinematics covered by current and
planned deep-inelastic scattering (DIS) facilities, we revisit the
formalism, practical implementation, and numerical impact of target
mass corrections (TMCs) for DIS on unpolarized nuclear targets.  An
important aspect is that we only use nuclear and later partonic
degrees of freedom, carefully avoiding a picture of the nucleus in
terms of nucleons.  
After establishing that formulae used for individual nucleon targets
$(p,n)$, derived in the Operator Product Expansion (OPE) formalism,
are indeed applicable to nuclear targets, we rewrite expressions for
nuclear TMCs in terms of \mbox{re-scaled} (or averaged) kinematic
variables. As a consequence, we find a representation for nuclear TMCs
that is approximately independent of the nuclear target. We go on to
construct a single-parameter fit
for all nuclear targets that is in good numerical agreement with full
computations of TMCs.
We discuss in detail qualitative and quantitative differences between
nuclear TMCs built in the OPE and the parton model
formalisms, as well as give numerical predictions for
current and future facilities.
\end{abstract} 

\begin{keyword}
{\small
DIS, Structure Functions, Target Mass Corrections, OPE, nuclear PDFs
}
\\
Journal: \href{https://doi.org/10.1016/j.ppnp.2023.104096}{Prog. Part. Nucl. Phys. 136 (2024) 104096}
\\
ArXiv: \href{https://arxiv.org/abs/2301.07715}{2301.07715}

\end{keyword}

\end{frontmatter}

\newpage
\tableofcontents
\newpage

\section{Introduction}
\label{sec:intro}

Deep-inelastic scattering (DIS) of high-energy leptons off nucleons and nuclei is a key process for 
studying the structure of hadrons in terms of their quark and gluon degrees of freedom
\cite{Taylor:1991ew,Kendall:1991np,Friedman:1991nq,Devenish:2004pb}.
For both nucleons 
\cite{Hou:2019efy,Accardi:2016qay,Bailey:2020ooq,Abramowicz:2015mha,Ball:2017nwa,Alekhin:2017kpj}
and nuclei
\cite{Hirai:2007sx,deFlorian:2011fp,Eskola:2009uj,Eskola:2016oht,Eskola:2021nhw,Schienbein:2009kk,Kovarik:2015cma,Kusina:2016fxy,Segarra:2020gtj,Kusina:2020lyz,Duwentaster:2021ioo,Duwentaster:2022kpv,Muzakka:2022wey,AbdulKhalek:2019mzd,AbdulKhalek:2020yuc,Khalek:2022zqe,Walt:2019slu,Khanpour:2020zyu},
present data from electron and neutrino DIS experiments form the backbone of modern determinations
of parton distribution functions (PDFs), which describe and encode the internal structure of hadrons.
This process will again be at the forefront of the future Electron-Ion Collider (EIC) \cite{Accardi:2012qut,AbdulKhalek:2021gbh}, 
where DIS off nucleons and various nuclear targets will be studied with high precision 
over a wide kinematic range.
Similar endeavors are also in discussion for the proposed Forward Physics Facility at CERN~\cite{Feng:2022inv}. 
Furthermore, 
high-precision fixed-target measurements from JLab 
provide complementary information~\cite{Arrington:2021alx}.
In addition to these facilities, DIS also plays an important role at current and future
accelerator-based neutrino oscillation facilities such as DUNE and the Short-Baseline Neutrino Program at Fermilab~\cite{DUNE:2015lol,Machado:2019oxb,Ruso:2022qes}, 
where experiments operate at somewhat lower energies compared to the EIC, 
and where the (quasi)elastic and resonant processes must also be taken into account.
Here, precise control over the contributions from each of these processes is essential, and duality-based arguments for the scaling of DIS cross sections at higher energies into the resonance region can play an important role~\cite{Bloom:1970xb,Melnitchouk:2005zr}.
However, due to the relative weakness of the effective neutrino-nucleon  interaction, neutrino experiments typically
use massive nuclear targets to enhance event rates.

In order to reach the ambitious physics goals of future experiments,charged lepton-nucleus and neutrino-nucleus interactions need to be precisely understood,
both experimentally
and theoretically.
On the theory side, there are broad efforts to improve the description of DIS in Monte Carlo simulation tools~\cite{Andreopoulos:2009rq,
Alwall:2014hca,Sherpa:2019gpd,Garcia:2019hze,Garcia:2020jwr,Bierlich:2022pfr,Isaacson:2022cwh}, 
increase the perturbative accuracy of matrix elements, particularly for more exclusive DIS channels,
and of course improve determinations of parton distribution functions of nuclei~\cite{Khanpour:2020zyu,Helenius:2021tof,Eskola:2021nhw,Kovarik:2015cma,Duwentaster:2022kpv,AbdulKhalek:2022fyi}.
Still, it is a fact that the different properties of  charged lepton-nucleus and neutrino-nucleus DIS  are not well understood,
particularly for extreme kinematics,
and is a topic of ongoing study~\cite{Muzakka:2022wey}.
The specific case in DIS when, simultaneously, momentum transfers are small while the scaling variable approaches unity
is the focus of this work.

\fhead{Operator Product Expansion (OPE):}
Along these lines, there are two major theoretical approaches to lepton-\textit{nucleon} DIS
that are standard material in textbooks and review articles; 
both have been employed to include the effects of  target nucleons
having nonzero masses.
One approach is based on the
operator product expansion (OPE) \cite{Wilson:1969zs,Brandt:1970kg,Christ:1972ms}.
The other is the QCD-improved parton model rigorously derived from first principles
of QCD in the context of the  collinear factorization theorems \cite{Bodwin:1984hc,Collins:1985ue,Collins:1988ig,Collins:1987pm,Collins:1989gx,Collins:1998rz,Collins:2011zzd,Berger:1987er}.
The OPE was first employed to derive target mass corrections (TMCs) to structure functions in DIS at leading order (LO) in QCD in the seminal paper by 
Georgi \& Politzer in 1976\,\cite{Georgi:1976ve,Georgi:1976vf}.
While the application of the OPE to DIS is widely accepted, 
proofs of the OPE have only been presented for simple scalar models~\cite{Muta:1998vi,Muta:2010xua}.
The question as to whether this result can be extended to QCD in general remains open to the best of our knowledge.

\fhead{Target Mass Corrections (TMCs):}
Qualitatively, TMCs in DIS accounts for the mass of the target nucleon (or target nucleus) in kinematical variables and structure functions, 
which are typically derived in the limit of a massless target.
In the work of Georgi \& Politzer it was argued that for each power,
or ``twist''\footnote{%
In the context of the OPE we define the twist $\tau=d-s$ of a field operator $\hat{\mathcal{O}}$ in the canonical fashion, i.e., the dimension $d$ of $\hat{\mathcal{O}}$ (in the usual power-counting sense) 
minus its spin $s$ (the number of un-contracted Lorentz indices). 
For structure functions in DIS, ``higher twist'' terms 
are generally power-suppressed by some positive power of a hard scale of the process, i.e., $Q$.},
of the OPE expansion of hadronic matrix elements in DIS there is a tower of (kinematical~\cite{Ellis:1982wd,Ellis:1982cd}) corrections of the form $x_A^{j}(M_A^2/Q^2)^{j'}$, and 
specifically $(x_A M_A^2/Q^2)^j$ at leading power, 
where $x_A$ is the Bjorken scaling variable,
$M_A$ is the mass of the target hadron, 
$Q^2>0$ is the (squared) virtuality of the intermediate exchange boson, and $j,j'$ are some positive powers.
Remarkably, the entire tower of 
$(x_A M_A^2/Q^2)^k$ corrections at leading power 
can be summed, leading to a closed-form solution for TMCs at this power~\cite{Georgi:1976ve,Georgi:1976vf}.
In the same year, Barbieri, et al., derived TMCs 
at LO including effects arising from non-zero quark masses~\cite{Barbieri:1976rd,Barbieri:1976bj}.
Later, the original work of Georgi \& Politzer was extended to next-to-leading order (NLO) in QCD for the structure functions $W_{1,2}$
in the context of local duality in electro-production and using off-shell regularization~\cite{DeRujula:1976baf}. 
More recently, Kretzer \& Reno presented TMCs
for charged current (CC) and weak neutral current (NC) neutrino-nucleon DIS,
including NLO QCD corrections and heavy quark mass effects using modern
conventions \cite{Kretzer:2003iu}.

\fhead{Parton Model:}
In the context of the parton model, 
Ellis, Furmanski, \& Petronzio
derived TMCs at LO in QCD in the framework of a non-collinear version of the parton model that includes the effects of partonic transverse momentum ($k_T$)~\cite{Ellis:1982wd,Ellis:1982cd}. 
Agreement with the OPE-based results~\cite{Georgi:1976ve,Georgi:1976vf,Barbieri:1976rd,Barbieri:1976bj} was found at leading power,
thereby demonstrating the equivalence of the OPE at LO in QCD
to a non-collinear parton approach, where
the incident parton is on-shell but not collinear with its parent nucleon.
In these works, the terminology ``kinematical'' operators was introduced to describe TMCs in the OPE 
in order to distinguish them 
from ``dynamical'' higher twist operators that remain in the OPE when $(x_A M_A^2/Q^2)\to0$~\cite{Ellis:1982wd,Ellis:1982cd}.

\begin{figure}[!t]
\begin{center}
\includegraphics[width=\textwidth]{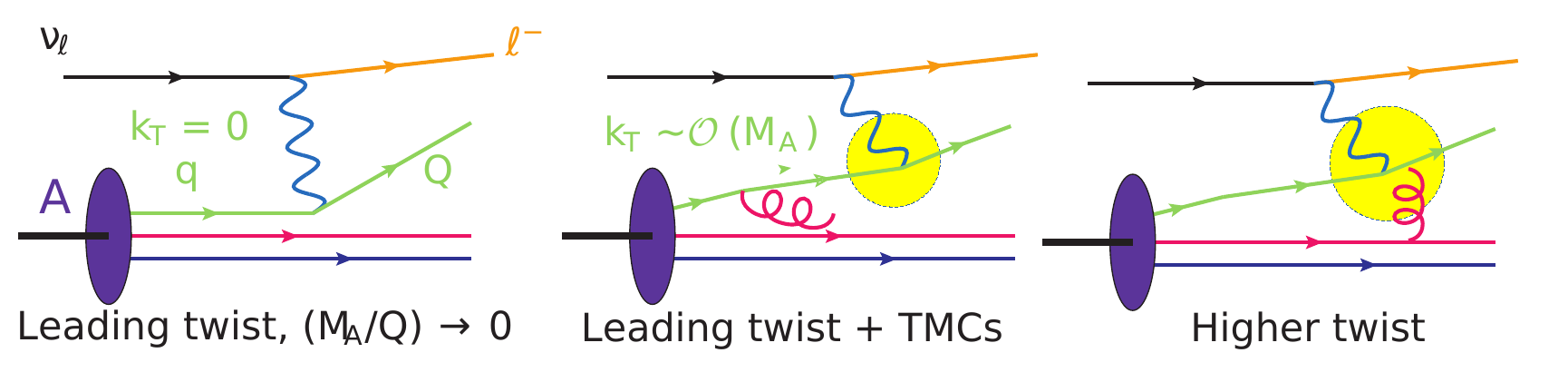}
\end{center}
\caption{Schematic of $\nu_\ell A \to \ell X$  for a lowest-order heavy quark production, $qW^*\to Q$.
a)~[left] Leading twist $\{M_A=0, k_T=0\}$.
b)~[center] Leading twist with TMCs $\{{\rm finite}\ M_A, \ k_T\sim \mathcal{O}(M_A)\}$.
The ``hard scattering'' is indicated by the yellow circle. 
The hadron dynamics gives the parton (green) a finite $k_T$, 
illustrated by the (red) gluon radiation; 
this is not a next-to-leading order (NLO) correction as the  soft (low energy) gluon radiation  is outside (before) the hard scattering process (yellow circle). 
c)~[right] Higher twist. A second parton (gluon) is exchanged between the hadron
and the hard scattering (yellow circle).
}
\label{fig:twist}
\end{figure}

\fhead{Higher Twist (HT):}
At this point, we stress the distinction between TMCs and (genuine) higher-twist corrections.
TMCs arise from the modification of kinematics due to  the presence of a hadron's mass $M_A$.
The behavior of amplitudes in the massless limit implies that 
mass terms are relatively suppressed by the hard scale of the process, leading to corrections that scale as powers of $(M_A^2/Q^2)$.
In contrast, higher-twist corrections can arise from exchanges of extra bosons (gluons and pions) between the hard  process and the hadronic remnant.
Hence, they are \textit{dynamical}.
Examples include double parton scattering and color (re)connection between active  and spectator quarks.
The presence of an ``extra'' boson exchange in higher-twist processes implies additional propagators, and therefore amplitudes that are also suppressed by the hard scale of the process.\footnote{For sufficiently inclusive observables, the contributions from soft gluons are subject to cancellations~\cite{Collins:1981ta,Collins:2011zzd}.} 
This leads to corrections in powers of $(\Lambda_{\rm NP}^2/Q^2)$, where  $\Lambda_{\rm NP}$ is the characteristic non-perturbative scale of a hadron.
Since, in practice, $\Lambda_{\rm NP}$ is often set to $M_A$,
the two corrections can be confused despite their distinct origins; in some sense, TMCs in DIS ``accidentally'' have the same characteristic  power-suppression as in higher-twist corrections.

We illustrate the differences between leading twist, TMCs, and higher twist in Fig.~\ref{fig:twist}
for leading-order heavy quark $Q$ production 
in the charged current process 
$\nu_\ell A \to \ell X$.
In the left panel~(a) is a depiction of the leading-twist process in the limit of vanishing TMCs, i.e., $k_T=0$ and $(M_A/Q)\to0$.
In the center panel~(b) is the leading-twist process with finite $(M_A/Q)$ which yields TMCs and generates $k_T\sim\mathcal{O}(M_A)$ via a gluon emission that occurs at a time $\tau_{\rm TMC}\sim1/k_T$  \textbf{before} the hard process (indicated with a yellow circle). 
In the right panel~(c) is the same process but with a typical higher-twist correction, i.e., an addition parton (gluon) exchange between the hadron and the hard scattering. 
The characteristic time of the interactions is inversely proportional to the energy scale,
${\rm (time)}\sim 1/({\rm Energy})$. For large $Q$, the hard scattering time scale $(1/Q)$ is short, 
while the characteristic time scale of the hadron dynamics ($1/\Lambda_{NP}$) is long. 
The probability of a second parton (gluon) participating in the hard interaction 
is proportional to the ratio of the time scales; 
thus, we expect higher twist contributions to be suppressed by powers of $(\Lambda_{NP}/Q)$.

\fhead{ACOT Formalism:}
Returning to the literature, we also note that TMC prescriptions based on the OPE and the factorization approach of Ellis, Furmanski, and Petronzio were similarly compared for semi-inclusive processes in Ref.~\cite{Accardi:2009md}.
In the collinear parton model ($k_T=0$), TMCs have been accounted for in the Aivazis-Collins-Olness-Tung (ACOT) formalism~\cite{Aivazis:1993kh,Aivazis:1993pi}, which is rigorously based on the factorization theorem including heavy quark masses~\cite{Collins:1998rz}. Furthermore, the full QCD framework for the evaluation of tau-neutrino deep-inelastic CC cross sections, including NLO corrections, charm production, tau-mass threshold, and target mass effects in the collinear approximation, was also presented in Ref.~\cite{Kretzer:2002fr}.

\fhead{Proton TMCs:}
An earlier review\,\cite{Schienbein:2007gr} of TMCs for nucleons in unpolarized DIS 
culminated 
in a set of so-called master formulae \cite{Kretzer:2003iu}, which
we extend and present below in Eq.~\eqref{eq:master} for \textit{nuclei}.
This set of equations is quite remarkable, 
organizing the rather complicated expressions in the aforementioned OPE-based papers
into a simple, easy-to-use, and modular form, valid at all orders in perturbative QCD  while still taking into account quark masses.
In principle, 
TMCs derived for DIS off nucleons\,\cite{Georgi:1976ve,Georgi:1976vf,Jaffe:1985je,Kretzer:2003iu,Schienbein:2007gr} 
apply also to DIS off nuclear targets since the OPE is (in theory) 
independent of target states. 
However, 
previous discussions~\cite{Schienbein:2007gr} do not address the subtle distinctions between a nucleon and nuclear target. 
Furthermore, the fact that established notation does not consistently distinguish between nucleons and nuclei in their respective  kinematics, 
the fact that the spin of a generic nucleus can be different from 1/2,
and the fact that the master equations in Eq.~\eqref{eq:master}
can be expressed conveniently in terms of ``averaged nucleon kinematics'' (as they are  in Eq.~\eqref{eq:master-rescaled}) 
have led 
to the question, why nucleon-like expressions could possibly be valid for nuclear targets?

\fhead{Nuclear TMCs:}
With an abundance of recent nuclear data from both electron and neutrino beams, and with the %
EIC and 3rd-generation long-baseline oscillations experiments on the horizon, 
there is a compelling need to 
rigorously revisit the derivation of TMCs to structure functions in charged-lepton and neutrino DIS with a particular focus on the {\em nuclear} case.
One of the main goals for this article is to  demonstrate the validity of the TMC master equations, as given in Eq.~\eqref{eq:master}, for the case of massive nuclear targets within the framework of the OPE.
Compared to past works on the subject\,\cite{Kretzer:2003iu,Schienbein:2007gr}, here
we present the derivation of TMCs from the OPE in much greater detail.
We hope that this material will be useful for students and researchers looking for a modern, in-depth discussion of the many technicalities that are frequently omitted in the literature and textbooks.

\fhead{Comparing OPE and Parton Model:}
In this work, we discuss and contrast the results for TMCs obtained in both the OPE and the collinear parton model;
throughout this review, we consider them on equal footing.
As discussed above, the collinear parton model is rigorously based on factorization theorems.
These provide field-theoretic definitions of PDFs and make statements about the error of the factorization approximation, which is generally inversely proportional to a positive power of a hard scale of the process.
It is generally believed that collinear factorization 
remains valid in lepton-nucleus and proton-nucleus collisions, possibly with nuclear-enhanced higher twist terms~\cite{Qiu:2003cg}.
However, the literature on factorization in the nuclear case is sparse and we consider this a working assumption.
Moreover, the non-collinear parton model (where the TMCs were shown to be equivalent to the OPE results at leading power) 
is not covered by the factorization proofs cited above.
Exploring these more theoretical questions is interesting and relevant but beyond the scope of this article.
Here, we take the collinear parton model and the OPE for granted, and explore extensively what happens when we transition from nucleons to nuclei in DIS.

\begin{figure}[!t]
\begin{center}
    \framebox{
\begin{minipage}[t]{0.90\textwidth}

\centering
\includegraphics[width=0.45\textwidth]{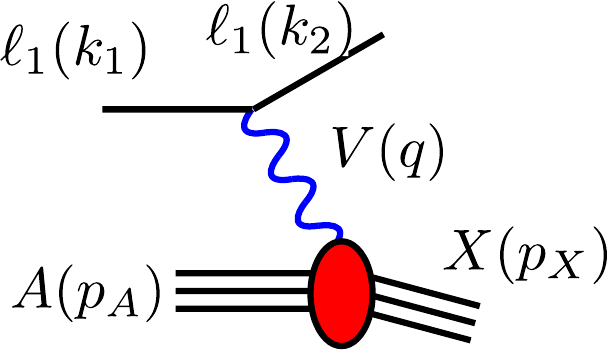}
\vspace*{0.5cm}\null
\renewcommand{\arraystretch}{2.0}
\begin{tabular}{|>{\centering}m{0.5\textwidth}|>{\centering}m{0.4\textwidth}|}
\hline 
\textbf{\large{}Kinematic variable} & 
\textbf{\large{}Description} \tabularnewline
\hline 
\hline 
$\nu_{A}=\dfrac{q\cdot p_{A}}{M_{A}}\;\stackrel{lab}{=}\;E_{\ell_1}-E_{\ell_2}$ & Lepton energy loss in the \\ nucleon rest frame (laboratory frame)\tabularnewline
\hline 
$y_{A}=\dfrac{q\cdot p_{A}}{k_1\cdot p_{A}}\;\stackrel{lab}{=}\;\dfrac{\nu_{A}}{E_{\ell_1}}$ & Inelasticity $y_A\in[0,1]$\tabularnewline
\hline 
$Q^{2}=-q^{2}>0$ & Squared boson momentum transfer\tabularnewline
\hline 
$x_{A}=\dfrac{Q^{2}}{2p_{A}\cdot q}=\dfrac{Q^{2}}{2M_{A}\nu_{A}}$ & Bjorken $x_A$ with $x_A\in[0,1]$\tabularnewline
\hline 
$W_A^{2}=(p_{A}+q)^{2}=M_{A}^{2}+Q^{2}\dfrac{1-x_{A}}{x_{A}}$ & Squared mass  of the recoil system\tabularnewline
\hline 
$s=(k_1+p_{A})^{2} = {\dfrac{Q^2}{x_A \, y_A}}  + M_A^2 + m_{\ell}^2$ & Squared center-of-mass System (CMS) energy\tabularnewline
\hline 
\end{tabular}
\renewcommand{\arraystretch}{1.0}

\caption{We consider the basic charged current ($V{=}W^\pm$) or neutral current ($V{=}\{\gamma,Z\}$)  lepton-nucleus DIS process 
\mbox{$\ell_1(k_1)+A(p_A)\to\ell_2(k_2)+X(p_{X})$} where
the incoming lepton can be a charged lepton ($e$, $\mu$)
or a neutrino ($\nu_e,\nu_\mu, \nu_\tau$).
Note, the energy loss symbol $\nu_A$ should not be confused with the neutrino symbols.
The 4-momentum of the exchange boson  is denoted $q=k_1-k_2$,
and $p_A^2=M_A^2$.
Here, the ``lab frame'' denotes the rest frame of $A$, meaning that $p_A=(M_A,0,0,0)$.
}
\label{fig:dis}
\end{minipage}\hfill
}  %
\end{center}
\end{figure}

\fhead{Overview:}
The starting point of our analysis is to consider the full nucleus as our target and apply only general symmetry principles, e.g., Lorentz invariance, in deriving nuclear structure functions and their TMCs. This means that until Sec.~\ref{sec:rescaling} 
there is no reference to (or dependence on) nucleon degrees of freedom.
Furthermore, until Sec.~\ref{sec:parton} there is no reference to (or dependence on) 
the individual partonic degrees of freedom.\footnote{Explicitly, until Sec.~\ref{sec:parton} we consider operators and matrix elements derived from quark, antiquark, and gluon fields but do not identify these as partons, nor identify structure functions as combinations of PDFs, i.e., the parton model.}
To do this,
we first outline in Sec.~\ref{sec:kin} key kinematic relations and definitions in DIS of a lepton off a nucleus, $\ell_1(k_1) + A(p_A) \to \ell_2(k_2) + X(p_X)$.
As depicted in Fig.~\ref{fig:dis}, $A$ is a nuclear target with mass number $A$, $\ell$ denotes  either a charged lepton~$(\ell^\pm )$ or neutrino~($\nu$), and $X$ represents all final-state hadrons.
In Sec.~\ref{sec:ope}, we discuss precisely the criteria for light-cone dominance in nuclear DIS, and then present in Sec.~\ref{sub:master} a formula for nuclear TMCs. The result is analogous to the nucleon case~\cite{Schienbein:2007gr}, 
but expressed in terms of the nuclear scaling variable $x_A$ and the mass of the nucleus $M_A$.

In Sec.~\ref{sec:rescaling}, we go on to perform a rescaling in order to express our formula for nuclear TMCs to the more familiar averaged nucleon quantities $x_N=Ax_A$ and $M_N=M_A/A$. This is the key step that allows us to compare 
structure functions across different nuclei, including the proton, in a meaningful way.
As an important consequence of rescaling, we obtain an alternative formula for nuclear TMCs that is universally 
applicable to all nuclei.
In Sec.~\ref{sec:parton} we introduce the QCD-improved parton model, both before and after rescaling.
We show that rescaling at the hadronic level and in the parton model
are consistent, and we comprehensively discuss the relationship between nuclear structure functions and nuclear PDFs (nPDFs).
We believe that the discussions in Secs.~\ref{sec:rescaling} and~\ref{sec:parton}, while relatively straight forward, are original
and have never been 
presented in full detail.
Our motivation for these later discussions is the importance of establishing
 proper theoretical definitions of nuclear structure functions and nPDFs as they are intuitively used in the literature.

In Sec.~\ref{sec:acot}, we derive nuclear TMCs in the ACOT light cone formalism, which has been used in past nCTEQ analyses~\cite{Duwentaster:2021ioo, Segarra:2020gtj, Kusina:2020lyz, Kovarik:2015cma, Schienbein:2009kk, Schienbein:2007fs, Kovarik:2010uv}, and compare them to the OPE results.
Having provided rigorous theoretical definitions for the physical observables, we perform numerical studies in Sec.~\ref{sec:num}. This includes a comparison to a selection of data, as well as cross section predictions for current and future DIS experiments.
We provide a parameterization of the TMCs that is accurate at the sub-percent level and can be used with any underlying set of nuclear structure functions available in the massless parton model to obtain the full TMCs for any nucleus in a simple way.
We hope that this parameterization will be useful for the community investigating  hadron structure.
Finally, in Sec.~\ref{sec:conclusons} we highlight the key observations of this analysis and conclude.

A detailed derivation of TMCs at twist $\tau=2$ for nuclear structure functions
using the OPE is provided in Appendix~\ref{app:nTMC_derivation}.
Finally, technical details on our parameterization of the TMCs have been relegated to Appendix~\ref{app:par}.
\section{Kinematics of lepton-nucleus DIS} 
\label{sec:kin}

Before proceeding with an in-depth treatment of nuclear structure functions,
we layout the definitions, notation, and baseline assumptions used throughout this work.
(However, assumptions related to light-cone dominance are
discussed in Sec.\,\ref{sec:light-cone-dominance}.)
While much of the following is standard material for modern textbooks,
many conventions are used in the literature. 
Therefore, this section serves 
(a) to fix the notation and conventions that we use,
and
(b) to make the work both self-consistent and self-contained. 
We start with definitions of kinematic variables in Sec.\,\ref{sec:variables}
and move onto the definitions of the leptonic and hadronic tensors for DIS in Sec.\,\ref{sec:kin_tensor}.

\subsection{Definitions of kinematic variables}
\label{sec:variables}

Throughout this work we consider the basic charged-current (CC) or neutral-current (NC) lepton-nucleus DIS process  involving a high-energy lepton $\ell_1$ of four-momentum $k_1$ scattering off an unpolarized nuclear target $A$, with an outgoing lepton $\ell_2$ of momentum $k_2$ in association with the inclusive hadronic remnant $X$ of momentum $p_X$ and mass $W_A=\sqrt{p_X^2}$.
We assume that $A$ has momentum $p_A$, mass $M_A=\sqrt{p_A^2}$, atomic number $Z$, mass number $A$, and neutron number $(A-Z)$.
As sketched in Fig.~\ref{fig:dis}, this is given by the hadron-level expression
\begin{align}
\label{eq:setup_dis_def}
 \ell_1(k_1)\ +\ A(p_A)\ \to\ \ell_2(k_2)\ +\ X(p_X)\ .
\end{align}
We consider that the incoming lepton 
can be a light charged lepton $\{e^\pm, \mu^\pm\}$,
a neutrino $\{\nu_e,\nu_\mu, \nu_\tau\}$, 
or an antineutrino 
$\{\overline{\nu_e},\overline{\nu_\mu}, \overline{\nu_\tau}\}$ 
in the Standard Model.
We work in the limit that all incoming leptons are massless but allow for possibility that the  outgoing lepton is a (massive) $\tau^\pm$ lepton.
$X(p_X)$, sometimes denoted with a subscript as $X_n(p_X)$,
represents the fragmentation of $A$. 
This fragmentation is an $n$-body final state with net quantum numbers corresponding to a colorless state carrying the same spin statistic, QED charge, and weak isospin charge as nucleus $A$, modulo differences between $\ell_1$ and $\ell_2$.
The momentum of $X$ can be parameterized by its $n$ constituents, with $p_X = \sum^n_{i=1} p_{Xi}$.

We work at lowest order in the electroweak theory, in the so-called  ``one-boson-exchange'' approximation. Under this assumption, the leptonic system $(\ell_2\ell_1)$ probes the hadronic system via the exchange of a single (off-mass-shell) electroweak boson $V$ ($V\in \{\gamma,Z\}$ for NC DIS
or $V=W^{\pm}$ for CC DIS) with time-like momentum $q=k_1-k_2 = p_X - p_A$. 
The kinematic invariants and scaling variables of DIS are given and defined by
\begin{subequations}
\label{eq:app_kinDef}
\begin{align}
k_1^2 =m_{\ell_1}^2 ,\quad k_2^2 = m_{\ell_2}^2, \qquad p_A^2 &= M_A^2 > 0, \qquad Q^2 = -q^2 > 0, \qquad  s=(k_1 + p_A)^2 = m_{\ell_1}^2 + M_A^2 + 2p_A\cdot k_1, 
\quad
\\
x_A = \frac{Q^2}{2 p_A \cdot q}, 
\qquad 
\nu_A &= \frac{p_A\cdot q}{M_A}\ , 
\qquad 
y_A = \frac{p_A\cdot q}{p_A\cdot k_1}\ ,
\qquad
Q^2 = (s-m_{\ell_1}^2 - M_A^2)x_Ay_A,
\\
W_A^2\ 
=\ p_X^2\ 
&=\ ( p_A + q)^2\ 
=\ M_A^2 - Q^2 + 2\nu_A M_A\ 
=\ M_A^2 + Q^2\frac{1-x_A}{x_A}\
\geq\ 
M_A^2\ .
\label{eq:app_kinDef_hadronicMass}
\end{align}
\end{subequations}
Importantly, the above  are defined at the level of hadrons (or nuclei).
While some are related to the kinematics of a hadron's   constituents, the quantities in Eq.~\eqref{eq:app_kinDef} are defined without referencing this internal structure.

Throughout this work we generally neglect the lepton masses $\{m_{\ell_1}, m_{\ell_2} \}$,
except for the case of $\tau$ leptons.
$\sqrt{s}$ is the hadron-level center-of-mass (cm) energy and the total energy available in the DIS process.
The (dimensionless) Bjorken scaling variable $x_A\in[0,1]$ 
is defined\footnote{Eventually in Sec.~\ref{sec:parton} we note that the quantity $p_i=x_A p_A$ can be identified in the parton model as the momentum of a massless parton of $A$. 
One can see this correspondence by
evaluating $W_A^2 = \left(p_i + (1-x_A)p_A + q\right)^2$, setting all $p_i^2=0$, and yet still recover Eq.~\eqref{eq:app_kinDef_hadronicMass}.
Similarly, other hadronic-level variables in
Eqs.~\eqref{eq:app_kinDef}-\eqref{eq:rma}
 have correspondence with partonic-level kinematics.
}
such that in the elastic scattering limit, i.e., where $X$ remains as an intact and on-shell $A$, 
the condition $W_A^2 = M_A^2$ leads to $2(p_A\cdot q)=Q^2$, or that $x_A=1$.
In the inelastic scattering limit, i.e., where $A$ disassociates, the condition $W_A^2 > M_A^2$ leads to $2(p_A\cdot q)>Q^2$, or that $x_A<1$. And in the no-scattering limit, the lepton momenta obey the condition $k_2=k_1$, which implies $Q^2=0$, or the $x_A=0$.
The quantity $\nu_A$, which has mass dimension 1, is a measure of the energy carried by the virtual state $V$. In the rest frame of $A$, it simplifies precisely to this: $\nu_A\vert_{lab} = k_1^0-k_2^0$.
Here and below, the ``lab frame'' denotes the rest frame of $A$, meaning that $p_A=(M_A,0,0,0)$.
The (dimensionless) inelasticity variable spans $y_A\in[0,1]$,
and in the ``lab frame'' we have  $y_A=\nu_A/E$.
In the rest frame of $A$, it is a measure of $\ell_2$'s energy; in other frames, $y_A$ is a measure of $\ell_2$'s outgoing angle.
(Note the inelasticity variable $y_A\in[0,1]$ 
should not be confused with the rapidity or pseudorapidity variables 
$y,\eta\in[-\infty,\infty]$
which measure of a particle's direction.)

For later use with TMCs to structure functions (Secs. \ref{sub:master} and \ref{sec:rescaling}), we also define the quantity
\begin{equation}
r_A\ =\ \sqrt{1 + 4 x_A^2 M_A^2/Q^2}\ =\ 
1 + \frac{2x_A^2 M_A^2}{Q^2} + \mathcal{O}\left(\frac{4x_A^4 M_A^4}{Q^4}\right)\, ,
\end{equation}
as well as the so-called Nachtmann scaling variable~\cite{Nachtmann:1973mr}
\begin{equation}
\xi_A\ =\ \frac{2 }{1 + r_A}\, x_A \equiv R_M x_A\ , \qquad {\rm where}
\qquad  R_M = \frac{2}{1+r_A} \quad . 
\label{eq:rma}
\end{equation}
$R_M$ is the target-mass-dependent factor relating the Bjorken scaling variable $x_A$  
to the Nachtmann scaling variable $\xi_A$.
In other words, $\xi_A$ is the true scaling variable in DIS when one   accounts for the mass of the hadronic target~\cite{Georgi:1976vf,Georgi:1976ve}.
The ``$M$'' subscript  of $R_M$ indicates this quantity depends on the hadron mass $M_A$.
In the limit that $(M_A/Q)\to 0$ or that $x_A\to 0$, we have 
$r_A\to 1$, $R_M\to 1$, and $\xi_A\to x_A$
since:
\begin{equation}
\xi_A 
=
x_A\ 
\left[
1 
- \frac{x_A^2 M_A^2}{Q^2}
+ \mathcal{O}\left(\frac{x_A^4 M_A^4}{Q^4}\right)
\right]\ .
\label{eq:nachtmann}
\end{equation}
We see in Eq.~\eqref{eq:nachtmann} that the 
Nachtmann variable $\xi_A$ is essentially the Bjorken scaling $x_A$ modified by the target mass $M_A$.
The $r_A$ and $R_M$ factors are ultimately kinematical in origin [this is shown below Eq.~\eqref{eq:deltaIdentity} in App.~\ref{app:opeMassive}] and take into account that $M_A$ is non-zero.
For example: in the rest frame of $A$, the energy of the time-like exchange boson $V$ is $q^0\vert_{lab}=Q^2/(2x_AM_A)$ and its speed is $\beta_q=\vert\vec{q}\vert/q^0\vert_{lab} = r_A$.

\begin{figure*}[tb]
    \centering
\subfigure[\label{fig:xi_a}]{\includegraphics[width=0.48\textwidth]{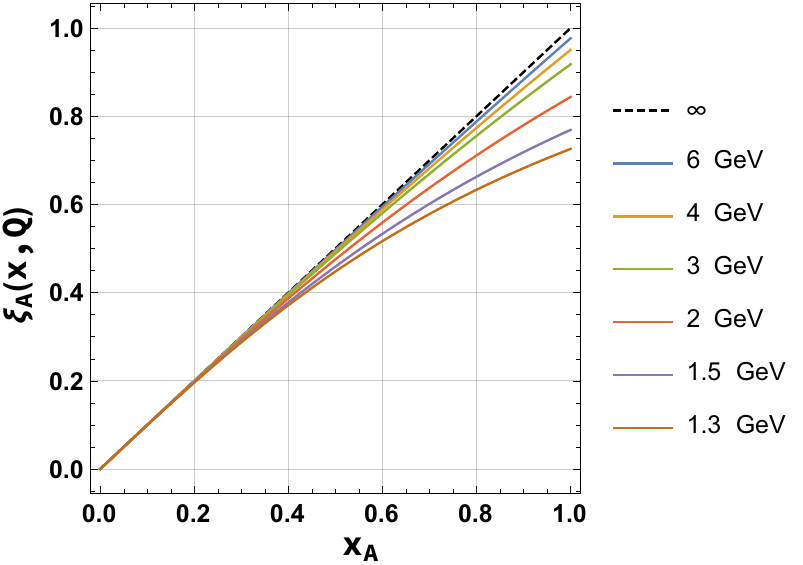}}
\subfigure[\label{fig:xi_b}]{\includegraphics[width=0.48\textwidth]{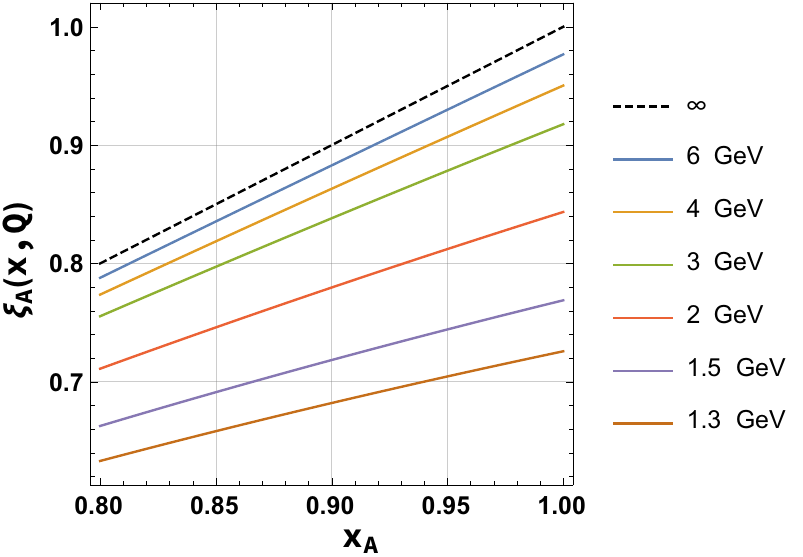}}
\caption{(a) The Nachtmann scaling variable $\xi_A(x_A,Q)$ as a function of the Bjorken scaling variable $x_A$ for a selection of $Q$ values [GeV], as indicated in the legend,  
        with $M_A=M_{\rm proton}$.
        (b) Same as (a) but for large $x_A$.
We observe the that for large $x_A$ and small $Q$ values, 
the target mass $M_A$ modifies the 
true (Nachtmann) scaling variable $\xi_A$ relative to the 
usual Bjorken scaling variable $x_A$, 
{\it c.f.}, Ref.~\cite{Schienbein:2007gr}.
}\label{fig:xi}
\end{figure*}

Qualitatively, the $R_M$ factor indicates that (slightly) less \textit{3-momentum} is available  than  suggested by $x_A$ and simplified arguments based purely on energy conservation. 
In other words, the leading kinematics in DIS are modified by  
$\mathcal{O}(x_A M_A^2/Q^2)$ terms that can be neglected at high momentum-transfers.
As discussed and demonstrated in later chapters of this work, 
this is often remedied in practice by the appropriate substitution of $x_A$ with $\xi_A$.
Quantitatively, the difference between $\xi_A$ and $x_A$ is illustrated in Fig.~\ref{fig:xi}, where $\xi_A$ is plotted as a function of $x_A$ 
for a selection of exchange-boson virtualities $Q$ with $M_A=M_{\rm proton}$.
Fig.~\ref{fig:xi_a} displays the full $\{\xi_A,x_A\}$ plane. 
Fig.~\ref{fig:xi_b} highlights the large~$x$ region, 
where the difference between $\xi_A$ and $x_A$ is more pronounced. 
Overall, $x_A$ and $\xi_A$ are mostly indistinguishable for $x_A\lesssim 0.4-0.5$ over a large range of $Q$ and $M_A$.
At around $x_A=0.8$, one has approximately $\xi_A\approx0.63~(0.71)~[0.76]$ for $Q=1.3~(2)~[3]\GeV$.
For the same $Q$ values, one has approximately $\xi_A\approx0.76~(0.78)~[0.84]$ at around $x_A=0.9$.

\subsection{Leptonic and hadronic tensors in DIS from 
experimentally observable kinematics}\label{sec:kin_tensor}

Deep-inelastic scattering is a powerful, elucidating probe of the internal structure of nucleons and nuclei, i.e., hadrons. This ability stems from the fact that under strong but general assumptions one can write the cross section for 
any sufficiently inclusive DIS process
\mbox{$\ell_1 + A \xrightarrow{V} \ell_2 + X$} 
as a combination of:
(a)~leptonic and hadronic kinematics, which can be measured, and 
(b) hadronic structure functions, which parameterize the internal dynamics of $A$. 
This discussion is based solely on kinematics and symmetries,
including Lorentz symmetry.
The only approximation of consequence that is made when constructing 
structure functions \textit{from experimentally observable kinematics} 
is the one where DIS is mediated by the exchange of only one electroweak boson.
(Relaxing this has been explored
elsewhere~\cite{Blunden:2003sp,Blunden:2005ew,Arrington:2007ux,Arrington:2011dn}.)
Apart from this assumption, expressions derived from kinematics are completely general:
they take into account all twist contributions,
at all orders in the strong coupling constant,
and all hadron- and quark-mass effects.
And in particular, the parton model does not need to be invoked.
\\

Schematically, the differential DIS cross section $d\sigma$ is proportional to the product of a leptonic tensor $L^{\mu\nu}(k_1,k_2)$, 
which is the square of the $\ell_1\to\ell_2 V$ splitting matrix element {(the matrix element of the leptonic current)}, and 
a hadronic tensor $\tilde{W}^A_{\mu\nu}(p_A,q)$, 
which is proportional to the square of the $A V\to X$ matrix element {(the matrix element of the hadronic current)}.
Subsequently, $d\sigma$ can be written as
\begin{equation}
d\sigma ~\propto ~ \tilde{W}^A_{\mu\nu}\ L^{\mu\nu}\quad ,   
\end{equation}
and is given more completely by Eq.~\eqref{eq:dxsecDef} in App.~\ref{app:xsec_intro}.
Often in the literature there are ambiguities as to whether a quantity in lepton-nucleus scattering is defined at the \textit{nucleus} level or the \textit{nucleon} level. 
Indeed, a key takeaway of this work is how the two levels are related in the DIS limit.
Therefore, to maximize clarity, we denote hadronic quantities, e.g., 
\textbf{structure functions, at the \textbf{nucleus} level using the 
tilde  notation $(\sim)$.} 
Hadronic quantities at the \textbf{nucleon} level 
 do not use the tilde notation.
We introduce this second notation in Sec.\,\ref{sec:rescaling},
where we perform a rescaling of 
nuclear momentum to average nucleon momentum.

The leptonic tensor can be constructed from Feynman diagrams for $\ell_1\to \ell_2 V$ splitting with standard methods, e.g., trace technology or helicity amplitudes.
For the case of massless charged leptons exchanging a photon with the nucleus $A$,
the spin-summed leptonic tensor for massless leptons is
\begin{equation}
\sum_{\{\lambda\}} L^{\mu\nu}\Big\vert_{\rm QED} = 4e^2  \Big\{k_1^\mu k_2^\nu + k_1^\nu k_2^\mu
 - (k_1\cdot k_2)g^{\mu\nu}\Big\} \quad .
\label{eq:lmunu}
\end{equation}
Here, $\{\lambda\}$ represents the sum over spins of external particles; spin-averaging is not yet performed.
Analogous expressions for the exchange of the $W$ and $Z$ bosons can be found in App.~\ref{app:matrixelement}.

In practice, the tensor $\tilde{W}^A_{\mu\nu}(p_A,q)$, 
which can only be a function of external momenta $p_A$ and $q$ for unpolarized targets, 
is decomposed into a sum over tensor-valued coefficient functions multiplied by
dimensionless, 
scalar-valued functions $\tilde{W}_i$. These $\tilde{W}_i$ are known in the literature as ``structure functions'' since they parameterize the structure of nucleons and nuclei.
Formally there are 
six\footnote{%
Since $\tilde{W}^A_{\mu\nu}(p_A,q)$ is a tensor made from the product of Dirac fermion spinors, it is a bilinear, and subsequently has six independent components.
For further details, see App.~\ref{app:ope}.
Separately, we will observe in Sec.~\ref{sec:acot} 
that using the helicity representation similar considerations 
(Lorentz invariance, angular momentum conservation) also yields six independent structure functions.
\\
} %
linearly independent $\tilde{W}_i$
into which $\tilde{W}^A_{\mu\nu}(p_A,q)$
can be decomposed.
Lorentz symmetry and Hermiticity dictate that only certain combinations of $p_A$ and $q$ are allowed in the coefficient functions.
For the case of an unpolarized (spin-averaged) nuclear target, the ``hadronic'' tensor $\tilde{W}^A_{\mu\nu}(p_A,q)$,
in terms of the square of the $A V\to X$ current $J(z)$ in coordinate space,
is given by 
\begin{align}
\tilde{W}^A_{\mu\nu}(p_A,q) 
\quad &\equiv
\frac{1}{4\pi}\sumint d^{4}z\ e^{iq\cdot z} 
\langle A(p_A) \vert\, J_{\mu}(z) \vert X(p_X)\rangle\langle X(p_X) \vert J_{\nu}(0)  \, |A(p_A)\rangle\ 
\label{eq:wmunu}\\
&=
-g_{\mu \nu}\tilde{W}_1 
+ \frac{p_{A\mu} p_{A\nu} }{ M_A^2}\tilde{W}_2
-i\epsilon_{\mu\nu\rho\sigma} \frac{p_A^{\rho}q^\sigma }{ M_A^2} \tilde{W}_3
\nonumber\\
&+  \frac{q_\mu q_\nu }{ M_A^2} \tilde{W}_4
+ \frac{p_{A\mu} q_\nu + p_{A\nu} q_\mu }{ M_A^2} \tilde{W}_{5}
+ \frac{p_{A\mu} q_\nu - p_{A\nu} q_\mu }{  M_A^2} \tilde{W}_{6}
\ .
\label{eq:wmunuExpand}
\end{align}
In Eq.~\eqref{eq:wmunu}, 
{the normalization factor $1/(4 \pi)$ is conventional} and
the sum and integral $(\sumint)$ run over all discrete and continuous configurations of $X$, implying that $\tilde{W}^A_{\mu\nu}(p_A,q)$ is inclusive with respect to $X$
(see also App.~\ref{app:matrixelement}).
We note that, although the targets considered in this work are unpolarized nuclei, 
we refer to quantities such as that given in Eq.~\eqref{eq:wmunu} as a hadronic tensors, in keeping with
convention.
For a polarized target, the decomposition into structure functions takes on a more complicated structure, see e.g.
\cite{Jaffe:1985je,Blumlein:1998nv,Accardi:2012qut,Aidala:2012mv,Deshpande:2005wd,ParticleDataGroup:2020ssz}.
In Sec.\,\ref{sec:buildingOPE} and App.~\ref{app:opeMassive}, we review the connection between the $\tilde{W}_i$ and the internal structure of $A$.

With further strong but reasonable assumptions, the number of structure functions appearing in Eq.~\eqref{eq:wmunuExpand} can be reduced in real-life calculations~\cite{Treiman:1972,Collins:1984xc,Sterman:1995fz}.
For instance: $\tilde{W}_{6}$ is only nonzero  if QCD violates time-reversal (or charge-parity) symmetry. (Whether $\tilde{W}_{6}$ is actually zero is an alternative formulation of the ``strong CP problem'' of QCD.)
The coefficients for $\tilde{W}_{4}$ and $\tilde{W}_{5}$ contain factors of $q_\mu$ and $q_\nu$. This implies that the contractions with the leptonic tensor $L^{\mu\nu}\cdot q_\mu$ and $L^{\mu\nu}\cdot q_\nu$ are proportional to the masses of $\ell_1$ and $\ell_2$, and therefore vanish when lepton masses are negligible.
$\tilde{W}_{3}$ is only nonzero if parity symmetry is violated; it therefore only appears for $W$ and $Z$ boson exchanges.
In summary, if one considers only electromagnetic interactions, standard QCD, and ignores lepton masses, then $\tilde{W}^A_{\mu\nu}(p_A,q)$ can be described entirely by  $\tilde{W}_{1}$ and $\tilde{W}_{2}$.

Modern notation calls for using the structure functions
$\tilde{F}_i$ rather than $\tilde{W}_i$. 
The mapping between the two sets of dimensionless 
structure functions is given by
\begin{align}\label{eq:WtoF}
&\left\{ \tilde{F}_1,\ \tilde{F}_2,\ \tilde{F}_3,
\ \tilde{F}_4,\ \tilde{F}_{5,  6}
\right\}\
\nonumber\\
& \qquad ~ \qquad =
\ \left\{\tilde{W}_1, 
\ \frac{Q^2}{2 x_A M_A^2}\tilde{W}_2 ,
\ \frac{Q^2}{x_A M_A^2}\tilde{W}_3 ,
\ \frac{Q^2}{2 M_A^2} \tilde{W}_4 ,
\ \frac{Q^2}{2 x_A M_A^2} \tilde{W}_{5, 6}
\right\}\ . 
\end{align}
The purpose of using $\tilde{F}_i$ is to factor out known dependence on $Q^2$ and make more manifest the phenomenon of scaling, i.e., that $\tilde{F}_i$ depend only on $x_A$, a dimensionless quantity, up to small logarithmic {QCD} corrections.
In the discussion that follows, we focus on
$\tilde{F}_1, \tilde{F}_2$, and $\tilde{F}_3$. 
However, we include a detailed discussion of $\tilde{F}_4$ and $\tilde{F}_5$ in Appendix~\ref{app:nTMC_derivation}.
The structure functions $\tilde{F}_4$ and $\tilde{F}_5$
enter into differential cross sections, but are suppressed by
a factor 
$\mathcal{O}(\frac{m_\ell^2}{M_A E_\ell})$, 
where $m_\ell^2$ is the lepton mass squared, 
$M_A$ is the mass of the hadronic target, and $E_\ell$ is the energy of one of the external leptons~\cite{Kretzer:2003iu}.
This suppression is a consequence of contracting the symmetric leptonic tensor $L^{\mu \nu}$ with $q_\mu$ (or $q_\nu$), which subsequently vanishes due to the conservation of weak currents by massless leptons.
Notably, finite lepton-mass effects could be measured in $\nu_\tau$-DIS, 
such as at the SHIP, FASER, or SND@LHC detectors at CERN~\cite{Alekhin:2015byh,FASER:2019dxq,SHiP:2020sos,Feng:2022inv}.
As for $\tilde{F}_6$, which signifies charge-parity violation,
the coefficient vanishes when contracted with Eq.~\eqref{eq:lmunu}.
Hence, it does not contribute to the cross section.

\subsubsection*{Considerations for spin-1 and greater}

Contrary to a nucleon target with spin-1/2, nuclei can have spin-1 or greater. 
The case of a spin-1 nuclear targets in NC DIS with charged leptons
has been discussed since the 80s~\cite{Hoodbhoy:1988am}.
At leading twist, i.e., twist $\tau=2$, 
the additional effects of scattering on a polarized spin-1 target reside in a 
single new structure function $\tilde{b}_1(x)$.
This structure function effectively measures the extent to which a target nucleus deviates from
a trivial bound state of protons and neutrons.
For the deuteron, it is expected that $\tilde{b}_1\approx 0$, but for other nuclei one could have 
$\tilde{b}_1\sim {\cal O}(\tilde{F}_1)$.

More generally,  using  gauge invariance and P- and T-invariance for the spin-1 case,
the hadronic tensor can be expressed in terms of eight independent structure functions,
$\{\tilde{F}_1, \tilde{F}_2, \tilde{b}_{1,2,3,4}, \tilde{g}_1, \tilde{g}_2\}$.
Similar results are found for the hadronic tensor of a (space-like) virtual photon target~\cite{Schienbein:2002wj}.
The functions $\tilde{F}_1$, $\tilde{F}_2$, $\tilde{g}_1$, and $\tilde{g}_2$ are analogous to the scaling structure functions  of a spin-1/2 target.
$\tilde{F}_1$ and $\tilde{F}_2$ can be measured in the DIS of unpolarized leptons off an 
unpolarized target whereas measurements of $\tilde{g}_1$ and $\tilde{g}_2$ require polarized lepton beams and a polarized target.
The four structure functions $\tilde{b}_{1,2,3,4}$ are new in the spin-1 case. 
$\tilde{b}_{1,2}$ are quantities that appear at leading twist, while
$\tilde{b}_{3,4}$ appear at higher twist.
$\tilde{b}_{1,2}$ are connected via  $\tilde{b}_{2}= 2x\tilde{b}_{1}$ in a manner analogous to the 
Callan-Gross relation, and will receive corrections beyond lowest order in QCD.
The structure functions $\tilde{b}_{1,2,3,4}$ can be measured using an unpolarized lepton
beam but require a polarized \mbox{spin-1} target.
The TMCs for a (spin-1) deuterium target have been calculated~\cite{Detmold:2005iz}
as well as those for a virtual photon target~\cite{Kitadono:2008iw}.

In this work we are studying unpolarized DIS of charged and neutral leptons off an unpolarized nucleus 
with any spin greater or equal to $1/2$.
For such circumstances, 
the hadronic tensor has the same decomposition into structure functions
as for a spin-$1/2$ target. 
For this reason we focus only on the $\tilde{F}_i$ structure functions in the remainder of this text;
the treatment 
of additional structure functions, such as $\tilde{b}_i$, can be dealt with in a similar manner,
but we shall not address them explicitly. 

\section{Nuclear structure functions in the OPE}  \label{sec:ope}

In this section, we discuss nuclear structure functions and their TMCs in the context of the OPE. We start in Sec.\,\ref{sec:light-cone-dominance} with a precise stipulation of the criteria for light-cone dominance involving nuclear targets, whose masses can readily exceed $\mathcal{O}(50-100)\GeV$.
In Sec.\,\ref{sec:buildingOPE}, we sketch the construction of nuclear structure functions and their TMCs in the OPE; a fuller
derivation is given in App.~\ref{app:nTMC_derivation}.
We then present in Sec.\,\ref{sub:master} the main formula for nuclear TMCs in terms of nuclear quantities $x_A$ and~$M_A$.
\subsection{Light-cone dominance of nuclear DIS}
\label{sec:light-cone-dominance}

A crucial step in deriving structure functions for nucleons and nuclei involves employing the OPE to expand the product of currents that comprise the hadronic tensor of Eq.~\eqref{eq:wmunu}.
This expansion, however, requires that one is in the limit of light-cone dominance, i.e., $z^2 \sim 0$, where $z$ is the Fourier conjugate of the DIS momentum $q$.
While this is well-established in the case of a nucleon target,\footnote{%
A demonstration of the light cone dominance of DIS can be found,
for example, in the textbook by Muta~\cite{Muta:1998vi,Muta:2010xua}, 
see (2nd edition) pages 228-229 and 262.
}  %
demonstrating light-cone dominance for nuclear targets requires care due to the relative sizes of $Q^2 = -q^2 > 0$ and $M_A^2$.

In many constructions of PDFs and related quantities in DIS, light cone-dominance is described as corresponding to the momentum configuration
\begin{equation}
    Q^2 \sim |p_N \cdot q| \gg M_N^2 \, ,
    \label{eq:dis-pQCD-validity}
\end{equation}
where $p_N=p_A/A$ and $M_N=M_A/A$ are respectively the four-momentum and mass of a single nucleon, {\it c.f.}, Table~\ref{tab:one}.
Under this condition, naively replacing $p_N$ and $M_N$ by the nuclear momentum $p_A$ and nuclear mass $M_A$
would imply (incorrectly) that light-cone dominance is only satisfied when
\begin{equation} \qquad \qquad 
    Q^2 \sim |p_A \cdot q| \gg M_A^2\, . \qquad {\rm [Incorrect]}
    \label{eq:dis-limit2}
\end{equation}
This incorrect condition suggests increasingly large $Q^2$ are needed for increasing $A$,
and downplays the validity of both perturbative QCD and the parton model. For example: 
Eq.~\eqref{eq:dis-limit2} implies that $Q\gg50\GeV$, which is more than half the $Z$ boson's mass,  would be needed to describe DIS data for $^{56}$Fe, or that $Q\gg180\GeV$, which is beyond the top quark's mass,  would be needed for $^{197}$Au.
Fortunately, this is not the case as Eq.~\eqref{eq:dis-pQCD-validity} 
is the correct condition.

To resolve this,
we apply the arguments of Refs.~\cite{Muta:1998vi,Muta:2010xua} for light-cone dominance in DIS with a nucleon to the case of an arbitrary \textit{nucleus}.
Our goal is to identify the dominant contribution to the
Fourier integral in  Eq.~\eqref{eq:wmunu} in the deeply inelastic limit:
\begin{equation}
    Q^2 \to \infty, \quad \nu_A \to \infty, \quad \text{such that}\quad
    \frac{Q^2}{\nu_A} = 2 M_A x_A\quad  \text{is fixed}.
    \label{eq:disCriteria}
\end{equation}
Note that in terms of averaged nuclear quantities, one has $Q^2/\nu_A = Q^2/\nu_N = 2 M_A x_A = 2 M_N x_N$. That is to say,
a bigger nuclear mass $M_A$ is compensated by a smaller Bjorken
variable $x_A$ such that the fixed quantity $Q^2/\nu$ is 
independent of the atomic number.

Following Refs.~\cite{Muta:1998vi,Muta:2010xua}, 
we  examine   $\tilde{W}^A_{\mu\nu}(p_A,q)$ in Eq.~\eqref{eq:wmunu}, 
and identify the integration regions that give rise to the  dominant contributions.
In the DIS limit of Eq.~\eqref{eq:disCriteria}, as $|q \cdot z| \to \infty$ the exponential of the Fourier integral 
oscillates without bound and thus makes a  vanishing contribution to the integral. 
Therefore, we need only to consider the region with finite $|q \cdot z|$ in the deeply inelastic regime.

In the target's rest frame, $\nu_A = q_0$.
Defining the quantity $\rr{}=\vec{q} \cdot \vec{z}/|\vec{q}|$, we have 
\begin{equation}\label{eq:qz}
    q \cdot z\ =\ q_0 z_0 - \vec{q} \cdot \vec{z}\ =\ 
    \nu_A \left(z_0 - \frac{|\vec{q}|}{\nu_A} \rr{}\right)\ 
    =\ \nu_A \left(z_0 - \sqrt{1+Q^2/\nu_A^2}\ \rr{}\right)\ ,
\end{equation}
since $\vert \vec{q}\vert^2 = \nu_A^2 + Q^2$.
We note again that this equation is independent of $A$.
In the DIS limit, $Q^2/\nu_A^2 = (2 M_A x_A)/\nu_A$ is small
and we can expand the square root:
\begin{equation}
    q \cdot z\ =\ \nu_A (z_0 - \rr{})\ - M_A x_A \rr{}\ +\ {\cal O}\left(\frac{M_A^2 \rr{}}{\nu_A}\right)\, .
    \label{eq:phase}
\end{equation}
Here, the target mass $M_A$ appears but only in combination with $x_A$ and always satisfies
$M_A x_A = M_N x_N$.
In order to keep $|q \cdot z|$ finite in the deeply inelastic limit, 
each term on the right hand side of Eq.~\eqref{eq:phase} must separately be finite. 
(Being separately infinite requires the scaling $(z_0-\rr{})\sim (\rr{}/\nu_A) \to \infty$, which cannot be consistently satisfied.)
Since $M_A x_A$ is fixed, $\rr{}$ itself must be finite. Therefore, for some constants $c>0$ and $d>0$, one has
\begin{equation}
    |z_0 - \rr{}| < c/\nu_A \,  \quad\text{and}\quad \, |\rr{}| < d/(x_A M_A)\, .
    \label{eq:bounds}
\end{equation}
The first inequality implies that $|z_0| < |\rr{}| + c/\nu_A$.
After squaring and using $\rr{}^2 = \vec{z}^2 - z_\perp^2 < \vec{z}^2$, where $z_\perp$ is the component of $z$ orthogonal to $\vec{q}$, we obtain the inequality chain
\begin{equation}
    z_0^2 ~ <~ (|\rr{}| + c/\nu_A)^2 = \rr{}^2 + \frac{2c |\rr{}|}{\nu_A} +
    {\cal O}\left(\frac{1}{\nu_A^2}\right) ~<~ \vec{z}^2 + \frac{2c d}{(x_A M_A \nu_A)}
    + {\cal O}\left(\frac{1}{\nu_A^2}\right)\, .
\end{equation}
To obtain the rightmost bound, we used the second inequality in Eq.~\eqref{eq:bounds}. 
Using the rightmost equality in Eq~\eqref{eq:disCriteria}, we obtain the final result:
\begin{equation}
    z^2 ~=~ z_0^2 - \vec{z}^2 ~<~ \frac{2cd}{(x_A M_A \nu_A)}  + {\cal O}\left(\frac{1}{\nu_A^2}\right)
    ~=~ \frac{4cd}{Q^2} + {\cal O}\left(\frac{1}{\nu_A^2}\right) \, .
\label{eq:lcdCriterion}
\end{equation}

We therefore find that the dominant region remains $0 \le z^2 \le (2cd/Q^2)$, and is independent of the target. This assumes, of course, that no extra $A$ dependence is hidden in the constants $c$ and $d$.
Such an assumption, however, is safe upon inspecting the above derivation.
For example: suppose there were separate (minimal) $c_A$ and $d_A$ for each $A$ that satisfied the inequality of Eq.~\eqref{eq:bounds}. Since there are finite many $A$, we can simply take the largest $c_A$ and $d_A$, and then define $c=\max\{c_A\}$ and $d=\max\{d_A\}$. This implies that there is a 
large enough 
$c$ and $d$ such that Eq.~\eqref{eq:bounds} holds independent of $A$.

In light of Eq.~\eqref{eq:lcdCriterion}, the requirement of Eq.~\eqref{eq:dis-limit2} 
is obviously too harsh.
Instead, we advocate that Eq.~\eqref{eq:dis-pQCD-validity} should be understood as requiring  $Q^2 \gtrsim$ a few GeV$^2$, independent of the hadronic target, in order to justify  applying perturbative QCD and the parton model. Then, given a non-perturbative (NP)
scale $\Lambda_{\rm NP} \lesssim 1\GeV$, which parameterizes the onset of hadronization, light-cone dominance in DIS for a nucleus $A$ occurs when 
\begin{equation}
    Q^2\ \sim\ |p_A \cdot q|\ \gg\ \Lambda_{\rm NP}^2.
    \label{eq:dis-pQCD}
\end{equation}

Under this criterion, Eq.~\eqref{eq:dis-pQCD-validity} is automatically satisfied since nucleon masses scale as
$M_N\sim \Lambda_{\rm NP}$. At smaller $Q^2$, but still satisfying $Q^2 \gtrsim \Lambda_{\rm NP}^2$, perturbative techniques and the parton model can still be employed
as evident by
the phenomenon 
of ``precocious scaling''~\cite{Georgi:1976vf,Georgi:1976ve,DeRujula:1976baf}. This denotes the onset of Bjorken scaling 
at moderate energies despite $\alpha_s(\mu_r=Q)\lesssim\mathcal{O}(1)$ being sizable, and follows from the fact that light quark masses $m_q$ are comparable to the scale of the QCD Landau pole, $\Lambda_{\rm QCD}\sim\mathcal{O}(0.1)\GeV$. 

Such a separation of scales implies the inequality
\begin{equation}
    Q^2\ \gtrsim\ \Lambda_{\rm NP}^2\ \gg\ \Lambda_{\rm QCD}^2\ \sim\ m_q^2\ ,
\end{equation}
{where we take $m_q$ to be the light-quark constituent mass at soft quark momenta.}
This demonstrates that, up to corrections of order $\mathcal{O}(\Lambda_{\rm QCD}^2/Q^2)$, light quarks can be approximated as free and massless partons,
even for low $Q^2$.
However, when such $Q^2$ is probing a nuclear target $A$ at large $x_A$,
it is paramount to stress that $x_A^2 M_A^2/Q^2$ is not guaranteed to be small or below unity. In this case, TMCs are important.
\subsection{Structure functions in the OPE}
\label{sec:buildingOPE}

In this section, we briefly outline the key elements of the OPE required to obtain expressions for structure functions with TMCs at leading twist $\tau$. Formally, TMCs are $\mathcal{O}(x_A M_A^2/Q^2)$ corrections to structure functions that are first generated by the \textit{same} operators that define the structure functions themselves at leading power.
For this reason they are sometimes called ``kinematical'' power corrections~\cite{Ellis:1982wd,Ellis:1982cd}. Such power corrections can be isolated from ``dynamical'' power corrections, 
by setting powers of $(M_A^2/Q^2)$ in the OPE to zero.
We generally follow the early literature of   Refs.~\cite{Nachtmann:1973mr,Georgi:1976ve,Georgi:1976vf,DeRujula:1976baf,Jaffe:1982pm,Collins:1984xc}, which shed light on the underlying dynamics of hadrons,
but note that TMCs can also be derived using diagrammatic methods and finite parton $k_T$~\cite{Ellis:1982wd,Ellis:1982cd}.
While we only outline the derivation here, a fuller derivation of nuclear structure functions and their TMCs at leading power in the OPE is provided in App.~\ref{app:nTMC_derivation}. 

Constructing TMCs for structure functions 
in the OPE provides a systematic
organization of short and long distance physics that does not manifestly rely on the perturbativity of QCD. 
(This means that it is possible to obtain results that are all orders in $\alpha_s$.)
Inspired by the program of Refs.~\cite{Georgi:1976ve,Georgi:1976vf}, the construction of structure functions follows by first recognizing that the inclusive hadronic tensor describing $\gamma^* A\to X$ scattering in the deeply inelastic limit,\footnote{The subscripts $XA$ on the vector currents $J_\mu,J_\nu$ are simply labels to denote the hadronic current that takes $A$ to $X$. The current is independent of states $A,X$ but depends on whether the current is electromagnetic or weak.}
\begin{equation}
\tilde{W}^A_{\mu\nu}(p_A,q)\ =\
 \frac{1}{4\pi} \int d^4z ~ e^{iq\cdot z}  
 \langle A \vert J_{XA\mu}^\dagger(z)\  J_{XA\nu}(0) \vert A\rangle\ ,
\end{equation}
which is equivalent to the expression in Eq.\,\eqref{eq:wmunu} of Sec.\,\ref{sec:kin_tensor} [for details, see below Eq.~\eqref{eq:hComplete}], and the time-ordered amplitude for virtual Compton scattering process $\gamma^* A \to \gamma^* A$ 
in the short-distance limit
\begin{align}
    \tilde{T}^A_{\mu\nu}(p_A,q)\  \equiv&   \int d^{4}z\ e^{iq\cdot z}\ \langle A\vert \mathcal{T}J_{A\mu}^\dagger(z)\ J_{A\nu}(0)\vert A\rangle\ ,
    \label{eq:virtComptonDef}
    \\
    =&
-g_{\mu\nu}                         \Delta \tilde{T}^{A}_{1} +
\frac{p_{A\mu}p_{A\nu}}{M_A^2}      \Delta \tilde{T}^{A}_{2}
-
i\epsilon_{\mu\nu\alpha\beta} \frac{ p^\alpha_A q^\beta}{M_A^2}   
                                    \Delta \tilde{T}^{A}_{3}
\nonumber\\
& +
\frac{q_\mu q_\nu}{M_A^2}           \Delta \tilde{T}^{A}_{4}                              
+          
\frac{(p_{A\mu}q_\nu  \pm  p_{A\nu}q_\mu)}{M_A^2}
                                    \Delta \tilde{T}^{A}_{5,6}\ ,
\label{eq:virtComptonDef_lorentz}                                    
\end{align}
are related by the following dispersion relationship~\cite{Christ:1972ms,Collins:1984xc}
\begin{subequations}
\label{eq:WFdispersion}
\begin{align}
\label{eq:WFdispersion_a}
\tilde{T}^A_{\mu\nu}(p_A,q)\Big\vert^{(1/x_A)+i\varepsilon}_{(1/x_A)-i\varepsilon} &= 4\pi\ \tilde{W}^A_{\mu\nu}(p_A,q), \quad\text{for}\quad x_A > 0\ ,
    \\
\label{eq:WFdispersion_b}    
\tilde{T}^A_{\mu\nu}(p_A,q)\Big\vert^{(1/x_A)-i\varepsilon}_{(1/x_A)+i\varepsilon} &= 4\pi\ 
 \left[ \tilde{W}^{A}_{\mu\nu}(p_A,-q)\right]^\dagger,
\quad\text{for}\quad x_A < 0\ .
\end{align}
\end{subequations}
In Eq.~\eqref{eq:virtComptonDef_lorentz}, $\tilde{T}^A_{\mu\nu}(p_A,q)$ is decomposed into coefficients of Lorentz structures in the same manner as $\tilde{W}^A_{\mu\nu}(p_A,q)$ in Eq.~\eqref{eq:wmunuExpand}. As $\tilde{T}^A_{\mu\nu}(p_A,q)$ is time ordered, the first (second) relation in Eq.~\eqref{eq:WFdispersion} is for DIS with (anti)particles.
We assume here and below that the hadronic currents $J_{A}^\mu$ and $J_{A}^\nu$ are always renormalized objects in QCD.
\begin{figure}[t]
\centering
\includegraphics[width=0.70\textwidth]{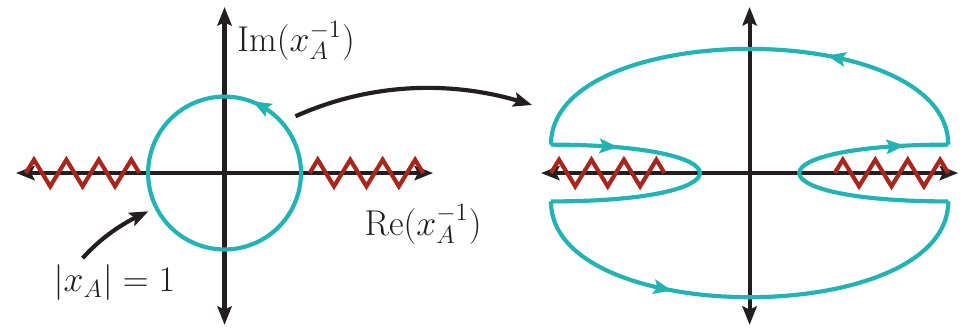}
\caption{
(Left) A contour (circle) along the radius of convergence of the time-ordered matrix element $\tilde{T}^A_{\mu\nu}(p_A,q)$ in the complex $(1/x_A)$ plane for $\gamma^* A \to \gamma^* A$ scattering. Branch cuts (sawtooth lines) are along the real axis at $x_A^{-1}<-1$ and $x_A^{-1}>1$.
(Right) A deformation of the contour along the branch cuts with vanishing arcs.
}
\label{fig:diagram_DIS_contour}
\end{figure}

The distinction between the short-distance limit and the DIS limit is important. 
In the DIS limit, \mbox{$(Q^2/M_A^2)\to\infty$} 
while $x_A=(Q^2/2p_A\cdot q)$ is fixed.
In the short-distance limit, \mbox{$(Q^2/M_A^2)\to\infty$} while $x_A/Q$ is fixed, meaning that $x_A$ grows with $(Q/M_A)$; equivalently, the quantity $(1/x_A)$ is small with increasing $(Q/M_A)$ in the short-distance limit.
Formally, $\tilde{T}^A_{\mu\nu}(p_A,q)$ and $\tilde{W}^A_{\mu\nu}(p_A,q)$ are defined in two different limits but can be related through analytic continuation. 
When analytically continued to complex values of $x_A$, 
$\tilde{T}^A_{\mu\nu}$ has desirable analytic properties for $\vert (1/x_A)\vert < 1$ (the short-distance limit), whereas 
$\tilde{W}^A_{\mu\nu}$ is meaningful for $\vert (1/x_A)\vert > 1$ (the DIS limit).
Hence the discontinuities (branch cuts) of $\tilde{T}^A_{\mu\nu}$  correspond to the kinematics of DIS.
Therefore, one can take $\tilde{T}^A_{\mu\nu}(p_A,q)$, decompose it into a contour integral over $x_A^{-1}$ using Cauchy's integral formula, and deform the contour around the discontinuities at $x_A^{-1}<1$ and $x_A^{-1}>1$ as shown in Fig.~\ref{fig:diagram_DIS_contour}.
(The sawtooth lines along $\vert x_A^{-1}\vert>1$ are the branch cuts of $\tilde{T}^A_{\mu\nu}$.)
Taking the difference of ${T}^A_{\mu\nu}$ when evaluated just below and above the branch cut, i.e., at the points  $x_A^{-1} \pm i\varepsilon$ for $x_A^{-1}>1$, gives $\tilde{W}^A_{\mu\nu}(p_A,q)$.
(For a fuller derivation, see above  Eq.~\eqref{eq:app_WFdispersion} in App.~\ref{app:ope}.)

An important consequence of Eq.~\eqref{eq:WFdispersion} is establishing a link between individual $\Delta \tilde{T}^{A}_{i}$ in Eq.~\eqref{eq:virtComptonDef_lorentz} and the Mellin moments of the structure functions $\tilde{F}_i^A$ defined in Eq.~\eqref{eq:WtoF}.
Following the steps below Eq.~\eqref{eq:cauchy} in App.~\ref{app:ope},
one can write for the 
case\footnote{See App.~\ref{app:ope} for the treatment of all six $\tilde{F}_i^A$.} of $\tilde{F}_2$~\cite{Collins:1984xc}
\begin{subequations}
\label{eq:taylorStrFn}
\begin{align}
    \left(\frac{Q^2}{2 x_A M_A^2}\right)\ 
    \Delta\tilde{T}_2^{A}(x_A, Q^2) 
    &= -4i\ 
    \sum_{N=0}^\infty\  
    \tilde{F}_2^{AN}(Q^2)\
    x_A^{-N}\ .
\end{align}
\end{subequations}
Here, $\tilde{F}_i^{AN}(Q^2)$ is the $N^{th}$ Mellin moment of the structure function $\tilde{F}_i^{A}(x_A,Q^2)$. The normalizations of the Mellin transformation and its inverse are set by the definitions
\begin{equation}
 F^N = \int_0^1 dz ~ z^{N-1} ~ F(z) \quad\text{with}\quad
 F(z) = \frac{1}{2\pi i}\int^{c+i\infty}_{c-i\infty} dN ~ z^{-N} ~ F^N \ ,
\end{equation}
{where $c$ has to be to the right of any poles of $F^N$ in the complex $N$-plane.}
As a brief comment, the Mellin transform is an integral over the positive reals, while the inverse transform is an integral over the complex plane. We therefore assume that structure functions can be analytically continued. (In practice, Mellin and inverse-Mellin transforms are implemented in the many computer codes that perform the DGLAP evolution equations in Mellin space, see e.g. Ref.~\cite{Vogt:2004ns}.)

The next step in deriving TMCs in the OPE formalism is to take the operator-product expansion of $\tilde{T}^A_{\mu\nu}$ itself in the short-distance limit.
After considerable reorganization (see App.~\ref{app:ope}), one obtains expressions (see App.~\ref{app:opeMixing}) for $\Delta\tilde{T}_i^{A}(x_A, Q^2)$ at leading power in the OPE. More specifically, one constructs a power series with respect to powers of $x_A^{-1}$, and therefore obtains expressions for $\tilde{F}_i^{AN}(Q^2)$, at leading power / leading twist.

In more detail: the OPE makes use of basic symmetries, e.g., Lorentz invariance and unitarity, 
to expand hadronic matrix elements into a complete set of local operators $\mathcal{O}$~\cite{Miramontes:1988fz}.
For forward Compton scattering, the OPE of $ \tilde{T}^A_{\mu\nu}$ is given by~\cite{Christ:1972ms}:
\begin{align}
\lim_{z\to0}\ 
{T}^A_{\mu\nu}(p_A,q)\  \overset{\rm OPE}{=} & 
-2i \sum_{\iota,n} ~ 
c_{\mu\nu\mu_1...\mu_n}^{\tau=2,\iota}(q)\
\langle A|\mathcal{O}_{\iota,\tau=2}^{\mu_{1}...\mu_{n}}|A\rangle 
\quad +\quad  \text{power corrections at }\ 
\mathcal{O}(\tau>2)
\label{eq:virtComptonOPE}
\\
= \ & 
-2i \sum_{k=k_{\min}}^\infty\Bigl[ 
-2g^{\mu\nu} q_{\mu_1}q_{\mu_2} C_1^{2k}
+ g^\mu_{\mu_1}g^\nu_{\mu_2} Q^2 C_2^{2k} 
- i\epsilon ^{\mu\nu\alpha\beta}
g_{\alpha\mu_1}q_\beta q_{\mu_2}C_3^{2k}
\nonumber \\ 
& \quad +
4\frac{q^\mu q^\nu }{ Q^2} q_{\mu_1}q_{\mu_2} C_4^{2k}
+2( g^\mu_{\mu_1}q^\nu q_{\mu_2} \pm  g^\nu_{\mu_1}q^\mu q_{\mu_2} )C_{5,6}^{2k}
\Bigr] \cdot 
\left.  %
\begin{cases}
1 \, ; & k=1
\\
\prod^{2k}_{m=3}q_{\mu_m}\,
; & k>1
\end{cases}
\right\} %
\nonumber\\
& \quad \times \frac{2^{2k} }{ Q^{4k}} A^{2k}_{\tau=2} \tilde{\Pi}^{\mu_1 ...
\mu_{2k}}\ 
\quad + \quad\mathcal{O}(\tau>2).
\label{eq:virtComptonTwist2}
\end{align}
In Eq.~\eqref{eq:virtComptonTwist2}, the summation over $n$ is set to $n=2k$.
The starting point for the summation over $k$ depends on the particular Wilson coefficient;
specifically, $k_{\min}=2$ for $C_2^{2k}$. while  $k_{\min}=1$ for all other $C_\iota^{2k}$'s.
In Eq.~\eqref{eq:virtComptonOPE}, the $\mathcal{O}^{\mu_1\dots\mu_n}_{\iota,\tau}$ are composite {local} operators, consisting of quark fields, 
QCD covariant derivatives, and/or gluon field strengths.
Operators with $n$ un-contracted Lorentz indices carry a spin of $n$.
Operators and their coefficients are organized according to their twist $\tau=d-n$, where $d$ is the dimensionality of $\mathcal{O}^{\mu_1\dots\mu_n}_{\iota,\tau}$ in the standard sense of dimension power counting in an effective field theory. 
In this study, we focus exclusively on the OPE at leading twist, i.e., $\tau=2$.
We neglect power corrections at $\tau > 2$ but attempt to keep track of their presence.
The index $\iota$ catalogs all operators $\mathcal{O}$ at a fixed spin and twist.
All numerical prefactors are sequestered into effective Wilson coefficients $c_{\mu\nu\mu_1...\mu_n}^{\tau,\iota}(q)$, which are only functions of the external momentum $q$.

The  Wilson coefficients\footnote{\label{foot:flavor_sum}%
In principle, the  Wilson coefficients $C_\iota^{2k}$ also carry a species index $f$, which runs over the gluon and quark flavors.
In this case, $C_\iota^{2k}$ should be replaced by $\sum_{f=u,g,\overline{d},\dots} C_{if}^{2k}$, but this dependence is implicit here for simplicity.
} 
$C_\iota^{2k}$, which are the same for all possible nuclei, can be identified and matched to quantities calculated in perturbative QCD.
The quantity $A^{2k}_{\tau=2}$ is the scalar-valued coefficient of the expectation value $\langle A|\mathcal{O}_{\iota,\tau=2}^{\mu_{1}...\mu_{2k}}|A\rangle = A^{2k}_{\tau=2}\times \tilde{\Pi}^{\mu_{1}...\mu_{2k}}$.
$A^{2k}_{\tau=2}$ is sometimes labeled as the ``reduced matrix element'' of $\langle A|\mathcal{O}_{\iota,\tau=2}^{\mu_{1}...\mu_{2k}}|A\rangle$. Importantly, the Lorentz structure of this expectation value $(\tilde{\Pi}^{\mu_{1}...\mu_{2k}})$ can only be a function of $p_A$ as the $q$ dependence is sequestered elsewhere. Consequentially, $\tilde{\Pi}^{\mu_{1}...\mu_{2k}}$ can be organized according to permutations and powers of momenta $p_A^{\mu_m}$ and metrics $g^{\mu_m\mu_n}$, and scales as~\cite{Georgi:1976ve,Georgi:1976vf}
\begin{align}
    \tilde{\Pi}^{\mu_{1}...\mu_{2k}} 
    &=
\sum_{j=0}^{k}(-1)^{j}\ \frac{(2k-j)!}{2^{j}(2k)!}\ 
{\underbrace{\{ g...g\}}_{\scriptstyle j\ g^{\mu_{n}\mu_{m}}{}'s}}
\quad
{\underbrace{\{ p_A...p_A\}}_{\scriptstyle (2k-2j)\ p_A^{\mu_{n}}{}'s}}
(p_A^{2})^{j}\
+\ \text{power corrections}
\label{eq:pi}
\\
&
    \sim 
    {\underbrace{\{ p_A...p_A\}}_{\scriptstyle 2k\ p_A^{\mu_{m}}{}}}
    ~+~
    {\underbrace{\{ p_A...p_A\}}_{\scriptstyle (2k-2)\ p_A^{\mu_{m}}{}}}
    {\underbrace{\{ g...g\}}_{~\scriptstyle ~1\ g^{\mu_{m}\mu_{n}}{}}} (p_A^2)^1
    ~+~
    {\underbrace{\{ p_A...p_A\}}_{\scriptstyle (2k-4)\ p_A^{\mu_{m}}{}}}
    {\underbrace{\{ g...g\}}_{~\scriptstyle ~2\ g^{\mu_{m}\mu_{n}}{}}} (p_A^2)^2
    ~+~ 
    \dots\, .
    \label{eq:piSimple}
\end{align}

Focusing on Eq.~\eqref{eq:piSimple}, one sees that the momentum factor $\tilde{\Pi}^{\mu_{1}...\mu_{2k}}$ scales as a tower of momentum and mass terms. 
For a fixed index $k$, the series ranges from the product of $2k$ instances of the nucleus' momentum $p_A^{\mu_m}$ and no instance of the nucleus' mass $p_A^2=M_A^2$ (as shown in the leftmost term), to no instances of $p_A^{\mu_m}$ but one instance of $(p_A^2)^k=(M_A^2)^k$. The brackets $\{...\}$ represent the collection of all possible permutations of Lorentz indices.  The dimension of each term is [mass]$^{2k}$.
Again, we neglect higher-order power corrections.

Whether one retains $(M_A^2)^j$ terms or neglects them outright is the distinction between obtaining structure functions with TMCs or obtaining structure functions 
without TMCs (which we shall later label as ``No-TMC'').
For example: after contracting all Lorentz indices in Eq.~\eqref{eq:virtComptonTwist2}, one can neglect all factors of $(M_A^2/Q^2)$. This is equivalent to truncating the summation in Eq.~\eqref{eq:pi} at $j=0$. Doing so and comparing the result to Eq.~\eqref{eq:taylorStrFn} recovers (see App.~\ref{app:opeMassive} for details): 
\begin{subequations}
 \label{eq:strFnXDefOPEmassless_all}
\begin{align}
   \tilde{F}_i^{AN}=\int_0^1 dy ~ y^{N-1} ~ \tilde{F}_i^A(y,Q^2) &= C_i^{N}(Q^2) A^{N}_{\tau=2} 
   +\ \text{power corrections}, 
   \quad\text{for}\quad i = 1,3,4,5,
 \label{eq:strFnXDefOPEmassless}
 \\
  \tilde{F}_2^{A(N-1)}=\int_0^1 dy ~ y^{N-2} ~ \tilde{F}^A_2(y,Q^2) &= C_2^{N}(Q^2) A^{N}_{\tau=2}
  +\ \text{power corrections}\ .
 \label{eq:strFn2DefOPEmassless}  
\end{align}
\end{subequations}
Non-trivially, Eq.~\eqref{eq:strFnXDefOPEmassless_all} states that one can identify the product of the Wilson coefficient and reduced hadronic matrix element, $(C_\iota^{2k}A^{2k}_{\tau=2})$, as integer  Mellin moments of structure functions, up to power corrections.
(We note that footnote~\ref{foot:flavor_sum} is still applicable.)

If one does not truncate the momentum factor $\tilde{\Pi}^{\mu_{1}...\mu_{2k}}$ in Eq.~\eqref{eq:pi} at $j=0$ and instead retains the sums over all $j$, and hence retains the sum over powers of $p_A^2=M_A^2$, then one obtains expressions that are analogous to Eq.~\eqref{eq:strFnXDefOPEmassless_all}. For example: one can identify the Mellin transformation of the structure function $F_2$ with a nonzero target mass
as the product $(C_{\iota=2}^{2k}A^{2k}_{\tau=2})$ with a coefficient 
proportional to
$(M_A^2/Q^2)^j$: 
\begin{equation}
\int_0^1 dx_A\, x_A^{N-2} \tilde{F}_2^{A,\TMC}(x_A,Q^2) =
\sum _{j=0}^{\infty}
\Biggl( \frac{M_A^2}{Q^2}\Biggr)^j\frac{(N+j)!}{j!\ (N-2)!}\
\frac{C_2^{N+2j}{A}^{N+2j}_{\tau=2}}{(N+2j)(N+2j-1)}\ .
 \label{eq:strFn2DefOPEmassive}  
\end{equation}
Similar expressions can be found in App.~\ref{app:opeMassive} for the other structure functions. Intuitively, the right hand side of Eq.~\eqref{eq:strFn2DefOPEmassive} states that structure functions with TMCs can be thought simply as the product (in moment space) 
of a structure function for a massless target, i.e., the $(C_iA_{\tau=2})$ factor, and a kinematical factor, i.e., everything else. 
For this reason, the TMCs under discussion are sometimes called ``purely kinematical''~\cite{Ellis:1982cd,Ellis:1982wd,Moffat:2017sha}.
It is not obvious but, with the use of generating functions, the inverse-Mellin transform of Eq.~\eqref{eq:strFn2DefOPEmassive} has a closed form~\cite{Georgi:1976ve,Georgi:1976vf,Detmold:2005iz}.
That is to say, one can obtain concise  expressions for structure functions with TMCs in terms of structure functions for a massless target in $x$-space. These results are summarized in the following section for $F_1^A, F_2^A$, and $F_3^A$, and in App.~\ref{app:opeMassive} for all $F_i$.

\subsection{Master formula for structure functions with TMCs in \texorpdfstring{$\ell A$}{lA} DIS}\label{sub:master}

Following the procedure outlined in Sec.\,\ref{sec:buildingOPE}, we obtain a set of master formulae for twist-2 target mass corrections to structure functions for nuclei that are similar to those in Eq.~(23) in Ref.~\cite{Schienbein:2007gr} for
nucleons. 
Using the notation of Refs.~\cite{Kretzer:2003iu,Schienbein:2007gr}, the general formula  for target mass-corrected structure functions $F_j^{A, \rm TMC}$ for $ j=1,\ldots,6$ reads:
\begin{align}
\tilde{F}_{j}^{{A,\TMC}}(x_A,Q^{2})=
\sum_{i=1}^{6}A_{j}^{i}\tilde{F}_{i}^{A,(0)}(\xi_A,Q^{2})
+B_{j}^{i}\tilde{h}_{i}^A(\xi_A,Q^{2})
+C_{j}\tilde{g}_{2}^A(\xi_A,Q^{2})\, .
\label{eq:master}
\end{align}
On the left-hand side, the $F_j^{A, \rm TMC}$ take as arguments the Bjorken scaling variable $x_A$ and scale $Q$; on the right-hand side, individual terms are given in terms of structure functions for massless nuclei $\tilde{F}_i^{A,(0)}$, which take as arguments the Nachtmann variable $\xi_A$ and scale $Q$. The coefficients $A_{j}^{i}$, $B_{j}^{i}$, $C_{j}$   are  derived in App.~\ref{app:nTMC_derivation},
and are the same that are given in Tables I, II, III in Ref.~\cite{Kretzer:2003iu} for $j=1,\dots,5$.
To our knowledge, TMCs for $\tilde{F}_6^{A}$ have not been previously published.

Suppressing the $Q^2$ dependence for brevity, one finds at twist $\tau=2$ the following:
\begin{subequations}
\label{eq:master_full}
\begin{align}
\tilde{F}_{1}^{A,\TMC}(x_A)  &=  \left(\frac{x_A}{\xi_A r_A}\right) \tilde{F}_{1}^{A,(0)}(\xi_A)
 +\left(\frac{M_A^2 x_A^{2}}{Q^2 r_A^{2}}\right) \, \tilde{h}_{2}^A(\xi_A)
+\left(\frac{2 M_A^4 x_A^{3}}{Q^4 r_A^{3}}\right) \, \tilde{g}_{2}^A(\xi_A)\, , 
\label{eq:master1}
 \\
\tilde{F}_{2}^{A,\TMC}(x_A)  &=  \left(\frac{x_A^{2}}{\xi_A^{2}r_A^{3}} \right) \tilde{F}_{2}^{A,(0)}(\xi_A)
 +\left(\frac{6 M_A^2 x_A^{3}}{Q^2 r_A^{4}}\right) \, \tilde{h}_{2}^A(\xi_A)
+\left(\frac{12 M_A^4x_A^{4}}{Q^4 r_A^{5}}\right) \, \tilde{g}_{2}^A(\xi_A)\, ,
\label{eq:master2} 
 \\
\tilde{F}_{3}^{A,\TMC}(x_A)  &=  \left(\frac{x_A}{\xi_A r_A^{2}}\right) \tilde{F}_{3}^{A,(0)}(\xi_A)
+\left(\frac{2 M_A^2x_A^{2}}{Q^2 r_A^{3}}\right) \, \tilde{h}_{3}^A(\xi_A)\, , 
\label{eq:master3}
\\
\tilde{F}_4^{A,\TMC}(x_A) &= 
\left(\frac{x_A}{\xi_A r_A}\right)\ \tilde{F}_4^{A, (0)}(\xi_A)\ 
-\ 
\left(\frac{2M_A^2 x_A^2}{Q^2 r_A^2}\right)\ 
\tilde{F}_5^{A, (0)}(\xi_A)\ 
+\ 
\left(\frac{M_A^4 x_A^3}{Q^4 r_A^3}\right)\ 
\tilde{F}_2^{A, (0)}(\xi_A)\ 
\nonumber \\ 
& 
\qquad\ +\   
\left(\frac{M_A^2 x_A^2}{Q^2 r_A^3}\right)\ 
\tilde{h}_5^{A}(\xi_A)\
-\ 
\left(\frac{2M_A^4 x_A^4}{Q^4 r_A^4}\right)\ 
\left(2-\xi_A^2 M_A^2/Q^2\right)\ 
\tilde{h}_2^{A}(\xi_A)\ 
\nonumber \\ 
& 
\qquad\ +\
\left(\frac{2M_A^4 x_A^3}{Q^4 r_A^5}\right)\ 
\left(1-2 x_A^2 M_A^2/Q^2\right)\ 
\tilde{g}_2^{A}(\xi_A)\ ,
\label{eq:master4}
\\
\tilde{F}_5^{A,\TMC}(x_A) &= 
\left(\frac{x_A}{\xi_A r_A^2}\right)\ \tilde{F}_5^{A, (0)}(\xi_A)\ 
-\ 
\left(\frac{M_A^2 x_A^2}{Q^2 r_A^3 \xi_A}\right)\ 
\tilde{F}_2^{A, (0)}(\xi_A)\ 
\nonumber \\ 
& 
\qquad\ +\   
\left(\frac{M_A^2 x_A^2}{Q^2 r_A^3}\right)\ 
\tilde{h}_5^{A}(\xi_A)\
-\ 
\left(\frac{2M_A^2 x_A^2}{Q^2 r_A^4}\right)\ 
\left(1-x_A \xi_A  M_A^2/Q^2\right)\ 
\tilde{h}_2^{A}(\xi_A)\ 
\nonumber \\ 
& 
\qquad\ +\
\left(\frac{6 M_A^4 x_A^3}{Q^4 r_A^5}\right)\ 
\tilde{g}_2^{A}(\xi_A)\  ,
\label{eq:master5}
\\
\tilde{F}_6^{A,\TMC}(x_A) &= 
\left(\frac{x_A}{\xi_A r_A^2}\right)\ 
\tilde{F}_6^{A, (0)}(\xi_A)\ 
+\
\left(\frac{2M_A^2 x_A^2}{Q^2 r_A^3}\right)\ \tilde{h}_6(\xi_A)\ .
\label{eq:master6}
\end{align}
\end{subequations}
We remind the reader that the tilde notation $(\sim)$ 
on the $\tilde{F}_i$ structure functions 
indicates these are for \textbf{nuclei}; 
we will consider \textbf{nucleon} structure functions $F_i$
in the following section. 
Here, the functions $\tilde{h}_{i}^A(\xi_A,Q^{2})$ and $\tilde{g}_{2}^A(\xi_A,Q^{2})$ are given by the integrals
\begin{subequations}
\label{eq:master_auxFns}
\begin{multicols}{2}
\setlength{\columnsep}{0.4\textwidth}
\begin{align}
\tilde{h}_{2}^A(\xi_A,Q^{2}) & =  \int_{\xi_A}^{1}du_A\ \frac{\tilde{F}_{2}^{A,(0)}(u_A,Q^{2})}{u_A^{2}}\ ,
\label{eq:defh2} 
\\
\tilde{h}_{5}^A(\xi_A,Q^{2}) & =  \int_{\xi_A}^{1}du_A\ \frac{2\tilde{F}_{5}^{A,(0)}(u_A,Q^{2})}{u_A}\ ,
\label{eq:defh5}
\\
\tilde{g}_{2}^A(\xi_A,Q^{2}) & =  \int_{\xi_A}^{1}du_A\ 
\tilde{h}_{2}^A(u_A,Q^{2})\ ,
\end{align}

\columnbreak

\setlength{\columnsep}{0.4\textwidth}

\begin{align}
\tilde{h}_{3}^A(\xi_A,Q^{2}) & =  \int_{\xi_A}^{1}du_A\ \frac{\tilde{F}_{3}^{A,(0)}(u_A,Q^{2})}{u_A}\ ,
\label{eq:defh3}
\\
\tilde{h}_{6}^A(\xi_A,Q^{2}) & =  \int_{\xi_A}^{1}du_A\ \frac{\tilde{F}_{6}^{A,(0)}(u_A,Q^{2})}{u_A}\ .
\label{eq:defh6}
\end{align}
\end{multicols}
\end{subequations}

The equation Eq.~\eqref{eq:master} does not assume or imply any Callan-Gross relation~\cite{Callan:1969uq} between $F_1$ and $F_2$. 
This relationship is typically presented as the equality $2xF_1 = F_2$
and is a prediction of the (massless) quark-parton model at 
leading order in the asymptotic limit, where $Q^2\to \infty$.
At this same order and without any mass contributions (either hadronic or partonic), 
the longitudinal structure function is given by
 $F_L = F_2 - 2 x F_1$.
Thus, Callan-Gross  is a statement regarding $F_L$.
This relation is modified 
if we include masses (either hadronic masses or partonic masses), e.g., via helicity inversion, 
or
if we include contributions beyond leading order in QCD.
Furthermore, 
 were there elementary constituents in hadrons with spin zero, i.e., scalar quarks or scalar gluons,
then these would contribute to the  longitudinal structure function $F_L$ at leading order.
Thus, the Callan-Gross relation also implies that there 
are no elementary scalar (spin zero) constituents of a hadron.

Using the above, one can also compute the longitudinal structure
function with TMCs and obtain: 
\begin{align}
\tilde{F}_{L}^{{A,\TMC}}(x_A) & =  r_A^{2}\tilde{F}_{2}^{{A,\TMC}}(x_A)-2x_A \tilde{F}_{1}^{{A,\TMC}}(x_A)
\nonumber \\
 & = 
 \frac{x_A^{2}}{\xi_A^{2}r_A}
 [\tilde{F}_{2}^{A,(0)}(\xi_A)-2\xi_A \tilde{F}_{1}^{A,(0)}(\xi_A)]
 +\frac{4M_A^{2}x_A^{3}}{Q^{2}r_A^{2}} \,  \tilde{h}_{2}^A(\xi_A)
 +\frac{8M_A^{4}x_A^{4}}{Q^{4}r_A^{3}} \, \tilde{g}_{2}^A(\xi_A)
 \nonumber \\
 & =  \frac{x_A^{2}}{\xi_A^{2}r_A}\tilde{F}_{L}^{A,(0)}(\xi_A)
 +\frac{4M_A^{2}x_A^{3}}{Q^{2}r_A^{2}} \, \tilde{h}_{2}^A(\xi_A)
 +\frac{8M_A^{4}x_A^{4}}{Q^{4}r_A^{3}} \, \tilde{g}_{2}^A(\xi_A) \ .
\label{eq:fl}
\end{align}
This general result is non-zero and thus violates the Callan-Gross
relation, as it should. The leading term $\tilde{F}_{L}^{A,(0)}$ will be
non-zero for finite quark masses, and the sub-leading terms 
$\tilde{h}_{i}^A,\tilde{g}^A_{2}$ contribute for finite hadron mass $M_A$.
Note that effects from heavy quark masses are separate 
from the $M_A$-modifications of the Nachtmann   variable $\xi_A$, 
and are taken into account in the $\tilde{F}_{i}^{A,(0)}$
as outlined in Sec.\,\ref{sec:rescaling} and Ref.~\cite{Schienbein:2007gr};
we address this in Sec.~\ref{sec:QuarkMasses}.
\subsection*{Remarks}

Note that the formula for TMCs in Eq.~\eqref{eq:master} was derived without invoking \textit{perturbative} QCD. In principle, one can start from matrix elements that describe collinear parton splitting at some fixed order in perturbative QCD, construct structure functions, 
compare to $\tilde{T}_{\mu\nu}^{A}$ in Eq.~\eqref{eq:virtComptonTwist2}, 
and obtain expressions for $C_\iota^{2k}$ at some power in the strong coupling constant $\alpha_{s}$.
However, by constructing structure functions in the OPE 
the Wilson coefficients $C_\iota^{2k}$ in Eq.~\eqref{eq:virtComptonTwist2} are exact quantities,
i.e., they are defined at all orders in $\alpha_{s}$.
(Expanding $C_\iota^{2k}$ in powers of $\alpha_{s}$ would then yield the perturbative QCD result.) 
Ultimately, this implies that all the $\alpha_s$ dependence of $\tilde{F}_i^{A, \rm TMC}$ is contained in $\tilde{F}_i^{A,(0)}$. This also means that the coefficients $A_{j}^{i}$, $B_{j}^{i}$ and $C_{j}$, and the variable $\xi_A$ are the same irrespective of the order (LO, NLO, NNLO,~\ldots{}) at which the structure functions $\tilde{F}_{i}^{A, (0)}$ are considered.

In the above expressions for $F_{i}^{{\TMC}}$, we emphasize that it is $(x_A,Q^{2})$  and not $(\xi_A,Q^{2})$ that is the correct point in phase space to evaluate structure functions. 
While on the surface it may appear strange to have the left-hand side of Eq.~\eqref{eq:master} to be a function of $x_A$ while the right-hand side be a function of $\xi_A$, this difference arises naturally in the calculation. In particular, final-state kinematics in $\ell A$ DIS are constrained by momentum conservation, with $\delta^{4}(q+P_A-P_{X})\sim\delta(x_A-\xi_A)$.
Thus, we can write schematically
\begin{equation}
 F_{i}^{\TMC}(x_A,Q^{2})\sim F_{i}^{(0)}(x_A,Q^2)\,
\delta(x_A-\xi_A)\sim F_{i}^{(0)}(\xi_A,Q^2).   
\end{equation}
Note that it would be incorrect to write $F_{i}^{\TMC}(\xi_A,Q^{2})\sim F_{i}^{(0)}(\xi_A,Q^2)$, as the Mellin and inverse-Mellin transformations in Sec.\,\ref{sec:buildingOPE} are defined with respect to $x_A$, not $\xi_A$.

Another feature of Eq.~\eqref{eq:master} is that $h_2$ and $g_2$ appear in the formulae for both $F_1^{\TMC}$ and $F_2^{\TMC}$, and is an example of the phenomenon of structure function mixing. This follows directly from the forms of $\tilde{T}_{\mu\nu}^{A}$ and the momentum factor $\tilde{\Pi}$ in Eq.~\eqref{eq:virtComptonTwist2}, in the massive target limit. For example: the terms proportional to $C_1^{2k}$ and to $C_2^{2k}$ both contribute to $\Delta \tilde{T}^{A,1}$, which multiplies the Lorentz structure  $-g_{\mu\nu}$ in Eq.~\eqref{eq:virtComptonTwist2}. As shown in Eq.~\eqref{eq:piSimple}, $\tilde{\Pi}$ in the massive-target limit  is proportional to terms that (a) consist exclusively of $p_A^{\mu_n}$ factors as well as to terms that (b) include factors of momentum $p_A^{\mu_n}$, metric $g^{\mu_m \mu_n}$, and mass $M_A$. Among the many possible contractions in Eq.~\eqref{eq:virtComptonTwist2} are 
the ``(a)'' terms contracting with the $C_1^{2k}$ term and 
the ``(b)'' terms contracting with the $C_2^{2k}$ term. Both sets of contractions generate a term proportional to $-g_{\mu\nu}$. However, the ``(b)'' set of contractions only enters $\Delta \tilde{T}^{A,1}$ as a TMC, i.e., it vanishes in the $(M_A^2/Q^2)\to0$ limit. Arguably, one can think of $C_1^{2k}$ feeding into  $F_1^{\TMC}$ as a helicity-conserving contribution, which survives in the massless limit, whereas the contribution from $C_2^{2k}$ is a helicity-inverting contribution, which vanishes in the massless limit.

\section{Rescaling}\label{sec:rescaling} 

\global\long\def\arraystretch{2}%

\newcommand\Tstrut{\rule{0pt}{28pt}}       %
\newcommand\Bstrut{\rule[-12pt]{0pt}{0pt}} %
\newcommand\BBstrut{\rule[-18pt]{0pt}{0pt}} %

\begin{table}[!p]
\framebox{\begin{minipage}[t]{0.90\textwidth} 
\begin{centering}
\vspace{0.2cm}
 \textbf{\Large{}Summary table of key relations}\\
 {\Large{}\vspace{0.2cm}
 }%
\begin{tabular}{|c|c|c|}
\hline 
\textbf{\large{}Nucleus A}  &  & \textbf{\large{}Nucleon N}\tabularnewline
\hline 
\hline 
$M_{A}=A\,M_{N}$  &  & $M_{N}=M_{A}/A$ \tabularnewline
\hline 
$p_{A}=A\,p_{N}$  &  & $p_{N}=p_{A}/A$ \tabularnewline
\hline 
$x_{A}=\mathlarger{\frac{Q^{2}}{2p_{A}\cdot q}}
\equiv{x_{N}}/{A}$  &  & $x_{N}=\mathlarger{\frac{Q^{2}}{2p_{N}\cdot q}}\equiv A\,x_{A}$ \tabularnewline
$x_{A}\in[0,1]$  &  & $x_{N}\in[0,A]$ \tabularnewline
\hline 
$W_{A}^{2}=(p_{A}+q)^{2}$  &  & $W_{N}^{2}=(p_{N}+q)^{2}$ \tabularnewline
\hline 
\Tstrut\BBstrut
$\nu_{A}=\mathlarger{\frac{(q\cdot p_{A})}{M_{A}}} 
=\mathlarger{\frac{Q^2}{2M_A x_A}}
\equiv\nu_{N}$  &  & 
$\nu_{N}=\mathlarger{\frac{(q\cdot p_{N})}{M_{N}}
=\mathlarger{\frac{Q^2}{2M_N x_N}}
\equiv\nu_{A}}$ \tabularnewline
\hline 
\Bstrut
$y_{A}=\mathlarger{\frac{\nu_{A}}{E}\equiv y_{N}}$  &  &
$y_{N}=\mathlarger{\frac{\nu_{N}}{E}\equiv y_{A}}$
\tabularnewline
\hline 
\multicolumn{3}{|c|}{}
\\[-10pt] %
\hline 
\hline 
\multicolumn{3}{|c|}{\textbf{\large{}Nachtmann variable \& hadronic mass}
\Bstrut}\tabularnewline
\hline 
\Tstrut \BBstrut
$r_{A}=\sqrt{1+\mathlarger{\frac{4x_{A}^{2}M_{A}^{2}}{Q^{2}}}}\equiv r_{N}$  &  & $r_{N}=\sqrt{1+\mathlarger{\frac{4x_{N}^{2}M_{N}^{2}}{Q^{2}}}}\equiv r_{A}$ \tabularnewline
\hline 
$\xi_{A}=R_{M}x_{A}\equiv\xi_{N}/A$  &  & $\xi_{N}=R_{M}x_{N}\equiv A\,\xi_{A}$ \tabularnewline
$\xi_{A}\in[0,1]$  &  & $\xi_{N}\in[0,A]$ \tabularnewline
\hline 
\multicolumn{3}{|c|}{Since\  $r_{A} = r_{N}$   }\\[-10pt]
\multicolumn{3}{|c|}{We\ define\ $r\equiv r_{A} = r_{N}$, such\ that 
$R_{M}=\mathlarger{\frac{2}{  1+r}}$ 
\Bstrut
}\tabularnewline
\hline 
\multicolumn{3}{|c|}{Also, \ 
$\xi_{A}/x_{A}=\xi_{N}/x_{N}=R_{M}=\mathlarger{\frac{2}{1+r}}$ 
\Bstrut
}\tabularnewline
\hline 
\multicolumn{3}{|c|}{ We also define $\varepsilon=(xM/Q)$. }\tabularnewline
\hline 
\end{tabular}
\par\end{centering}
\begin{centering}
 
\par\end{centering}
\vspace{+0.25cm}
 \caption{ We summarize the key relations for a nucleus ($A$) and nucleon ($N$).
We also find it convenient to define $\varepsilon=(xM/Q)$; we omit
the subscripts on $\varepsilon$ for brevity as $(x_{A}M_{A}/Q)=(x_{N}M_{N}/Q)$.
We caution that $W_{A}$ and $W_{N}$ are \textbf{not} simply related,
\textit{c.f.}, Sec.~\ref{sec:wcut}. Note, the target mass modifies
the scaling variable via $\xi_{A}=R_{M}\,x_{A}$. Additionally, we
introduce the shorthand notation $r=r_{A}=r_{N}$.
\label{tab:one} }
\end{minipage}

} 
\end{table}

\global\long\def\arraystretch{1}%

In this section we introduce a rescaling 
between the nuclear and averaged nucleon kinematics.
Many of the key relations are summarized in Table~\ref{tab:one}. 
We typically identify the nuclear variables with an ``$A$'' subscript $(x_A, \xi_A, r_A)$,
and those of the nucleon with an ``$N$'' subscript $(x_N, \xi_N, r_N)$.
Similarly, the original nuclear structure functions are identified with a tilde $(\tilde{W},\tilde{F})$,
while those re-scaled to the kinematics of an averaged nucleon
are without $(W,F)$.
\subsection{Nuclear and nucleon kinematics}

From Eqs.~\eqref{eq:master_full} and \eqref{eq:fl},
we see, perhaps with some minor algebraic  manipulation,
that $M_A^{2j}$ terms are always accompanied by $x_A^{2j}$ factors in $x$-space. 
Furthermore, since structure functions $\tilde{F}_2^{A,(0)}(x_A)$ have their main support at\footnote{%
Note that we have introduced the nucleon scaling variable $x_N=A x_A$; this will be discussed  in detail throughout this section.
Although  $x_N$ can in principle extend to $x_N=A$, the dominant 
range of the kinematics is $0 \le x_N \lesssim 1$.  
Discussion on the $x_N>1$ region are given in Sec.~\ref{sec:largex}.
} %
$x_A = x_N/A \sim 1/A$,
we have effectively
\begin{equation}
    \frac{M_A^{2j} x_A^{2j}}{Q^{2j}} = 
    \frac{(M_N^{2j} A^{2j})\, x_A^{2j}}{Q^{2j}}  =
    \frac{M_N^{2j} x_N^{2j}}{ Q^{2j}} \sim 
    \frac{M_N^{2j}}{Q^{2j}}\, ,
\end{equation}
where $M_N=M_A/A$ is defined as the (average) nucleon mass such that the terms with $j>0$ in Eq.~\eqref{eq:piSimple} 
are suppressed 
for $Q^2 > M_N^2$ independent of the nuclear target.
Subsequently, we introduce the {\em average nucleon momentum} $p_N := p_A/A$ inside the nucleus
and define a nucleon scaling variable 
\begin{equation}
x_N = \frac{Q^2}{2p_N\cdot q}  = A\, x_A\ , \qquad {\rm where} \quad x_N \in [0,A]\ ,
\end{equation}
in contrast to the original Bjorken variable 
\begin{equation}
x_A = \frac{Q^2}{2p_A\cdot q}  = x_N/A\ , \qquad {\rm where} \quad x_A \in [0,1]\ .
\end{equation}
Note that  the original $x_A$ variable can be constructed directly from the external momenta of the particles
whereas the ``averaged quantities'' $x_N$ (and $p_N$) are not directly observable for nuclear targets with $A>1$.

Using $M_A = A M_N$, we find for the Nachtmann scaling variable
\begin{equation}
\xi_A = \frac{2 x_A}{1 + \sqrt{1+4 x_A^2 M_A^2/Q^2}} = \frac{1}{A} \xi_N,
\quad\text{with}\quad
\xi_N = \frac{2 x_N}{1 + \sqrt{1+4 x_N^2 M_N^2/Q^2}} \, .
\end{equation}
Conversely, 
the quantity $R_M$, 
which relates the Bjorken scaling $x_A$ variable to the Nachtmann scaling variable $\xi_A$  via $\xi_A=R_M x_A$, 
is the same for both the nucleus and (averaged) nucleon case: 
\begin{equation}
R_M=
\frac{2}{1+r_A} = 
\frac{2}{1+r_N}\ .
\end{equation}
This is because 
the factor $r_A$, which is kinematical in origin (see Sec.~\ref{sec:variables})
and appears throughout the master formulae of Eq.~\eqref{eq:master},
is also the same for nuclei and the average nucleon:
\begin{equation}
r_A = \sqrt{1 + 4 x_A^2 M_A^2/Q^2} = \sqrt{1+ 4 x_N^2 M_N^2/Q^2} = r_N\, .
\end{equation}
Hence, we have  $\xi_{A} = R_M x_{A}$ and $\xi_{N} = R_M x_{N}$, 
where $R_M$ takes the same value in both equations.

\subsection{Rescaled structure functions}
\label{sec:rescaled_str_fn}

Turning to the structure functions, 
we define a rescaled hadronic tensor via the relation
\begin{equation}\label{eqs:rescaledWmunu}
A\, W^A_{\mu\nu}(p_N,q) := \tilde{W}^A_{\mu\nu}(p_A,q) \ ,
\end{equation}
expressed in terms of rescaled structure functions:
\begin{equation}
F_2^A(x_N,Q^2) := \tilde{F}_2^A(x_A,Q^2)
\ , \qquad 
x_N F_{1,3-6}^A(x_N,Q^2) := x_A \tilde{F}_{1,3-6}^A(x_A,Q^2) \, .
\label{eq:sfs}
\end{equation}
The general relations in Eq.~\eqref{eq:sfs} hold for 
both with and without TMCs and for any value of   twist.
As we show below, they are consistent with the master formula for the target-mass-corrected structure functions in Eq.~\eqref{eq:master}.
Hence, the pattern of the rescaled  equations is consistent between the nucleus and nucleon.

Indeed, one easily finds from Eq.~\eqref{eq:master_auxFns}:
\begin{equation}
\tilde h_2^A(\xi_A) = \int_{\xi_A}^1 du_A \frac{\tilde F_2^{A(0)}(u_A)}{u_A^2}  = A \int_{\xi_N}^A du_N \frac{F_2^{A(0)}(u_N)}{u_N^2} =: A h_2^A(\xi_N)\, ,
\end{equation}
\begin{equation}
\tilde h_3^A(\xi_A) = \int_{\xi_A}^1 du_A \frac{\tilde F_3^{A(0)}(u_A)}{u_A}  = A \int_{\xi_N}^A du_N \frac{F_3^{A(0)}(u_N)}{u_N} =: A h_3^A(\xi_N)\, ,
\end{equation}
\begin{equation}
\tilde h_5^A(\xi_A) = \int_{\xi_A}^1 du_A \frac{2\tilde F_5^{A(0)}(u_A)}{u_A}  = A \int_{\xi_N}^A du_N \frac{2F_5^{A(0)}(u_N)}{u_N} =: A h_5^A(\xi_N)\, ,
\end{equation}
\begin{equation}
\tilde h_6^A(\xi_A) = \int_{\xi_A}^1 du_A \frac{\tilde F_6^{A(0)}(u_A)}{u_A}  = A \int_{\xi_N}^A du_N \frac{F_6^{A(0)}(u_N)}{u_N} =: A h_6^A(\xi_N)\, ,
\end{equation}
\begin{equation}
\tilde g_2^A(\xi_A) = \int_{\xi_A}^1 du_A \tilde h_2^{A}(u_A)  = 
\frac{1}{A} \int_{\xi_N}^A du_N A h_2^A(u_N) =: g_2^A(\xi_N)\, .
\end{equation}

With these expressions, and using $x_N {=} A x_A$, $\xi_N {=} A \xi_A$, $r_A{=}r_N$, $M_A {=} A M_N$, we obtain the  TMC formula for the
rescaled nuclear structure functions:
\begin{subequations}
\label{eq:master-rescaled}
\begin{align}
F_{1}^{A,\TMC}(x_N,Q^{2}) & =  \left(\frac{x_N}{\xi_N r_N}\right) F_{1}^{A,(0)}(\xi_N,Q^{2})
+\left(\frac{M_N^2 x_N^{2}}{Q^2 r_N^{2}}\right) \, h_{2}^A(\xi_N,Q^{2})
+\left(\frac{2 M_N^4 x_N^{3}}{Q^4 r_N^{3}}\right) \, g_{2}^A(\xi_N,Q^{2})\, , 
\nonumber \\
\label{eq:master1-rescaled}
\\
F_{2}^{A,\TMC}(x_N,Q^{2}) & =  \left(\frac{x_N^{2}}{\xi_N^{2}r_N^{3}}\right) F_{2}^{A,(0)}(\xi_N,Q^{2})
+\left(\frac{6 M_N^2 x_N^{3}}{Q^2 r_N^{4}}\right) \, h_{2}^A(\xi_N,Q^{2})
+\left(\frac{12 M_N^4 x_N^{4}}{Q^4 r_N^{5}}\right) \, g_{2}^A(\xi_N,Q^{2})\, ,
\nonumber \\
\label{eq:master2-rescaled} 
\\
F_{3}^{A,\TMC}(x_N,Q^{2}) & =  \left(\frac{x_N}{\xi_N r_N^{2}}\right) F_{3}^{A,(0)}(\xi_N,Q^{2})
+\left(\frac{2 M_N^2x_N^{2}}{Q^2 r_N^{3}}\right) \, h_{3}^A(\xi_N,Q^{2})\ ,
\label{eq:master3-rescaled}
\\
{F}_4^{A,\TMC}(x_N,Q^{2}) &= 
\left(\frac{x_N}{\xi_N r_N}\right)\ {F}_4^{A, (0)}(\xi_N,Q^{2})\ 
-\ 
\left(\frac{2M_N^2 x_N^2}{Q^2 r_N^2}\right)\ 
{F}_5^{A, (0)}(\xi_N,Q^{2})\ 
+\ 
\left(\frac{M_N^4 x_N^3}{Q^4 r_N^3}\right)\ 
{F}_2^{A, (0)}(\xi_N,Q^{2})\ 
\nonumber \\ 
& 
\qquad\ +\   
\left(\frac{M_N^2 x_N^2}{Q^2 r_N^3}\right)\ 
h_5^{A}(\xi_N,Q^{2})\
-\ 
\left(\frac{2M_N^4 x_N^4}{Q^4 r_N^4}\right)\ 
\left(2-\xi_N^2 M_N^2/Q^2\right)\ 
h_2^{A}(\xi_N,Q^{2})\ 
\nonumber \\ 
& 
\qquad\ +\
\left(\frac{2M_N^4 x_N^3}{Q^4 r_N^5}\right)\ 
\left(1-2 x_N^2 M_N^2/Q^2\right)\ 
g_2^{A}(\xi_N,Q^{2})\ ,
\label{eq:master4-rescaled}
\\
{F}_5^{A,\TMC}(x_N,Q^{2}) &= 
\left(\frac{x_N}{\xi_N r_N^2}\right)\ {F}_5^{A, (0)}(\xi_N,Q^{2})\ 
-\ 
\left(\frac{M_N^2 x_N^2}{Q^2 r_N^3 \xi_N}\right)\ 
{F}_2^{A, (0)}(\xi_N,Q^{2})\ 
\nonumber \\ 
& 
\qquad\ +\   
\left(\frac{M_N^2 x_N^2}{Q^2 r_N^3}\right)\ 
h_5^{A}(\xi_N,Q^{2})\
-\ 
\left(\frac{2M_N^2 x_N^2}{Q^2 r_N^4}\right)\ 
\left(1-x_N \xi_N  M_N^2/Q^2\right)\ 
h_2^{A}(\xi_N,Q^{2})\ 
\nonumber \\ 
& 
\qquad\ +\
\left(\frac{6 M_N^4 x_N^3}{Q^4 r_N^5}\right)\ 
g_2^{A}(\xi_N,Q^{2})\  ,
\label{eq:master5-rescaled}
\\
{F}_6^{A,\TMC}(x_N,Q^{2}) &= 
\left(\frac{x_N}{\xi_N r_N^2}\right)\
{F}_6^{A, (0)}(\xi_N,Q^{2})\ 
+\
\left(\frac{2M_N^2 x_N^2}{Q^2 r_N^3}\right)\ h_6(\xi_N,Q^{2})\ .
\label{eq:master6-rescaled}
\end{align}
\end{subequations}
The above rescaled equations for the 
\textbf{nucleon} structure functions  $F_i$ (without the ``$\sim$'' notation)
take the same form as the master equations 
for the \textbf{nuclear} structure functions $\tilde{F}_i$ 
in  Eq.~\eqref{eq:master} but are entirely expressed
in terms of averaged nucleon variables. This is one of the main
results of this rescaling demonstration.
\subsection{Impact of quark masses \label{sec:QuarkMasses}}
\renewcommand{\arraystretch}{2}

\begin{table}[!t]
\centering{}%
\begin{tabular}{|c||c|}
\hline 
\multicolumn{2}{|c|}{\textbf{\large{}Partonic masses in scaling variables}}\tabularnewline
\hline 
\hline 
\multicolumn{2}{|c|}{$\bar{\xi}_{A}=R_{ij}\xi_{A}=R_{ij}R_{M}x_{A}$ \quad Nucleus }\\[-10pt]
\multicolumn{2}{|c|}{$\bar{\xi}_{N}=R_{ij}\xi_{N}=R_{ij}R_{M}x_{N}$ \quad Nucleon }\tabularnewline
\hline 
\hline 
\multicolumn{2}{|l|}{${\xi}_{A,N}$ :\quad \  TMC \ corrections (no overbar)   }\\[-10pt]
\multicolumn{2}{|l|}{$\bar{\xi}_{A,N}$ : \quad TMC + parton mass corrections }\tabularnewline
\hline 
\hline 
\multicolumn{2}{|c|}{{\Large{}For\ \ $q_{i}+V\to q_{j}$}}\tabularnewline
 \hdashline
\multicolumn{2}{|c|}{$\ensuremath{R_{ij}=
\mathlarger{\mathlarger{\frac{1}{{ 2Q^{2}}}[Q^{2}-m_{i}^{2}+m_{j}^{2}+\Delta(Q^{2},m_{i}^{2},m_{j}^{2})]}}}
$}\tabularnewline
\multicolumn{2}{|c|}{$\ensuremath{\Delta(a,b,c)=\mathlarger{\sqrt{a^{2}+b^{2}+c^{2}-2(ab+bc+ca)}}}$}\tabularnewline
\hline 
\hline 
\multicolumn{2}{|c|}{{\Large{}For\ \ $\ensuremath{g+V\to q_{j}+q_{k}}$}}\tabularnewline
 \hdashline
\multicolumn{2}{|c|}{$\ensuremath{R_{jk}=1+
\mathlarger{\mathlarger{\frac{{\displaystyle (m_{j}+m_{k})^{2}}}{{ Q^{2}}}}}}
$}\tabularnewline
\hline 
\end{tabular}\caption{Summary of partonic mass relations in scaling variables. 
The target mass modifies the Nachtmann scaling variable 
via $\xi_{A}=R_{M}\,x_{A}$,
where $R_{M}$ is defined in Table\,\ref{tab:one}.
This is further
modified by the parton masses $\overline{\xi}_{A}=R_{ij}\xi_{A}=R_{ij}R_{M}x_{A}$;
hence these multiplicative corrections factorize between the hadron
and parton masses~\cite{Schienbein:2007gr}.
In the limit the initial-state parton mass $m_{i}\to0$, we obtain
the ``slow rescaling'' result: \mbox{$R_{ij}=\left(1+m_{j}^{2}/Q^{2}\right)$.}
Thus, $\xi_A$ contains modifications due to the hadron mass $M_A$,
while  $\bar{\xi}_A$ (with the overbar) contains modifications due to \textbf{both} the hadron mass $M_A$
and the parton masses~$\{m_{i},m_{j}\}$.
}
\label{tab:PartonMass}
\end{table}

\renewcommand{\arraystretch}{1}
We stress that all of the above derivations have made no mention of the parton model. 
This can be understood from the fact that the quark and gluon 
sub-structure of nuclei enters only implicitly in intermediate steps of the derivation of
the master and rescaled formulae since the hadronic currents appearing in the hadronic tensor are composed 
 of quark and gluon fields and we have relied on the validity of the OPE for the
products of these currents.
Thus, due to the nature of the factorization, all details related to quark masses is contained entirely
inside the $F_i^{A,(0)}$ structure functions, and the forms of the rescaled formulae in Eq.~\eqref{eq:master-rescaled} 
are unchanged.

The treatment of the quark masses was reviewed in detail in Ref.~\cite{Schienbein:2007gr},
but it is instructive to briefly mention how the quark masses effect the rescaling of the 
kinematic variables as summarized in  Table~\ref{tab:PartonMass}.
Just as the hadron mass modifies the scaling variable via the relation $\xi=R_M x$ due to momentum conservation, 
the quark masses further modify the scaling variable via the relation~\cite{Olness:1986mv,Aivazis:1993pi,Aivazis:1993kh} 
\begin{align}
 \bar{\xi}\ =\ R_{ij}\ \xi\ =\ R_{ij}\ R_M\ x \ .
\end{align}
In our notation, $\xi_A$ contains modifications due to the hadron mass $M_A$,
while  $\bar{\xi}_A$ (with the overline) contains modifications due to \textbf{both} the hadron mass $M_A$
and the parton masses $m_{i},m_{j}$.
The $R_M$ factor, which is given in Eq.~\eqref{eq:rma}, depends on the hadron (nucleus) mass, and the $R_{ij}$ factor depends on the 
incoming and outgoing quark masses  $\{m_i,m_j\}$, respectively 
({\it c.f.}, Table\,\ref{tab:PartonMass}).
Note that the hadronic and partonic correction factors factorize and are subsequently applied multiplicatively.
Intuitively, this factorization signifies that hadronic kinematics can be considered separately 
from partonic kinematics, up to momentum conservation.
Additionally, in the limit where the mass of the initial-state quark vanishes $m_i=0$, we have the familiar 
``slow rescaling'' limit $R_{ij}\approx(1+m_j^2/Q^2)$. 
For further details, see Refs.~\cite{Olness:1986mv,Aivazis:1993pi,Aivazis:1993kh,Schienbein:2007gr}. 

\section{Parton model} 
\label{sec:parton}

We now turn to a discussion of structure functions in the context of the parton model
originally proposed by Feynman in 1969 \cite{Feynman:1969ej} and applied to electron-proton DIS 
by Bjorken and Paschos \cite{Bjorken:1969ja}. In the years following these works, partons were identified with
  quarks and gluons, and the heuristic parton model including QCD effects was rigorously derived 
from first principles of QCD in the context of factorization theorems \cite{Bodwin:1984hc,Collins:1985ue,Collins:1988ig,Collins:1987pm,Collins:1989gx,Collins:1998rz,Collins:2011zzd,Berger:1987er}.
In this section we summarize the main elements of the conventional ``massless'' QCD-improved parton model, 
also known as Zero Mass Variable Flavor Number Scheme (ZM-VFNS), 
where quark-mass effects in the hard scattering cross sections and target-mass effects are neglected.
There is an extensive body of literature on computations with heavy partons; 
Collins, for example, extended the factorization proof~\cite{Collins:1985ue} to the case of heavy partons in Ref.~\cite{Collins:1989gx}, and implementations can be found in 
Refs.~\cite{Aivazis:1993pi,Olness:1997yc,Thorne:2000zd,Alekhin:2020edf,Alekhin:2009ni}.
A discussion of the parton model including such mass effects will be given in    Sec.~\ref{sec:acot}.

In the QCD-improved parton model the nuclear structure functions $({\cal \tilde{F}}^A_k)$ are given as
convolutions of target-independent, short-distance, Wilson coefficients ($C_{k,i}$) with 
universal nuclear parton distribution functions ($\tilde{f}_i^A$):
\begin{equation}
    {\cal \tilde{F}}^A_k(x_A,Q^2) 
    = \int_{x_A}^1 \frac{dy_A}{y_A} \  \tilde{f}_i^A(y_A,Q^2) \  C_{k,i}(x_A/y_A) \  + \ldots
    \label{eq:pm}
\end{equation}
where the ellipses represent higher twist contributions ${\cal \tilde{F}}_k^{A,\tau \ge 4}(x_A,Q^2)$,
and a sum over $i=q,g$ species is understood.
As in the previous section, we use the tilde notation to identify the nuclear PDFs $\tilde{f}_i^A(x_A,Q^2)$,
and the rescaled  PDFs are without the tilde $f_i^A(x_N,Q^2)$.
Here, we use the shorthand
\begin{equation}
   {\cal \tilde{F}}^A_k(x_A) = [\tilde{F}_1(x_A),\tilde{F}_2(x_A)/x_A,\tilde{F}_3(x_A)] \qquad\text{for}\qquad  k=1,2,3\, . 
\end{equation}
The nuclear PDFs  $(\tilde{f}_i^A)$ are defined as Fourier transforms of matrix elements of local, twist-2 operators,
which we denote\footnote{The notation for field operators, which typically carry indices throughout this work,   should not be confused with the ``Big $O$'' notation $\mathcal{O}(x)$, which  we also use in this work (without indices) to denote an uncertainty or expansion to order $x$.} as ${\cal O}_i(z)$ and which are defined initially in coordinate space.
These operators are composed of quark and gluon fields, and are acted upon by the hadronic, i.e., nuclear, state $A(p_A)$:
\begin{equation}
 \tilde{f}_i^A(x_A,Q^2) \sim \langle A(p_A) |{\cal O}_i| A(p_A) \rangle   \, ,
\end{equation}
where $x_A \in [0,1]$ and the factorization and renormalization scales have been identified with the exchange boson virtuality $\mu_F=\mu_R=Q$.

In this paper we focus on unpolarized DIS. However, 
the application to polarized DIS are relatively straightforward. 
For a spin-$j$ target there are $2(2j+1)$ independent helicity-dependent quark  distributions
$\tilde{q}_{+}^m(x)$ and $\tilde{q}_{-}^m(x)$ with $-j \le m \le j$. 
Here $\tilde{q}_{+}^m(x)$ $[\tilde{q}_{-}^m(x)]$ describes the number density for finding a quark with momentum fraction $x$ and
spin $m$  parallel [anti-parallel] to the the \mbox{$z$-axis} 
of a hadron moving with infinite momentum along the $z$-axis.
Due to parity invariance of the strong interaction, 
we have a relation between the up (parallel) and down (anti-parallel) spin ($\tilde{q}_{+}^m = \tilde{q}_{-}^{-m}$) 
such that there remain only $2j+1$ independent distributions.
However, this symmetry is broken once weak interaction effects are taken into account.
For a spin-$j$ target, the unpolarized quark densities are related to the helicity-dependent distributions
as follows:
 \begin{equation}
     \tilde{q}_i^A(x_A,Q^2) := \frac{1}{2 j+1} \sum_{m=-j}^j (\tilde{q}^{m}_{+} + \tilde{q}^{m}_{-})\, .
 \end{equation}
For a spin-$1/2$ target (and using $\tilde{q}_{+}^{1/2} = \tilde{q}_{-}^{-1/2}$), this reduces to the well-known expression:
\begin{equation}
    \tilde{q}_i^A(x_A,Q^2) := \tilde{q}^{1/2}_{+} + \tilde{q}^{1/2}_{-}\, .
\end{equation}
Similar definitions can be written down for the unpolarized gluon distribution.

\subsection{Nuclear DGLAP evolution}\label{sub:dglap}

The unpolarized nuclear PDFs (for any target spin) satisfy the usual 
Dokshitzer-Gribov-Lipatov-Altarelli-Parisi (DGLAP)
evolution equations~\cite{Gribov:1972ri,Altarelli:1977zs,Dokshitzer:1977sg}:
\begin{eqnarray}
\frac{d \tilde{f}_i^A(x_A,Q^2)}{d \ln Q^2} & = &
\frac{\alpha_s(Q^2)}{2 \pi} \int_{x_A}^1 \frac{dy_A}{y_A} \, 
P_{ij}(y_A) \, \tilde{f}_j^A(x_A/y_A,Q^2)
\\ 
& = & 
\frac{\alpha_s(Q^2)}{2 \pi} \int_{x_A}^1 \frac{dy_A}{y_A} \, 
P_{ij}(x_A/y_A) \, \tilde{f}_j^A(y_A,Q^2) \quad .
\end{eqnarray}
Furthermore, the following sum rules 
due to charge, baryon number, and momentum conservation 
are satisfied:
\begin{subequations}
\label{eq:sumrules_A}
\begin{align}
 \int_0^1 &dx_A\ \tilde{u}_v^A(x_A,Q^2) = 2Z + N \, ,
 \\
 \int_0^1 &dx_A\ \tilde{d}_v^A(x_A,Q^2) = Z + 2N \, ,
 \\
 \int_0^1 &dx_A\ x_A \sum_i \tilde{f}_i^A(x_A,Q^2) = 1\, , 
\end{align}
\end{subequations}
where $Z$ is the electric charge of the nucleus
with baryon number $A=Z+N$.

\subsection{Relation to the OPE}

A Mellin transformation can turn a convolution integral   $(A \otimes B)(x) = \int_x^1 dy/y A(y) B(x/y)$
into an ordinary product  of Mellin moments: 
\begin{equation}
    \int_0^1 dx x^{n-1} (A \otimes B)(x) \equiv (A \otimes B)_n = A_n \cdot B_n \  ,
\end{equation}
where
\begin{equation}
    A_n \equiv \int_0^1 dx x^{n-1} A(x)\,  \qquad\text{and}\qquad  B_n \equiv \int_0^1 dx x^{n-1} B(x) \  .
\end{equation}
Applying this to structure functions of the QCD-improved parton model in Eq.~\eqref{eq:pm}, we obtain:
\begin{equation}
\int_0^1 dx_A x_A^{n-1} \, {\cal \tilde{F}}^A_k(x_A,Q^2) 
= 
\sum_n
\tilde{f}_{i}^{A,n}(Q^2) \, C_{k,i}^n\, .
\label{eq:Mellin-pm}
\end{equation}
where on the RHS the Mellin transformation has yielded a simple sum of Mellin moments. 
(Note that we indicate the Mellin moment by the ``$n$'' superscript, as the tilde-notation 
has already been used for the pre-scaled PDFs, which we discuss in the following section, and $N$ already denotes averaged nucleon quantities.)

The relation between the structure functions in the parton model and the 
OPE can now be easily seen by comparing the Mellin-moment of Eq.~\eqref{eq:Mellin-pm} with
the $j{=}0$ term in Eq.~\eqref{eq:strFn2DefOPEmassive}:
\begin{equation}
 {\cal \tilde{F}}_{k}^{(0),n}
 = C_{k,i}^n \, \tilde{A}_{i}^n 
 = C_{k,i}^n \, \tilde{f}_{i}^{A,n}(Q^2)\ , \quad\text{for}\quad k=1,2,3\, ,
 \label{eq:mellinCF}
\end{equation}
where a sum over $i=q,g$ is understood.
From this equation we see that the matrix elements $\tilde{A}^n$ of Eq.~\eqref{eq:strFn2DefOPEmassive}
are simply the Mellin moments of the  parton distribution functions:
\begin{equation}
    \tilde{A}_{i}^n(Q^2) =  \tilde{f}_{i}^{A,n}(Q^2) 
    \equiv
    \int_0^1 dx_A x_A^{n-1} \, \tilde{f}_i(x_A,Q^2) \ .
\end{equation}
Similarly, the Wilson coefficients  $C_{k,i}^n$  of  Eq.~\eqref{eq:mellinCF}
are directly  identified with those of Eq.~\eqref{eq:strFn2DefOPEmassive} in the OPE.

\subsection{Rescaling}\label{subs:rescaling}

The parton  distributions $\tilde{f}_i^A(x_A)$ are number densities defined on the interval $x_A\in[0,1]$ 
such that $\tilde{f}_i^A(x_A)dx_A$ can be interpreted as
the number of partons ``$i$'' carrying a fraction
of the parent hadron momentum in the interval $[x_A, x_A+dx_A]$.
We define rescaled parton densities 
$f_i^A(x_N,Q^2)$ in the variable $x_N\in[0,A]$
with $x_N = A x_A$
by imposing that the number of partons remains unchanged
in the corresponding momentum intervals:
\begin{equation}\label{eqs:rescpdf}
    f_i^A(x_N,Q^2)dx_N := \tilde{f}_i^A(x_A,Q^2)dx_A\  .
\end{equation}
This equation may appear a bit odd as  $x_N\in[0,A]$ on the LHS,
while  $x_A\in[0,1]$  on the RHS; 
we will provide additional clarification in the following discussion. 
Additionally, 
it should be noted that the rescaled nuclear PDF 
$f_i^A(x_N,Q^2)$ are the ones usually used in the literature 
as $f_i^A(x_N,Q^2)$ can be meaningfully compared across different nuclei,
in contrast to  $\tilde{f}_i^A(x_A,Q^2)$. 

It can be easily checked that the rescaling defined
here at the level of PDFs is consistent with
the rescaling introduced above at the hadronic level.
Correspondingly, the rescaled objects satisfy analogous DGLAP evolution
equations:
\begin{eqnarray}
\frac{df_i^A(x_N,Q^2)}{d \ln Q^2} & = &
\frac{\alpha_s(Q^2)}{2 \pi} \int_{x_N}^A \frac{dy_N}{y_N}\ 
P_{ij}\left(\frac{x_N}{y_N} \right)  \  f_j^A(y_N,Q^2)\, .
\end{eqnarray}
Furthermore, the sum rules take the following form
after the rescaling:
\begin{subequations}
\begin{eqnarray}
 \int_0^A &dx_N& u_v^A(x_N,Q^2) = 2Z + N \, ,
\label{eq:nucSumRulesi}
 \\
 \int_0^A &dx_N& d_v^A(x_N,Q^2) = Z + 2N \, ,
\label{eq:nucSumRulesii}
 \\
 \int_0^A &dx_N& x_N \sum_i f_i^A(x_N,Q^2) = A\, ,
\label{eq:nucSumRulesiii}
\end{eqnarray}
\end{subequations}
which match those in Eq.~\eqref{eq:sumrules_A}.
However, note the above integrals extend to $A$ since $x_N\in[0,A]$.

While the rescaled nuclear PDFs are formally defined on the interval $x_N\in[0,A]$, 
it is important to identify how the bulk of the PDFs are distributed. 
For a proton with three valence quarks, we generally expect the valence PDFs to be peaked 
in the region of $x_A=x_N\sim 1/3$. 
In a similar manner, for nuclei of mass $A$, which carries $A$ nucleons, 
we generally expect the valence PDFs to be peaked 
in the region of $x_A\sim 1/(3A)$,
or equivalently $x_N\sim 1/3$,
which is independent of $A$ and thus facilitates  a meaningful comparison between different nuclei.
Therefore, even though $x_N$ can in principle span the range $[0,A]$, 
we expect the dominant support of the PDFs to be in the region $x_A\leq 1/A$, or $x_N\leq 1$.
For $x_N$ to exceed unity, 
 one \textit{parton} alone would essentially need to acquire more momentum than an average \textit{nucleon}; this is 
highly unlikely.
Thus, it is common to assume the probability for $x_N>1$  to be small,
and global analyses of nPDFs typically implement the condition that $f_i^A(x_N)=0$ for $x_N>1$. 
This induces only a very small error~\cite{Segarra:2020gtj},
and we discuss the possibility of $x_N>1$ further  in Sec.~\ref{sec:largex}.

Setting $f_i^A(x_N,Q_0^2)=0$ for $x_N\geq 1$ at the initial scale $Q_0$
yields an important computational benefit since
the DGLAP evolution framework consistently leads to
$f_i^A(x_N,Q^2)=0$   
for $x_N \geq 1$
at all larger scales $Q$.
Consequently, the advantage of setting $f_i^A(x_N{\geq}1,Q^2)=0$ is that
the same evolution equations can be used for all nuclei, including nucleons:
\begin{equation}
 \frac{df_i^A(x_N,Q^2)}{d \ln Q^2} = 
\begin{cases}
\quad
\frac{\alpha_s(Q^2)}{2 \pi} \int_{x_N}^1 \frac{dy_N}{y_N}
\ P_{ij}\left(\frac{x_N}{y_N} \right) \ f_j^A(y_N,Q^2)\,  \quad &: 0 < x_N \le 1\ ,
\\[5pt]
 \qquad 0 &: 1 < x_N \le A\, .
\end{cases}
\label{eq:xlimit}
\end{equation}

\begin{figure}[!tb]
    \centering
    \includegraphics[angle=0,width=0.98\textwidth]{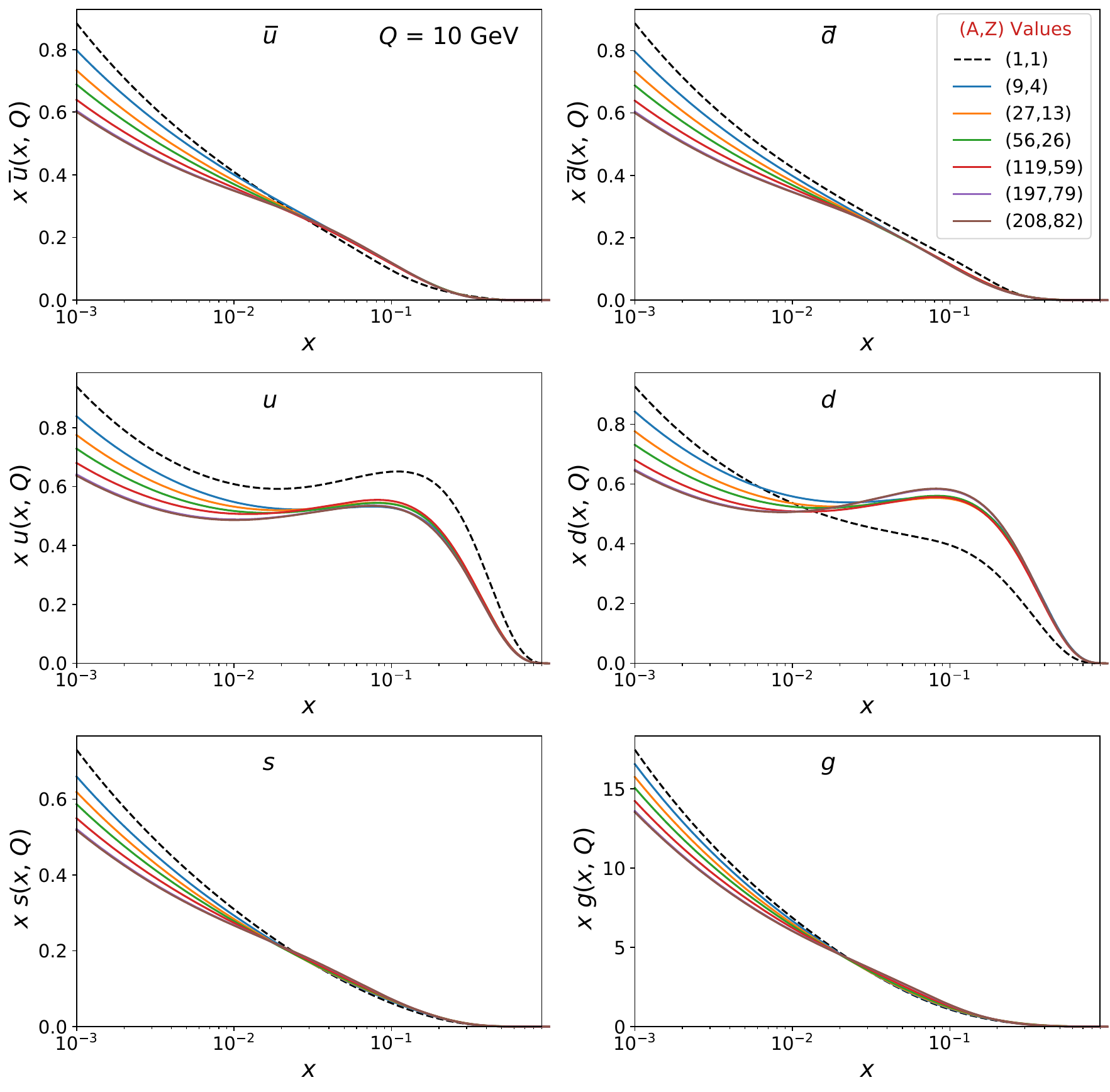} 

    \caption{We display the nCTEQ15 nPDFs~\cite{Kovarik:2015cma} 
    as a function of $x$ at $Q=10$~GeV 
    for selected nuclear $(A,Z)$ values
    as indicated in the legend. 
    These are the scaled nPDFs of Eq.~\eqref{eq:fnuc}, and this feature is most evident 
    when comparing the $xu(x,Q)$ and $xd(x,Q)$  of the proton ($A=1$).
    \label{fig:nucpdf} }
\end{figure}

With the above framework for  nuclear PDFs, we can perform  DGLAP evolution on the interval $x\in[0,1]$
in a manner similar to proton evolution. 
However, the sum rules of Eq.~\eqref{eq:nucSumRulesi}, \eqref{eq:nucSumRulesii}, and~\eqref{eq:nucSumRulesiii}
differ from the proton case. 
Thus, it is common (but not necessary) to further decompose the nuclear PDF $f_i^A$ 
as
\begin{equation}
    f_i^{A}(x,Q) = 
    \frac{Z}{A} f_i^{p/A}(x,Q)
    + \frac{A-Z}{A} f_i^{n/A}(x,Q)
    \, ,
    \label{eq:fnuc}
\end{equation}
where  $f_i^{p/A}$  and $f_i^{n/A}$ represent effective ``bound proton and neutron'' PDFs
for finding a parton ``$i$'' inside a nucleon. 
The ``bound proton'' PDF $f_i^{p/A}$  satisfies identical sum rules as the (free) proton PDF, 
so this quantity can be computed using standard DGLAP programs. 
The ``bound neutron'' PDF is commonly obtained from the ``bound 
proton'' one  by assuming isospin symmetry.\footnote{%
While isospin symmetry is typically used to relate $f_i^{n/A}$ to $f_i^{p/A}$, 
in principle one could add isospin violating contributions. 
The recent Marathon data~\cite{MARATHON:2021vqu} suggests there may be isospin violations in the 
very large $x$ region~\cite{Cocuzza:2021rfn}. 
} %

For comparison of nPDFs across different  nuclei, it is typically the scaled 
nPDF $f_i^{(A,Z)}(x,Q)$ of Eq.~\eqref{eq:fnuc} which can be meaningfully compared,
as we illustrate in Fig.~\ref{fig:nucpdf}. This quantity is defined on $x\in[0,1]$ 
and normalized so that the integrated momentum is unity.
Conversely, although the effective ``bound proton and neutron'' PDFs of Eq.~\eqref{eq:fnuc} are convenient for 
computing the DGLAP evolution, these are {\em not} physically measurable objects.  
Caution is required when comparing them. 
 
In particular, if we try to invert the relations of Eq.~\eqref{eq:fnuc}
for the up and down quarks, 
we find:
\begin{eqnarray}
 u^{p/A} &=& \frac{Z}{2Z-A} \, u^{A} - \frac{A-Z}{2Z-A} \,  d^{A}  
 \qquad {\rm and} \qquad 
 d^{p/A} = \frac{Z}{2Z-A} \, d^{A} - \frac{A-Z}{2Z-A} \, u^{A}  \quad . 
 \label{eq:nucIspin}
\end{eqnarray}
This relation has the expected limits that for a proton: with $A=p$ and $A=Z=1$ we find
$\{ u^{p/A} {=} u^{A=p};\   d^{p/A}{=}d^{A=p}\}$.
Also, for a neutron with $A=n$ and $\{A,Z\}=\{1,0\}$ we find
$\{d^{p/A}{=}u^{A=n};\  u^{p/A}{=}d^{A=n}\}$.
However, for an isoscalar target $(A=2Z)$, Eq.~\eqref{eq:nucIspin} is singular 
because in this limit it is impossible to differentiate the $u^{A}$ and $d^{A}$ distributions 
and the decomposition into 
$u^{p/A}$ and $d^{p/A}$ becomes arbitrary. 
Since the majority of the nuclear data is taken on nuclear targets which lie along the 
nuclear stability line in the $\{A,Z\}$ plane where $A\sim 2Z$ (except for very heavy nuclei), 
our ability to separately determine the nuclear $u^A$ and $d^A$ distributions is limited.
\subsection{Kinematic \texorpdfstring{$\pmb{W}$}{W} cut}\label{sec:wcut}

Determinations of PDFs are performed by global analyses using a wide array of data sets that can be described by the parton model within the factorization framework. 
It is therefore important to exclude data in the kinematic region that are not reliably 
predicted within this model. 

In lepton-nucleon scattering, there are different types of scattering scenarios depending on the value of the hadronic invariant mass, $W^2=(p+q)^2$, and the  virtuality of the exchanged boson $Q$ \cite{SajjadAthar:2020nvy}.  
We can divide these cases into four categories: 
\begin{enumerate}
\itemsep-0em  %
\item elastic scattering ($W=M_N$),
\item shallowly inelastic scattering (SIS) ($W\leq 2$ GeV), 
\item soft DIS ($W>2$ GeV, $Q< 1$ GeV), and
\item DIS ($W>2$ GeV and $Q\geq 1$ GeV). 
\end{enumerate} 
Among these scenarios,  DIS and soft DIS (dominated by non-resonant pion production) 
are the most reliably described by the parton model.
Therefore, a $W$ cut is traditionally imposed on DIS data to extract PDFs.  
While imposing a $W$ cut  is straightforward for the case of lepton-proton scattering, 
it is  more subtle for lepton-nucleus scattering when we compute in terms of the rescaled variables. 
For $\ell_1 +A \rightarrow \ell_2 + X$ scattering, we have:
\begin{equation}
   W_A^2 = (p_A+q)^2 
   = M_A^2+ Q^2\ \frac{1-x_A}{x_A}
   =
   A^2 M_N^2+ Q^2 \ \frac{1-x_N/A}{x_N/A} \quad . 
   \label{eq:wa}
\end{equation}
The average $W_A$ per nucleon, $W_{\rm aver}=W_A/A$, is then given by
\begin{eqnarray}
 W_{\rm aver}^2 &=& 
\left(\frac{W_A}{A}\right)^2 = M_N^2+Q^2 \  \frac{A-x_N}{A^2 x_N}\quad . 
\label{eq:waver}
\end{eqnarray}
Additionally, we observe that $W_{\rm aver}^2$ and $W_N^2$ are not the same:
\begin{eqnarray}
 W_{N}^2 &=& 
(p_N+q)^2 = M_N^2+Q^2 \  \frac{1-x_N}{ x_N}\quad . 
\label{eq:wn}
\end{eqnarray}
For $x_N\in[0,A]$, we find $W_{\rm aver}^2$ is always positive while
$W_N^2$ can be negative for $x_N>1$. On the other hand, $W_N$ is independent of $A$
and can be used to compare  nuclei as long as $x_N$ is restricted to values below unity.

If we impose a lower bound on $W_A$, this translates to the well-known upper bound on $x_A$ of
\begin{equation}
0 \leq x_A \leq   
\left[ 1+ \frac{W_A^2-M_A^2}{Q^2}\right]^{-1}
\leq 1\, .
\end{equation}
In terms of the rescaled variable $x_N = A x_A\ \in[0,A]$
this translates to
\begin{equation}
0 \leq x_N \leq   
A\left[ 1+ \frac{W_A^2-M_A^2}{Q^2}\right]^{-1}
\leq A\, .
\end{equation}

Consequently, when we impose a $W$ kinematic cut for the nuclear case ($A>1$), 
we need to be cognizant of these issues. 
One could try to limit the resonance region with a mass cut ($m_{\rm cut}$) by setting 
$W_N^2 =M_N^2 +m_{\rm cut}^2/A^2$. For $A>1$, the resulting high-$x$ cut turns out to be more relaxed than the usual $W$ cut. Thus, using the traditional $W$ cut here is much safer and hence, despite its pathological nature at $x>1$, this cut can still be used. While cutting on $W_A$ or $(W_A/A)$ is more natural, the value of $W_N$, which is $A$ dependent, needs further investigations. 
\begin{figure*}[!tb]
    \centering
\includegraphics[width=0.65\textwidth]{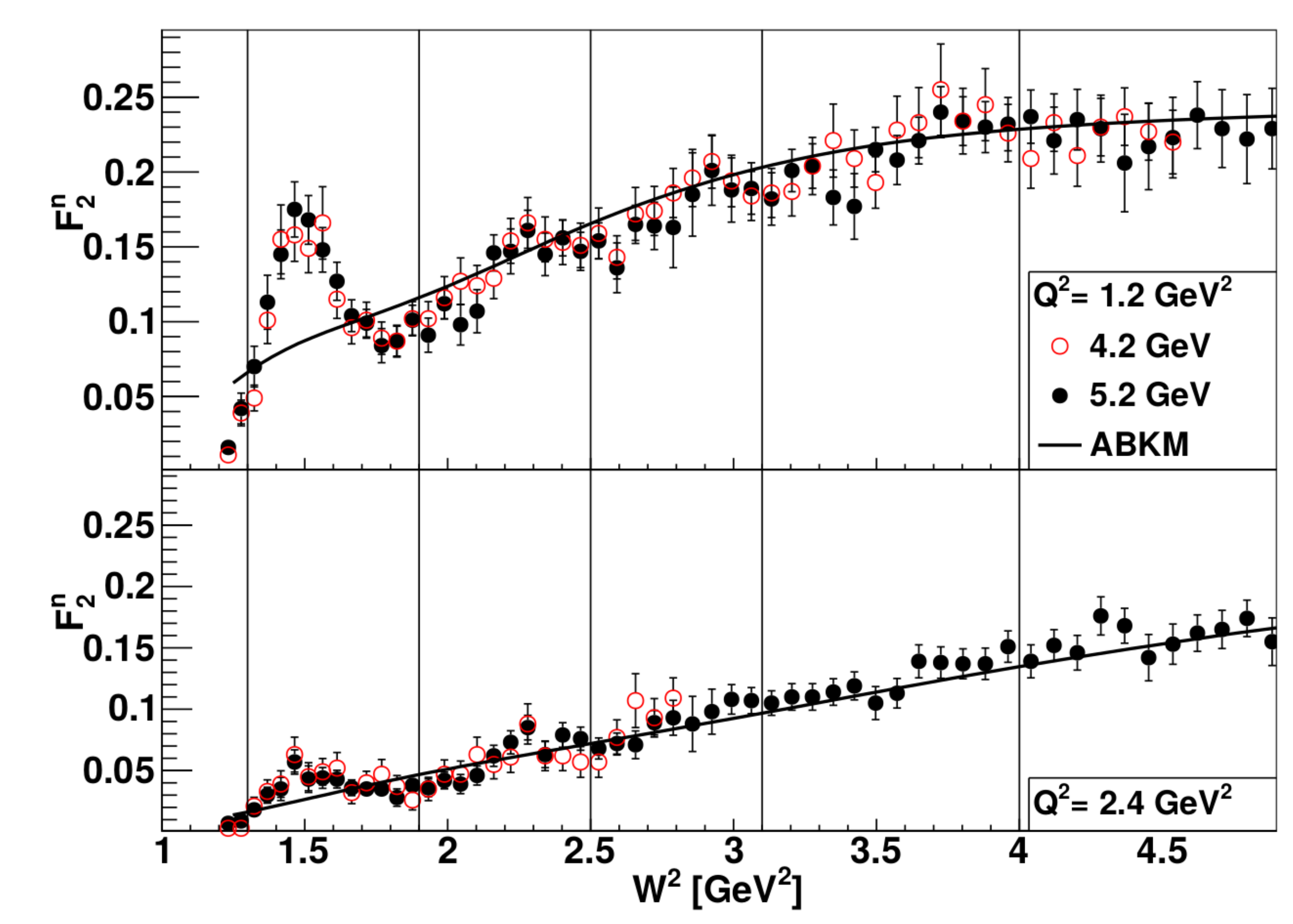}        
\caption{  
We display the neutron structure function $F_2^n$ as measured by the BONuS data~\cite{Niculescu:2015wka} at selected $Q^2$ values.
The open (filled) circles represent data for a beam energy of {$E$ {=} 4.223}
(5.262)~GeV. 
This is compared with ABKM PDF predictions~\cite{Alekhin:2009ni} 
including higher twist effects and target mass corrections.
We observe the resonance structure at low $W^2$ values. 
Figure taken from Ref.~\cite{Niculescu:2015wka}. 
}
\label{fig:bonusW}
\end{figure*}
\begin{figure*}[!tb]
    \centering
\includegraphics[width=0.99\textwidth]{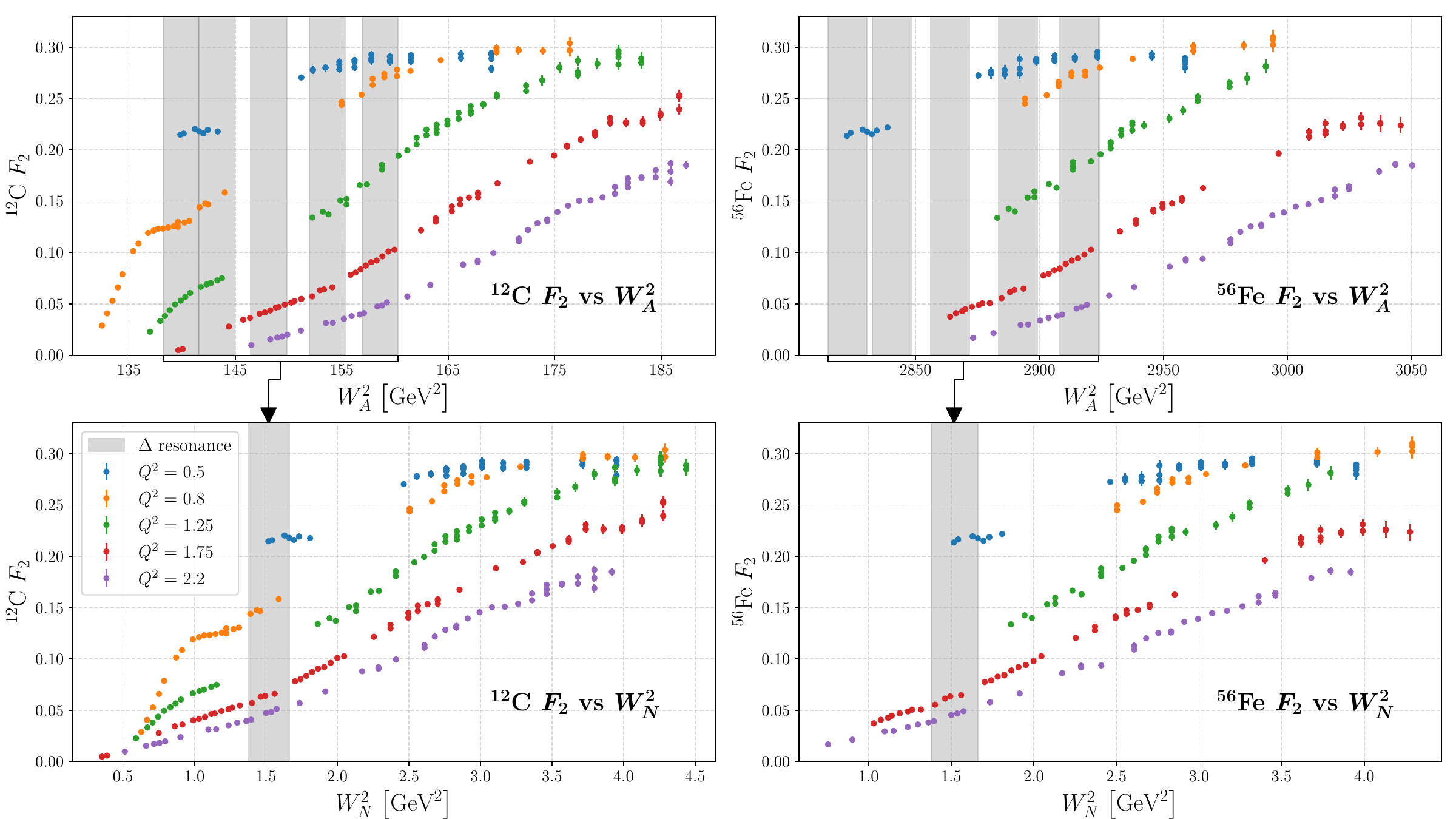}        
\caption{  
We plot $F_2^A$ for carbon and iron for selected $Q^2$ values (in GeV${}^2$) 
as a function of both $W_A^2$ (upper panels) and $W_N^2$ (lower panels). 
The first ($\Delta$) resonance region is highlighted with the vertical shaded bands. 
For $W_N^2$, this region is ${\sim} M_\Delta^2$. 
For $W_A^2$, there is a more pronounced $Q^2$ dependence so that the 
resonance region for the lowest $Q^2$ is at the lowest $W_A^2$, 
and the one for the highest $Q^2$ is at the highest $W_A^2$ values. 
The data are taken from Ref.~\cite{Alsalmi:2019sie}. 
}
\label{fig:jlabWplot}
\end{figure*}

\subsection*{Structure function data for nucleons and nuclei}
In Fig.~\ref{fig:bonusW} we display $F_2^n$ as measured by the BONuS experiment 
and compared with the ABKM global fit~\cite{Alekhin:2009ni}. 
BONuS studied electron-deuteron scattering with two electron energies ($E=4.2$ and $5.2$~GeV) 
and used a novel spectator tagging technique to extract the free neutron cross section.
The resonance structure is  evident in the upper panel with $Q^2=1.2$~GeV${}^2$
where we observe the first ($\Delta$) resonance in the region of $W^2\sim 1.5$~GeV${}^2$.
As we go to larger $Q^2$ values, the effect of the resonances are reduced as seen
in the lower panel ($Q^2=2.4$~GeV${}^2$).

Changing our focus  from a nucleon to a nucleus, in Fig.~\ref{fig:jlabWplot} we display  
the $F_2^A$ structure functions for carbon and iron at selected $Q^2$ values 
as measured in  Ref.~\cite{Alsalmi:2019sie}. 
The region of the first resonance is highlighted by the shaded vertical bands. 
We immediately note that the characteristic resonance structure  is 
more difficult to observe for the nuclear case as compared to the nucleon (neutron) case above. 
Furthermore, this is complicated by the (comparably) limited data in the low $W^2$ region. 

We start by considering the bottom panels of Fig.~\ref{fig:jlabWplot}, which plot $F_2^A$ 
as a function of $W_N^2$. 
Here, we have highlighted the resonance region  $m^2_{\Delta}\sim W_N^2$.
Due to the limited data in the low $W^2$ region, it is difficult to identify the resonances;
however, at larger $Q^2$ values it is clear that any resonance structure again appears to be reduced. 
We now draw attention to the upper panels of  Fig.~\ref{fig:jlabWplot}, which are plotted versus $W_A^2$.
The $W_A^2$-axes of the neutron, carbon and iron all differ as they are now proportional to $A$.
Additionally, there is a relative shift of the curves due to the different $Q^2$ dependence
as compared with $W_N^2$.
Specifically, we have converted using:
$W_A^2 = A W_N^2 + A(A-1) M_N^2 + (A-1)Q^2$.
This conversion also shifts the resonance region in a manner dependent on $Q^2$;
in the upper panels, the resonance region for the lowest $Q^2$ is at the lowest $W_A^2$, 
and the one for the highest $Q^2$ is at the the highest $W_A^2$ values. 

We now summarize our main observations in the following. 
\begin{itemize}

    \item 
    The resonance structure is most evident at lower $Q^2$ values.
As $Q^2$ increases, the resonance peaks are reduced as the phase space for the DIS continuum is 
growing and becomes dominant. 

    \item 
    As we move to heavier nuclei, the resonance structure  is also comparably reduced.

    \item 
    The $W_N^2$ variable has the advantage that this identifies a uniform kinematic region 
    largely independent of $Q^2$ or $A$. This simplifies the task of placing cuts on the resonance region.

    \item 
    The $W_A^2$ variable has the disadvantage that the kinematic region is dependent on $A$ and $Q^2$, 
    and this complicates the task  of placing cuts on the resonance region.

\end{itemize}

\subsection{nPDFs for \texorpdfstring{$x_N\geq 1$}{xN>=1} }\label{sec:largex}

The rescaled PDFs allow us to compare distributions from different nuclei in a meaningful 
manner as illustrated in Fig.~\ref{fig:nucpdf}. Operationally, all current global nuclear PDF analyses 
work on the interval $x_N\in[0,1]$ even though the technically allowed range is $x_N\in[0,A]$.
While details of the $x_N>1$ region are beyond the scope of this study, 
we briefly mention experimental measurements that provide insights into this 
extreme kinematic region.

\subsubsection*{The Fermi region}

\begin{figure}
    \centering
    \includegraphics[angle=0,width=0.85\textwidth]{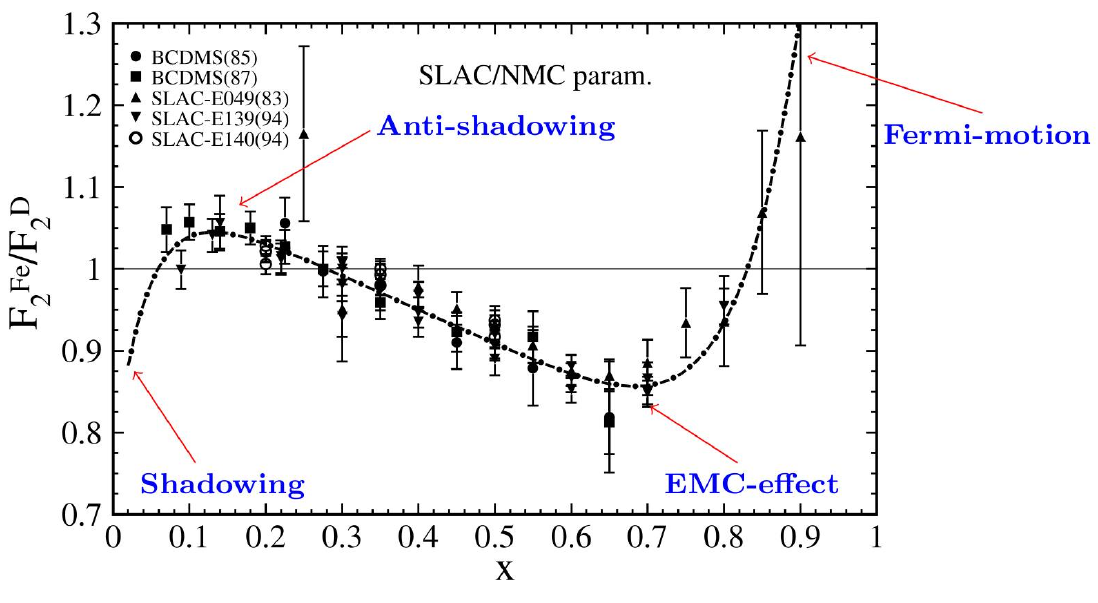}
    \caption{We display a characteristic nuclear correction ratio 
    $F_2^{Fe}/F_2^{D}$ for iron over deuterium. 
    We observe the shadowing region \mbox{($x\lesssim 0.1$)},
    the anti-shadowing region \mbox{($x\sim 0.2$)},
    the EMC region \mbox{($x\sim 0.5$)},
    and the Fermi region \mbox{($x\gtrsim 0.8$)}.
    The figure is taken from Ref.~\cite{Schienbein:2007fs}
    which contains the details and references for   the data sets.
    \label{fig:f2emc}
}
\end{figure}

The effects of nuclear binding can modify the partonic momentum, 
and thus dramatically impact the resulting structure functions~\cite{Kopeliovich:2012kw}.
Fig.~\ref{fig:f2emc} shows the characteristic form of the nuclear correction factor $F_2^{Fe}/F_2^{D}$ 
with\footnote{The nomenclature is historical:
The term ``shadowing'' refers to the front of a large-$A$ nucleus casting a metaphoric shadow over its back since a leptonic probe is likelier to scatter off the front (and not ``see'' the back) when partons have larger wavelengths (smaller $x_N$). This leads to a suppression of the $F_2^A/F_2^D$ ratio at $x_N\lesssim0.1$; for further details of this geometric picture, see Ref.~\cite{Jaffe:1985je,Kopeliovich:2012kw}.
The term ``anti-shadowing'' refers to the experimental observation that the ratio is enhanced at $x_N\sim0.1$.
The term ``EMC effect'' refers to the suppression of the $F_2^A/F_2^D$ ratio at moderate $x_N$ discovered by the EMC collaboration~\cite{EuropeanMuon:1983wih}.
The term ``Fermi region'' refers to the high-$x_N$ region when Fermi motion becomes relevant enabling $x_N \geq 1.0$ for bound nucleons 
}
shadowing at small $x_N$ ($x_N < 0.1)$,
anti-shadowing just beyond the shadowing region ($x_N\sim 0.1)$,
the EMC region for intermediate $x$,
and the Fermi region for large $x$.
The  nuclear binding effects are especially evident at larger momentum fractions (large $x$) 
in the Fermi region where the PDFs are steeply falling. 
There are a variety of theoretical approaches to describe this region, 
as well as experimental measurements of these nuclear binding effects. 
We provide only a brief overview below and refer to the reference for additional details.

\subsubsection*{Challenges of the Fermi region}

The limitations of constraining the PDFs in the $x_N\sim 1$ region  are evident
when  examining the nuclear ratio $F_2^A/F_2^p$
in the large $x_N$ region. 
As $x$ increases, the ${\cal R}=F_2^A/F_2^p$ ratio 
transitions from ${\cal R}\lesssim 1$ in the EMC region,
to ${\cal R}\gtrsim 1$ as $x_N\to 1$ in the Fermi region. 
We find that $F_2^p(x)$ with $A=1$ for the proton must vanish at $x_N=1$, 
whereas $F_2^A(x)$ can be finite since $x_N\in[0,A]$.
This means that 
the denominator of ${\cal R}$ is vanishing while the numerator is finite.
Hence, ${\cal R}$ will rapidly increase at  $x\sim 1$, and is consistent with experimental 
measurements. 
More generally, we expect $F_2^{A_1}>F_2^{A_2}$ for $A_1>A_2$,
and this is the case for $F_2^{{}^{12}C}/F_2^D$ as shown in Fig.~\ref{fig:epsscan}.

\subsubsection*{Theoretical implementations}

\begin{figure}[t]
\centering
\includegraphics[width=0.95\textwidth]{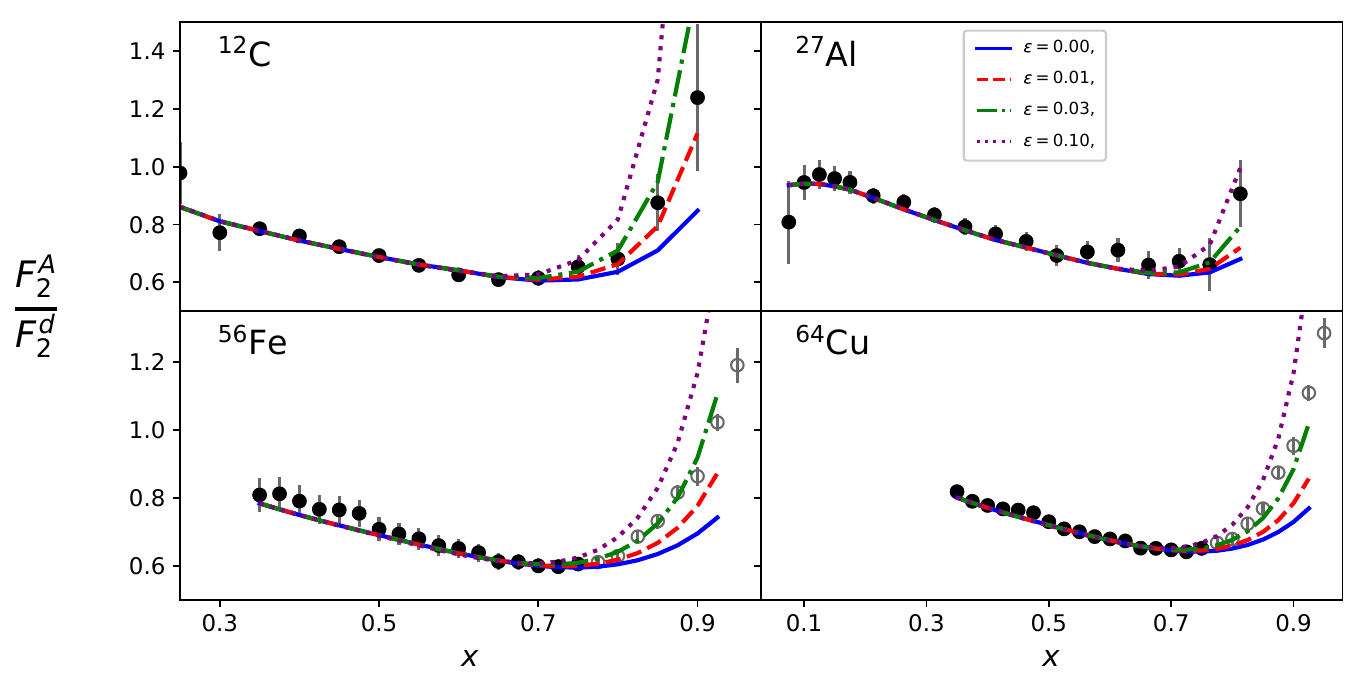}
\caption{We display the ratio $F_2^A /F_2^d$  for selected nuclei data sets in the large~$x$ region. 
The different curves show the impact of the  $\varepsilon$ parameter 
for the  rescaling 
\mbox{$x'_N\to  [x_N - \varepsilon x_N^\kappa \log_{10} A]$} 
with $\kappa= 10$, 
{\it c.f.}, Ref.~\cite{Segarra:2020gtj} for details. 
The data with solid circles satisfy the cuts $Q>1.3$~GeV and $W>1.7$~GeV;
the open circles do not satisfy the cuts and are not included in the fit. 
This figure is taken from Ref.~\cite{Segarra:2020gtj}. 
\label{fig:epsscan}
} %
\end{figure}
On the theoretical side, there are a number of approaches that have been proposed to 
model nPDFs in the $x_N>1$ region, and we point to a selection of references~\cite{Frankfurt:1988nt,Frankfurt:1993sp,Frankfurt:1981mk,Sargsian:2002wc,Rozynek:1988xr,Niculescu:2005rh,Melnitchouk:2005zr,Kondratyuk:1984qj,Sargsian:2007gd,Hirai:2007sx,Freese:2014zda}.
There is also a LO solution  of DGLAP equation for quarks taking into account the full support 
(including $x_N>1$ region)  which is detailed in  Ref.~\cite{Freese:2015ebu}.

One technique to describe the $x_N>1$ region is to 
use a convolution of the  nucleon PDFs $f(x,Q)$ defined on the interval $x\in[0,1]$ 
with a smearing kernel $S_A$ which will shift the PDFs out to larger $x$ values:
\begin{equation}
    f_A(x,Q) = 
    \int_x^A \frac{dy}{y} S_A(y,Q)\, f(x/y,Q) \quad .
\end{equation}
Here, the smearing kernel $S_A$ is typically a Gaussian  with an $A$-dependent width; 
larger nuclei have a larger smearing width, 
and hence a larger proportion of the $f_A$ distribution will populate  the $x>1$ region. 
The behavior of the PDF convolution can also be approximated using 
an $A$-dependent rescaling: \mbox{$x'_N\to  [x_N - \varepsilon x_N^\kappa \log_{10} A]$}, 
{\it c.f.}, Ref.~\cite{Segarra:2020gtj}. 
This   mimics the PDF convolution method described above, 
but the rescaling avoids the convolution integral, so it is computationally fast to evaluate.

In Fig.~\ref{fig:epsscan} we display $F_2^A/F_2^d$ for selected nuclei and data sets,
and indicate various levels of rescaling  (controlled by the $\varepsilon$ parameter).
The data points are described in Ref.~\cite{Segarra:2020gtj}, and the 
solid curve (blue) with $\varepsilon{=}0$ is the default theory with no $x$-rescaling.
The solid points are within the kinematic cuts of the global fit ($Q>1.3$~GeV, $W_N>1.7$~GeV),
and the hollow points are outside these cuts. 
As we increase the $\varepsilon$ parameter, we see this increases the ratio in the large~$x$ region,
and can improve the comparison of the data and theory.

Judging from this  sample of data, 
it appears that a detailed modeling of the Fermi region  can improve the 
description of the data in this extreme kinematic region  (low $W_N$, low $Q$, large $x_N$).
These corrections could be a combination of TMCs, $x_N\,{>}\,1$ effects, 
higher twist, or other non-perturbative corrections. 
Thus, a complete analysis must include and balance all these contributions.

\color{black} %

\subsubsection*{Experimental investigations}

There are a number of studies that have explored the structure functions 
in the Fermi region at very large $x_N$,
and we mention a few examples below. 

For example: Refs.~\cite{Bodek:1980ar,Bodek:1981wr} 
examined SLAC DIS data from a variety of nuclear targets and 
found that in the larger $x_N$ region ($x_N\, {\gtrsim}\, 0.5$)
the Fermi motion effects are similar to those in the deuteron, and increase for heavier nuclei.
Measurements in this region are particularly challenging, and are subject to a variety of
uncertainties as outlined in the Appendix of Ref.~\cite{Bodek:1980ar}.

For the CC neutrino DIS process, CCFR~\cite{CCFR:1999kbf}
measured ${\sim}2000$ events in the region $x_N>0.75$, and used this to study $F_2$ at very large $x_N$.
They fit the high $x_N$ behavior to a decaying exponential $\propto \exp(-s x_N)$ normalized at $x_N=0.65$,
and find a slope of $s=8.3\pm 0.7$. They note this compares favorably with E133 at SLAC~\cite{Rock:1986te}, 
which observed $s\,{\sim}\,[7,8]$.

Hall C at Jefferson Lab measured the NC structure function $F_2$ on a variety of nuclei ranging 
from ${}^2$H and ${}^3$He up to ${}^{197}$Au in the large $x_N$ region~\cite{disProposal2006,Fomin:2010ei} 
out to $x_N\sim 1.4$.  
This work extends an earlier study of Ref.~\cite{Arrington:1995hs},
and related measurements performed at SLAC~\cite{Filippone:1992iz}.
Taking account of the kinematic pre-factors and the $\{h_2, g_2 \}$ contributions, they extract 
$F_2^{(0)}(\xi_N,Q^2)$, where $\xi_N$ is the Nachtmann variable, and fit this to a decaying exponential of the form $\exp(-s \xi_N)$ in the region $\xi_N\geq 0.75$.
The large $\xi_N$ fits finds  the slope  is in the general range $s\sim 15\pm 2$.
The slope has  a mild increase with increasing $Q^2$, 
and a mild decrease with increasing nuclear $A$.
These results also compare favorably with the BCDMS muon scattering data~\cite{BCDMS:1994ala}
which finds $s=16.5\pm 0.6$ for $x_N\,{\sim}\,[0.75,1.05]$.

\subsubsection*{Parton momentum in the Fermi region}

Based on the above large $x_N$ measurements, 
we can ask the question: How much parton momentum is in the region $x_N>1$?

The experimental measurements discussed above parameterize $F_2$ at large $x_N$ as $\exp[-s x_N]$, 
where $s$ can range from  about ${\sim}8.3$  to  ${\sim}14$.
With $s$=8.3, the fraction of $F_2$ in the range $x_N\in[1,\infty]$ 
is less than ${\sim} 6\%$ than that in the range  $x_N\in[0.65,1]$. 
Given that the average momentum fraction in the interval $x_N\in[0.65,1]$
is  $\lesssim 2\%$ of the total nucleon momentum, this gives a rough estimate of 
the potential momentum fraction  beyond $x_N>1$ of $6\%\times 2\% \lesssim 0.1\%$ of the total
momentum fraction. 
For $s\sim 14$, the decrease at large $x_N$ is even steeper. 
Here, the $F_2$ ratio of 
$x_N\in[0.65,1]$ compared to $x_N\in[1,\infty]$ drops to $\lesssim0.1\%$, and 
the potential momentum fraction  beyond $x_N>1$ estimate becomes $1\%\times 2\% \lesssim 0.02\%$ of the total. 
In both of the above cases, the integrated parton momentum is very small in the $x_N>1$ region, but 
the impact on the $F_2^{A_1}/F_2^{A_2}$ ratio still yields 
the characteristic increase in the  $x_N\sim 1$ Fermi region.

\subsection{Threshold problem and higher twist contributions}  \label{sec:ht}

On the basis of the correspondence between the limits $x_N \to 1$
and $W_N^2 \to M_A^2$, the very high-$x_N$ behavior of DIS structure functions 
at fixed $Q^2$ is sensitive to the presence of power-suppressed
corrections beyond leading twist (twist-4 for unpolarized processes).
It has been conjectured that the problematic threshold behavior in
target-mass corrected structure functions in the free nucleon case
are at least partly due to the absence of higher-twist contributions.
As an approximate test of the inclusion of twist-4 contributions, Ref.~\cite{Kulagin:2004ie}
proposed a $1/Q^2$ expansion of the OPE mass-corrected structure functions,
which indeed suppresses the problematic threshold behavior. These calculations
were done for $F_2$, and later, for other tensor structure functions in
Ref.~\cite{Brady:2011uy}.
While this was done for the free nucleon, a similar exercise might be
carried out for nuclear structure functions as another TMC
prescription.
In particular, Kulagin and Petti~\cite{Kulagin:2004ie} showed that by expanding the
target mass corrected structure functions to leading order in $1/Q^2$,
the resulting functions have the correct $x_N\to 1$ limits.
While avoiding the threshold problem, this prescription, however, raises
the question of whether the $1/Q^2$ approximation is sufficiently
accurate for structure functions near $x_N\approx 1$ at moderate $Q^2$.

To test the convergence of the $1/Q^2$ expansion at large $x$,
Ref.~\cite{Brady:2011uy}  further expands the OPE result(s) to include
${\cal O}(1/Q^4)$ corrections.
In fact, one can demonstrate that for a structure
function that behaves at large $x$ as $(1-x)^n$, the target mass
corrected result will vanish in the $x_N\to 1$ limit up to order
$1/Q^{2n-2}$ in the expansion.  For $n \approx 3$, as is typical for
nucleon structure functions, the threshold problem will therefore
appear only at order $1/Q^6$.

\section{The parton model with quark and hadron masses
}\label{sec:acot}

We now turn to a discussion of the parton model, including the effects of quark and hadron mass.
Such mass effects can be rigorously included in the QCD-improved parton model, as shown in a general factorization proof with quark masses by Collins \cite{Collins:1998rz}.
This work extended the factorization proof to the case of massive partons 
and demonstrated that all contributions proportional to the parton mass 
can be fully factorized. 
The underlying scheme is also known in the literature as the Aivazis-Collins-Olness-Tung (ACOT) scheme
\cite{Aivazis:1993kh,Aivazis:1993pi,Aivazis:1990pe}, which is an example of what is called 
a General Mass Variable Flavor Number Scheme (GM-VFNS) in the modern 
literature~\cite{Aivazis:1993pi,Olness:1997yc,Thorne:2000zd,Kniehl:2005mk,Alekhin:2020edf,Alekhin:2009ni}.
In the following subsections we summarize DIS structure functions in the ACOT scheme, including
quark and hadron (nuclear) mass effects, and compare and contrast these expressions to the structure functions in the OPE.
As discussed in the introduction (Sec.~\ref{sec:intro}), we consider the theoretical status of both approaches, the OPE
and the collinear-factorization-based QCD-improved parton model, on equal footing.

In the na\"ive, massless parton model, we have a hadron $A$ of momentum $p_A$ that emits
a collinear parton of momentum $k=x_A \, p_A$. If we try and extend this simple picture
to the case of massive partons, the presence of the parton's mass $m_i$ violates collinear kinematics and we encounter ambiguities proportional to the parton mass. 
We work in the 
ACOT formalism~\cite{Aivazis:1993kh,Aivazis:1993pi,Aivazis:1990pe} 
to compute the structure functions in the helicity basis.
This has the advantage that the polarization vectors are boost invariant between the hadron and parton frame,
and  provides advantages for incorporating both the hadron and parton masses, as we  illustrate below. 
We observed in Sec.~\ref{sec:rescaling} that  hadron masses modify the scaling variable via the relation $\xi_A = R_M x_A$, where the $\xi_A$ is the Nachtmann variable.
Within the context of the OPE and the parton model, the effect of the quark masses factorizes from the hadronic modifications. The rescaling variable is modified by an additional factor $R_{ij}$, which is a function of the quark masses $\{m_i,m_j \}$ and is defined in Table~\ref{tab:PartonMass}. Thus, we have $\bar{\xi}_A =R_{ij} R_M x_A$.
Recall the Nachtmann variable $\xi_A$ includes the TMCs due to $M_A$, 
and the variable  $\bar{\xi}_A$ (with the overline) includes both the TMCs  and the parton mass corrections.

In the following, we also review the structure functions in the light-front formalism,
and compare with the TMC expressions obtained from the OPE. 
There has been extensive discussion in the literature about computations with massive 
partons~\cite{Aivazis:1993pi,Olness:1997yc,Thorne:2000zd,Alekhin:2020edf,Alekhin:2009ni}.
The present discussion  is focused on the detailed 
organization of the massive parton contributions in the formalism outlined by 
Collins~\cite{Collins:1998rz}. 
Thus, we include all terms of order ${\cal O}(m_i/Q)^2$, and the non-factorizable terms are 
suppressed by powers of $(\Lambda_{\rm NP}/Q)^2$.

\subsection{Helicity formulation of the parton model}
\label{sec:aot}

When we generalize the massless QCD parton model to the massive
case, there are potential ambiguities proportional to the parton mass $m_i$, 
which may enter if one is not careful about organizing the perturbative expansion.
This organization is more easily dealt with in the helicity basis\footnote{For clarity, this is the same  technology used to compute helicity amplitudes and can be found in popular textbooks~\cite{Halzen:1984mc,Peskin:1995ev}.} as we outline in this section;
however, helicity and more common tensor formulation 
lead to identical results.

The original ACOT formalism for massive DIS was derived in the helicity
basis, and this provides some advantages in working with both hadron and parton masses. 
For example: in the
helicity formalism, we can define polarization vectors (Sec.~\ref{sec:aot_decomposition}) in
a boost-invariant manner  so that they are the same for both the hadron
and parton reference frames (Sec.~\ref{sec:boost_defs}). This has the advantage that there are
no mixing of terms between the partonic and hadronic structure functions\footnote{ %
The partonic structure functions describe the theoretically calculable 
interaction of the partons with a vector boson. 
} %
in the helicity basis that could otherwise originate from making a boost (helicity eigenstates, e.g., polarization vectors and spinors, are not boost-invariant quantities).
Additionally, by working with light-front momenta, $\{p^\pm, k^\pm\}$, we can take advantage of the fact that
$p^+ \gg p^-$ and $k^+ \gg k^-$, and define our scaling variable $\xi$ as 
 $k^{+}=\xi p^{+}$, which is valid for both massless and massive partons. 
In this Section, we will initially use $\xi$ (un-subscripted) as a generic scaling variable, 
and we will connect this to   $\xi_A$ and $\bar{\xi}_A$ in Sec.~\ref{sec:pm2ope}.

For the present discussion we  focus on the target mass corrections
that enter the hadronic structure functions. We  show that the helicity
formalism matches the leading terms of the OPE master formula.
We compare
these result both analytically in Sec.~\ref{sec:pm2ope} and numerically in Sec.~\ref{sec:pm2ope_pt2} (and Sec.~\ref{sec:num}). 

\subsection{Helicity decomposition}
\label{sec:aot_decomposition}

In analogy to what was done in Sec.~\ref{sec:kin}, one can also decompose the hadronic tensor $\tilde{W}^A_{\mu\nu}$
to obtain the six structure functions $\tilde{F}^A_{mn}$ in a basis of helicity polarizations by projecting out the Lorentz tensors~\cite{Aivazis:1993kh,Aivazis:1993pi,CTEQ:1993hwr}. 
Conventionally, this basis is defined using the polarization vectors $\epsilon_{\mu}(q,\lambda)$  of the intermediate exchange boson $V$, with momentum $q$ and helicity $\lambda$.
Explicitly, the decomposition for nucleus $A$ is given by
\begin{align}
\label{eq:helicity_decomposition}
    \tilde{W}^A_{\mu\nu}(p_A,q)\ &=\ \sum_{m,n} \varepsilon_\mu^*(q,m)\ 
    \tilde{F}^A_{mn}(p_A,q)\ \varepsilon_{\nu}(q,n)\ ,\ 
\end{align}
where polarization vectors according to the ACOT convention are
\begin{align}
   \varepsilon_\mu(q,m=\pm) =
      \frac{1}{\sqrt{2}}\ (0,\mp 1, -i,0)\ ,
\quad
   \varepsilon_\mu(q,m=0) =
      \frac{(-q^2)p_{\mu} + (p\cdot q)q_{\mu}}{\sqrt{(-q^2)[(p\cdot q)^2- q^2 p^2]}}\ ,
      \quad
  \varepsilon_\mu(q,m=s)  = \frac{q_\mu}{\sqrt{Q^2}}\ .
\end{align}
(Note the $\mp$ Condon-Shortley convention for the transverse polarizations.)
Here, the transverse polarization vectors are defined in a collinear frame where $q_z>0$.
The longitudinal polarization vector is defined with respect to a reference vector $p$ (usually taken to be $p_A$).
The scalar polarization vector $\varepsilon_\mu(q,m=s)$, which is sometimes called the ``auxiliary polarization vector''~\cite{Halzen:1984mc,BuarqueFranzosi:2019boy}, ensures 
the completeness relationship 
for off-shell and $t$-channel exchanges of the photon:
\begin{align}
        -g_{\mu\nu}\ =\
        \varepsilon_\mu(q,+)\varepsilon^*_\nu(q,+)\ +\ 
        \varepsilon_\mu(q,-)\varepsilon^*_\nu(q,-)\ -\ 
        \varepsilon_\mu(q,0)\varepsilon^*_\nu(q,0)\ +\
        \varepsilon^\gamma_\mu(q,s)\varepsilon^{\gamma*}_\nu(q,s)\ .
\end{align}

\begin{table}[!t]
\begin{doublespace}
\begin{center}
\begin{tabular}{|c||c|c|c|:c|c|c|}
\hline\hline
 Tensor $ \Leftrightarrow $ Helicity
& $\tilde{F}^A_1$ & $\tilde{F}^A_2$ & $\tilde{F}^A_3$ 
& $\tilde{F}^A_4$ & $\tilde{F}^A_5$ & $\tilde{F}^A_6$
\tabularnewline
\hline
\hline
$\tilde{F}_{+}^A \equiv \tilde{F}_{++}^A$ & $1$ &  & $ -r $ &  &  & 
\tabularnewline
\hline
$\tilde{F}_{-}^A \equiv \tilde{F}_{--}^A$ & $1$ &  & $ +r $ &  &  & 
\tabularnewline
\hline
$\tilde{F}_{0}^A \equiv \tilde{F}_{00}^A$ & $-1$ & $\frac{r^2}{2x_A}$ &  &  &  &  
\tabularnewline
\hline
\hdashline
$\tilde{F}_{S}^A \equiv \tilde{F}_{SS}^A$ & $1$ & $\frac{1}{2x_A}$ & & $2\, \frac{Q^2}{M_A^2}$ & $\frac{-1}{x_A}\, \frac{Q^2}{M_A^2}$ & 
\tabularnewline
\hline
$\tilde{F}_{0S}^A + \tilde{F}_{0S}^A$ & & $\frac{r}{x_A}$ & & & $\frac{-r}{x_A} \,\frac{Q^2}{M^2_A}$ &
\tabularnewline
\hline
$\tilde{F}_{0S}^A - \tilde{F}_{0S}^A$ & & & & & & $\frac{-r}{x_A}\, \frac{Q^2}{M_A^2}$
\tabularnewline
\hline
\hline 
\end{tabular}
\end{center}
\end{doublespace}
\caption{The  transformation between the tensor and helicity basis.
The dotted lines highlight the scalar-based modes which are 
suppressed by the lepton mass.
This table demonstrates there is a one-to-one 
relation between the helicity and tensor 
structure functions.
This table matches the ACOT convention  of Table~III in Ref.~\cite{Aivazis:1993kh} with appropriate conversions:
(i) we replace  $r\to \rho$;
(ii) Ref.~\cite{Aivazis:1993kh} includes an additional factor of $\nicefrac{1}{2}$
in the definitions of $\{W_3, W_5,W_6 \}$, so this table
includes a relative factor of 2 in the  columns for 
$\{ \tilde{F}^{A}_{3}, \tilde{F}^{A}_{5}, \tilde{F}^{A}_{6}\}$.
Additionally, note there is a relative  factor of $2x$ between
the  helicity $F_0$ and longitudinal $F_L$ structure function 
definitions $F_L=2xF_0$ as noted in Eq.\,(\ref{eq:FLF0}).
} %
\label{tab:helicity_expansion}
\end{table}

The indices $m,n$ 
in Eq.~\eqref{eq:helicity_decomposition}
span the helicities plus ($+$), minus ($-$),
longitudinal ($0$), and scalar ($s$). Angular momentum conservation
reduces these to six combination $\{++,--,00,ss,0s,s0\}$, just as
we have six hadronic structure functions in the tensor basis $F_{i}$
with $\{i=1...6\}$. Additionally, the three combinations $\{F_{ss},F_{s0},F_{0s}\}$
are suppressed by the lepton mass in a manner analogous to $\{F_{4},F_{5},F_{6}\}$,
so typically these are ignored. 
We indicate the 
remaining combinations,  $\{F_{++},F_{00},F_{--}\}$, 
by the short-hand notation $F_{\lambda}$
with $\lambda=\{+,0,-\}$.

Using the orthogonality and normalization conditions of the polarization vectors, 
the decomposition of Eq.~\eqref{eq:helicity_decomposition} can be inverted, leading to the  expression
\begin{align}
\label{eq:helicityStrFnDef}
    \tilde{F}^{A}_{mn}(p_A,q)\ =\ 
    \varepsilon^\mu(q,m)\ 
    \tilde{W}^A_{\mu\nu}(p_A,q)\ 
    \varepsilon^{\nu *}(q,n)\ ,
\end{align}
which can be used to construct all six structure functions in the helicity basis. 
Explicit calculation gives
\begin{align}
\label{eq:helicity_str_fns}
    \tilde{F}^A_{mn}(x_A) &= \sum_{i=1}^6\ R_{i}^{A(m,n)}(x_A,Q^2)\ \times\ \tilde{F}_i^A(x_A)\ ,\
\end{align}
where the coefficient functions $R_i^{A(m,n)}(x_A,Q^2)$ are given in Table~\ref{tab:helicity_expansion}.
In the DIS limit, the $R_i^{A(m,n)}$ coefficients  simplify since $\sqrt{1+ Q^2/(q^{0})^2}=\sqrt{1+ (2x_A M_A)/q^{0}}\to1$ in the high energy limit.
Alternative helicity bases
can lead to relative sign differences in Table~\ref{tab:helicity_expansion}, particularly for the scalar polarizations.
For example: in the HELAS basis~\cite{Murayama:1992gi},
Eq.~\eqref{eq:helicityStrFnDef} is augmented by factors of 
$1/(\varepsilon\cdot\varepsilon^*)$ to account for the normalization differences.
Moreover, defining Eq.~\eqref{eq:helicityStrFnDef} using $\tilde{W}_i$, as done in the original ACOT works~\cite{Aivazis:1993kh,Aivazis:1993pi} can also lead to normalization differences.

We obtain the hadron-level cross section by contracting the hadronic and leptonic
tensors:
\begin{eqnarray}
d\sigma & \sim & L^{\mu\nu}(q,k) \ W_{\mu\nu}(p,q) 
\nonumber \\
&\equiv& 
L^{m'n'}(q,k) \ d_{m'}^{m}(\psi)^{-1}\ d_{n'}^{n}(\psi) \ F_{mn}(p,q)\ , \label{eq:sig}
\end{eqnarray}
where we show the corresponding relations in both the tensor and helicity basis.
Here the Wigner rotation matrix
$d_{n'}^{n}(\psi)$
is the $SO(2,1)$ analog of the familiar $3\times3$ ``rotation''
matrix (since the exchanged particle is spin-1 with 3 polarization
states), and $\psi$ is the boost angle that transforms between the
hadronic Breit (``brick wall'') frame, where $p\|q$, and the leptonic
frame where $k \|k'$, with
\begin{equation}
\cosh\psi
=\frac{2p\cdot(k+k')}{\Delta[-Q^{2},p^{2},p_{x}^{2}]}
=\frac{\xi^{2}M^{2}-Q^{2}+2\xi(s-M^{2})}{\xi^{2}M^{2}+Q^{2}} \quad
\xrightarrow[M\to0]{}
\quad\frac{2-y}{y} \quad .
\end{equation}
Here, 
$\Delta^2[a,b,c]=a^2+b^2+c^2-2(ab+bc+ca)$,
and $p_x$ is the the 4-momentum 
of the outgoing proton remnant~($X$), {\it c.f.}, Ref.~\cite{Aivazis:1993pi}.
An elegant feature of Eq.~\eqref{eq:sig} is that 
the leptonic ($L^{mn}$) and hadronic ($F_{mn}$) helicity structure functions are simple and 
all the kinematic complications due to the target mass are
contained in the $d(\psi)$ rotation matrices. This simplicity is
a consequence of the underlying group-theoretic approach to the factorized
structure.\footnote{While the helicity formalism elegantly exhibits the underlying symmetries of the interaction, as we go to higher-order the computation of the  Wigner rotations for multiple intermediate particles can become more complex~\cite{Aivazis:1993pi}.}

\subsection{Boost-invariant polarizations}\label{sec:boost_defs}

If we work in light-front coordinates $\{x^{+},\vec{x},x^{-}\}$
with $x^{\pm}=(x^{0}\pm x^{3})/\sqrt{2}$ and $\vec{x}=\{x^{1},x^{2}\}$,
then 
\begin{eqnarray}
p^{\mu}_A & = & \left\{ p_A^{+},\vec{0},\frac{M_A^{2}}{2p_A^{+}}\right\} 
\qquad \text{and} \qquad 
q^{\mu} =  \left\{ -\xi p_A^{+},\vec{0},\frac{Q^{2}}{2\xi p_A^{+}}\right\}\ ,
\end{eqnarray}
such that $p_A^{2}=M_A^{2}$ and $q^{2}=-Q^{2}$. 
Additionally, $2(p_A{\cdot} q) = Q^2/x_A = Q^2/\xi \,{-} \xi M_A^2$.
We now choose to define
the parton momentum for a massive state as 
\begin{eqnarray}
k^{\mu} & = & \left\{ \xi p_A^{+},\vec{0},\frac{m^{2}}{2\xi p_A^{+}}\right\} \ ,
\end{eqnarray}
such that $k^{2}=m^{2}$ and $k^{+}=\xi p_A^{+}$,
with parton mass $m$.
This scaling relation holds for both massless and massive partons. 

We require
two reference vectors to define the polarizations. 
For the hadron polarizations $\epsilon_{n}^{\nu}(p,q)$ we choose the vectors $\{p,q\}$,
while for the parton polarizations $\epsilon_{n}^{\nu}(k,q)$ we
choose the vectors $\{k,q\}$. For both the hadron and parton polarizations,
the scalar polarization is given by $\epsilon_{0}^{\mu}=q^{\mu}/Q$,
and the transverse polarizations are defined as usual to be $\epsilon^{\pm}=(0,\mp1,-i,0)/\sqrt{2}$.
The longitudinal polarizations are constructed from the reference
vectors using a completeness relation~\cite{Aivazis:1993kh}. The polarization vectors only
depend on the reference vectors $p^{\mu}$ and $k^{\mu}$ to the extent
that they define the $t-z$ plane in conjunction with $q^{\mu}$; see Ref.~\cite{Aivazis:1993kh} for details. 

Thus, the polarization vectors are invariant
between the hadron and parton reference frame. 
Additionally, there is a direct relation between the hadronic helicity structure functions
$F_{\lambda}$ and the partonic helicity structure functions $\omega_{\lambda}$:
\begin{eqnarray}
F_{\lambda} & = & \delta^{\lambda\lambda'}f\otimes\omega_{\lambda'}
\quad .
\end{eqnarray}
Here, $f$ is the PDF, $\otimes$ represents a convolution, $\delta^{\lambda \lambda'}$ is the usual Kronecker $\delta$-function, and $\lambda$ sums over the six helicity configurations. 
Specifically, we note there is no mixing as indicated by the $\delta^{\lambda \lambda'}$ function.
This is in contrast to the
structure functions in the tensor basis: 
\begin{eqnarray}
F_{i} & = & c^{ij}\,f\otimes\omega_{j}
\end{eqnarray}
where the mixing coefficients $c^{ij}$ contain off-diagonal elements, with indices $i,j$ summing over the six tensor structure functions~\cite{Aivazis:1993kh}.

\begin{table}[t]
\begin{doublespace}
\begin{centering}
\begin{tabular}{|c|c|c|c|}
\hline 
 & (OPE) TMC &  ACOT-TMC & $  \frac{\mathrm{ACOT-TMC}}{\mathrm{(OPE)\ TMC}} $
 \tabularnewline
\hline 
\hline 
$F_{1}$ & $\frac{x}{\xi\,r}$ & $1$ & 
$\frac{\xi\,r}{x} \sim 1+\varepsilon^2 $\tabularnewline
\hline 
$F_{2}$ & $\frac{x^{2}}{\xi^{2}\,r^{3}}$ & $\frac{x}{\xi \, r^{2}}$ &
$\frac{\xi\,r}{x}  \sim 1+\varepsilon^2$ \tabularnewline
\hline 
$F_{3}$ & $\frac{x}{\xi\,r^{2}}$ & $\frac{1}{r}$ & 
$\frac{\xi\,r}{x}  \sim 1+\varepsilon^2$\tabularnewline
\hline 
\end{tabular}
\par\end{centering}
\end{doublespace}
\caption{Tabulation of prefactors from the  formulae 
for the  TMC (obtained using the OPE) of Eq.~\eqref{eq:master_full} 
and  {ACOT-TMC} light-front derivations of Eq.~\eqref{eq:master_acot}, where $\varepsilon=(x_AM_A/Q)=(x_NM_N/Q)$, and 
$r=r_A=r_N = \sqrt{1+4 \varepsilon^2}$.
While the $F_i$'s are generally evaluated at $x_N$, 
using Table~\ref{tab:one} we have $x_N/\xi_N = x_A/\xi_A$
and these ratios are equivalent. 
The TMC and ACOT-TMC prefactors 
differ by a uniform factor $\xi r/x$ and 
match to ${\cal O}(\varepsilon^2)$,
and this is plotted in Fig.~\ref{fig:relNormACOT}.
The source of this factor is discussed in the text. 
}
\label{tab:acot}
\end{table}
\subsection{Relationship between TMCs on the light-front and in the OPE} \label{sec:pm2ope}

Finally, coming back to the hadronic level, we can relate the hadronic
helicity structure functions $F_{\lambda}$ to the tensor structure
functions $F_{i}$ using 
Eq.~\eqref{eq:helicity_str_fns}
and obtain:\footnote{%
Note that while the $F_i$ structure functions of Eq.~\eqref{eq:master-rescaled} are 
evaluated at $x_N$, the relations of this section hold at the nucleus and averaged nucleon level; hence, we omit here the 
super/subscripts on $F_i$ and $x_N$. 
}  %
\begin{eqnarray}
F_{1} & = & \frac{1}{2}\left\{ F_{+}+F_{-}\right\} \ ,  \qquad
F_{2} = \frac{x}{r^{2}}\left\{ F_{+}+F_{-}+2F_{0}\right\} \ , \qquad 
F_{3} = \frac{1}{2r}\left\{ -F_{+}+F_{-}\right\} \ ,
\end{eqnarray}
where
$r=r_A=r_N = \sqrt{1+4 \varepsilon^2}$ 
and   $\varepsilon=(xM/Q)$.
The $F_0$ helicity structure function is 
related to the longitudinal structure $F_L$ of Eq.~\eqref{eq:fl}
via
\begin{eqnarray}
F_L  &=& 2x \, F_0 = r^2 F_2 - 2 x F_1 \ .
\label{eq:FLF0}
\end{eqnarray}
Note there is a relative factor of $2x$ connecting 
the helicity $F_0$ and the longitudinal $F_L$ structure functions. 

Translating the above relations into the  formulae for the TMCs in the light-front
approach yields:
\begin{subequations}
\label{eq:master_acot}
\begin{alignat}{3}
F_{1}^{\mathrm{ACOT-TMC}}(x) & =&  & \, F_{1}^{(0)}(\xi)\ , \\
F_{2}^{\mathrm{ACOT-TMC}}(x) & =&  \  \frac{x}{\xi \, r^{2}} & \, F_{2}^{(0)}(\xi)\ ,\\
F_{3}^{\mathrm{ACOT-TMC}}(x) & =&  \frac{1}{r} & \, F_{3}^{(0)}(\xi)\ ,
\end{alignat}
\end{subequations}
and $F_{L}^{\mathrm{ACOT-TMC}}(x)$ can be constructed using Eq.\,\eqref{eq:fl}.
The proof of collinear factorization for heavy quarks is
detailed in Ref.~\cite{Collins:1998rz}.

The correspondence between the  {ACOT-TMC} and {(OPE) TMC} expressions are summarized in Table~\ref{tab:acot}.
In the right-most column, we display the relative conversion factor  ($\xi r/x$), 
which is the same for all three $F_{i}^{\mathrm{ACOT-TMC}}(x)$ results.
Further manipulation of the helicity structure functions in Eq.~\eqref{eq:helicity_str_fns} reveals 
that $F_{4,5,6}$ also follow this pattern.
For $F_{i}$,  we see that the {TMC} and {ACOT-TMC} approach align   to 
${\cal O}(\varepsilon^2)$, 
where $\varepsilon=(xM/Q)$ vanishes outside the dominant TMC region of large $x$ and small $Q$.
Note, these comparisons are for the leading term $F_i^{\mathrm{Leading-TMC}}$ only (see below and Sec.~\ref{sec:num}), as the  higher order
terms in $(x M/Q)^2$ with the $h_i$ and $g_i$ functions are not included in the ACOT expressions.

\begin{figure}[!t]
\centering
\includegraphics[width=0.65\textwidth]{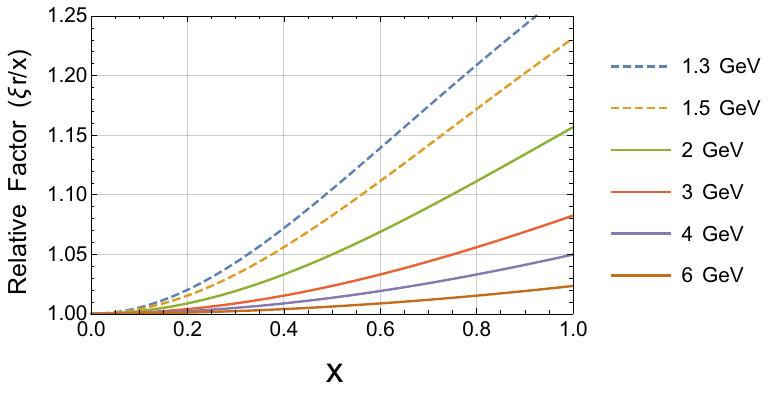}
\caption{We display the relative factor $(\xi r/x)$
between the  ACOT and leading OPE terms
as tabulated in  Table~\ref{tab:acot}  
for $Q=\{1.3,1.5,2,3,4,6\}$~GeV 
as a function of $x$ with $M=M_{proton}$. 
The lower two $Q$ values $\{1.3, 1.5\}$ are dashed to indicate these values 
are  below typical kinematic cuts ($Q_{cut} \sim 2$~GeV) for global PDF fits. 
Recall $(\xi_A r_A /x_A) = (\xi_N r_N /x_N)$.
}
\label{fig:relNormACOT}
\end{figure}
In the above Eq.~\ref{eq:master_acot}, we have used generic (un-subscripted) variables
$\{x, \xi\}$, and this equation is valid for both \textbf{nuclear} variables $\{x_A, \xi_A\}$,
and \textbf{nucleon}  variables $\{x_N, \xi_N\}$, (recall, $r_A=r_N$).
On the LHS of Eq.~\ref{eq:master_acot} we have the Bjorken scaling variable $x$
and on the RHS this is modified by the presence of the target mass 
to be the Nachtmann variable $\xi$.
Note that  the nuclear structure function $F_i^{(0)}$ was derived without knowing 
any details of the partonic structure; 
this is the essence of the factorization between the partonic $(R_{ij})$ and hadronic $(R_M)$
corrections. 
Therefore, 
the correct scaling variable in  Eq.~\ref{eq:master_acot} is $\xi$,
and not   $\bar{\xi}=R_{ij}\, \xi$.

Separately, when we compute  $F_i^{(0)}$ in the parton model, 
we will have individual partonic contributions   $f(\bar{\xi}, m_i,m_j)$ that will depend on the parton masses $\{ m_i, m_j \}$. 
We can incorporate the parton masses
by using the full rescaling variable  $\bar{\xi} =R_{ij}\, \xi =R_{ij} R_M\, x$ (with overline).
This scaling of course cannot be done at the hadron level as the scaling for the 
light quarks $\{u,d,s\}$ will be different from the heavy quarks $\{ c,b,t\}$.

\subsection{Relation of the parton model to the OPE}\label{sec:pm2ope_pt2}

The parton model {ACOT-TMC} results differ from the TMC results obtained with OPE in two respects:
i)~there is an overall factor of $(\xi r/x)$, and 
ii)~the OPE result contains additional  $(x M/Q)^2$ and  $(x M/Q)^4$ terms. 
We will discuss the origin of these differences below. 

\paragraph{Correspondence of the leading factor $(\xi r/x)$:}
The TMC (OPE) results of Eq.\,\eqref{eq:master-rescaled} contain an overall factor of $(\xi r/x)$ 
as compared with the ACOT-TMC results of Eq.\,\eqref{eq:master_acot}.
The source of this factor comes from 
fully including the hadron mass in the calculation as outlined in~\ref{app:opeMassive}
which computes both the massless and massive results. 
The relative factor is most obvious in the case of $F_1$, 
where  the ACOT-TMC result is simply given by $F_1^{(0)}(\xi)$.

In contrast, the massive OPE results of Eq.\,\eqref{eq:appF1Atilde} 
include the additional  $(\xi r/x)$  factor
which arises from the delta function of Eq.\,\eqref{eq:deltaDecom} 
using the identities of Eq.\,\eqref{eq:deltaDecom2}. 
This delta function enforces the relationship between the Bjorken and Nachtmann variables, and is a manifestation of  momentum conservation.
The same factor is present in all the OPE  results for $F_i$ as computed
in~\ref{app:opeMassive}, and is summarized in Table~\ref{tab:acot} for $F_{1,2,3}$.
Additionally, this factor is also present in Ref.~\cite{Ellis:1982wd};
{\it c.f.}, Eq.\,(1.24) where, matching notation,
$(1+\mu \xi^2)=(r\xi/x)$ with $\mu=(M/Q)^2$.

In Fig.~\ref{fig:relNormACOT} we show the size of this  relative factor 
as a function of $x$ for selected $Q$ values. 
For small $x$ and large $Q$ this factor is close to unity, but 
grows for large $x$ and small $Q$ values 
where the TMCs are typically substantial.

\paragraph{Correspondence of the $(M/Q)$ powers:}
The OPE master equation, Eq.\,\eqref{eq:master_full}, includes higher powers of $(xM^2/Q^2)$, i.e., the $\tilde{h}_i^A$ and $\tilde{g}_2^A$ terms, 
which are formally twist-2 (leading twist). The source of these is the intrinsic transverse momentum of  
partons $(k_T)$ and manifests through the conservation of momentum~\cite{Ellis:1982cd,Ellis:1982wd}. 
For such configurations, partonic cross sections are evaluated with 
(initial) parton momenta being on-shell
but not collinear with 
the parent hadron. 
Details are provided in Refs.~\cite{Ellis:1982cd,Ellis:1982wd},
which include the transverse momentum in the description of the partons, 
[{\it c.f.}, $f(x,k_T^2)$ in Eq.\,(1.22) of Ref.~\cite{Ellis:1982wd}].
As the transverse momentum  is of order $M$, $k_T$-dependent contributions generate
the $(M^2/Q^2)$ and $(M^4/Q^4)$ terms of  Eq.\,\eqref{eq:master_full}.
Thus, if the transverse momentum is accounted for in the parton model, one matches the OPE result~\cite{Ellis:1982cd,Ellis:1982wd}. 

\paragraph{Recap:}

We have derived the structure function results for the parton model using the ACOT formalism 
in a helicity basis as summarized in Eq.\,\eqref{eq:master_acot}.
We can obtain a complete correspondence between the parton model and OPE structure function results if
i)~we fully account for the hadron mass in the delta function (c.f., Eq.\,\eqref{eq:deltaDecom}), which  ensures the relationship between the Bjorken and Nachtmann variables,
and 
ii)~we account for the transverse momentum of the parton (c.f., Ref.~\cite{Ellis:1982wd}).

\section{Numerical results} \label{sec:num}

Having discussed various theoretical and phenomenological aspects
of TMCs for nuclei from an analytic perspective, we now turn our focus to the quantitative aspect of TMCs.
In this section, we explore the numerical impact of TMCs on  first the 
proton structure functions (Sec.~\ref{sec:numbers_proton}), and then the 
nuclear structure functions (Sec.~\ref{sec:numbers_nuclei}).
We will examine both the leading-TMC and sub-leading-TMC contributions, and explore the variation across the nuclear $A$ range.
These observations will allow us to design a simple parameterization of the TMCs which are computationally efficient 
 (Sec.~\ref{sec:parm}).
Finally, we will compute the impact of the TMCs for 
DIS reduced cross sections with sample kinematics from HERA, JLab, EIC, and LBNF (Sec.~\ref{sec:numbers_xsec}).

In order to build intuition for  numerical results throughout this section, it is useful to recall several relations between structure functions and quark and antiquark PDFs, but which hold only at LO in the absence of CKM mixing and masses. In particular, for neutrino and antineutrino DIS on an arbitrary nuclear target $(A)$, where $A$ may also be just a proton, the charged current structure functions $(F^{W^\pm}_i)$ and PDFs are related by:
\begin{subequations}
\label{eq:nuDIS_strFn_Identities}
\begin{alignat}{7}
 F_1^{\nu A} &=& (d+s + \bar{u}+ \bar{c}), \qquad &  F_1^{\bar{\nu} A} &=& \quad (u+c + \bar{d}+ \bar{s}),& 
 \label{F1nu}\\
F_2^{\nu A} &=& 2x\left( d+s + \bar{u}+ \bar{c}\right), \qquad & F_2^{\bar{\nu} A} &=& 2x\left( u+c + \bar{d}+ \bar{s}\right) &, 
\label{F2nu}\\
F_3^{\nu A} &=& +2\left( d+s -\bar{u} -\bar{c}\right), \qquad & F_3^{\bar{\nu} A} &=& {-}2\left( u+c -\bar{d} - \bar{s}\right) & \ ,
\label{F3nu}
\end{alignat}
\end{subequations}
in the limit of four quarks. 
Here $\{u, d, ...\}$ are the PDFs of a full nucleus $A$.
Likewise, when the exchange of $Z$ bosons can be neglected in charged lepton DIS, the neutral current structure functions $(F^{\gamma}_i)$ are related to PDFs at LO by
\begin{equation}
    F_2^{l^\pm A} = x\frac{1}{9} \left[4(u+\bar{u})+ (d+\bar{d}) + 4(c+\bar{c})+ (s+\bar{s}) \right].
    \label{eq:strFnPDFs_F2NC}
\end{equation}
While the LO relations shown above are intuitively useful, 
our calculations are performed at full NLO in QCD including the quark mass contributions;
specifically, we use the S-ACOT($\chi$) scheme~\cite{Kramer:2000hn}.

Since we will compare separate components of the TMC contributions, to reduce ambiguity  we introduce the nomenclature 
which we will use throughout our presentation.
Schematically, the OPE TMC terms are related as  follows:
{
\def\dobig#1{\scaleto{\color{blue}\bf #1 }{8pt}}
\def\dobigg#1#2{\scaleto{\color{blue}\bf  #2 }{#1pt}}
\begin{equation}
  \overbrace{F_{2}^{A,{\rm  TMC}}(x_{N},Q^{2})}^{\dobig{TMC}}
  =
  \underbrace{\left(\frac{x_{N}^{2}}{\xi_{N}^{2}r_{N}^{3}} \right)
    \overbrace{F_{2}^{A,(0)}(\xi_{N},Q^{2})}^{
    {\dobig{No-TMC}  \atop \dobigg{8}{  with\  \xi\to x   }
    }}}_{\dobigg{10}{Leading-TMC} }
  +  
  \underbrace{\left(\frac{6M_{N}^{2}x_{N}^{3}}{Q^{2}r_{N}^{4}}\right)\,h_{2}^{A}(\xi_{N},Q^{2})}_{\dobig{h-term}}
  +
  \underbrace{\left(\frac{12M_{N}^{4}x_{N}^{4}}{Q^{4}r_{N}^{5}}\right)\,g_{2}^{A}(\xi_{N},Q^{2})}_{\dobigg{10}{g-term}}
  \label{eq:Fnotation}
\end{equation}
}
We provide additional details below. 

\begin{itemize}

    \item[\textbf{TMC:}] We identify the full set of contributions to the structure function,  given by the rescaled OPE equation of Eq.~\eqref{eq:master-rescaled}, as the ``TMC'' result;
    this label is without additional qualifiers. 
    
    \item[\textbf{No-TMC:}] 
    As we take the   $(x M^2/Q^2)\to 0$ limit, 
    the pre-factors of $F_i^{A,(0)}$ become unity, and the higher-order 
    $(x M/Q)^2$  and $(x M/Q)^4$ terms containing the $\{h_i, g_i\}$ functions vanish. 
    Additionally, we have $\xi\to x$ so that 
    the expressions in Eq.~\eqref{eq:master-rescaled} reduces to $F_i^{A,(0)}(x_N,Q^2)$ alone. We refer to this as the ``No\nobreakdash-TMC'' result.\footnote{%
Note, we always retain the full quark mass dependence in all calculations
    as this is factorized from the hadron-level kinematics. {\it C.f.},  
    Ref.~\cite{Schienbein:2007gr} for details.
    For this reason, we \textbf{do not} 
    refer to the \textbf{No-TMC} term as a  ``massless" result. }  

    \item[\textbf{Leading-TMC:}] 
    The ``Leading-TMC'' structure functions are obtained from Eq.~\eqref{eq:master-rescaled} 
    by only keeping the first term on the RHS which is proportional  to $(x M/Q)^0$.
    Specifically, we are neglecting  the terms proportional to 
    $(x M/Q)^2$ and $(x M/Q)^4$ which contain the $\{h_i, g_i\}$ functions.
    
    \item[\textbf{ACOT-TMC:}] 
    We obtain the ``ACOT-TMC'' structure functions from the ACOT TMC equation
    of Eq.\,\eqref{eq:master_acot}. This result is similar to the ``Leading-TMC,''
    but the pre-factors differ by   $(\xi r/x)$ as detailed in Table~\ref{tab:acot}.
    Importantly, the ``ACOT-TMC''  result does not include  the higher-order 
    $(x M/Q)^2$  and $(x M/Q)^4$ terms containing the $\{h_i, g_i\}$ functions.

    \item[\textbf{h-terms \& g-terms:}] The ``$h$'' and ``$g$'' terms are those terms in the full (OPE) TMC result that are proportional to the 
    $h_i$ and $g_i$ functions.
    We observe that 
    the ``h'' contributions are proportional to $(M_N^2/Q^2)$, 
    while 
    the ``g'' contributions are proportional to $(M_N^4/Q^4)$.

\end{itemize}
\subsection{Proton structure functions with TMCs}\label{sec:numbers_proton}

In comparison to structure functions for massless protons,
structure functions with TMCs for nuclei contain two additional layers of complexity.
The first, of course, is the larger nucleon content; the second are the TMCs themselves.
Therefore, in order to establish a baseline intuition of TMCs for nuclear structure functions, we consider briefly TMCs for proton structure functions. For more comprehensive studies of TMCs for protons, see Ref.~\cite{Schienbein:2007gr} and references therein.

\begin{figure}[!t]
    \centering
    \subfigure[\label{fig:ntmc_AbsStr_vs_x_proton_a}]{\includegraphics[width=0.48\textwidth]{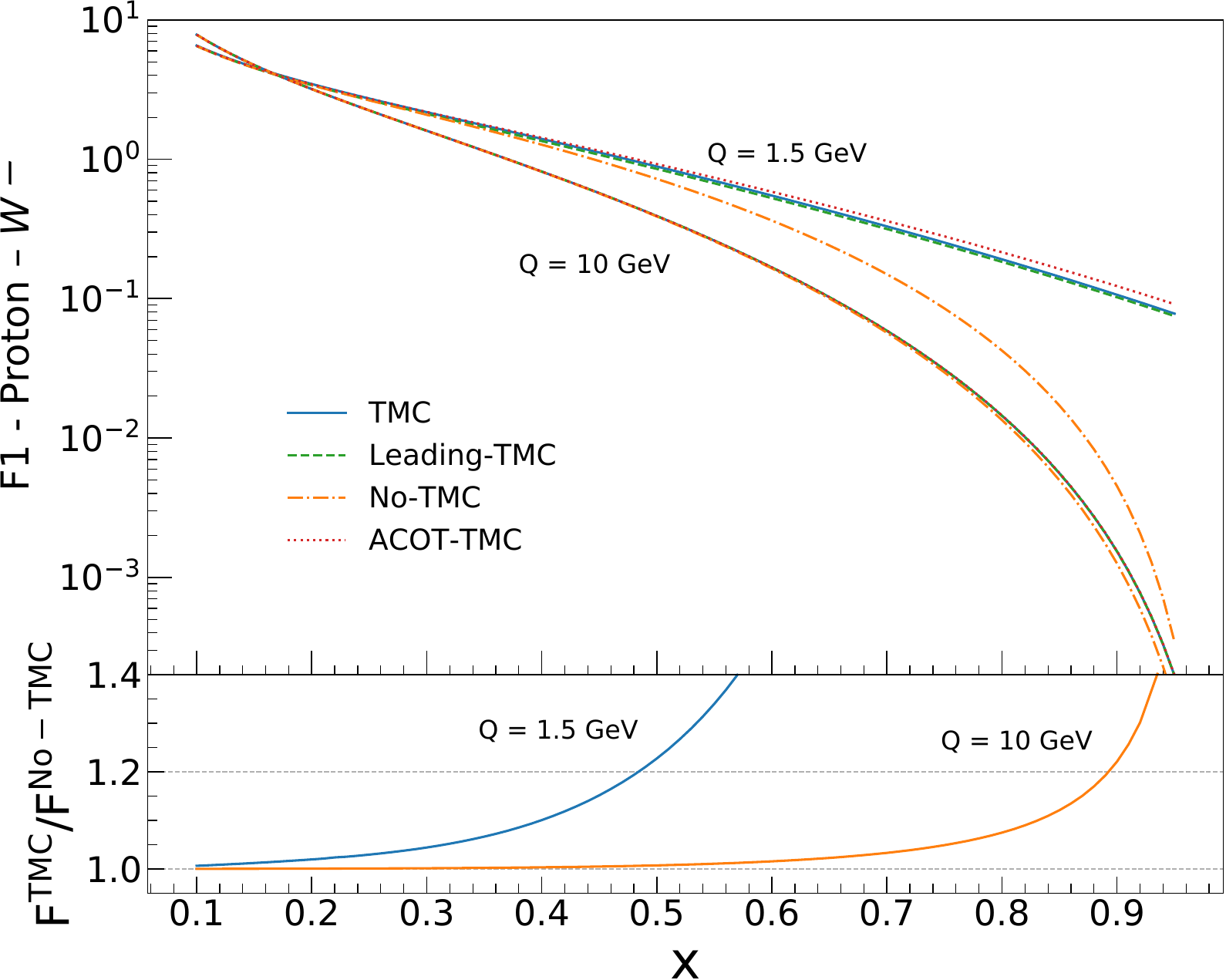}}   
    \hfill
    \subfigure[\label{fig:ntmc_AbsStr_vs_x_proton_b}]{\includegraphics[width=0.48\textwidth]{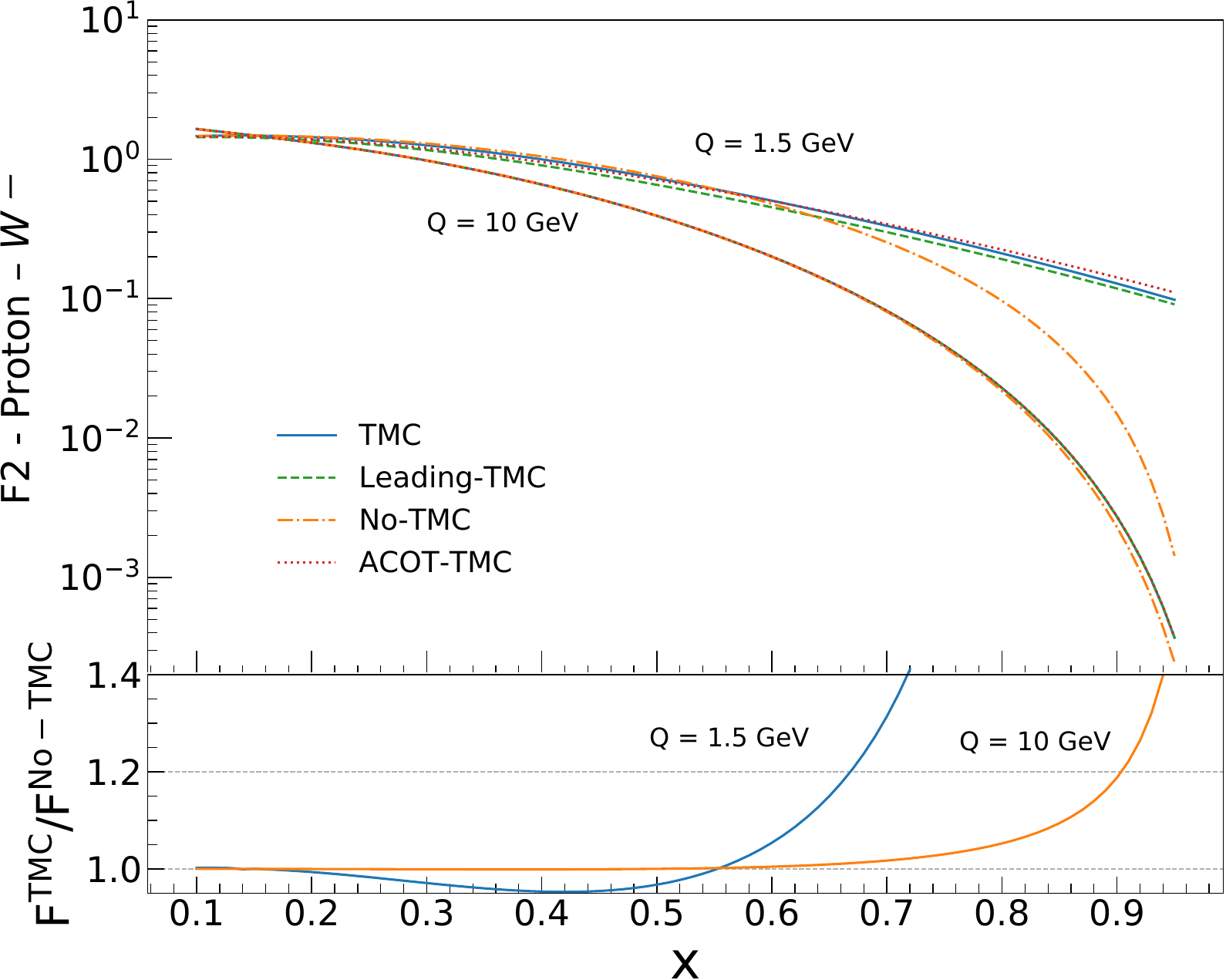}}  
    \\
    \subfigure[\label{fig:ntmc_AbsStr_vs_x_proton_c}]{\includegraphics[width=0.48\textwidth]{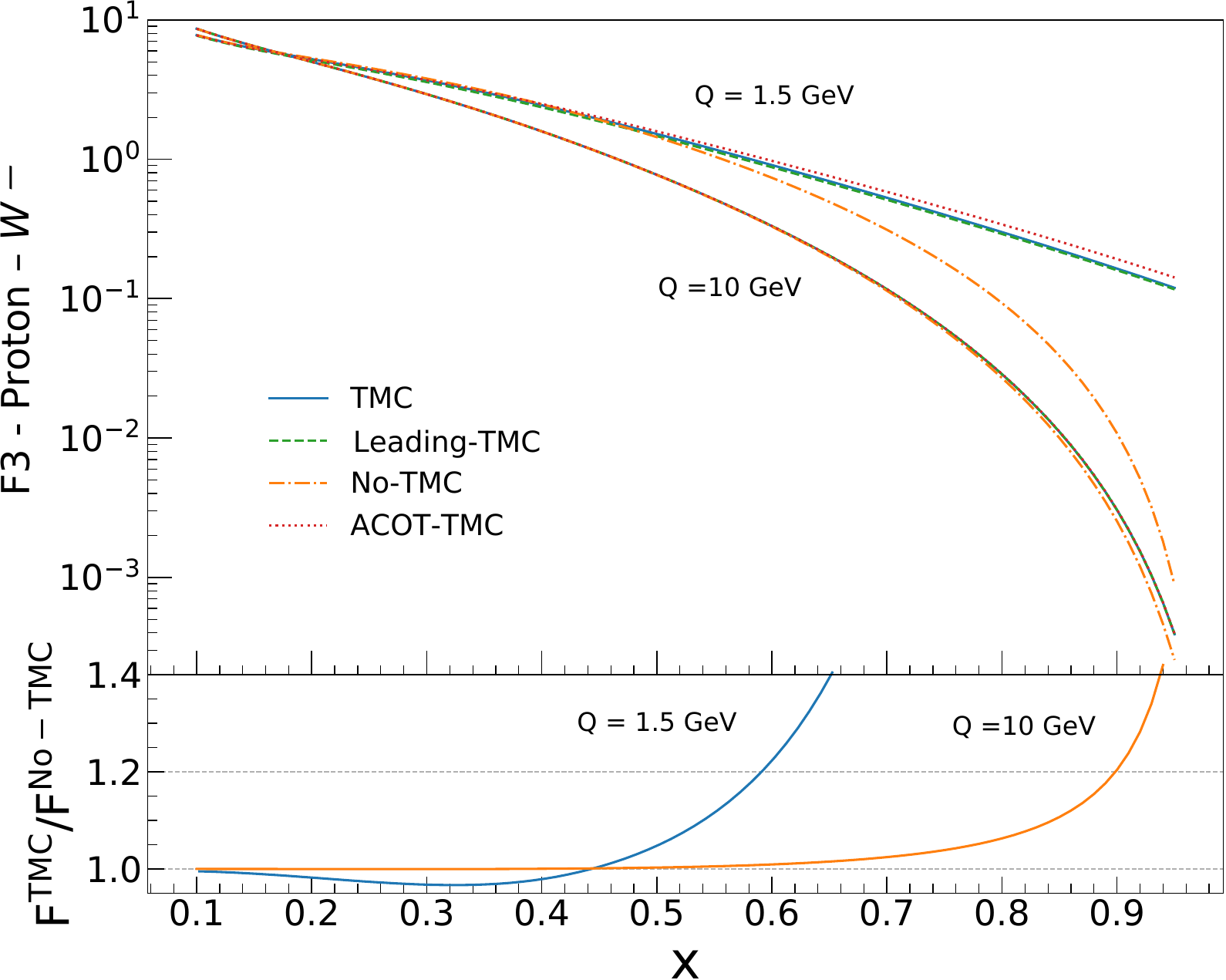}}   
    \hfill 
    \subfigure[\label{fig:ntmc_AbsStr_vs_x_proton_d}]{\includegraphics[width=0.48\textwidth]{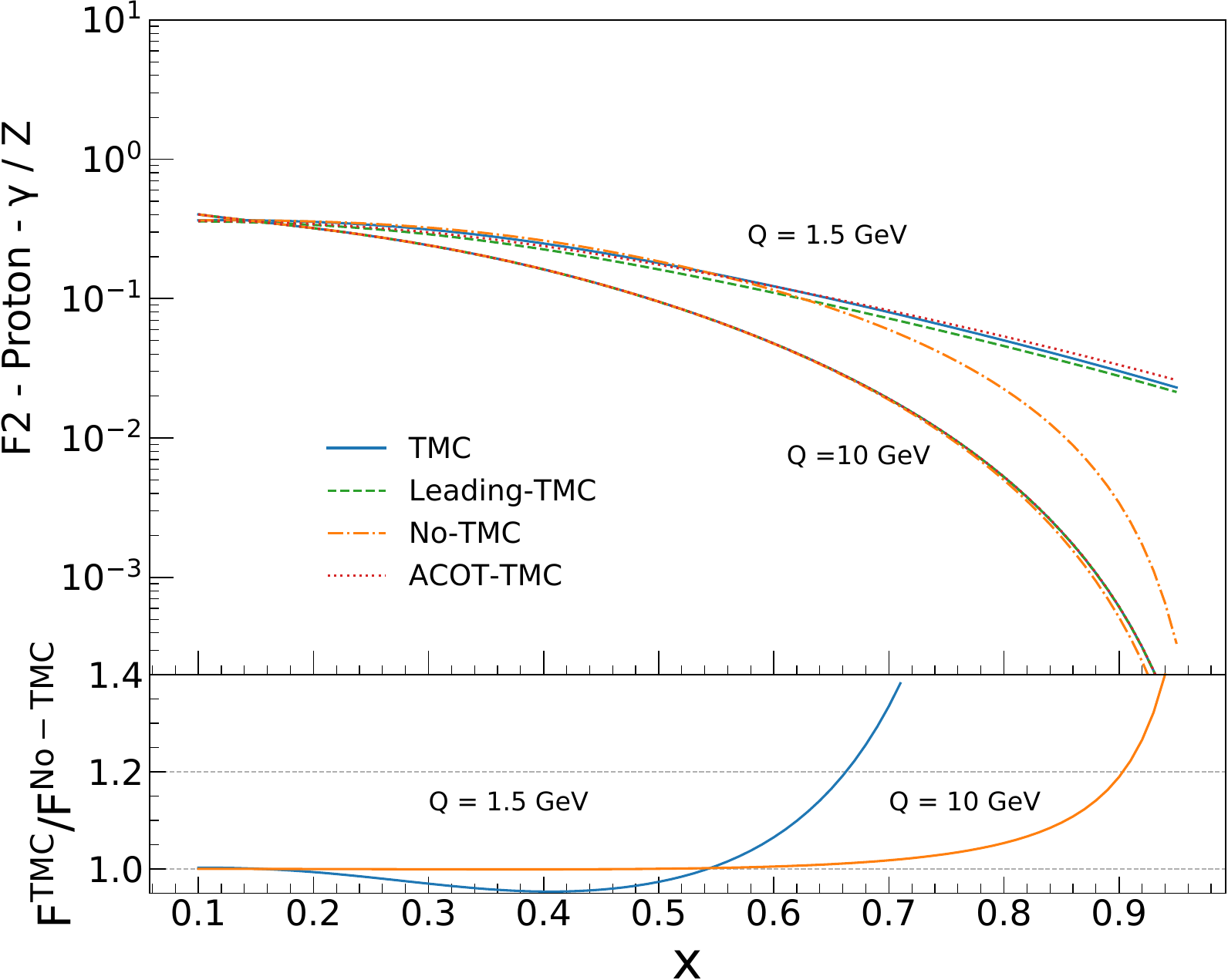}}
        \caption{
        {\bf Upper panels:} Proton structure functions 
        for charged current $W^-$ (a)~$F_1$, (b)~$F_2$, and (c)~$F_3$  
        and (d)~neutral current $\gamma/Z$  $F_2$ 
        as a function of momentum fraction $x_N$.
        We display results for the 
        full ``TMC'' (solid blue), 
        ``Leading-TMC'' (dashed green),  
        \mbox{``No-TMC''} (dot-dash orange),
        and
        ``ACOT-TMC'' (dotted red) 
        at $Q=1.5\GeV$ (upper lines) and $10\GeV$ (lower lines).
        {\bf Lower panels:} Ratio of $F_i$ with full ``TMC'' to ``No-TMC'' at $Q=1.5\GeV$ and $10\GeV$.
}
    \label{fig:ntmc_AbsStr_vs_x_proton}
\end{figure}

We will show results for a range of $Q$ values. 
For a typical global analysis, 
the PDF evolution uses an initial scale $Q_0$
in the range of $1.3$ to $1.5$~GeV, and a typical 
kinematic cut is $Q\gtrsim  2$~GeV.
For example, in Fig.~\ref{fig:ntmc_AbsStr_vs_x_proton} 
we display $Q=1.5$~GeV, which is near the initial evolution scale
but below the typical kinematic cut.
We also display  $Q=10$~GeV, which is above the kinematic $Q$ cut
(the $W$ cut is more complex).

We begin with Fig.~\ref{fig:ntmc_AbsStr_vs_x_proton}, where in the upper panel we plot 
the absolute structure functions 
$F_i$ for charged current $(W^-)$  and neutral current $(\gamma,Z)$ exchange 
showing 
full ``TMC'' (solid), 
``Leading-TMC'' (dash), 
``No-TMC'' (dash-dot), 
and ``ACOT-TMC'' (dotted),
at $Q=1.5\GeV$ and $10\GeV$ as a function of the Bjorken scaling variable $x$.
In the lower panel we show the ratio of the full ``TMC'' to the ``No-TMC'' results. 
\paragraph{$F_i$ structure functions: 
\rm (Fig.~\ref{fig:ntmc_AbsStr_vs_x_proton} Upper Panels)} %
We focus first on the upper panels of Fig.~\ref{fig:ntmc_AbsStr_vs_x_proton} which shows 
the proton structure functions vs.\  $x.$
All the structure functions are steeply decreasing for large $x$. 
This  reflects the underlying  PDF structure  of 
Eqs.~\eqref{eq:nuDIS_strFn_Identities} and \eqref{eq:strFnPDFs_F2NC}:
PDFs, and hence structure functions,
must decrease at large $x$ due to momentum conservation. 

For both $Q=1.5$~GeV and $Q=10$~GeV, we observe that 
the full ``TMC,''  ``Leading-TMC'' and ``ACOT-TMC'' 
results roughly coincide.
The good agreement between the full ``TMC'' and   ``Leading-TMC'' 
allows us to infer that the sub-leading contributions containing the 
$\{h_i, g_i\}$ terms are small; this is to be expected as they are 
suppressed by powers of $(M/Q)^2$ and  $(M/Q)^4$. 
We will further examine the impact of these sub-leading terms 
in Sec.~\ref{sec:numbers_nuclei_subleading}.
The approximate agreement of  the   ``Leading-TMC'' and the ``ACOT-TMC'' 
reflects the impact of the relative factor of $(\xi r/x)$ discussed in Sec.~\ref{sec:pm2ope}.

In contrast, at $Q=1.5$~GeV, the ``No-TMC'' results are dramatically different 
from the other three curves. 
The ``No-TMC'' expression is most similar to the ``Leading-TMC'' 
as illustrated below for the case of $F_1$:
\begin{subequations} \label{eq:nTMC_F_expansion}
\begin{alignat}{7}
 &F_1^{\rm TMC-leading}(x_N) \  &=& \ \frac{x_N}{\xi_N r_N} &&  F_1^{(0)}(\xi_N) \ ,
 \\
 &F_1^{\rm No-TMC}(x_N) &=& \null   &&   F_1^{(0)}(x_N)  \quad ,
\end{alignat}\end{subequations}
where $r_N=\sqrt{1+4\varepsilon^2}$. 
There are two differences between these expressions:
i)~for  ``No-TMC''  the pre-factor  is set to unity, and 
ii)~the argument of $F_i^{(0)}$ is set to $x$ instead of $\xi$.
Given that $\xi= 2x/(1+r) \simeq x(1-x^2 M^2/Q^2)$, 
the expression for $F_1^{\rm No-TMC}$ is neglecting the $(M/Q)$ mass effects.

\begin{figure*}[!t]
    \centering
    \includegraphics[width=\textwidth]{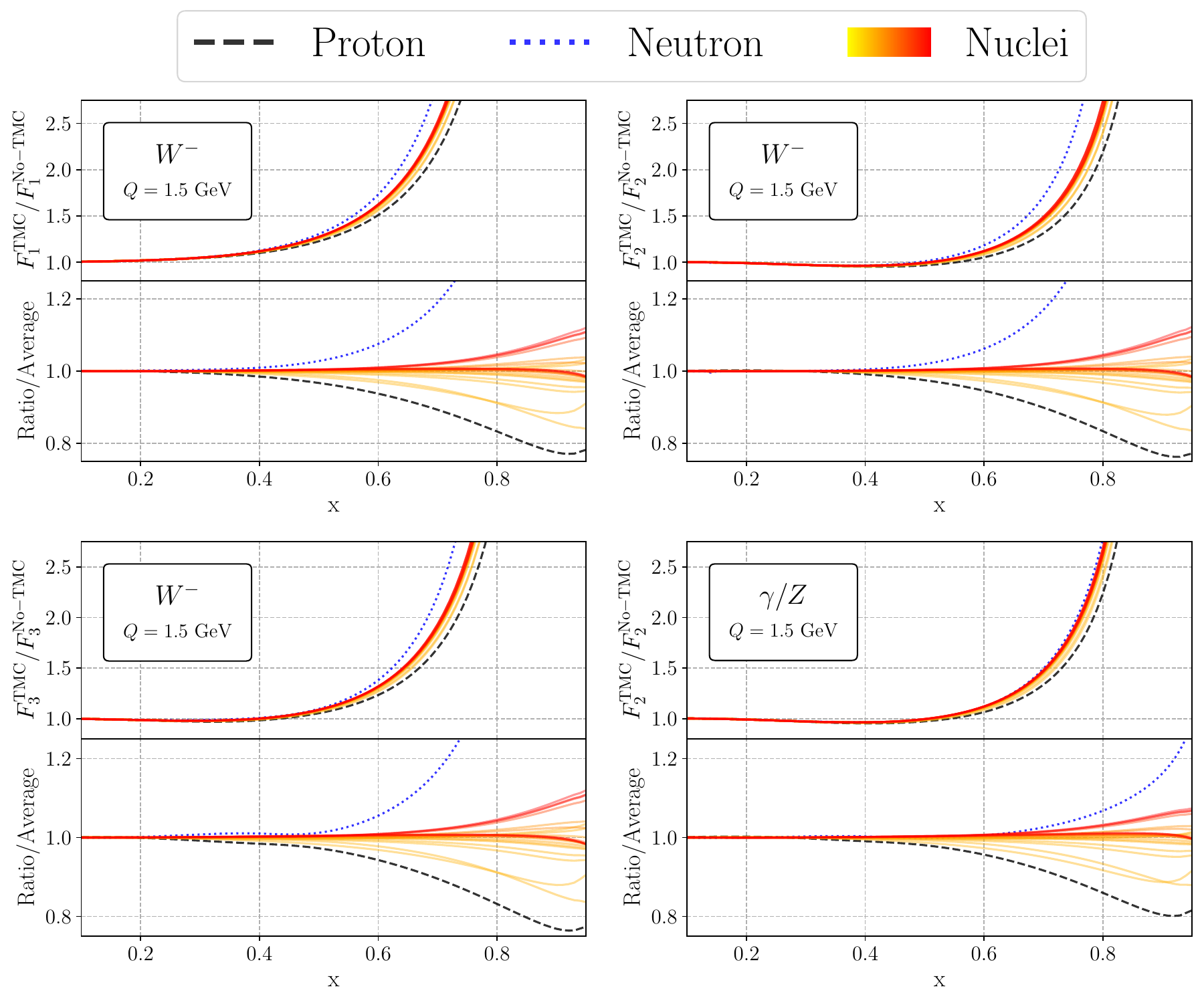}
        \caption{  
        Ratio of nuclear structure functions $F_i$ for the 
        full ``TMC'' over ``No-TMC'' vs.\  $x_N$  at $Q=1.5\GeV$.
        We display ratios for charged current $W^-$ (a)~$F_1$, (b)~$F_2$, and (c)~$F_3$  
        and (d)~neutral current $\gamma/Z$  $F_2$ 
        for the nuclei listed in Table~\ref{tab:isotopeList}.
The proton is indicated with the black dashed line, and the neutron with the blue dotted line. 
The narrow band of colored lines shows the various nuclear results.
}
    \label{fig:ntmc_StrFn_TMC_vs_x_MultiA}
\end{figure*}

We can  separately explore the impact of these two components and determine 
which is dominant. 
For $F_1$ the relative factor $x/(\xi r)$ is precisely the inverse of what is displayed in 
Fig.~\ref{fig:relNormACOT}. 
Thus, we observe that for $Q=1.5$~GeV and $x=0.9$, this factor yields a shift of ${\sim}20\%$.
Comparing with Fig.~\ref{fig:ntmc_StrFn_TMC_vs_x_MultiA}, we see this is clearly not sufficient 
to explain the observed large difference between the  ``No-TMC'' and  ``Leading-TMC'' results. 
While $F_2$ and $F_3$ have slightly different prefactors, 
we obtain the same conclusion that these prefactors do not generate 
the sizable differences observed between the ``No-TMC'' and the other results.

Thus, the dominant variation 
of $F_i^{\rm No-TMC}$ and $F_i^{\rm TMC-Leading}$  must be due to 
the different arguments: \mbox{$x$ vs.\  $\xi$.}
This correspondence of  $x$ and $\xi$ is displayed in Fig.~\ref{fig:xi} where we observe 
that  $x\sim \xi$ for small $x$ and large $Q$, 
but for large $x$ and small $Q$ we find $\xi<x$. 
Because the PDFs are steeply falling at large $x$ 
 (where $x$ and $\xi$ have the largest deviation),
it is the difference of $\xi$ vs.\  $x$ in the arguments on the RHS of Eq.~\eqref{eq:nTMC_F_expansion}
that is driving the variation observed in Fig.~\ref{fig:ntmc_StrFn_TMC_vs_x_MultiA}.
While we have detailed the comparison of $F_1$ in 
Fig.~\ref{fig:ntmc_AbsStr_vs_x_proton}, %
the same rationale applies to the other structure functions displayed.

In contrast to the sizable differences between the ``No-TMC'' result and the other three curves
for the lower $Q$ value ($1.5$~GeV),  
the ``No-TMC'' result is roughly comparable to the others for  $Q=10$~GeV.  
In total, these comparisons affirm that TMCs are negligible for small-$x$ 
or for intermediate to large $Q$.

\vspace{1cm} %

\paragraph{$F_i$ structure function ratios: 
\rm (Fig.~\ref{fig:ntmc_AbsStr_vs_x_proton} Lower Panels)} %
To highlight the importance of including the TMCs, 
we now examine the lower panels of  Fig.~\ref{fig:ntmc_AbsStr_vs_x_proton}
which display the ratio of $F_i^{\rm TMC}$ to $F_i^{\rm No-TMC}$. 
The ratio plots accentuate subtle features that are not evident
on the log-scale plots of the upper panels. 

If we first focus on the $Q=1.5$~GeV results, we observe the deviation 
between $F_i^{\rm TMC}$ and $F_i^{\rm No-TMC}$ differs markedly 
for the separate $F_{1,2,3}$ structure functions. 
These differences arise from:
i)~the different prefactors, see e.g. Eq.~\eqref{eq:master-rescaled}, 
ii)~the different argument of $F_i^{(0)}$ (as noted previously),
and also 
iii)~the sub-leading $\{h_i, g_i\}$ contributions. 
It is the combination of these components that cause 
the variation among the Fig.~\ref{fig:ntmc_AbsStr_vs_x_proton} sub-figures. 
For example, 
we see the $F_1$ curves deviate at smaller $x$ values as
compared with the $F_2$ and $F_3$ results. 
Additionally, the $F_2$ and $F_3$ ratios dip  below unity for intermediate $x$ values 
($x\sim 0.3 - 0.5$) while the $F_1$ ratio is above unity.

\begin{table}[!t]
\begin{centering}
\begin{tabular}{|c|c|c|c||c|c|c|c||c|c|c|c|}
\hline 
\rule[-1mm]{0mm}{16pt}%
\textbf{Symbol} & \textbf{$A$} & \textbf{$Z$} & \textbf{$J^{P}$} & \textbf{Symbol} & \textbf{$A$} & \textbf{$Z$} & \textbf{$J^{P}$} & \textbf{Symbol} & \textbf{$A$} & \textbf{$Z$} & \textbf{$J^{P}$}\tabularnewline
\hline 
\hline 
H & 1 & 1 & $\frac{1}{2}^{+}$ & N & 14 & 7 & $1^{+}$ & Ag$_{{\rm \,iso}}$ & 108 & 54 & -\tabularnewline
D & 2 & 1 & $1^{+}$ & Ne & 20 & 10 & $0^{+}$ & Sn$_{{\rm \,iso}}$ & 119 & 59.5 & -\tabularnewline
$^{3}$He & 3 & 2 & $\frac{1}{2}^{+}$ & Al & 27 & 13 & $\frac{5}{2}^{+}$ & Xe & 131 & 54 & $\frac{3}{2}^{+}$\tabularnewline
He & 4 & 2 & $0^{+}$ & Ar & 40 & 18 & $0^{+}$ & W & 184 & 74 & $0^{+}$\tabularnewline
Li & 6 & 3 & $1^{+}$ & Ca & 40 & 20 & $0^{+}$ & Au & 197 & 79 & $\frac{3}{2}^{+}$\tabularnewline
Li & 7 & 3 & $\frac{3}{2}^{-}$ & Fe & 56 & 26 & $0^{+}$ & Au$_{{\rm \,iso}}$ & 197 & 98.5 & -\tabularnewline
Be & 9 & 4 & $\frac{3}{2}^{-}$ & Cu$_{{\rm \,iso}}$ & 64 & 32 & - & Pb$_{{\rm \,iso}}$ & 207 & 103.5 & -\tabularnewline
C & 12 & 6 & $0^{+}$ & Kr$_{{\rm \,iso}}$ & 84 & 42 & - & Pb & 208 & 82 & $0^{+}$\tabularnewline
\hline 
\end{tabular}
\par\end{centering}
\caption{List of nuclear PDFs considered in this work. We use the nCTEQ15 nPDF
set~\cite{Kovarik:2015cma}. The nuclei indicated with \textquotedblleft iso\textquotedblright{}
subscript are isoscalar corrected; thus, the $Z$ value can be half-integer.
We will also show results with a neutron PDF computed using isospin symmetry. 
\label{tab:isotopeList}
}%
\end{table}

\subsection{Nuclear structure functions with TMCs}\label{sec:numbers_nuclei}

Having examined the proton structure functions in Fig.\,\ref{fig:ntmc_AbsStr_vs_x_proton},
we now consider the corresponding nuclear structure functions 
of  Fig.\,\ref{fig:ntmc_StrFn_TMC_vs_x_MultiA}.
The list of nuclear isotopes that we consider are summarized in Tab.\,\ref{tab:isotopeList}.

\subsubsection*{Full TMCs for nuclei}\label{sec:numbers_nuclei_full}

In  Fig.~\ref{fig:ntmc_StrFn_TMC_vs_x_MultiA} we plot the ratio of the 
full ``TMC'' structure functions $F_i^{\rm TMC}$  
to the  ``No-TMC'' structure functions $F_i^{\rm No-TMC}$  
for charged current $(W^-)$  and neutral current $(\gamma,Z)$ exchange 
for  the nuclei of Table~\ref{tab:isotopeList}.
This ratio highlights the impact of the TMCs.

We  observe the general behavior of the nuclear ratios
displayed in  Fig.~\ref{fig:ntmc_StrFn_TMC_vs_x_MultiA}
are very similar to the corresponding proton results in the lower panels of Fig.~\ref{fig:ntmc_AbsStr_vs_x_proton}.
Because the nuclear and proton results are so similar, 
we only plot a single $Q$ value ($1.5~$GeV)
in  Fig.~\ref{fig:ntmc_StrFn_TMC_vs_x_MultiA}
as other $Q$ values (e.g., $Q=10$~GeV) will be similar to the proton results. 
We will study the detailed $Q$ dependence in the following.

The results of Fig.~\ref{fig:ntmc_StrFn_TMC_vs_x_MultiA} 
clearly demonstrate that the TMCs  
for the  nuclei are effectively independent of $A$. 
We attribute this curious finding to the  re-scaling property ($x_A M_A = x_N M_N$) 
as shown in Sec.~\ref{sec:rescaling}.
This re-scaling allows us to identically rewrite the TMCs for structure functions in Eq.~\eqref{eq:master}, which are functions of nuclear-level quantities $x_A$ and $M_A$, in terms of   nucleon-level quantities $x_N$ and $M_N$.
As a result, the variable nucleon content of nuclei, which is the defining characteristic of nuclei, is averaged out, resulting in near universal behavior. 

To illustrate the emergence of this universality, 
let us Taylor-expand the nuclear structure function ratio 
$F_1^{\rm TMC-Leading}/F_1^{\rm No-TMC}$ 
to show this is independent of the nuclear $A$ value 
up to corrections $\varepsilon^2=(xM/Q)^2$.
We use the expressions of  Eq.~\eqref{eq:nTMC_F_expansion} 
and computing the  derivative as  
$F_i^{(0)}(\xi_N)\approx F_i(x_N) + \delta x_N [dF_i(y)/dy]_{y=x_N}$,
where $\delta x_N = (\xi_N-x_N)\approx (x^3 \pMass^2/Q^2) \ll1$,
If we apply this approximation to the  $F_1$ ratio, for example, 
we then obtain:
\begin{align}
& \frac{F_1^{\rm TMC-Leading}(x_N)}{F_1^{\rm No-TMC}(x_N)}  \sim 
\left(1- \varepsilon^2 \right) 
 \frac{F_1^{(0)}(\xi_N)}{F_1^{(0)}(x_N)} 
 \sim 
 \left(1-\varepsilon^2\right) 
 \frac{F_1^{(0)}(x_N) +\mathcal{O}(\varepsilon^2)}{F_1^{(0)}(x_N)} 
\sim 
  \left(1-\varepsilon^2\right) 
  + \mathcal{O}\left(\varepsilon^4\right)
\ .
\end{align}
where we have defined $\varepsilon=(xM/Q)$.
Remarkably, we find that the ratio 
$F_1^{\rm TMC-Leading}(x_N)/F_1^{\rm No-TMC}(x_N)$
is independent of $A$ up to $\varepsilon^2$.
Recall the full TMC ratio 
$F_1^{\rm TMC}(x_N)/F_1^{\rm No-TMC}(x_N)$
differs only by the additional $\{h_i, g_i\}$ terms
which are suppressed by powers of $(M/Q)^2$.
This implies that nuclear TMCs themselves, when defined in terms of averaged quantities, are essentially universal. 

We now explore this approximate universal property of the TMCs, 
and determine how we can exploit this property to simplify 
certain nuclear TMC calculations.

\begin{figure}[tb]
\centering
\includegraphics[width=0.99\textwidth]{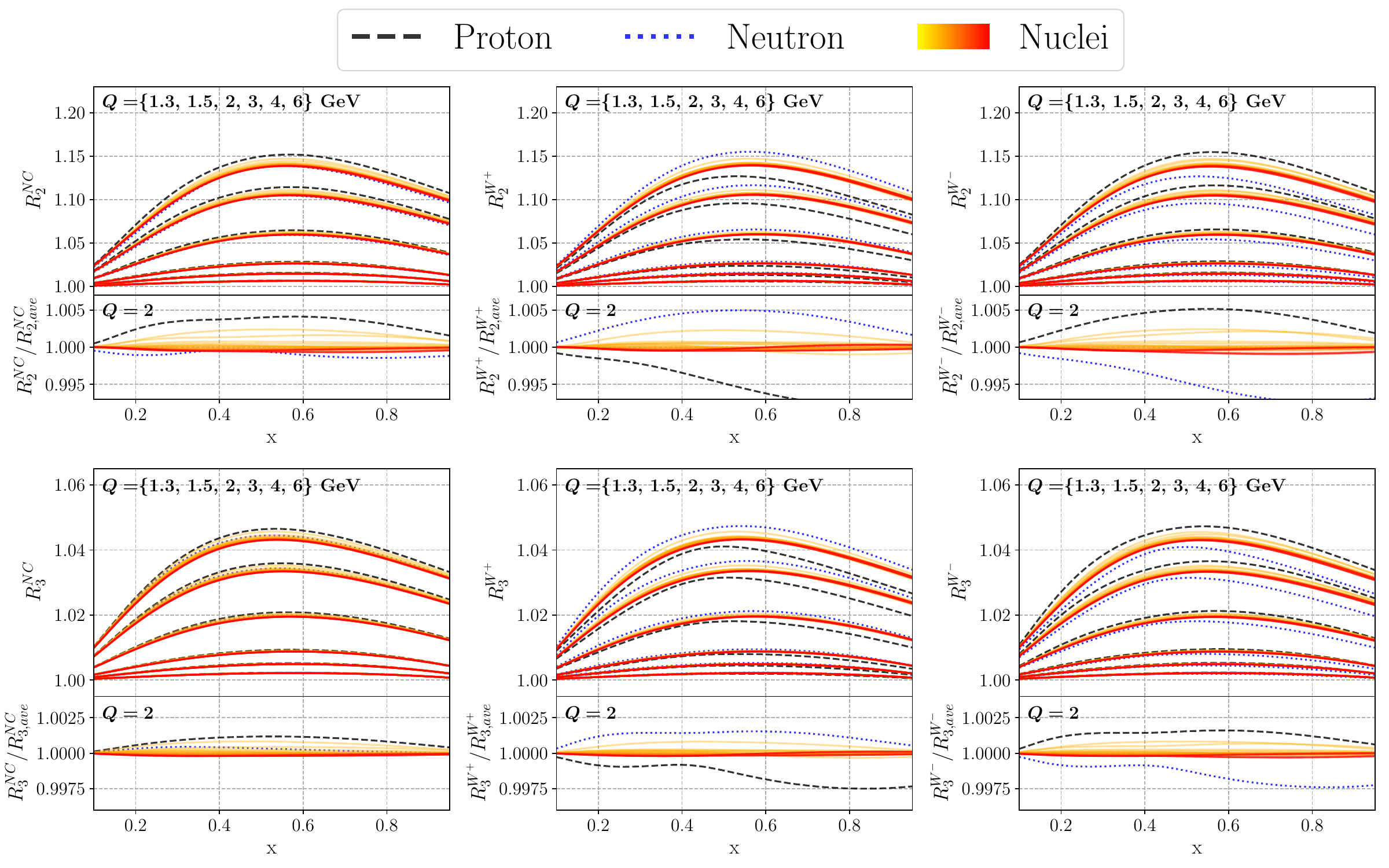}
\caption{
{\bf Upper panels:} 
We display the nuclear structure function ratios of $F_i^{\rm TMC}/F_i^{\rm TMC-Leading}$
for selected CC and NC processes for $Q=\{1.3,1.5,2,3,4,6\}\GeV$ (from top to bottom)
{\it vs.} $x_N$. 
{\bf Lower panels:}
We display the variation of this ratio compared to the
average variation as computed in
Eq.\,\protect\eqref{eq:numbers_averageDef} for $Q=2$~GeV.
In the top row we show $F_2$ results for 
(a)~NC $\gamma/Z$ exchange, (b)~CC $W^+$ exchange, and (c)~CC $W^-$ exchange, 
and the bottom row shows $F_3$ results for the same processes.
In all panels, 
the proton is indicated with the black dashed line, 
and the neutron with the blue dotted line. 
The narrow band of colored lines shows the various nuclear results.
}
\label{fig:f23nuc} 
\end{figure}
\subsubsection*{Comparison of full vs.\  leading-TMCs for nuclei}\label{sec:numbers_nuclei_leading}

In  Fig.~\ref{fig:ntmc_AbsStr_vs_x_proton} we observed that 
the ``Leading-TMC'' yielded a reasonable approximation to the 
full ``TMC'' result, even at low $Q$ and large $x$ values. 
Computationally, the ``Leading-TMC'' result is much simpler to 
compute as the full ``TMC'' result include contributions from the 
$\{h_i, g_i\}$ terms, each of which includes an integral. 
Additionally, if we can take advantage of the approximate $A$ independence 
observed in the previous section, this may allow us to greatly simplify 
the calculation of TMCs for the many different nuclei present in a typical 
nPDF fit. 

\paragraph{Magnitude of ratios: (Fig.~\ref{fig:f23nuc} upper panels)} 
We start by  comparing the ratio of the  ``Leading-TMC'' to the  full ``TMC'' result 
as shown in the upper panels of Fig.~\ref{fig:f23nuc}.
The results are shown for selected CC and NC processes with  representative values $Q=\{1.3,1.5,2,3,4,6\}\GeV$.

We display results for all the nuclei of Table~\ref{tab:isotopeList}
In these plots, the individual nuclear~$A$ values are not labeled, but the ratios coalesce into clearly defined bands for each $Q$ value. This coalescence holds for all  permutations of structure functions and boson exchanges. 
The different nuclei are shown as colored solid lines in the figure.
We also include results for the proton (dashed black) and neutron (dotted blue).

Focusing first on the bands of the top panels, we see that for $F_2^A$ the differences between the full  TMCs and Leading-TMCs are  as large as $\mathcal{O}(10\%-15\%)$ at $x\sim0.5$ for $Q=1.3\GeV$,
and  $\mathcal{O}(4\%-5\%)$ for $F_3^A$. 
For larger $Q$ values we expect these differences to be reduced as they are proportional to 
powers of $(M/Q)^2$. 
At $Q=2$~GeV, which is a typical $Q_{cut}$ value for many global PDF analyses, 
the ratio for $F_2^A$ is  $\lesssim 5\%$, and the ratio of $F_3^A$ is $\lesssim 2\%$. 
For $Q\gtrsim3-4\GeV$, we find that these differences reduces to the sub-percent level, 
and effectively vanish at $Q=6\GeV$.

\paragraph{Nuclear $A$ dependence of ratios:} %

Having discussed the magnitude of the  $F_i^{\rm TMC}/F_i^{\rm TMC-Leading}$ ratios, 
we now examine the $A$ dependence of these ratios. 
A distinctive feature of  Fig.~\ref{fig:f23nuc} is the coalescence of the results into individual bands, 
and this suggests that the nuclear $A$ dependence of this ratio is minimal. 
[The proton and neutron ratios (shown as dashed and dotted black lines, respectively), 
do not necessarily lie within the bands, and we will discuss these separately.] 
This apparent  universality of the full TMC/Leading-TMC ratios can be traced back to:
i)~the fact that this ratio has only mild dependence on the underlying PDF, and 
ii)~the fact that for nuclei, in the large $x$ region it is the average $u+d$ (isoscalar) PDF that dominates this result.

One approach to understand these features is to consider the analytically computed upper bound
for these ratios.
If we assume the structure functions are  monotonically decreasing
(an entirely reasonable assumption in the large $x$ region),
it is possible to obtain the following constraints  on these ratios~\cite{Schienbein:2007gr}:
\begin{subequations}
\begin{align} \label{eq:f2limit}
    \frac{F_2^{\rm TMC}}{F_2^{\rm TMC-leading}} (x, Q^2) &\ \leq\  1+\left(\frac{M}{Q}\right)^2 \frac{6x \xi}{r}(1-\xi) + \left(\frac{M}{Q}\right)^4  \frac{12 x^2 \xi^2}{r^2} (-\ln\xi -1+\xi)\\
    \frac{F_3^{\rm TMC}}{F_3^{\rm TMC-leading} }(x, Q^2) &\ \leq\  1-\left(\frac{M}{Q}\right)^2 \frac{2x \xi}{r}\ln\xi
        \quad .
        \label{eq:f3limit}
\end{align}\end{subequations}
Note, these bounds have absolutely no dependence on the PDF.
Here, we also explicitly see the powers of $(M/Q)^2$ which drive the ratios to unity for large $Q$.

\begin{figure}[!t]
    \centering
    \subfigure[]{\includegraphics[width=0.48\textwidth]{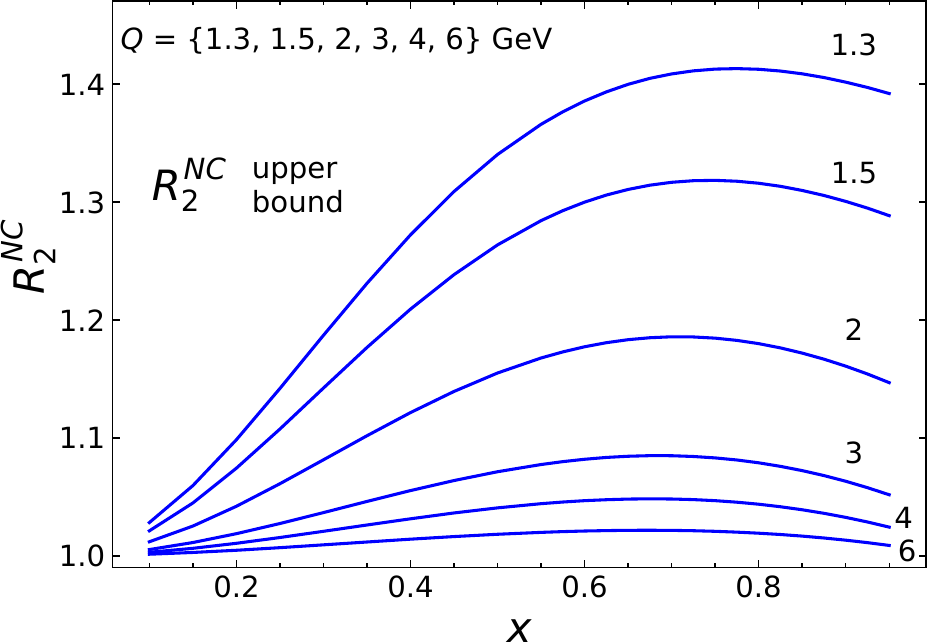}}
\hfill
    \subfigure[]{\includegraphics[width=0.48\textwidth]{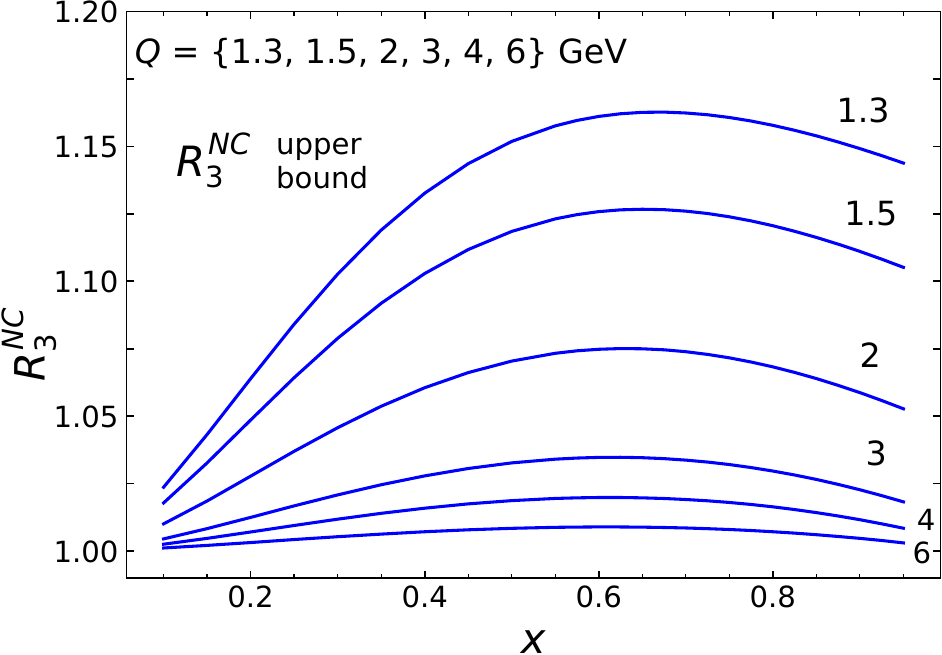}}
\caption{We show the upper bounds of $R_i^{\rm NC}=F_i^{\rm TMC}/F_i^{\rm TMC-Leading}$ for a)~$F_2$ and b)~$F_3$ 
    with selected \mbox{$Q=\{1.3, 1.5, 2,3,4,6\}$~GeV,} (from top to bottom)
    {\it vs.} $x_N$
    as computed with Eqs.~\eqref{eq:f2limit} and \eqref{eq:f3limit}.
    }
    \label{fig:RFUB}
\end{figure}

In Fig.\,\ref{fig:RFUB}, we plot these bounds as a function of $x$ for selected $Q$ values. 
Comparing the magnitudes of the bounds with the values of Fig.~\ref{fig:f23nuc},
we see the bounds are quite conservative.
Note that these bounds are valid for any nPDFs in the large $x$ region 
(where $F_i^A$ is decreasing monotonically),  and that is the relevant region of 
interest for the TMC effects. 
Consequently, the independence of the bound on the PDF
helps us understand the comparable nuclear $A$ independence of the  $F_i^{\rm TMC}/F_i^{\rm TMC-Leading}$ ratios
observed in Fig.~\ref{fig:f23nuc}.

\paragraph{Nuclear $A$ variation of ratios:} %

The observation that these ratios are relatively insensitive to the nuclear $A$ value  
suggests a ``short cut'' which can be used to efficiently implement TMCs into numerical calculations. 

More specifically, computing the full TMCs for each nucleus $A$ requires a calculation of the $h_i^A(\xi_N,Q^2)$ and $g_i^A(\xi_N,Q^2)$ terms in Eq.~\eqref{eq:master-rescaled}. Each term requires a separate integration over a nuclear structure function. Clearly, such steps would require significant CPU time if they were included inside the fitting loop of a global PDF analysis. If it is true that the size of TMCs to nuclear structure functions are relatively insensitive to the nuclear $A$ value, as our works suggests, then TMCs can be estimated through the application of a ``universal'' correction factor, $R_{i}^{V}(x,Q^2)$.
Such a factor can be constructed by taking an average TMC summed over the various $n_A$ nuclei: 
\begin{equation}
 R_i^{V}(x_N,Q^2) \ = \ 
 \frac{1}{n_A}\sum_k^{n_A} 
 \frac{F_i^{ A_k, \rm TMC}(x_N,Q^2)}{F_i^{ A_k, {\rm TMC-Leading}}(x_N,Q^2)} 
 \ = \ 
 \frac{1}{n_A} \sum_k^{n_A} R_i^{k, {V}}
\label{eq:numbers_averageDef}
 \end{equation}
where ``$V$'' labels the type of the exchanged gauge boson ($\gamma,Z,W^\pm$), 
and the sum over $k$ excludes the  proton and neutron. 

The utility of the ``universal'' correction factor $R_{i}^{V}(x,Q^2)$ 
is that we can compute this once at the beginning of a fit, and then apply this correction
factor inside the fitting loop without the need for recomputing additional integrals over $h_i$ and $g_i$ functions.

To determine the potential precision of such an approximation, 
we show in the lower panels of Fig.~\ref{fig:f23nuc}  a ratio of ratios 
constructed by taking  $(F_{i}^{\rm TMC}/F_{i}^{\rm TMC-Leading})$ 
over $R_i^{V}$ for $Q=2$~GeV, which is a typical $Q_{\rm cut}$ value for global fits.
It is remarkable how narrow the band of nuclear PDFs lie in comparison to the 
universal  $R_i^{V}$ function. 
For $F_2$ we see the nuclear bands lie within $0.2\%$ of the universal curve, 
while for $F_3$ the  bands lie within $0.1\%$.
For larger $Q$ values, the precision is further increased. 
This is clearly sufficient accuracy for any TMC calculation given the typical nPDF uncertainty 
in the large $x$ region.

\paragraph{Variation of proton \& neutron ratios:} %

In Fig.~\ref{fig:f23nuc} we observed that all the nuclear PDFs were uniformly within a narrow band.
In contrast,  the proton and neutron results displayed a wider variation lying outside these nuclear bands.
We will explain the source of this variation, 
and understand why the universal correction factor $R_{i}^{V}(x,Q^2)$ previously introduced,
still works well in the case of nuclei. 

Specifically, in Fig.~\ref{fig:f23nuc} we see that the proton ratios lie above the nuclear band 
for the NC and CC $W^-$ plots, and below for the CC $W^+$ plots; 
furthermore, the behavior of the neutron ratios is exactly the opposite of the proton.
We can understand these features by identifying  the dominant contribution
for each process. 
As $x\to1$, we neglect the gluon and sea quarks to obtain:
$F_2(\gamma/Z) \sim \frac{x}{9}[4 u(x) + d(x)]$
and 
$F_2(W^-) \sim 2 x u(x)$,
while
$F_2(W^+) \sim 2 x d(x)$.
Recall that for a proton PDF, the $d/u$ ratio in the large $x$ region 
can  typically be as small as $\sim 0.1$; 
that is, the up quark is an order of magnitude larger than the down quark.

We now understand the pattern of Fig.~\ref{fig:f23nuc}. 
The structure function for the NC and CC $W^-$ processes are driven by the up quark,
while the  CC $W^+$ is driven by the down quark. 
Thus, the  proton results are above the others for NC and CC $W^-$ 
but below for  CC~$W^+$ in plots of Fig.~\ref{fig:f23nuc}.
Since we obtain the neutron PDFs by isospin symmetry ($u\leftrightarrow d$),
so we also understand why the neutron results are the opposite of the proton. 

This exercise also demonstrates why the results for the nuclear PDFs lie together in a uniform band.
Most of the nuclei are closer to the isoscalar state $(A\sim 2Z)$ 
than either the proton  $(Z=A)$ or neutron state  $(Z=0)$. 
For an isoscalar nuclei we have $u=d$, and it is effectively the average
of the up and down PDFs $(u+d)/2$ 
which provides the dominant contributions to Fig.~\ref{fig:f23nuc}.
Consequently,  we find that the nuclear band is effectively an average of 
the proton and neutron result, and thus lies between the two in  Figs.~\ref{fig:f23nuc}.
The one slight variation of this pattern is the parity-violating $F_3$ NC
structure function; this effect must come from the $Z$ boson contribution because
the parity-conserving photon contribution vanishes.

\paragraph{Approximate nuclear $A$ independence:} %
\begin{figure}[p]
    \centering
    \subfigure[\label{fig:f2ccnc_lightA_a}]{\includegraphics[width=0.5\textwidth]{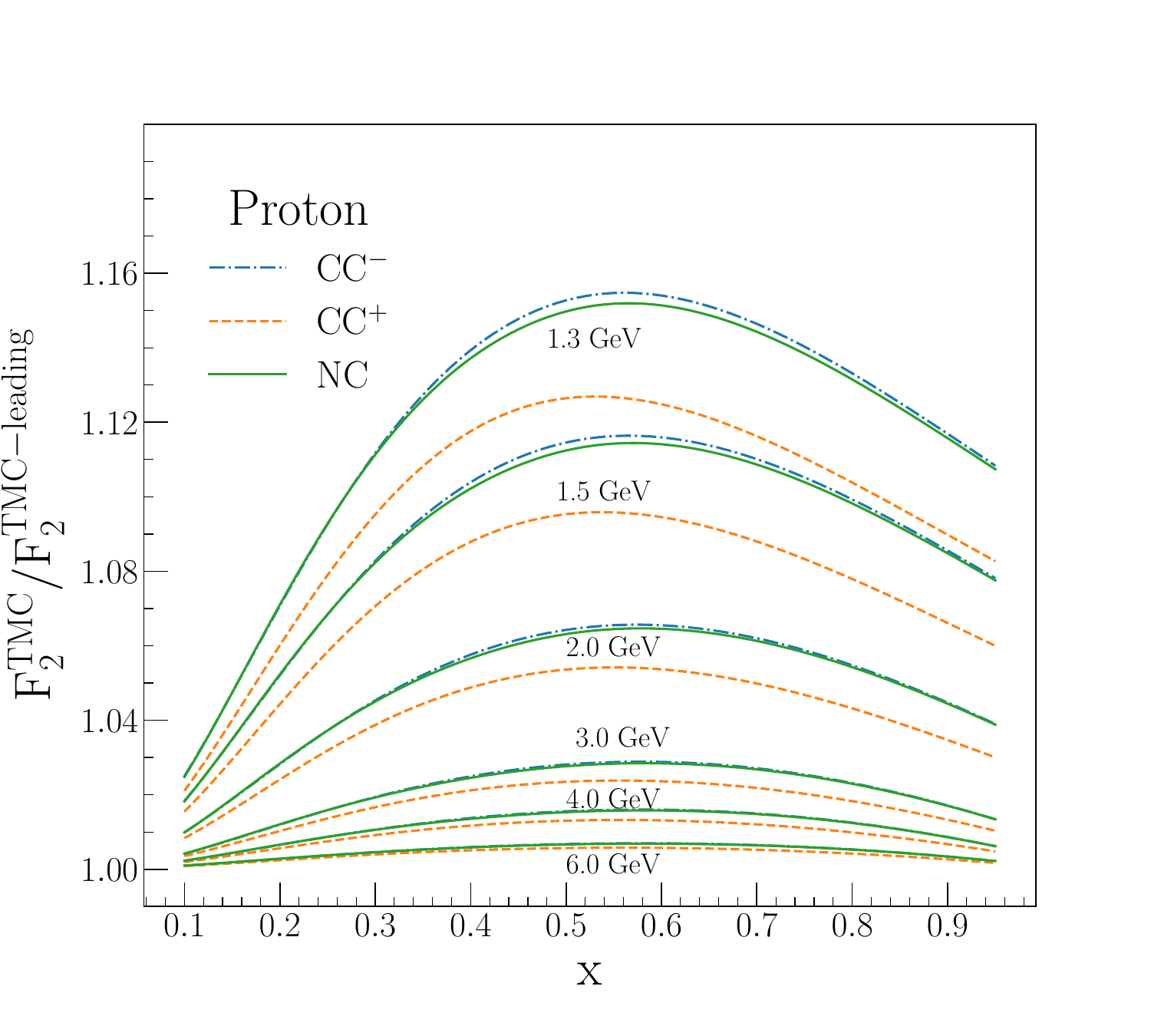}}%
    \subfigure[\label{fig:f2ccnc_lightA_b}]{\includegraphics[width=0.5\textwidth]{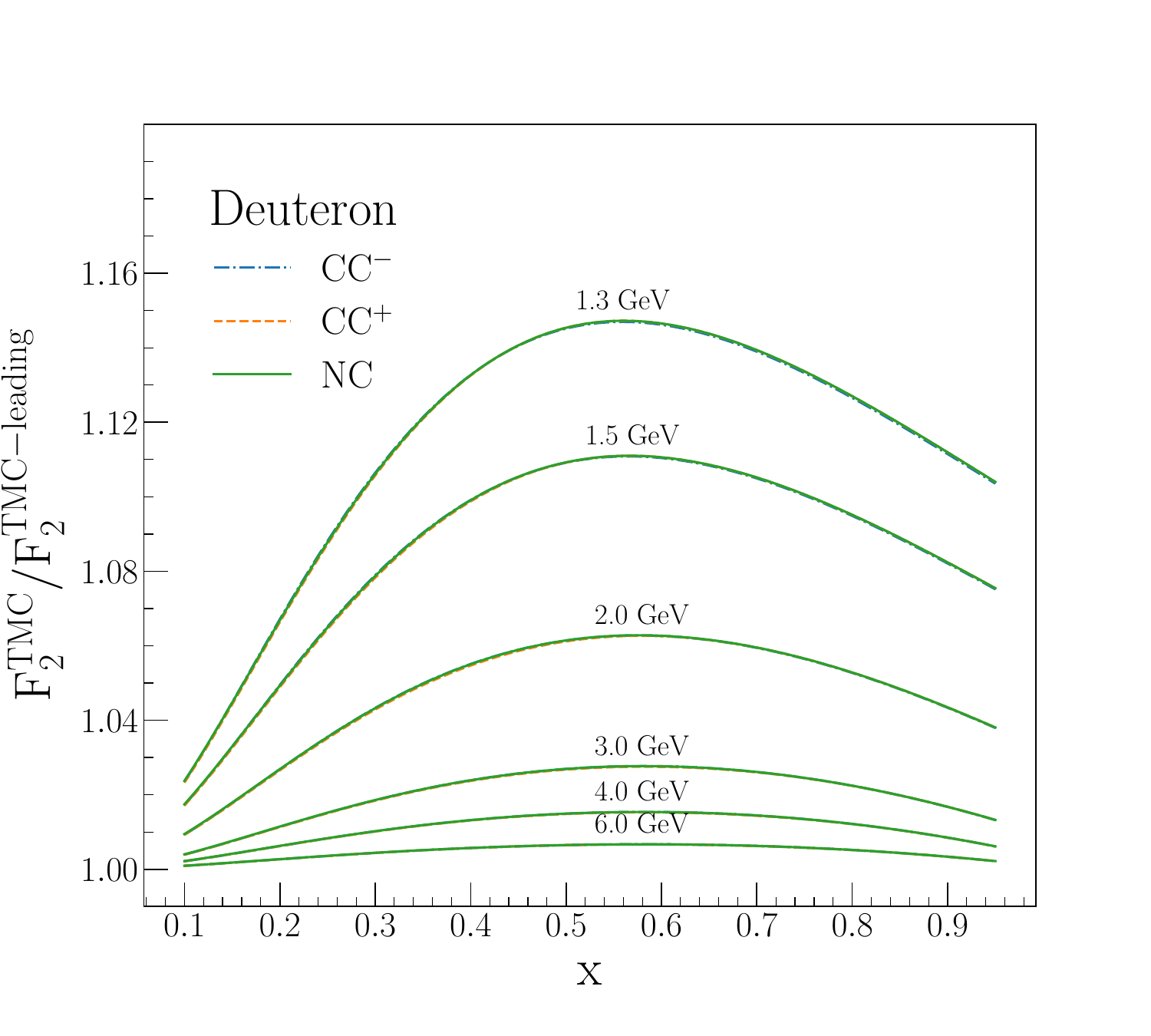}}
    \\
    \subfigure[\label{fig:f2ccnc_lightA_c}]{\includegraphics[width=0.5\textwidth]{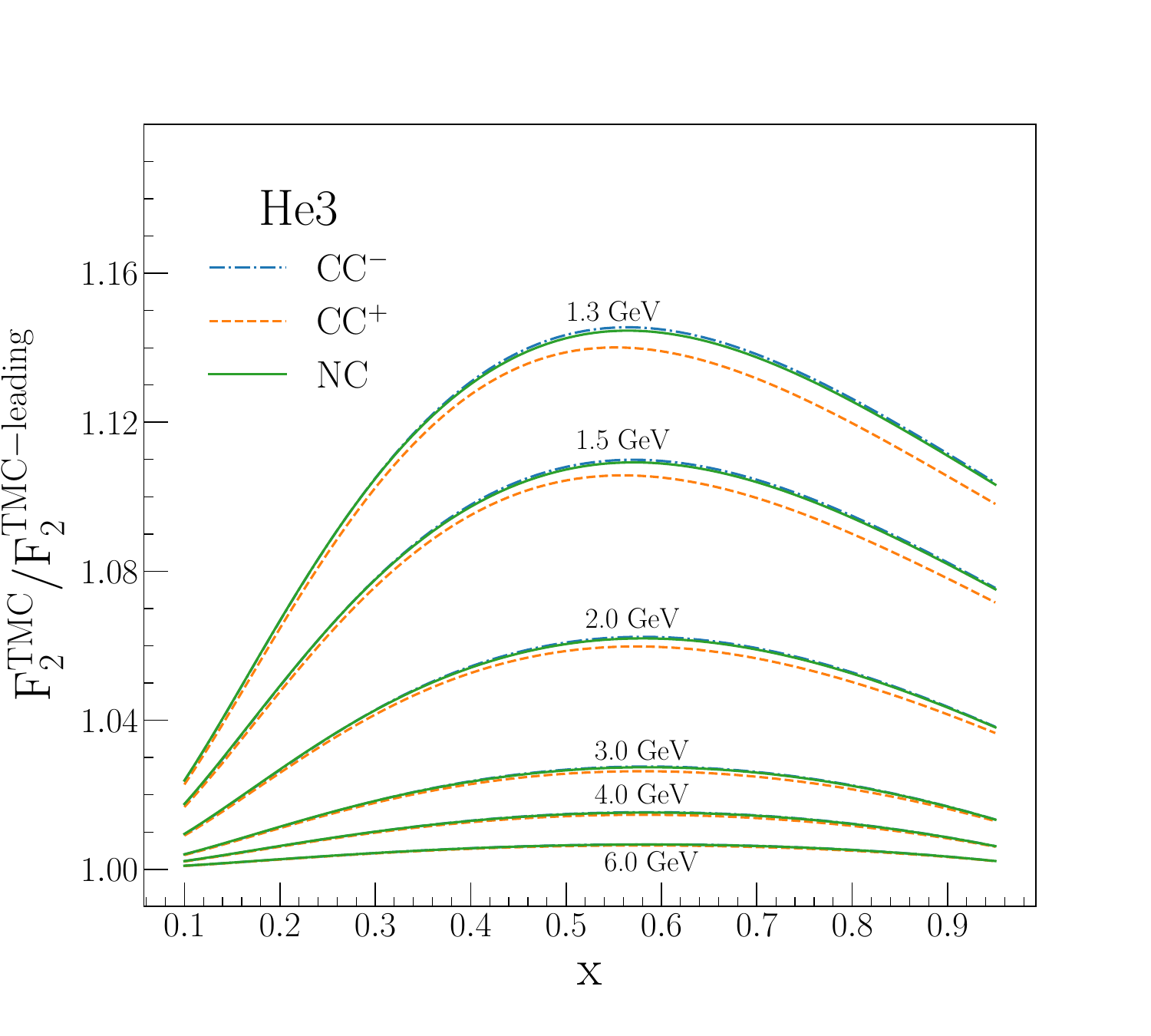}}
\caption{
The nuclear structure function ratio $F_2^{\rm TMC}/F_2^{\rm TMC-Leading}$ 
as function of $x_N$ for $Q=\{1.3,1.5,2,3,4,6 \}\GeV$
for  (a) the proton, (b) deuteron, and (c) ${}^3$He. 
For each nucleus, we overlay the result
for NC exchange (solid green), 
CC $W^+$ exchange (dotted orange), 
and 
CC $W^-$ exchange (dot-dashed blue).
Note that for the deuteron in Fig.~(b), the curves  for the three CC/NC processes coincide.
\label{fig:f2ccnc_lightA}}
\end{figure}

We illustrate this further in Fig.~\ref{fig:f2ccnc_lightA} which shows the ratios 
for the NC and CC processes for the a)~proton, b)~deuteron, and c)~${}^3$He.
We choose these examples because the deuteron is an isoscalar, and
${}^3$He is one of the most ``non-isoscalar'' nuclei (with the exception of hydrogen/proton).

The pattern in Fig.~\ref{fig:f2ccnc_lightA} is reminiscent of Fig.~\ref{fig:f23nuc}.
In Fig.~\ref{fig:f2ccnc_lightA_a} for the proton, we see the 
($u$\nobreakdash-quark dominated) NC and CC $W^-$ processes 
lie above the ($d$\nobreakdash-quark dominated) CC $W^+$ process.
In Fig.~\ref{fig:f2ccnc_lightA_b} for the isoscalar deuteron, 
we see all the curves coincide since $u=d$ for this nuclei. 
Finally, in Fig.~\ref{fig:f2ccnc_lightA_c} for the ${}^3\mathrm{He}$,
we see the NC and CC $W^-$ processes 
lie slightly above the CC $W^+$ process,
but the difference is not as dramatic as for the proton case. 
Note in these cases, the $u:d$ ratio is 
$2:1$ in the proton,
$1:1$ in the deuteron,
$5:4$ in ${}^3$He.
For any heavier nuclei, the $u:d$ ratio will be comparatively closer to unity,
and the results of the individual NC and CC processes will tend to coincide.

The outcome of these findings is that given the typical uncertainty of nuclear PDFs in the large $x$ region, applying an average, $A$-independent set of TMCs yields precision which is sufficient for a wide variety of real-world calculations.  
We shall exploit this observation in Sec.~\ref{sec:parm} and derive an approximate parameterization 
for these TMCs which can be applied to the general nuclear case.

Setting aside the proton and neutron cases, 
at $Q=2$~GeV we find the deviation of individual nuclei from the average to be 
less than $\sim0.2\%$ for $F_2$, 
and less than $\sim0.1\%$ for $F_3$.
And of course, these differences  decrease for larger $Q$.

\subsubsection*{Sub-leading TMCs for nuclei}\label{sec:numbers_nuclei_subleading}

\begin{figure}[t!]
    \centering
    \subfigure[]{\includegraphics[width=0.48\textwidth]{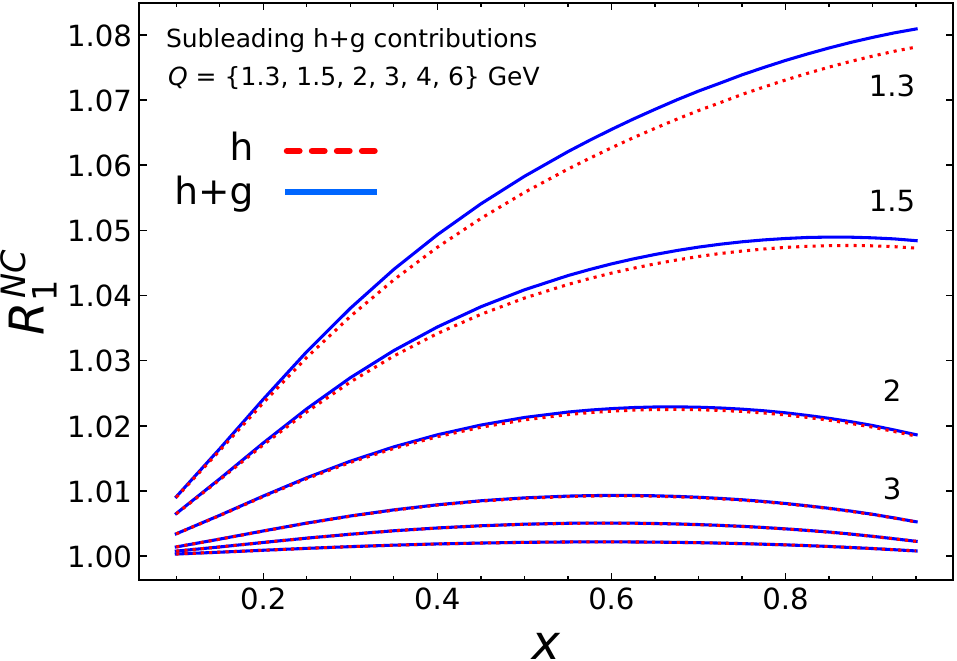}} \hfill
    \subfigure[]{\includegraphics[width=0.48\textwidth]{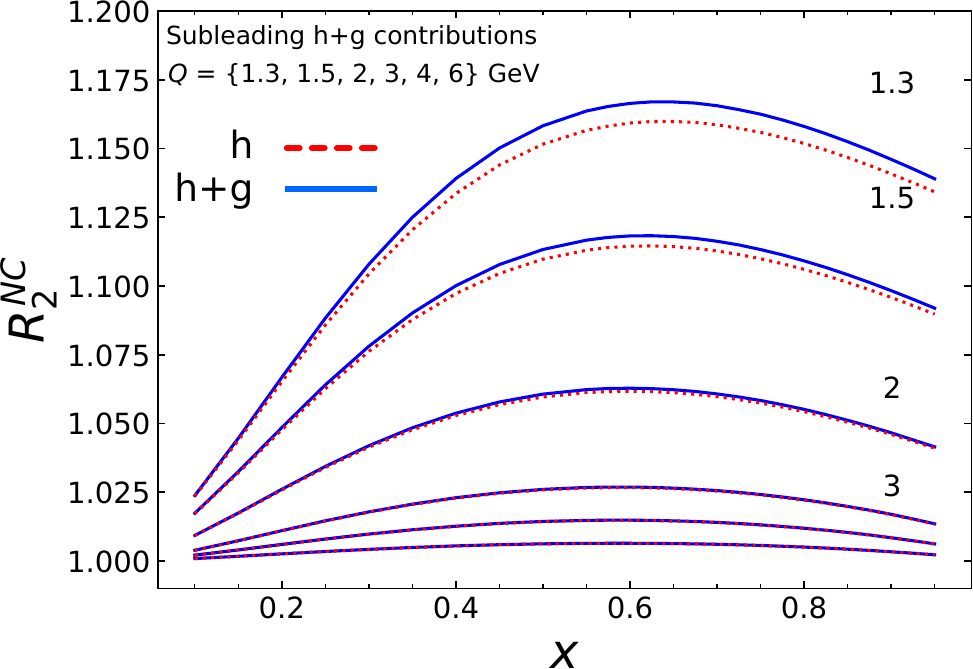}}
    \subfigure[]{\includegraphics[width=0.48\textwidth]{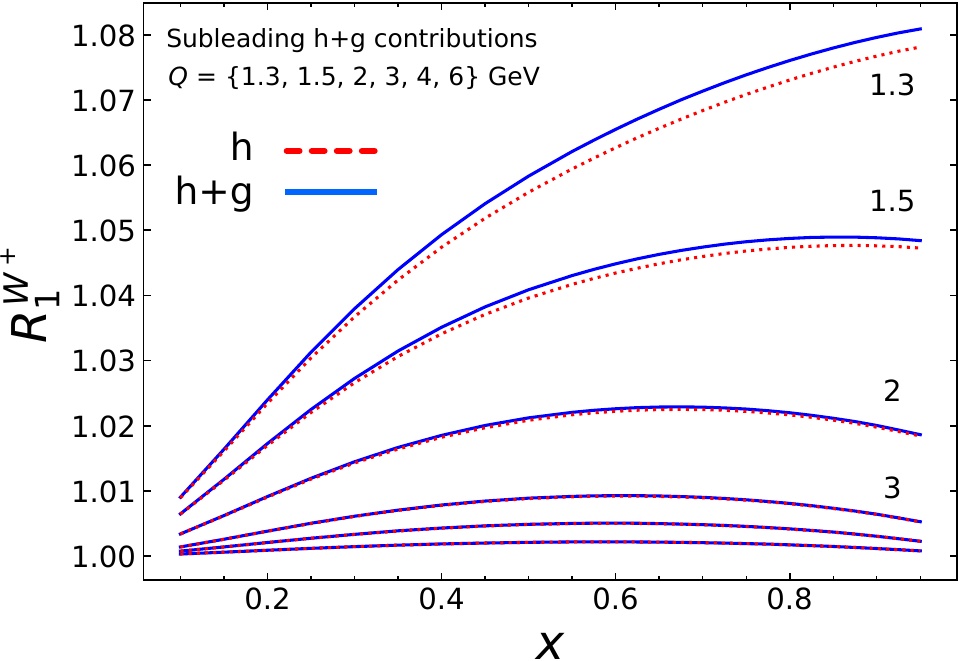}} \hfill
    \subfigure[]{\includegraphics[width=0.48\textwidth]{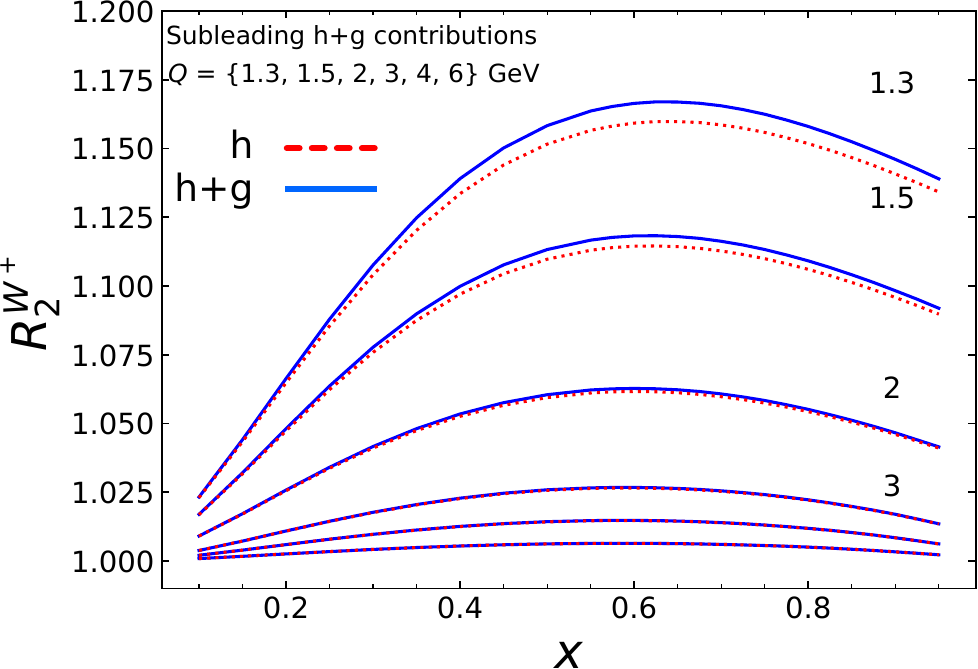}}
\caption{To highlight the effect of the sub-leading terms $\{h_i,g_i\}$,
we display the ratios $R_i^{Z,W^\pm}$.
Specifically, the ``h'' curve (dashed red) is the result for   
$[F_i^{\rm TMC-Leading}+F_i^{\rm h-term}]$,
and  the ``h+g'' curve (solid blue) is 
$[F_i^{\rm TMC-Leading}+F_i^{\rm h-term}+F_i^{\rm g-term}]$,
both compared to $[F_i^{\rm TMC-Leading}]$.
The notation is detailed in Eq.~\eqref{eq:Fnotation}.
We display selected \mbox{$Q= \{1.3,1.5, 2, 3,4,6\}$\GeV} (from top to bottom)
as a function of $x_N$ using the nCTEQ15  PDF for  $^{12}C$.
We observe for $Q\geq 2$~GeV (a typical kinematic cut), 
the  ``h+g'' contributions are less than~${\sim}2\%$.
    \label{fig:h2g2} }
\end{figure}

As a final study, we want to demonstrate the individual size of the  sub-leading $h_2^A$ and $g_2^A$ components of the full TMCs in Eq.~\eqref{eq:master-rescaled}. 
In Fig.~\ref{fig:h2g2} we present the results for NC ($\gamma/Z$) and CC ($W^\pm$) processes,
and we display a pair of ratios 
$F_2^{\rm TMC+h2} / F_2^{\rm TMC-Leading}$ and 
$F_2^{\rm TMC} / F_2^{\rm TMC-Leading}$,
for each $Q$ value.

Recall that $F_2^{\rm TMC-Leading}$ neglects the sub-leading $h_2^A$ and $g_2^A$ contributions,  where
$h_2^A$ is suppressed by $(M/Q)^2$, and 
$g_2^A$ is suppressed by $(M/Q)^4$.
The $F_2^{\rm TMC+h2}$ expression includes the $h_2^A$ contribution, but neglects $g_2^A$. 
Finally, 
the full $F_2^{\rm TMC}$ expression includes both $h_2^A$ and $g_2^A$; this could equivalently be called $F_2^{\rm TMC+h2+g2}$.

Since the $g_2^A$ term is suppressed by $(M/Q)^4$, we expect this will only contribute for very large $x$ at low $Q^2$ values. Therefore, we expect $F_2^{\rm TMC+h2}$ and $F_2^{\rm TMC}$ to coincide 
throughout most of the parameter space. 
This expectation is validated in Fig.~\ref{fig:h2g2} where we do see that $F_2^{\rm TMC+h2}$ and $F_2^{\rm TMC}$
differ only for the lowest few $Q$ values in the large $x$ region. 
For example, at $Q=2$~GeV the differences are barely discernible in the figure, 
and the effect is even smaller as $Q$ increases. 

Thus, $h_2^A$ provides the dominant contribution throughout the kinematic range, 
but $g_2^A$ does contribute for low $Q$ values. 
The pattern for the other structure functions $F_{1,3}^A$ is similar.

\subsection{Parameterizing \texorpdfstring{$F_i^{\rm TMC}/F_i^{\rm Leading-TMC}$}{Fi(TMC)/Fi(TMC-Leading)}}\label{sec:parm}
\begin{table}[t]
\begin{centering}
\begin{onehalfspacing}
\begin{tabular}{|c|cc|cc|cc|}
\hline 
 & \multicolumn{2}{c|}{$F_{1}^{TMC}/F_{1}^{leading}$} & \multicolumn{2}{c|}{$F_{2}^{TMC}/F_{2}^{leading}$} & \multicolumn{2}{c|}{$F_{3}^{TMC}/F_{3}^{leading}$}\tabularnewline
\hline 
nPDFs   & $\lambda_{1}$  & $\delta_{1}$  & $\lambda_{2}$  & $\delta_{2}$  & $\lambda_{3}$  & $\delta_{3}$ \tabularnewline
\hline 
\hline 
nCTEQ15 &2.275  & -0.014  & 2.144 & 0.100  & 2.183  & 0.029 \tabularnewline
\hline 
\hline 
EPPS16 &  2.226 & -0.026  & 2.086  & 0.095  & 2.142 & 0.022\tabularnewline
\hline 
\hline 
nNNPDF2.0  & 2.226 & -0.025 & 2.086  & 0.103  & 2.197  & 0.023\tabularnewline
\hline 
\hline 
TUJU19 & 2.290 & -0.0170 & 2.167 & 0.099 & 2.200 & 0.031 \tabularnewline
\hline 
\hline 
\end{tabular}
\end{onehalfspacing}
\par\end{centering}
\caption{We approximate the $\{h_{i}(\xi),g_{i}(\xi)\}$ contributions to  $F_{i}^{\rm TMC}$ using $F_{i}^{{\rm TMC-Leading}}$ together 
with a \mbox{2-parameter} function $\gamma_a(Q)$.
Here, $\gamma_a(Q)$ implicitly depends on the parameters $\{\lambda_a, \delta_a\}$ 
which are fitted for each structure function $\{F_{1,2,3}\}$ and for selected PDFs displayed in the table. 
The TMC corrections are accurate to within $\sim 0.3$\% for $Q>1.3$~GeV. 
The $\{\lambda_{i},\delta_{i}\}$ parameters are
independent of the exchanged boson ($\gamma,Z,W^{\pm}$), and are
relatively insensitive to the specific underlying nPDF. 
Once the $\{\lambda_{i},\delta_{i}\}$ parameters are determined for a particular nPDF, 
the full TMC corrections can be efficiently computed (without convolution integrations) to
high precision as illustrated in Fig.~\ref{fig:approx}. 
\label{tab:gamma} }
\end{table}
\begin{figure}[tb]
    \centering
    \includegraphics[width=0.99\textwidth]{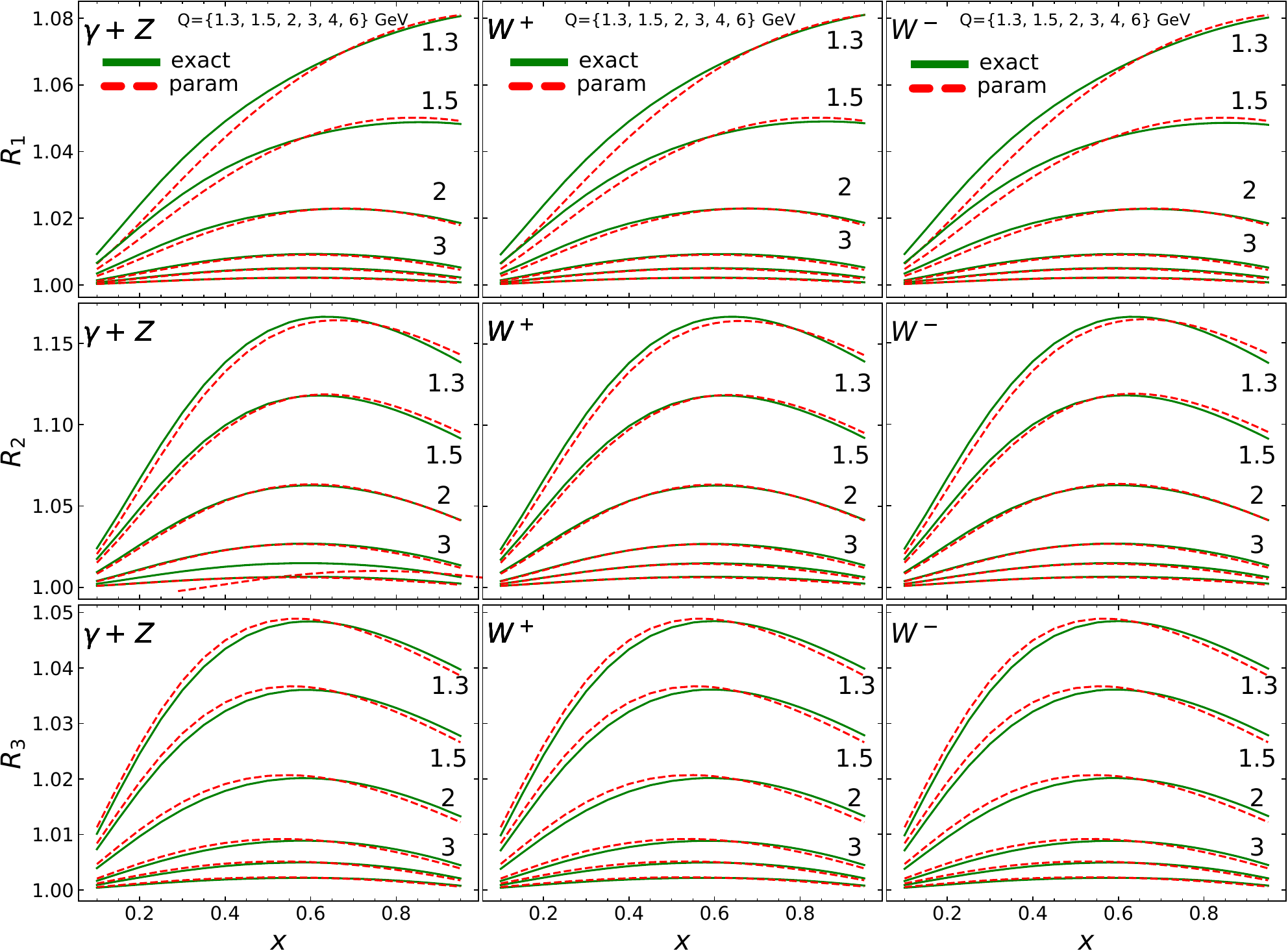}
\caption{We present the nuclear structure function ratio 
$R_i = F_i^{\rm TMC}/F_i^{\rm TMC-Leading}$   averaged over the 
nuclei of Table~\ref{tab:isotopeList} as defined in Eq.~\eqref{eq:numbers_averageDef}.
The ratios are shown as a function of $x_N$
for  $Q= \{1.3, 1.5, 2, 3, 4, 6 \}$~GeV (from top to bottom) 
using the nCTEQ15 nPDFs.
We compare the exact results (green solid) of Eq.~\eqref{eq:master-rescaled} 
to the approximate parameterizations (red dashed) based on 
Eq.~\eqref{eqs:approx} with parameters given in Table~\ref{tab:gamma}.
For $Q>2$~GeV, we observe the parameterization can match the exact results to better than 0.2\%.
} %
\label{fig:approx}
\end{figure}

In Sec.~\ref{sec:numbers_nuclei_leading} we demonstrated that the full TMC/Leading-TMC ratios were effectively insensitive to the nuclear $A$ value. In this section, we provide a parameterization for these ratios which allows us to efficiently compute the full TMCs given the Leading-TMC structure functions. 

\goodbreak
Starting from Eq.~\eqref{eq:master-rescaled}, we divide $F_i^{\rm TMC}(x)$ by the leading term $F_i^{\rm Leading-TMC}(x)$:
\begin{subequations}
\begin{align}
\frac{F_1^{\rm TMC}(x)}{F_1^{\rm Leading-TMC}(x)} &=  1+ \frac{M^2}{Q^2} \frac{x\xi}{r} \frac{h_2(\xi)}{F_1^{(0)}(\xi)} + \frac{M^4}{Q^4}\frac{2x^2\xi}{r^2} \frac{g_2(\xi)}{F_1^{(0)}(\xi)} \ , \\
\frac{F_2^{\rm TMC}(x)}{F_2^{\rm Leading-TMC}(x)} &=  1+ \frac{M^2}{Q^2} \frac{6x \xi^2}{r} \frac{h_2(\xi)}{F_2^{(0)}(\xi)} + \frac{M^4}{Q^4}\frac{12 x^2 \xi^2}{r^2} \frac{g_2(\xi)}{F_2^{(0)}(\xi)} \ , \\
\frac{F_3^{\rm TMC}(x)}{F_3^{\rm Leading-TMC}(x)} &=  1+\frac{M^2}{Q^2} \frac{2\xi x}{r} \frac{h_3(\xi)}{F_3^{(0)}(\xi)} \quad .
\end{align}\end{subequations}
The goal of this section is to parameterize these ratios. 

As explained in detail in~\ref{app:par}, 
if we assume the structure functions vanish at $x=1$ and expand them in a Taylor series,
we can obtain simple expressions for the ratios in terms of the parameter function $\gamma_i(Q)$, 
which characterizes the first derivative of the structure function. 
This allows us to obtain the approximate expressions for  the ratios $h_i/F_j^{(0)}$ and $g_2/F_j^{(0)}$:
\begin{subequations}\label{eqs:approx}
\begin{equation}
\frac{h_2(\xi)}{F_2^{(0)}(\xi)} = \frac{1-\xi}{\xi} +\gamma_2(Q)\frac{1-\xi}{\xi^2}  \sum_{j=1}^{j_{\rm max}} \frac{(-1)^j}{j! (j+1)} \,\, _2F_1\left(2, j+1, j+2, 1-\frac{1}{\xi}\right) \ , \label{h2of2}
\end{equation}
\begin{equation}
    \frac{h_3(\xi)}{F_3^{(0)}(\xi)}  = -\ln(\xi) +\gamma_3(Q) \frac{1-\xi}{\xi}\sum_{j=1}^{j_{\rm max}} \frac{(-1)^j}{j! (j+1)}  \,\, _2F_1\left(1, j+1, j+2, 1-\frac{1}{\xi}\right) \ , \label{h3of3}
\end{equation}
\begin{equation}
    \frac{h_2(\xi)}{F_1^{(0)}(\xi)} = 2\xi r^2 \left[\frac{1-\xi}{\xi} +\gamma_1(Q)\frac{1-\xi}{\xi^2}  \sum_{j=1}^{j_{\rm max}} \frac{(-1)^j}{j! (j+1)} \,\, _2F_1\left(2, j+1, j+2, 1-\frac{1}{\xi}\right) \right]  \ , \label{h2of1}
\end{equation}
\begin{equation}
    \frac{g_2(\xi)}{F_2^{(0)}(\xi)} = -\ln(\xi) -(1-\xi) +\gamma_2(Q)  \frac{(1-\xi)^2}{\xi^2} \sum_{j=1}^{j_{\rm max}} \frac{(-1)^j}{j! (j+2)}  \,\,_2F_1\left(2, j+2, j+3, 1-\frac{1}{\xi}\right)  \ , \label{g2of2}
\end{equation}
\begin{equation}
    \frac{g_2(\xi)}{F_1^{(0)}(\xi)}  = 2\xi r^2 \left[ -\ln(\xi) -(1-\xi) +\gamma_1(Q) \frac{(1-\xi)^2}{\xi^2} \sum_{j=1}^{j_{\rm max}} \frac{(-1)^j}{j! (j+2)}  \,\,_2F_1\left(2, j+2, j+3, 1-\frac{1}{\xi}\right)  \right]  \ ,
    \label{g2of1}
\end{equation}
\end{subequations}
where ${}_2F_1(a,b,c,z)$ is a hypergeometric function. 
Although the summation over the index $j$ can, in principle, go to infinity, we truncate the series at $j=j_{\rm max}$.
The $1/(j!)$ prefactor of the hypergeometric function helps ensure the series converges quickly, and we find  
$j_{\rm max}=4$ yields results that are $\lesssim 1\%$ accuracy.

The function $\gamma_a(Q)$ is given by
\begin{equation}
    \gamma_a(Q) =\lambda_a\ln(Q)^{\delta_a}\quad . 
\end{equation}
The values of $\{ \lambda_a, \delta_a\}$ are obtained by fitting the parameterization to the exact results for each $F_i^{\rm TMC}$ structure function. Note that the values of $\{ \lambda_a, \delta_a\}$ are independent of the type of the exchanged bosons; they have only a mild dependence on the specific structure function $F_{i}^A$ as displayed in  Table~\ref{tab:gamma}.

Additionally, while we have computed these results assuming an NLO DGLAP evolution, 
these results should be similar for an NNLO evolution. 
This is because: 
i)~we are  working at relatively low $Q$ values where the effects of the 
DGLAP evolution from $Q_0$ to $Q$ will be minimal, 
and ii)~since this quantity is a ratio of structure functions, 
the modification of the ratio is expected to be less than the impact on 
the individual structure functions.

We have also computed the results for a variety of different PDF sets to ensure 
this parameterization is largely insensitive to the specific PDF. 
This also is displayed in   Table~\ref{tab:gamma}, where we observe 
a mild variation of the parameters between PDF sets.

In Figure \ref{fig:approx}, we show the comparison between our parameterization with the exact results obtained using nCTEQ15 nPDFs. We can see that our parameterization works very well to reproduce the exact results. 
The parameterization has the advantage that it can be computed efficiently whereas the full $F_i^{\rm TMC}$ results require
additional convolution integrations for the $\{h_i(\xi), g_i(\xi)\}$ functions. 
For values of $Q\geq 2$~GeV, which is a typical cut for the global nPDF fits, we see 
the parameterization describes the full TMC results to better than 0.2\%.

\paragraph{Summary:}
In the previous sections we established several results which will facilitate 
efficient calculation of TMCs for nuclear structure functions. 
We briefly summarize the key observations below. 

\begin{itemize}

    \item{} TMCs can yield large corrections in the double limit of large $x$ and small $Q$.
    
    \item  The ``Leading-TMC'' result provides an excellent approximation to the  
    full ``TMC'' result as the $h_{2,3}$ contributions are suppressed by additional 
    $(M/Q)^2$, and the $g_2$ contributions by additional $(M/Q)^4$ powers.

    \item  The nuclear dependence of the ``TMC'' result 
    is approximately independent of $A$.  
    This observation suggests we can determine an $A$-independent ``universal''
    correction factor which can be used to quickly calculate 
    the nuclear structure function TMCs for any nuclei. 
    As illustrated in the bottom panels of Fig.~\ref{fig:f23nuc}, 
    this ``universal'' approximation holds to  
    ${\lesssim 0.1\%}$ for \mbox{$Q\geq 2$\GeV}.
    \item  As demonstrated in Fig.~\ref{fig:approx},
    we can parameterize this ``universal'' correction factor with Eq.~\eqref{eqs:approx} 
    into a 2-parameter form that can be easily incorporated in a efficient nuclear PDF fitting program. This parameterization matches the exact results 
    to better than 0.2\% for $Q\geq 2\GeV$.
    
\end{itemize}

Given the ``Leading-TMC'' result, the above allows us to efficiently
approximate the full ``TMC'' result for any nuclear structure function 
by applying ``universal'' correction factor. This avoids  the  computation of 
the numerical integrals contained in the $\{h_2, h_3, g_2\}$ functions.

\subsection{Reduced cross sections with TMCs for nuclei}\label{sec:numbers_xsec}
\begin{figure*}[!t]
\includegraphics[width=0.99\textwidth]{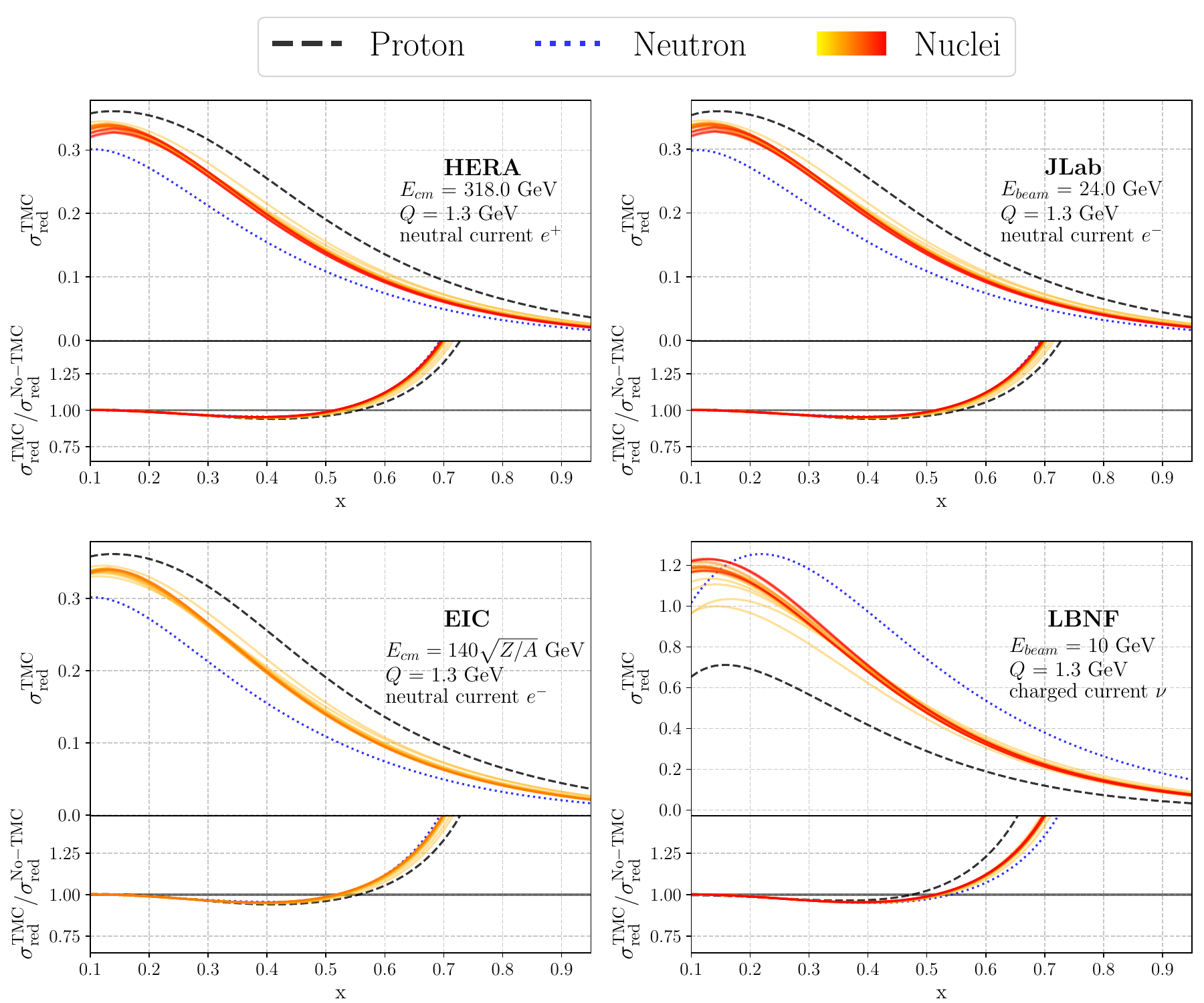}
\caption{ 
We present  reduced cross sections for the following scenarios:
(a)~HERA inspired $e^+p$ neutral current scattering with $E_{cm}=318$~GeV for  $Q=1.3$~GeV;
(b)~JLab inspired $e^- A$ charged current scattering with $E_{beam}=24$~GeV for $Q=1.3$~GeV;
(c)~EIC inspired $e^- A$ neutral current scattering with $E_{cm} = 140\sqrt{Z/A}$~GeV for $Q=1.3$~GeV;
(d)~LBNF inspired $\nu  A$ charged current scattering with $E_{beam}=10$~GeV for $Q=1.3$~GeV.
{\bf Upper panels:}
The upper panels display the  full ``TMC'' results.
{\bf Lower panels:}
The lower panels display the ratio of the full ``TMC'' to ``No-TMC''. 
The proton is indicated with the black dashed line, and the neutron with the blue dotted line. 
The narrow band of colored lines shows the various nuclear results.
}
    \label{fig:ntmc_ReducedXSec}
\end{figure*}

When extracting structure functions from DIS data or comparing theoretical expectations for DIS to data, one often works with the ``reduced cross section'' $\sigma_{\rm Red.}$. 
These quantities are derived from double differential cross sections 
scaled by kinematic factors, 
and provide a direct connection between DIS kinematics and structure functions.

The general cross section for neutral and charged current DIS  
can be expressed as:\cite{ParticleDataGroup:2020ssz}
\begin{eqnarray}
\frac{d^2\sigma^i}{dx \, dy} 
&=&
x(s-M^2) \, 
\frac{d^2 \sigma^i}{dx dQ^2} 
=
\frac{4\pi \alpha^2}{x y Q^2} \, \eta^i
\left\{
\left( 1 - y - \frac{x^2 y^2 M^2}{Q^2}\right) F_2^i
+
\left(y^2 x \right) F_1^i
\mp
\left( y-\frac{y^2}{2} \right) x F_3^i
\right\} 
\ ,
\label{eq:dsigDISxy}
\end{eqnarray}
where i=NC,~CC corresponds to neutral-current or  charged-current processes.
Additionally, 
$\eta^{NC}=1$ and 
$\eta^{CC}=4 \etaW $, with 
$\etaW=\frac{1}{2}[\frac{G_F M_W^2}{4\pi \alpha} \frac{Q^2}{Q^2+M_W^2}]^2$.
In the last term of Eq.~\eqref{eq:dsigDISxy}, 
the ``--''~sign is taken for an incoming anti-lepton $\{e^+,\bar{\nu}\}$ 
and the ``+''~sign for an incoming lepton $\{e^-,{\nu}\}$.

\paragraph{Neutral current cross section:} %

We find it convenient to define the reduced cross section by dividing out 
the leading factors. 
\begin{eqnarray}
\frac{d^2\sigma^{\rm NC}}{dx \, dy} 
&=&
x(s-M^2) \, 
\frac{d^2 \sigma^{\rm NC}}{dx dQ^2} 
=
\frac{4\pi \alpha^2}{x y Q^2}  
\
\left[ 
\frac{Y_+}{2} \ \sigma_{\rm Red.}^{NC}
\right] 
\ ,
\label{eq:dsigNCxy}
\\[10pt]
\sigma_{\rm Red.}^{NC} &=&
\left( 1 +  \frac{2 y^2 \varepsilon^2}{Y_+}\right) F_2^{\rm NC}
\mp 
\frac{Y_-}{Y_+} \, x  F_3^{\rm NC}
-
\frac{y^2}{Y_+} F_L^{\rm NC}
\ ,
\label{eq:dsigNCxyRED}
\end{eqnarray}
with $F_L = r^2 F_2 - 2 x F_1$,
$r=\sqrt{1+4\varepsilon^2 }$, 
$\varepsilon=(xM/Q)$
and $Y_\pm = 1\pm (1-y)^2$;
the $r^2$ term in $F_L$ is essential to obtain the correct sign for the $\varepsilon^2$ term
of Eq.~\eqref{eq:dsigNCxyRED}.
Note that at LO ($F_L\sim 0$) for parity-conserving photon exchange ($F_3=0$),
so that we have $\sigma_{\rm Red.}^{NC}$ simply reduces to $F_2^{\rm NC}$.

\paragraph{Charged current cross section:} %
For the charged current process, we have:
\begin{eqnarray}
\frac{d^2\sigma^{\rm CC}}{dx \, dy} 
&=&
x(s-M^2) \, 
\frac{d^2 \sigma^{\rm CC}}{dx dQ^2} 
=
\frac{G_F^2}{2\pi x} \ 
\frac{Q^2}{y} \ 
\left( \frac{M_W^2}{Q^2+M_W^2}\right)^2 \
\left[ 
\rule[-4mm]{0mm}{0mm}  %
\sigma_{\rm Red.}^{CC} 
\right]
\ ,
\label{eq:dsigCCxy}
\\[10pt]
\sigma_{\rm Red.}^{CC} &=&
\frac{Y_+}{2} \
\left( 1 +  \frac{2 y^2 \varepsilon^2}{Y_+}\right) F_2^{\rm CC}
\mp 
\frac{Y_-}{2} \, x  F_3^{\rm CC}
-
\frac{y^2}{2} F_L^{\rm CC}
\label{eq:dsigCCxyRED}
\end{eqnarray}
Here we have followed the HERA convention\cite{H1:2015ubc} 
where the  CC differs from NC by  a factor of $Y_+/2$.

For  both the NC and CC cases, 
the factor of $2y^2\varepsilon^2/Y_+$ 
with $\varepsilon=(xM/Q)$
multiplying $F_2$ 
arises directly from the DIS kinematics; however, this is often neglected 
as it is small for typical HERA kinematics.

\subsubsection*{TMCs for nuclear reduced cross sections}

As described in the preceding subsections, TMCs can have a large numerical impact on nuclear structure functions in the low $Q^2$ and large $x$ double limit. 
To illustrate the TMC impact,
in  Fig.~\ref{fig:ntmc_ReducedXSec} 
we present results for experimental scenarios including HERA, JLab, EIC, and LBNF. 
The upper panels of  Fig.~\ref{fig:ntmc_ReducedXSec} display the 
reduced cross sections, and the lower panels
show the ratio of the full ``TMC'' to the ``No-TMC'' result. 
As before, all the nuclear isotopes  
of Table~\ref{tab:isotopeList} are within the colored band, 
and the proton and neutron are displayed separately. 

In  Fig.~\ref{fig:ntmc_ReducedXSec}-(a) we show 
the HERA-inspired $e^+ p$ neutral current  $\sigma_{Red}^{NC}$ at $\sqrt{s}=318\GeV$
at \mbox{$Q=1.3$~GeV}  for  the full ``TMC''  results.
Although HERA ran proton beams, we also show the nuclear results which may be 
useful for a future LHeC/FCCeh facility. 
The TMCs yield a slight suppression
at intermediate $x$ values ($\sim0.4$), and an enhancement at larger $x$ values 
($\gtrsim 0.6$) which can be significant. 
The behavior of $\sigma^{NC}_{Red.}$ strongly resembles  that of $F_2$ in Fig.~\ref{fig:ntmc_AbsStr_vs_x_proton}; this is expected  as  
given our choices of $Q$ and $E_{\rm CM}$ we find
$y\ll 1$ and $Y_-\sim0$,
so we see from Eq.~\eqref{eq:dsigNCxyRED} that the $F_2$ contribution dominates.
For the higher $Q$ values (not shown)  
TMCs are reduced as for $F_2$ in Fig.~\ref{fig:ntmc_AbsStr_vs_x_proton},
and are generally  negligible except in the very large $x$ region.

In  Fig.~\ref{fig:ntmc_ReducedXSec}-(b) we show 
a  JLab-inspired $e^- A$ neutral-current $\sigma_{Red}^{NC}$
with $E_{\rm Beam}=24\GeV$ at $Q=1.3\GeV$.
The character of the curves for $E_{\rm Beam}=12\GeV$ (not shown)
is quite similar to the displayed results. 
The results here are qualitatively similar to the 
$F_2$ result of Fig.~\ref{fig:ntmc_AbsStr_vs_x_proton} and the 
HERA case above 
with a reduction for intermediate $x$ ($\sim0.4$), and an enhancement at larger $x$.

In  Fig.~\ref{fig:ntmc_ReducedXSec}-(c) we show 
an  EIC-inspired $e^- A$ neutral-current $\sigma_{Red}^{NC}$
with $E_{cm} = 140\sqrt{Z/A}\GeV$ at $Q=1.3\GeV$.
As with the above HERA and JLab cases, these results 
largely follow the $F_2$ structure function.

In  Fig.~\ref{fig:ntmc_ReducedXSec}-(d) we show 
an LBNF-inspired $\nu- A$ charged-current scattering with $E_{\rm beam}=10\GeV$ 
at $Q=1.3$\GeV. In the low $x$ region, we see the various nuclear curves 
separate due to the differing nuclear mass values.

From this exercise, one can see that for large $\sqrt{s}$,
we generally have small $y$, and $\sigma_{Red}$ closely follows $F_2$. 
For smaller $\sqrt{s}$, we have larger $y$, 
and the $F_3$ and $F_L$ structure functions can now contribute.

\section{Conclusions}  \label{sec:conclusons}

The structure of the proton has been extensively investigated, and
recent work has extended the precision to
next-to-next-to-leading-order (NNLO) and beyond.  With the upcoming
Electron Ion Collider (EIC) and other experiments discussed in the
introduction, we have the opportunity to build upon the proton
analyses and also determine the nuclear structure with extreme precision.

To exploit the current and future nuclear measurements, and to obtain
the desired accuracy of theory calculations, it is essential to
include TMCs in the analyses of structure functions.
However, extending the TMCs from the proton to nuclei requires a
rigorous theoretical foundation.  This foundation is provided by the
OPE.  Additionally, as we have shown, the equivalence of the OPE with
the parton model (with appropriate $M_A$ and $k_T$ contributions) 
provides a familiar framework to implement TMC in analyses.
Some of the tools from the proton analyses can be applied directly to the nuclear case,
while other elements (such as the nuclear DGLAP equations with $x_N\in[0,A]$
and $W_A\not= W_N$) must be adapted.  
Ultimately, the answer to the question,
\textit{Why nucleon-like expressions could possibly be valid for nuclear targets?}, 
is simply that the consequences of asymptotic freedom, 
e.g., light-cone dominance and the validity of perturbative methods,
begin to emerge when momentum transfers $Q$ surpass the non-perturbative scale $\Lambda_{\rm NP}\lesssim1\GeV$ (see Sec.~\ref{sec:light-cone-dominance}).
While this is (accidentally) comparable to the scale of free nucleons,   masses of target hadrons in DIS 
(albeit nucleons or nuclei) 
are a sub-leading ingredient and contribute as powers of $\mathcal{O}(x_A^2 M_A^2/Q^2)$.
The goal of this presentation is to
explicitly delineate the proper ingredients and adaptations required 
for a precision nuclear analysis.

The present  investigation  provides  both pedagogical material and 
a practical reference for computing and investigating 
TMCs for nuclear processes.
We briefly outline the key components and results in the following:

\begin{itemize}
    \item \fhead{OPE Formalism:}
    We extend the OPE formalism to the case of interacting nuclei
    and explicitly provide the correspondence between TMC expressions for a nucleon 
    and a nucleus, given in Eq.~\eqref{eq:master_full}. To derive these relations (see Sec.~\ref{sec:buildingOPE} and Appendix~\ref{app:nTMC_derivation}), we use only fundamental symmetry principles
    of the nuclei, and make no assumptions about the parton or other models.
    This is the essence of the factorization between the hadron and parton level processes. 

 \item \fhead{TMC Power Expansion:}
 We investigate analytically and numerically the magnitude of TMCs and show it is     the quantity $(x_A M_A/Q)=(x_N M_N/Q)$ that enters the TMC formula for \textbf{nuclei} (Eq.~\eqref{eq:master}) and \textbf{nucleons} (Eq.~\eqref{eq:master-rescaled}).
 This result is important for heavy nuclear $A$ targets such as lead. 
While $M_A$ can be large (${\sim} A\, M_N$), it is always combined with $x_A \lesssim 1/A$
(since an individual parton momentum fraction will typically be less than $1/A$ of the total 
\textbf{nuclear} momentum).
Thus, the OPE  expansion  for TMCs is effectively in powers of $(M_N/Q)$ and not $(M_A/Q)$, so 
it is equally accurate for both protons and nuclei.

    \item \fhead{Nuclear TMCs:}
    In Sec.~\ref{sub:master}, we extend the proton results of Ref.~\cite{Schienbein:2007gr}
    and derive a set of master equations (Eq.~\eqref{eq:master_full}) for nuclear structure functions with TMCs
    that  are the same for all $A$, including the proton,  neutron, and nuclear cases.
    This allows us to easily apply the tools of the OPE to nuclei as well as protons.

    \item \fhead{Nucleon TMCs:}
    In Sec.~\ref{sec:rescaled_str_fn}, 
    we rescale our \textbf{nuclear} structure functions of Eq.~\eqref{eq:master_full}
    to construct \textbf{nucleon}    structure functions in Eq.~\eqref{eq:master-rescaled}.
    These  rescaled structure functions are the ones commonly studied and presented in comparisons between different nuclear data. 
    This facilitates  extraction of the  ``nuclear correction factors'' 
    by taking ratios of  rescaled structure functions, as illustrated in 
    Fig.~\ref{fig:f2emc}. Additionally, this allows for investigations of collective nuclear 
    effects ({\it e.g.}, shadowing, anti-shadowing, EMC, Fermi, etc.).

    \item \fhead{The Parton Model:}
    In Sec.~\ref{sub:dglap}, we extend the rescaled results to  the parton model and 
    obtain the corresponding set of DGLAP equations and associated sum rules,
    including isospin symmetry.
    We also identify the complication of separately extracting $u(x)$ and $d(x)$ nPDFs.
    This correspondence allows us (with the reasonable assumption  $x_N\leq 1$)
    to  re-use our proton DGLAP evolution code for the case of nucleons.
    
    \item  \fhead{Nuclear Kinematics:}
    We establish the relations between the kinematic variables including 
    the \textbf{nuclear} scaling variables $\{x_A, \xi_A\}$ and 
    the \textbf{nucleon} scaling variables $\{x_N, \xi_N\}$. 
    Furthermore, we indicate in Sec.~\ref{sec:wcut} some subtle complications we encounter such as defining an appropriate $W^2$ cut  for the nucleon case. 
    Specifically, as demonstrated in Fig.~\ref{fig:jlabWplot}, 
    the dependence of $W_A^2$  on $A$ and $Q^2$,  complicate the task of placing cuts on the resonance region.

    \item  \fhead{The OPE and the Parton Model:}
    We obtain a complete correspondence
    between the parton model and OPE structure function results,
    and demonstrate these are equivalent \textbf{if} we include 
    i)~target mass effects and 
    ii)~$k_T$     transverse momentum    contributions in the parton model. 
    Additionally, in Sec.~\ref{sec:acot} we show the utility 
    of the  ACOT helicity approach with light-cone kinematics, 
    and demonstrate the correspondence to the traditional tensor basis
    as shown in Table~\ref{tab:helicity_expansion}.

    \item \fhead{Numerical Comparisons:}
    In Sec.~\ref{sec:num}, we provide a variety of numerical comparisons to demonstrate the  size of the TMCs in various kinematic regions. 
    These comparisons are shown for both structure functions, and the reduced cross section for some realistic machine kinematics. 
    We found that for typical kinematic cuts used in global analyses 
    $\{Q\gtrsim 2~{\rm GeV}, W \gtrsim 1.7~{\rm GeV},x\lesssim 0.7 \}$,
    the $F_i^{\rm TMC-Leading}$ was within $\sim 5\%$ or better of the full $F_i^{\rm TMC}$ result, and improves quickly to $\sim 1\%$ or better for 
    $Q\gtrsim 6$~GeV.

    \item \fhead{TMC Computation:}
    We also present numerically the dependence of the TMCs on the nuclear $A$ value and the exchanged vector boson $\{\gamma,Z,W^\pm\}$.
    We find the relative magnitude of the TMCs is  weakly dependent on the exchanged boson, and \textbf{very} weakly dependent on the nuclear $A$.
    This latter observation allows us to derive an approximate (and computationally expedient) parameterization for the nuclear TMCs 
    suitable for including inside a global fit function (see Sec.~\ref{sec:parm}). 
    For $Q>2$~GeV, the parameterized result matches the exact results 
    to better than $0.2\%$, which is well below the typical nPDF precision. 
    
\end{itemize}

In the near future, there will be ambitious new facilities  and experiments  that will investigate 
lepton-nucleus and neutrino-nucleus interactions
in extreme kinematic regions with high precision.
Just as Ref.~\cite{Schienbein:2007gr} provided the details for the proton case, 
this report offers foundational and reference material that can facilitate
incisive investigations of nuclear phenomena.

We can use the tools from the  precision proton analyses as an archetype, and extend them into the nuclear dimension. 
Furthermore, the nuclear interactions open up a whole new class of phenomena (saturation, shadowing)
and new states of matter (quark-gluon plasma,  Color Glass Condensates) to explore. 
Discovering and analyzing these phenomena will require precision on both the experimental and theoretical fronts;
a rigorous foundation for theoretical calculations of structure functions with TMCs is an essential step toward this goal.

\vspace{1cm}
\goodbreak
\newpage
\section*{Acknowledgments}
The authors would like to thank 
Eric~Christy,
John,~C.~Collins,
Stefan~Krezer, 
Pavel Nadolsky, 
Voica~Radescu,
and
Mary~Hall~Reno
for many useful comments and discussions. 

A.K. and R.R. acknowledge the support of  under
Grant No. 2019/34/E /ST2/00186.
R.R also acknowledges the support of the Polska Akademia Nauk
(grant agreement PAN.BFD.S.BDN. 613. 022. 2021 - PASIFIC 1, POPSICLE).
This work has received funding from the European Union's Horizon 2020 research and innovation program under the Sk{\l}odowska-Curie grant agreement No.  847639, and from the Polish Ministry of Education and Science.
The work at WWU M{\"u}nster was funded by the DFG through the Research Training Group 2149 ``Strong and Weak Interactions - from Hadrons to Dark Matter'' and the SFB 1225 ``Isoquant,'' {project\nobreakdash-id} 273811115. 
M.K.~thanks the School of Physics at the University of New South Wales in Sydney, Australia for its hospitality and financial support through the Gordon Godfrey visitors program.
The work of A.A. was supported by the U.S. Department of Energy under Grant No. DE-SC0008791.
The work of T.J.H.~was supported by the U.S.~Department of Energy under contract DE-AC02-06CH11357.
The research of P.~D. was funded as a part of the Center of Excellence in Quark Matter of the Academy of Finland (project 346326).
The work of  C.K.~was supported by the  U.S.\ Department of Energy contract DE-AC05-06OR23177, under which Jefferson Science Associates LLC manages and operates Jefferson Lab. 
J.G.M. has been supported by Fermi Research Alliance, LLC under Contract No. DE-AC02-07CH11359 with the U.S. Department of Energy, Office of Science, Office of High Energy Physics.
The work of F.O.~was supported by the U.S.~Department of Energy under Grant No.~DE-SC0010129, 
and  by the  Office of Science, Office of Nuclear Physics, within the framework of the Saturated Glue (SURGE) Topical Theory Collaboration.
The work of I.S.\ was supported by the French CNRS via the IN2P3 project GLUE@NLO.
\clearpage
\appendix
\renewcommand*{\thesection}{\Alph{section}} %

\section{Nuclear structure functions with TMCs in the OPE}\label{app:nTMC_derivation}

In this appendix, we build structure functions for massive, unpolarized nuclear targets following a modified prescription of Ref.~\cite{Georgi:1976ve,Georgi:1976vf}.
The purpose of this is two fold.
First is to establish that the TMC formulae used for free nucleons are applicable to nuclei, which is a key conclusion of this work.
Second is to provide a explicit derivation of the TMC master formulae, which are summarized in 
Eqs.~\eqref{eq:master}
and \eqref{eq:app_summary}, with details beyond what is generally available in the literature. 
It is well-documented that constructing TMCs to structure functions is a tedious endeavor, with many intermediate steps omitted from the literature. Some notable exceptions exist~\cite{Georgi:1976ve,Georgi:1976vf,Kretzer:2003iu,Detmold:2005iz,Schienbein:2007gr,Steffens:2012jx}. For example: Refs.~\cite{Georgi:1976ve,Georgi:1976vf} give a largely complete treatment of $F_2$ but implicitly employ
derivative, integral, and summation identities.
Some of these intermediate steps are more completely documented in Refs.~\cite{Detmold:2005iz,Steffens:2012jx},
but other steps may not be fully justified under their assumptions~\cite{Collins:1984xc}.
Similarly, while the structure functions $W_4$ and $W_5$ (or equivalently $F_4$ and $F_5$) are documented in Refs.~\cite{Kretzer:2003iu,Schienbein:2007gr}, neither works out the many permutations 
of contractions with
the momentum factor $\Pi^{\mu_1\dots\mu_{2k}}$ from Eq.~\eqref{eq:pi}.
Keeping track of these permutations requires care, as demonstrated for $F_1$ and $F_2$ in Ref.~\cite{Steffens:2012jx}. Notably, Ref.~\cite{Schienbein:2007gr} documents some differences and similarities between the above treatments. The treatment in this Appendix builds on the different steps employed throughout  
Refs.~\cite{Georgi:1976ve,Georgi:1976vf,Collins:1984xc,Kretzer:2003iu,Detmold:2005iz,Muta:2010xua,Muta:1998vi,Steffens:2012jx}.

We first build a formula for the inclusive DIS cross section
in App.~\ref{app:xsec_intro}
based on the notation defined in section~\ref{sec:variables}.
In App.~\ref{app:matrixelement}, we derive the matrix element, squared matrix element, and inclusive hadronic tensor in DIS.
In App.~\ref{app:ope}, the OPE is applied to the inclusive hadronic tensor, with the leading-power result organized according to Wilson coefficients.
In App.~\ref{app:opeMixing}, 
OPE terms are reorganized by Lorentz structures, resulting in structure-function mixing.
Structure functions for massless and massive nuclear targets are then constructed and summarized in App.~\ref{app:opeMassive}.
\subsection{Preliminaries and the inclusive DIS cross section formula}\label{app:xsec_intro}

To build nuclear structure functions with TMCs in the OPE formalism, we consider the DIS process as defined in Eq.~\eqref{eq:setup_dis_def} of section~\ref{sec:variables} and depicted in Fig.~\ref{fig:dis}.
Given this configuration, the total cross section for a generic $2\to(n+1)$ scattering process is given by the formula
\begin{subequations}
 \label{eq:xsecDef}
\begin{align}
 \sigma(\ell_1 + A \to \ell_2+X_n) &= \int dPS_{n+1} \frac{d\sigma}{dPS_{n+1}}\ ,
 \qquad  {\rm with } \qquad
 \frac{d\sigma}{dPS_{n+1}} = 
 \frac{1}{\mathcal{F}}
 \frac{1}{\mathcal{S}_{\ell_1}\mathcal{S}_A} \sum_{\rm \{\lambda\}}\sum_{\rm dof} \vert \mathcal{M} \vert^2\ ,
 \\
 \text{and}\quad
\mathcal{F} &= 4 \sqrt{(p_A\cdot k_1)^2 - p_A^2 k_1^2}~\ 
\overset{k_1^2=m_1^2\to 0}{\approx} \
2(s-M_A^2)\ .
\end{align}
\end{subequations}
Here, $\mathcal{F}$ is the Moller flux factor and reduces to $\mathcal{F}\approx 2(s-M_A^2)$ when the mass $k_1^2 = m_1^2$ vanishes.
$d\sigma/dPS_{n+1}$ is the totally differential, nucleus-level cross section.
$\mathcal{S}_{\ell_1}$ and $\mathcal{S}_A$ are the spin multiplicities for $\ell_1$ and $A$ in unpolarized scattering. The double summations in the differential cross section run over external helicities $\{\lambda\}$ and other discrete quantum numbers/degrees of freedom (dof). 
We introduce the subscript $n$ on $X_n$ to make explicitly that it is an $n$-body state.
In addition, $\mathcal{M}$ is the scattering amplitude for the $2\to(n+1)$  process. The associated $(n+1)$-body phase space measure is given by
\begin{align}
 dPS_{n+1}(k_1+p_A; k_2, p_X) =&  (2\pi)^4 \delta^4\left(k_1+p_A - k_2 - \sum_{i=1}^n p_{Xi}\right)
~ \frac{d^3k_2}{(2\pi)^3 2k^0_2}
 \times \prod_{i=1}^n \frac{d^3 p_{Xi}}{(2\pi)^3 2p^0_{Xi}}
 \label{eq:dPSdef}
 \\
 = &\  dPS_{n}(q+p_A; p_X)\
 \times\ \frac{d^3k_2}{(2\pi)^3 2k^0_2}
 \\
  = &  \int d^4z\  e^{i(q+p_A-p_X)\cdot z}\  
 \prod_{i=1}^n \frac{d^3 p_{Xi}}{(2\pi)^3 2p^0_{Xi}}\ 
 \times 
 \frac{d^3k_2}{(2\pi)^3 2k^0_2}\ ,
\end{align}
where  $k^0_2$ and $p^0_{Xi}$ are the energy components of the 
$k_2$ and $p_{Xi}$ 4-vectors, respectively.

In the above lines, we employed momentum conservation and $\delta$-function identities to rewrite the measure in ways that will be used below. Specifically in the third line, we used 
\begin{equation}
 (2\pi)^4\ \delta^4(q-p)\ =\ 
 \int d^4z\  e^{i(q-p)\cdot z}\ ,
 \label{eq:deltaFnID}
\end{equation}
where $q, p$ and $z$ are 4-vectors.
The point of these identities is to identify the phase space integral over the hadronic matrix element also as the Fourier transformation 
over composite operators. 
The modeling of $X_n$'s phase space is not precisely correct in many 
presentations of structure functions in DIS. 
Often, $p_X$ is treated as a one-body configuration despite
being 
an $n$-body configuration. While this mis-modeling is ultimately a technicality, it does impact the formal definition of $W_{\mu\nu}$; see, for instance, Ref.~\cite{Becher:2013xia}.
We reiterate that $X_n$ here (or $X$ in the main text) is an $n$-body state. 

Using the above phase space decomposition, we can write 
an expression for the  DIS cross section  
that (1)~is differential with respect to the kinematics of $\{ \ell_2, k_2 \}$,
and (2)~is explicitly inclusive over \textit{all} hadronic activity.
This expression is given by
\begin{align}
k^0_2 \times \frac{d\sigma}{d^3k_2} &= 
\frac{1}{(16\pi^2)} \frac{1}{\pi}
\frac{1}{\mathcal{F}}
\frac{1}{\mathcal{S}_{\ell_1}\mathcal{S}_A} ~
\sum_{n=1}^\infty 
\int d^4z ~ e^{i(q+p_A-p_X)\cdot z}  
  \prod_{i=1}^n \frac{d^3 p_{Xi}}{(2\pi)^3 2p^0_{Xi}}
 \sum_{\rm \{\lambda\}}\sum_{\rm dof} \vert \mathcal{M} \vert^2\ .
 \label{eq:dxsecDef}
\end{align}
Here, $\mathcal{S}_{\ell_1}=2$ for an unpolarized $\ell_1$, 
and $\mathcal{S}_{A}=2$ for an unpolarized, spin-1/2 nuclei~$A$. 
Additionally, $\mathcal{S}_{\ell_1}=1$ when $\ell_1$ is an incoming neutrino or antineutrino because neutrino beams are effectively 100\% polarized. 
More broadly, for unpolarized nuclear targets with spin $s_A > 1/2$, the spin-averaging factor $\mathcal{S}_{A}$ generalizes to $\mathcal{S}_{A}=(2 s_A+1) > 2$.

In Eq.~\eqref{eq:dxsecDef}, we inserted a sum over the multiplicity $n$ of $X_n$.
Conceptually, this is equivalent to the relationship
\begin{equation}
\label{eq:inclusive_def_app}
     \ell_1 ~+~ A ~\to~ \ell_2 ~+~ \text{any hadronic activity} = 
     \sum_{n=1}^\infty \left[ \ell_1 + A \to \ell_2 + X_n\right]\ .
\end{equation}
For example,  when $n=2$ this sums over the configurations $pp, pn, p\pi, p K, \pi \pi, \ldots$, which are all included in the inclusive sum.

\subsection{The DIS matrix element  and inclusive hadronic tensor} \label{app:matrixelement}

The hadronic tensor $W_{\mu\nu}^A$ of Eq.~\eqref{eq:wmunu} is built from the (squared) matrix element $\mathcal{M}$.
For $\ell_1 A\to\ell_2 X$ scattering via an arbitrary electroweak boson $V(q)$,
we parameterize this by
\begin{align}
 -i\mathcal{M} &=\ 
 \langle X \ell_2 \vert J^\mu_{\ell_2\ell_1}(0) 
 \cdot \Delta_{\mu\sigma}^V(q) \cdot 
 J^\sigma_{XA}(0) \vert A\ell_1\rangle
 \\
 &=\
 \langle \ell_2 \vert J^\mu_{\ell_2\ell_1}(0) \vert \ell_1\rangle
 \cdot \Delta_{\mu\sigma}^V(q) \cdot 
 \langle X \vert J^\sigma_{XA}(0) \vert A\rangle,
 \label{eq:appMEDef}
\end{align}
where the leptonic part of the matrix element $L^\mu$ and current $J^\mu_{\ell_2\ell_1}$  
in terms of 4-component helicity spinors, a generic coupling normalization $\tilde{g}$, and chiral couplings $g_V,g_A,$ 
are 
\begin{equation}
 L^\mu ~\equiv~ \langle \ell_2 \vert J^\mu_{\ell_2\ell_1}(0) \vert \ell_1\rangle
 ~=~ -i\tilde{g} ~ \overline{u}(k_2,\lambda_2)\left[
 g_V^{\ell}\gamma^\mu + g_A^{\ell}\gamma^\mu\gamma^5 \right]u(k_1,\lambda_1).
\end{equation}
We assume arbitrary vector and axial-vector couplings between the $(\ell_2\ell_1)$ system and $V$ in order to accommodate the possibility of $V$ being a photon or weak boson. The conversion between the different possibilities is given in Tab.~\ref{tab:coup}. Similarly, we write the hadronic part of the matrix element in terms of an arbitrary hadronic current $J^\sigma_{XA}(0)$: 
\begin{equation}
 H^{A \sigma} ~\equiv~  \langle X \vert J^\sigma_{XA}(0) \vert A\rangle ~\equiv~ \overline{u}_X(p_X,\lambda_X)[\dots]^\sigma u(p_A,\lambda_A).
\end{equation}
We assume here and below that $J^\sigma_{XA}$ is always a renormalized object in QCD.

\begin{table}[!t]
\begin{center}
\resizebox{\textwidth}{!}{
\renewcommand*{\arraystretch}{0.95}
\begin{tabular}{c|c|c|c|c|c}
\hline\hline
\multirow{2}{*}{Vertex} & Coupling  & \multirow{2}{*}{$g_R^\ell$}  & \multirow{2}{*}{$g_L^\ell$}  & \multirow{2}{*}{$g_V^\ell$}  & \multirow{2}{*}{$g_A^\ell$} \\
	& strength	&	& &  & \\
\hline\hline
$V-\ell_1-\ell_2$	& $\tilde{g}$				& $(g_V^\ell+g_A^\ell)$		& $(g_V^\ell-g_A^\ell)$		& $\frac{(g_R+g_L)}{2}$ 	& $\frac{(g_R-g_L)}{2}$ \\
$\gamma-\ell-\ell$	& $e Q^\ell$				& $1$				& $1$				& $1$				& $0$\\
$Z-\ell-\ell$	& $\frac{g_W}{\cos\theta_W}$	& $-Q^\ell\sin^2\theta_W$	& $(T_3^\ell)_L-Q^\ell\sin^2\theta_W$				& $\frac{1}{2}(T_3^\ell)_L-Q^\ell\sin^2\theta_W$ 			& $-\frac{1}{2}(T_3^\ell)_L$ \\
$W-\ell_1-\ell_2$	& $\frac{g_W}{\sqrt{2}}$			& $0$				& $1$				& $\frac{1}{2}$ 			& $-\frac{1}{2}$ \\
\hline\hline
\end{tabular}
}
\caption{Electroweak chiral couplings and coupling strength normalizations used for fermions $\ell_1,\ell_2$ with weak isospin charge $(T_3^\ell)_L=\pm1/2$ and electric charge $Q^\ell$, with normalization $Q^e=-1$.
}
\label{tab:coup}
\end{center}
\end{table}

In Eq.~\eqref{eq:appMEDef}, $\Delta_{\mu\sigma}^V(q)$ is the propagator of $V$
which is typically of the form 
$ -i (g_{\mu\sigma}-q_\mu q_\nu/M_V^2)/(q^2-M_V^2)$.
When we assume the $\ell_i$ are massless, the Dirac equation and SU$(2)_L$ invariance ensure that the longitudinal modes of $\Delta_{\mu\sigma}^V$ do not contribute to the scattering process.\footnote{%
We typically neglect the mass of the leptons, except for the tau-lepton.
} %
More specifically, 
the longitudinal component of $\Delta$ scales as $\Delta_{\mu\sigma}(q)\vert_{\rm long} \propto q_\mu q_\sigma$. 
Since $\not k_i u(k_i) = m_i u(k_i) = 0$
and $q_\mu = (k_1 - k_2)_\mu$,
one finds
\begin{align}
L^\mu \cdot \Delta_{\mu\sigma}^V \vert_{\rm long} 
& ~\propto~ q_\mu \cdot \overline{u}(k_2,\lambda_2)\left[
 g_V^{\ell}\gamma^\mu + g_A^{\ell}\gamma^\mu\gamma^5 \right]u(k_1,\lambda_1)  
 \\
 & ~=~ \overline{u}(k_2,\lambda_2)\left[
 g_V^{\ell}(\not k_1 - \not k_2) + g_A^{\ell}(\not k_1 - \not k_2)\gamma^5 \right]u(k_1,\lambda_1) =0.
\end{align}

The above implies that only the transverse components of $\Delta_{\mu\sigma}^V$ contributes to DIS for massless leptons. For $V=\gamma,W,Z$ with mass $M_V$ (where $M_\gamma=0$), one can then write
\begin{equation}
\label{eq:appMESimple}
-i\mathcal{M} = 
 L^\mu(k_2,k_1) \Delta_{\mu\sigma}^V(q) H^{A \sigma}(p_X,p_A) = 
 L^\mu \frac{-i g_{\mu\sigma}}{q^2 - M_V^2} H^{A \sigma} = \frac{-i}{q^2 - M_V^2} L^\mu  H_\mu^A.
\end{equation}
After squaring and summing over external helicity states, one obtains
\begin{equation}
 \sum_{\{\lambda\}} \vert \mathcal{M}\vert^2\ 
 =\  
 \frac{1}{(q^2-M_V^2)^2}\ 
 \sum_{\{\lambda\}} L^{\mu\nu}\ \cdot\  \sum_{\{\lambda\}}H_{\mu\nu}^A\ .
 \label{eq:squaredMEDIS}
\end{equation}
The leptonic tensor $L^{\mu\nu}$ denotes the square of $L^\mu$ and can be solved exactly using trace technology, helicity amplitudes, or other standard technology. It is given by
\begin{align}
 \sum_{\{\lambda\}} L^{\mu\nu} = & 
 \sum_{\{\lambda\}}  \left(L^\dagger\right)^\mu L^{\nu} 
 = 
 \sum_{\{\lambda\}} 
  \langle \ell_1 \vert \left(J_{\ell_2\ell_1}^\dagger(0)\right)^\mu \vert \ell_2\rangle
  \langle \ell_2 \vert J^\nu_{\ell_2\ell_1}(0) \vert \ell_1\rangle
  \\
  = & 4\tilde{g}^2 \Big\{(g_V^2+g_A^2)(k_1^\mu k_2^\nu + k_1^\nu k_2^\mu)
 - \left[(g_V^2+g_A^2)(k_1\cdot k_2) -  (g_V^2-g_A^2)m_1m_2\right]g^{\mu\nu}
\nonumber\\
& \qquad 
- 2i (g_V g_A) k_{1\alpha}k_{2\beta}\epsilon^{\mu\nu\alpha\beta}
\Big\}
 \label{eq:leptonicTensorFull}
 \\
  \overset{m_1,m_2\to0}{=}
&  4\tilde{g}^2 \Big\{(g_V^2+g_A^2)(k_1^\mu k_2^\nu + k_1^\nu k_2^\mu)
 - (g_V^2+g_A^2)(k_1\cdot k_2)g^{\mu\nu}
\nonumber\\
& \qquad 
- 2i (g_V g_A) k_{1\alpha}k_{2\beta}\epsilon^{\mu\nu\alpha\beta}
\Big\}
  \label{eq:leptonicTensorMassless}
\end{align}

For completeness, we provide the expression for $\sum_{\{\lambda\}} L^{\mu\nu}$ with massive leptons. If using this formula, note that  Eq.~\eqref{eq:appMESimple} must be modified since $\not k_i u(k_i) = m_i u(k_i) \neq 0$. For the remainder of this work, we keep $m_1,m_2 = 0$. Note that the expression here for the leptonic tensor only sums over initial- and final-state polarizations; a symmetry factor of $
1/\mathcal{S}_{\ell_1}=1/2$ for unpolarized initial-states has not yet been applied. Following the conventions of Table~\ref{tab:coup}, for the case of QED, where $g_V=1,~g_A=0$, and $\tilde{g}=e$, we recover the usual expression:
\begin{equation}
     \sum_{\{\lambda\}} L^{\mu\nu}\Big\vert_{\rm QED} = 4e^2  \Big\{k_1^\mu k_2^\nu + k_1^\nu k_2^\mu
 - (k_1\cdot k_2)g^{\mu\nu}\Big\}.
 \label{eq:app_lepTensor_qed}
\end{equation}
For the case of $W$ boson exchange, where $g_V=1/2,~g_A=-1/2,$ and $\tilde{g}=g_W/\sqrt{2}$, we have
\begin{equation}
    \sum_{\{\lambda\}} L^{\mu\nu}\Big\vert_{W} = 
    g_W^2\Big\{k_1^\mu k_2^\nu + k_1^\nu k_2^\mu
 - (k_1\cdot k_2)g^{\mu\nu} + i k_{1\alpha}k_{2\beta}\epsilon^{\mu\nu\alpha\beta}
\Big\}.
\label{eq:app_lepTensor_ew}
\end{equation}

The \textit{exclusive} hadronic tensor $H_{\mu\nu}^A$ for an $n$-body final state $X$ denotes the square of $H_\mu$. 
For a fixed $n$, it is given by
\begin{align}
 \sum_{\{\lambda\}} H_{\mu\nu}^A = \sum_{\{\lambda\}}  \left(H^{A\dagger}\right)_\mu H_{\nu}^A 
& = \sum_{\{\lambda\}}
 \langle A \vert \left(J_{XA}^\dagger(0)\right)_\mu \vert X\rangle 
 \langle X \vert J_{XA\nu}(0) \vert A\rangle\ . 
\end{align}
From $H_{\mu\nu}^A$ we can build the \textit{inclusive} hadronic tensor $W_{\mu\nu}^A$ that sums over all final states.
\subsection*{Building the inclusive hadronic tensor}

In inclusive DIS, one measures the kinematics of $\ell_2$ and remains totally inclusive regarding the hadronic system~$X$. (Though in practice, the invariant mass of $X$ is often measured.) The inclusiveness criterion means summing over all possible final-state multiplicities $n$ (where $n$ is same $n$ in Eq.~\eqref{eq:dPSdef}), discrete dof, and continuous dof, i.e., momenta. 
Performing this summation allows us to define the \textit{inclusive} hadronic tensor as
\begin{align}
 W_{\mu\nu}^A ~\equiv~ &  \frac{1}{4\pi}\sum_{n=1}^\infty \int dPS_n(q+p_A; p_X)\  \sum_{\rm dof}  \sum_{\{\lambda\}} H_{\mu\nu}^A
 \\
 ~=~ & \frac{1}{4\pi}
\int d^4z ~ e^{i(q+p_A-p_X)\cdot z}\ \sum_{n=1}^\infty   \int \prod_{i=1}^n \frac{d^3 p_{Xi}}{(2\pi)^3 2p^0_{Xi}}
\nonumber\\
&
\qquad ~ \qquad ~ \qquad ~ \quad  \times  \sum_{\rm dof}  \sum_{\{\lambda\}}
 \langle A \vert J_{XA\mu}^\dagger(0) \vert X\rangle 
 \langle X \vert J_{XA\nu}(0) \vert A\rangle
 \\
 ~=~ & \frac{1}{4\pi}
\int d^4z ~ e^{i(q+p_A-p_X)\cdot z}\
\sumint 
 \langle A \vert J_{XA\mu}^\dagger(0) \vert X\rangle 
 \langle X \vert J_{XA\nu}(0) \vert A\rangle.  
 \label{eq:translation}
\end{align}
In the second line, we used the phase space identities of Eq.~\eqref{eq:dPSdef}.
Following Ref.~\cite{Becher:2013xia} (and related references), we denote the $3^n$-dimensional integration over momenta and triple summation by $\sumint$. Importantly, the integration and summation constitute a summation over all possible configurations spanning the space of $\{\vert X \rangle\}$. Completeness dictates that
\begin{equation}
\sumint \vert X\rangle  \langle X \vert = \sum_{n=1}^\infty
 \int \prod_{i=1}^n \frac{d^3 p_{Xi}}{(2\pi)^3 2p^0_{Xi}} 
 ~ \sum_{\rm dof}  \sum_{\{\lambda\}} ~ \vert X\rangle  \langle X \vert = 1.
 \label{eq:hComplete}
 \end{equation}

Before employing Eq.~\eqref{eq:hComplete}, we use translation invariance  to build a spacetime separation between $J_{A\mu}^\dagger$ and $J_{A\nu}$ that is constrained by momentum conservation. 
Such separations in the DIS limit are restricted to be near the light cone, i.e., with
$z^2\sim0$~({\it c.f.}, Ref.~\cite{Muta:2010xua}). 
Applying translation invariance to  $J_{XA\mu}^\dagger(0)$ gives
\begin{align}
  \langle A \vert J_{XA\mu}^\dagger(0) \vert X\rangle
  & =  \langle A \vert 
  \left[e^{-i \hat{P}\cdot z} 
  J_{XA\mu}(z) 
    e^{i \hat{P}\cdot z} \right]^\dagger
  \vert X\rangle
\\
 & =  \langle A \vert 
  e^{-i \hat{P}\cdot z} 
  J_{XA\mu}^\dagger(z) 
    e^{i \hat{P}\cdot z} 
  \vert X\rangle
    =   e^{-i (p_A-p_X)\cdot z}   \langle A \vert 
  J_{XA\mu}^\dagger(z) 
  \vert X\rangle,
\end{align}
where $\hat{P}^\alpha$ is the four-momentum operator that generates the eigenvalue equation $\hat{P}^\alpha \vert Y(p_Y)\rangle = p_Y^\alpha \vert Y(p_Y)\rangle$. 
Combining the preceding expressions leads to familiar expression for $W_{\mu\nu}^A$:
\begin{align}
 W_{\mu\nu}^A &=
 \frac{1}{4\pi} \int d^4z ~ e^{iq\cdot z}  
 \langle A \vert J_{XA\mu}^\dagger(z)\  J_{XA\nu}(0) \vert A\rangle
 \ .
 \label{eq:hadTensorDagger}
\end{align}
In electromagnetism, currents are Hermitian. This leads to the omission of the $\dagger$ in many texts.
In other texts, Eq.~\eqref{eq:hadTensorDagger} is written using a commutator of currents~\cite{Muta:1998vi,Muta:2010xua}. However, Refs.~\cite{Collins:1984xc,Collins:2011zzd} note that commutators are not necessary to define
$W_{\mu\nu}^A$. And indeed, Refs.~\cite{Muta:1998vi,Muta:2010xua} show that the commutator-based result reduce to Eq.~\eqref{eq:hadTensorDagger} in certain limits.

We are now in position to write the differential cross section for the DIS process $\ell_1 + A \to \ell_2 +$ anything in terms of the inclusive hadronic tensor $W_{\mu\nu}^A$. To do this, we take the differential cross section formula of  Eq.~\eqref{eq:dxsecDef} and insert the squared matrix element  in  Eq.~\eqref{eq:squaredMEDIS}. We then substitute the summations and integration over the exclusive hadronic tensor $H_{\mu\nu}$ with the inclusive hadronic tensor $W_{\mu\nu}$
to obtain the formula,
\begin{align}
k^0_2 \times \frac{d\sigma}{d^3k_2} &= 
\frac{1}{(16\pi^2) s} \frac{2}{\mathcal{S}_{\ell_1}\mathcal{S}_A} ~
  \frac{1}{(q^2-M_V^2)^2}
  \left(\sum_{\{\lambda\}} L^{\mu\nu}\right) \cdot  W_{\mu\nu}^A
  \ .
 \label{eq:dxsecDef_simplified}
\end{align}
This formula holds for arbitrary $A$, including unbound nucleons. Expressions for the leptonic tensor (summed over all external spins) are found in Eqs.~\eqref{eq:leptonicTensorMassless}-\eqref{eq:app_lepTensor_ew}.
The factor of $2$ in the numerator originates from the adopting the $(1/4\pi)$ normalization of $W_{\mu\nu}$; when $A$ is a spin-1/2 object, it cancels the spin-averaging factor $\mathcal{S}_A$. The inclusive hadronic tensor is given by Eq.~\eqref{eq:hadTensorDagger}, but now we relate it to the time-order matrix element for virtual Compton scattering, which has a known expansion in the OPE at leading twist.
\subsection{Nuclear structure functions from the OPE I: organization}\label{app:ope}

As summarized in Sec.~\ref{sec:buildingOPE}, building structure functions in the OPE for massless and massive hadronic targets is a multi-step process. The advantage, however, is clear: the OPE, if it holds for QCD, facilitates an all-orders, operational definitions for parton densities and structure functions. 
Moreover, the power counting within the OPE formalism allows one to organize
contributions that are not clearly captured in fixed-order perturbation theory.

To do this in the manner of Refs.~\cite{Georgi:1976ve,Georgi:1976vf}, one starts with the inclusive hadronic tensor $W_{\mu\nu}^A$ in the DIS limit (large $Q$, fixed $x_A$) as defined above and relates it to the time-ordered matrix element 
$T_{\mu\nu}^A$
for virtual Compton scattering\footnote{$\gamma^*(q) A(p_A) \to \gamma^* A$ for electromagnetic currents,
$W^*(q) A(p_A) \to W^* A$ for charged currents, etc.}
 in the short-distance limit (large $Q$, fixed $x_A/Q$). 
The OPE of $T_{\mu\nu}^A$ is then organized by Lorentz structures. By comparing the Lorentz structures of the expansions of $W_{\mu\nu}^A$ and $T_{\mu\nu}^A$, one can identify the structure functions of $W_{\mu\nu}^A$ in terms of the OPE.
When the mass of a nucleus $A$ is neglected, the OPE of $T_{\mu\nu}^A$ simplifies to familiar expressions.
The differences between the fully massive and simplified massless expressions are the TMCs to the structure functions of $W_{\mu\nu}^A$.

Following this outline, we start by taking the expression for $W_{\mu\nu}^A(p_A,q)$ in Eq.~\eqref{eq:hadTensorDagger} and decomposing it into a sum of tensor-valued coefficients multiplied by dimensionless, scalar-valued functions $\tilde{W}_{i}^A$. The $\tilde{W}_{i}^A$ are the structure functions. Lorentz symmetry and hermiticity dictate that only certain combinations of $p_A$ and $q$ are allowed as coefficients. The most general combination allowed by symmetries for unpolarized $A$ is 
\begin{align}
W^A_{\mu\nu}(p_A,q) 
\quad &
=
 \frac{1}{4\pi}  \int d^4z ~ e^{iq\cdot z}  
 \langle A \vert J_{\mu}^\dagger(z)\  J_{\nu}(0) \vert A\rangle 
 \\
&=
-g_{\mu \nu}\tilde{W}_1^A ~+~ \frac{p_{A\mu} p_{A\nu} }{ M_A^2}\tilde{W}_2^A
~-~i\epsilon_{\mu\nu\rho\sigma} \frac{p_A^{\rho}q^\sigma }{ M_A^2}\tilde{W}_3^A
~+~\frac{q_\mu q_\nu }{ M_A^2}\tilde{W}_4^A
\nonumber\\
& ~ \qquad ~\quad ~ \quad
+\frac{p_{A\mu} q_\nu + p_{A\nu} q_\mu }{ M_A^2} \tilde{W}_{5}^A
~+~\frac{p_{A\mu} q_\nu - p_{A\nu} q_\mu }{ M_A^2} \tilde{W}_{6}^A
\label{eq:strFnExpandW}
 \\
&=
-g_{\mu\nu}\tilde{F}_1^A ~+~ 
\frac{p_{A\mu}p_{A\nu}}{Q^2}  2x_A \tilde{F}_2^A 
~-~ i\epsilon_{\mu\nu\alpha\beta} \frac{ p^\alpha_A q^\beta}{Q^2}   
{x_A}
\tilde{F}_3^A ~+~ \frac{q_\mu q_\nu}{Q^2}
{2}
\tilde{F}_4^A
\nonumber\\
& ~ \qquad ~ \quad ~ \quad
+\frac{(p_{A\mu}q_\nu  +  p_{A\nu}q_\mu)}{Q^2}2x_A \tilde{F}_{5}^A
~+~\frac{(p_{A\mu}q_\nu  -  p_{A\nu}q_\mu)}{Q^2}2x_A \tilde{F}_{6}^A\ .
\label{eq:strFnExpandF}
\end{align}
The presence of six structure functions follows from the fact that
(i) electromagnetic and weak currents are vector currents and 
(ii) the state $X$ in the process
$\ell_1(k_1) + A(p_A) \to \ell_2(k_2) + X(p_X)$
must have \textit{net} quantum numbers corresponding to a fermion. 
($VA$ scattering cannot, for example, convert nucleus $A$, 
which is a fermion, into a state with net scalar or vector boson quantum numbers due to Lorentz invariance / angular momentum conservation.)
The first point implies that the hadronic current $\tilde{W}^A_{\mu\nu}(p_A,q)$, which is built by squaring the $AV\to X$ current is a rank-2 tensor.
Now, since Dirac fermions have, in general, four independent components,  
$(4\times4)$ fermion bilinears have 16 components.
These can be arranged into five different ways:
a one-component scalar current,
a one-component pseudoscalar current,
a four-component vector current,
a four-component axialvector current,
and a six-component tensor current~\cite{Halzen:1984mc,Peskin:1995ev}.
\color{black} %
Since $\tilde{W}^A_{\mu\nu}(p_A,q)$ is made from the product of Dirac fermion spinors, it is a bilinear, and subsequently has six independent components.

Formally, the normalizations and organization of the $\tilde{W}_i^A$ and $\tilde{F}_i^A$ in Eqs.~\eqref{eq:strFnExpandW} and \eqref{eq:strFnExpandF} are conventional. Following Ref.~\cite{Schienbein:2007gr}, the $\tilde{W}_i^A$ are normalized such that each is dimensionless and has at most a prefactor of $M_A^{-2}$.
The $\tilde{F}_i^A$ are normalized 
to factor out known $Q^2$ dependence and are related to $\tilde{W}_i^A$ by
\\
\begin{subequations}
\label{eq:app_FW_normalization}
\begin{minipage}{.5\textwidth}
\begin{align}
    \tilde{F}_1^A &= \tilde{W}_1^A,
    \\
\tilde{F}_i^A &= 
\left(\frac{Q^2}{2x_A M_A^2}\right)\ \tilde{W}_i^A,\quad \text{for}\quad i=2,5,6
\end{align}
\end{minipage}
\begin{minipage}{.4\textwidth}
\begin{align}
    \tilde{F}_3^A &= \left(\frac{Q^2}{x_A M_A^2}\right)\ \tilde{W}_3^A,
    \\
     \tilde{F}_4^A &= \left(\frac{Q^2}{2 M_A^2}\right)\ \tilde{W}_4^A \ .
\end{align}
\end{minipage}
\end{subequations}
\\
The $\tilde{F}_i^A$ normalization makes scaling with respect to $x_A$ more manifest.
Under this normalization, $\tilde{W}_i^A$ and $\tilde{F}_i^A$ are real for $i=1-5$, whereas
$\tilde{W}_6^A$ and $\tilde{F}_6^A$ are imaginary~\cite{Collins:1984xc}.
Moreover, $\tilde{W}_6^A$ and $\tilde{F}_6^A$ are only nonzero if charge-parity symmetry is violated in QCD.
However, the coefficient of $\tilde{W}_{6}^A$ and $\tilde{F}_{6}^A$  will vanish when contracted
with the symmetric $L^{\mu\nu}$ tensor.
Choosing $\tilde{W}_i^A$, $\tilde{F}_i^A$, or other normalizations does not change the underlying physics. 
However, 
seemingly innocuous differences in their definitions can 
impact the final form of TMCs due
to structure function mixing, i.e., the off-diagonal terms in $A^i_j$ and $B^i_j$ in Eq.~\eqref{eq:master}. 

The next  step is to recognize that $\tilde{W}^A_{\mu\nu}$ in the DIS limit can be related to the time-ordered matrix element $T_{\mu\nu}^A$ for $A V^* \to A V^*$ scattering in short-distance limit.
This is given by~\cite{Christ:1972ms}
\begin{align}
\label{eq:app_compton_def}
    {T}^A_{\mu\nu}(p_A,q) \quad &\equiv   \int d^{4}z\ e^{iq\cdot z}\ \langle A\vert \mathcal{T}J_{\mu}^\dagger(z)\ J_{\nu}(0)\vert A\rangle
\\
&= \int d^{4}z\ e^{iq\cdot z}\ \langle A\vert J_{\mu}^\dagger(z)\ J_{\nu}(0)\vert A\rangle\
\theta(z^0 > 0)
\nonumber\\
&\qquad\ +\int d^{4}z\ e^{iq\cdot z}\ \langle A\vert  J_{\nu}(0)\ J_{\mu}^\dagger(z) \vert A\rangle\
\theta(z^0 < 0)\ 
\\
&\equiv T_{z^0>0}(p_A,q) + T_{z^0<0}(p_A,q),
\label{eq:app_compton_split}
\end{align}
where $\theta$ is the usual Heaviside step function normalized to unity.
The matrix elements under the integral can be identified as the (inverse) Fourier transforms (FT) of $W_{\mu\nu}^A$ over $q$:
\begin{subequations}
\label{eq:app_fourierTransform}
\begin{align}
\label{eq:app_fourierTransform_W}
 {\rm FT}[W^A_{\mu\nu}(p_A,q)](x) 
&=
 \int \frac{d^4q}{(2\pi)^4}\
 e^{-iq\cdot x}\  
 W^A_{\mu\nu}(p_A,q)\
=\frac{1}{4\pi} 
 \langle A \vert J_{\mu}^\dagger(x)\  J_{\nu}(0) \vert A\rangle\ , 
\\
\label{eq:app_fourierTransform_Wbar}
 {\rm FT}[\overline{W}^{A}_{\nu\mu}(p_A,q)](x) 
&=
 \int \frac{d^4q}{(2\pi)^4}\
 e^{-iq\cdot x}\  
 \overline{W}^{A}_{\nu\mu}(p_A,q)\
=\frac{1}{4\pi} 
 \langle A \vert J_{\nu}(0)\ J_{\mu}^\dagger(x) \vert A\rangle\ .
\end{align}
\end{subequations}
Equation \eqref{eq:app_fourierTransform_Wbar} defines the ``conjugate'' hadronic tensor $\overline{W}_{\nu\mu}^A$. It is related to ${W}_{\mu\nu}^A$ by
\begin{align}
\overline{W}^{A}_{\nu\mu}(p_A,q)\ =\ 
 \frac{1}{4\pi}  \int d^4z ~ e^{-i(-q)\cdot z}  
 \langle A \vert J_{\nu}(0)\ J_{\mu}^\dagger(z) \vert A\rangle\
=\ \left[ W^{A}_{\mu\nu}(p_A,-q)\right]^\dagger \ .
\label{eq:app_incHad_conjugate}
\end{align}
Note that the Hermitian conjugation operator $\dagger$ should be applied to both the structure functions and the tensor-valued coefficients of $W^{A}_{\mu\nu}$.
For the case of electromagnetic currents, these are Hermitian and Eq.~\eqref{eq:app_incHad_conjugate} reduces to $\overline{W}^{A}_{\nu\mu}(p_A,q)\big\vert_{\rm EM} = W^{A}_{\mu\nu}(p_A,-q)$.

After exchanging the order of integration, the first term in Eq.~\eqref{eq:app_compton_split} is
\begin{align}
   T_{z^0>0}(p_A,q) &=  
4\pi \int \frac{d^4k}{(2\pi)^4}\
 W^{A}_{\mu\nu}(p_A,k)\   
\int d^{4}z\ e^{i(q-k)\cdot z}\ 
\theta(z^0 > 0)   
\\
&=
4\pi \int \frac{d^4k}{(2\pi)^4}\
 W^{A}_{\mu\nu}(p_A,k)\   
 (2\pi)^3\delta(\vec{q}-\vec{k})\
\left[\pi\delta(q^0-k^0)+\frac{i}{q^0-k^0}\right].
\label{eq:app_compton_posTime}
\end{align}
The three-dimensional 
$\delta(\vec{q}-\vec{k})$ comes from integrating over all of $\vec{z}$-space and the bracketed factor comes from integrating only over $z^0>0$. 
Following the argument of Ref.~\cite{Collins:1984xc},
momentum conservation dictates that $(2p_A\cdot q)>Q^2$, which implies that ${W}^A_{\mu\nu}$ is zero for $ (2p_A\cdot q)  < Q^2$, or equivalently that ${W}^A_{\mu\nu}$ is zero for $x_A>1$ and $x_A<0$.
Fixing $Q^2$ and taking ${W}_{\mu\nu}^A$ to be analytic
on the plane $\omega_A\equiv x_A^{-1}$
further implies branch cuts along $\vert \omega_A \vert \geq 1$ as depicted in Fig.~\ref{fig:diagram_DIS_contour}.
Therefore, since the pole in Eq.~\eqref{eq:app_compton_posTime} can be decomposed into its principal $\mathcal{P}$ and singular parts when deformed, i.e.,
\begin{align}
    \frac{1}{q^0- k^0\mp i\varepsilon}\ =\ \mathcal{P}\ \pm\ i\pi \delta(q^0-k^0),
\end{align}
the tensor $W^{A}_{\mu\nu}$ can be identified as the discontinuity of $T_{z^0>0}$. Specifically, one finds:
\begin{align}
 {\rm disc}\ T_{z^0>0}\ &= \lim_{\varepsilon\to0}\ \left[T_{z^0>0}(q^0+i\varepsilon)-T_{z^0>0}(q^0-i\varepsilon)\right]
 \\
 &=
 4\pi\ \int \frac{d^4k}{(2\pi)^4}\
 W^{A}_{\mu\nu}(p_A,k)\   
 (2\pi)^3\delta(\vec{q}-\vec{k})\
 \nonumber\\
 & \qquad 
\times \ 
\left\{
\left[
i\mathcal{P} - i^2 \pi\delta(q^0-k^0)
\right]
-
\left[
i\mathcal{P} + i^2 \pi\delta(q^0-k^0)
\right]
\right\}
\\
&= 4\pi\  W^{A}_{\mu\nu}(p_A,q)\ .
\end{align}
A similar result holds for $T_{z^0<0}$.
Subsequently, one can write in terms of $\omega_A = x_A^{-1}$ the relations~\cite{Christ:1972ms,Collins:1984xc}
\begin{subequations}
\label{eq:app_WFdispersion}
\begin{align}
\label{eq:app_WFdispersion_a}
{T}^A_{\mu\nu}(p_A,q)\Big\vert^{\omega_A+i\varepsilon}_{\omega_A-i\varepsilon} &= 4\pi\ {W}^A_{\mu\nu}(p_A,q), \quad\text{for}\quad \omega_A > 0\ ,
    \\
\label{eq:app_WFdispersion_b}    
{T}^A_{\mu\nu}(p_A,q)\Big\vert^{\omega_A-i\varepsilon}_{\omega_A+i\varepsilon} &= 4\pi\ 
 \left[ W^{A}_{\mu\nu}(p_A,-q)\right]^\dagger,
\quad\text{for}\quad \omega_A < 0\ .
\end{align}
\end{subequations}
and with ${W}^A_{\mu\nu}$ and ${W}^{A\dagger}_{\nu\mu}$ vanishing for  $\omega_A < 1$ and $\omega_A > -1$, respectively. Intuitively, Eq.~\eqref{eq:app_WFdispersion} states that ${T}^A_{\mu\nu}$ and $W^A_{\mu\nu}$, which are defined in different kinematic limits (the short-distance and DIS limits, respectively), are nevertheless related through analytic continuation.
Therefore, one can take ${T}^A_{\mu\nu}(p_A,q)$, decompose it into a contour integral over $x_A^{-1}$ using Cauchy's integral formula, and deform the contour around the discontinuities at $x_A^{-1}<-1$ and $x_A^{-1}>1$ as shown in Fig.~\ref{fig:diagram_DIS_contour}.

\subsection*{Coefficient functions of $T_{\mu\nu}^A$ and moments of structure functions}

The relationship in Eq.~\eqref{eq:app_WFdispersion} can be refined by
decomposing ${T}^A_{\mu\nu}$ in three different ways.
The first is according to Lorentz structures as was done for ${W}^{A}_{\mu\nu}$ in Eq.~\eqref{eq:strFnExpandW}: 
\begin{align}
    {T}^A_{\mu\nu} 
&=
-g_{\mu\nu}\Delta \tilde{T}^{A}_{1} ~+~ \frac{p_{A\mu}p_{A\nu}}{M_A^2}   \Delta \tilde{T}^{A}_{2} 
~-~i\epsilon_{\mu\nu\alpha\beta} \frac{ p^\alpha_A q^\beta}{M_A^2}   \Delta \tilde{T}^{A}_{3}
~+~\frac{q_\mu q_\nu}{M_A^2}\Delta \tilde{T}^{A}_{4}
\nonumber\\
&  ~ \qquad ~ \quad ~ \qquad 
+\frac{(p_{A\mu}q_\nu  +  p_{A\nu}q_\mu)}{M_A^2}\Delta \tilde{T}^{A}_{5}
~+~\frac{(p_{A\mu}q_\nu  -  p_{A\nu}q_\mu)}{M_A^2}\Delta \tilde{T}^{A}_{6}.
\label{eq:forwardFull}
\end{align}
Component-by-component, the dispersion relationship of 
Eq.~\eqref{eq:app_WFdispersion} becomes 
\begin{subequations}
\label{eq:app_StrFndispersion}
\begin{align}
\Delta{\tilde{T}}^A_{i}(p_A,q)\Big\vert^{\omega_A+i\varepsilon}_{\omega_A-i\varepsilon} &= 4\pi\ \tilde{W}^A_{i}(p_A,q), \quad\text{for}\quad \omega_A > 0\ ,
    \\
\Delta{\tilde{T}}^A_{i}(p_A,q)\Big\vert^{\omega_A-i\varepsilon}_{\omega_A+i\varepsilon} &
= 4\pi\ (-1)^{b_i}\ {\tilde{W}}^{A}_{i}(p_A,-q), \quad\text{for}\quad \omega_A < 0\ ,
\\ 
\text{and}\quad b_i &= 0\ (1), \quad\text{for}\quad i=1,2,3,4,6\ (5)\ .
\end{align}
\end{subequations}
The factor of $(-1)^{b_i}$ comes from applying the $\dagger$ operator in Eq.~\eqref{eq:app_WFdispersion_b} to $\tilde{W}_i^A$ and its prefactor in combination with propagating the argument $-q$ into the prefactor of $\tilde{W}_i^A$. 
We also take into account whether $\tilde{W}_i$ is real or imaginary, as discussed below Eq.~\eqref{eq:app_FW_normalization}; 
recall ${\tilde{W}}^{A}_{6}$ is imaginary.

The second way of decomposing $T_{\mu\nu}^A$  is by the OPE (done below) and  gives expressions for $\Delta \tilde{T}_i^A$.

The third decomposition is by Taylor expansion in the short-distance limit. 
The distinction between this limit and the DIS limit is important:
In the DIS limit, $(Q^2/M_A^2)\to\infty$ while $x_A=(Q^2/2p_A\cdot q)$ is fixed.
In the short-distance limit, $(Q^2/M_A^2)\to\infty$ while $x_A/Q$ is fixed, meaning that $x_A$ grows with $(Q/M_A)$.
Formally, the short-distance limit corresponds to the limit $\omega_A = 2(p_A\cdot q)/Q^2\to0$.
We therefore assume that we can write 
\begin{subequations}
    \label{eq:opeDTtaylor2}
\begin{align}
    \zeta_i^A\ \times\ \omega_A^{a_i}\ \times\
    \Delta\tilde{T}_i^{A}(p_A,q)\
     &=\ \sum_{k=0}^\infty\  t_{i(k)}^{A}(Q^2)\  \omega_A^{k}\ , \quad\text{where}
    \\ 
    a_i = 0~(1)\ \text{for}\ i=1,4~(2,3,5,6)\ ,
    \quad\text{and}\quad
    \zeta_1^A &= 1\ ,\quad\ 
    \zeta_{2,4,5,6}^A = \left(\frac{Q^2}{2M_A^2}\right)\ ,\quad\ 
    \zeta_{3}^A = \left(\frac{Q^2}{M_A^2}\right)\ . 
\end{align}
\end{subequations}
Here, $t_{i(k)}^A$ is the $k^{th}$ coefficient of the expansion and is only a function of $Q^2$.
The factors $\zeta_i^A$ and $\omega_A^{a_i}$,
which are fixed in the short distance limit, are the same factors relating the structure functions $\tilde{W}_i$ and $\tilde{F}_i$ in Eqs.~\eqref{eq:WtoF} and \eqref{eq:app_FW_normalization}. 
Following Ref.~\cite{Collins:1984xc} and using Cauchy's integral formula
\begin{align}
    F(y) = \frac{1}{2\pi i}\int_{\mathcal{C}}\ \frac{dz}{z-y}\ F(z),
    \label{eq:cauchy}
\end{align}
where $\mathcal{C}$ is an appropriately chosen contour, 
it is possible to 
analytically continue $\Delta\tilde{T}_i^{A}$:
\begin{align}
    \zeta_i^A\ \times \omega_A^{a_i}\ \times
     \Delta\tilde{T}_i^{A}(p_A,q) 
    &= \frac{\zeta_i^A}{2\pi i}\int_{\mathcal{C}}\ \frac{d \omega_A'}{(\omega_A' - \omega_A)}\ 
    \left(\omega_A'\right)^{a_i}\
    \Delta\tilde{T}_i^{A}(p_A,q')
    \\
    &= \frac{\zeta_i^A}{2\pi i}\int_{\mathcal{C}}\ \frac{d \omega_A'}{\omega_A'(1 - \frac{\omega_A}{\omega_A'})}\ 
    \left(\omega_A'\right)^{a_i}\
    \Delta\tilde{T}_i^{A}(p_A,q')
    \\
    &= \sum_{k=0}^\infty\ 
    \frac{\zeta_i^A}{2\pi i}\int_{\mathcal{C}}\ 
    d \omega_A'\
    \left(\frac{\omega_A}{\omega_A'}\right)^k\
    \left(\omega_A'\right)^{a_i-1}\
    \Delta\tilde{T}_i^{A}(p_A,q').
    \label{eq:opeDTtaylor1}
\end{align}
The geometric series is obtained by fixing the integration variable $\omega_A'$, which can be large, but taking the external variable  $\omega_A\to0$. 

Comparing Eq.~\eqref{eq:opeDTtaylor1} to Eq.~\eqref{eq:opeDTtaylor2} and using Eqs.~\eqref{eq:app_StrFndispersion}, 
one obtains
\begin{align}
t_{i(k)}^{A}(Q^2) &= 
\frac{\zeta_i^A}{2\pi i}\int_{\mathcal{C}}\ 
d \omega_A'\ \left(\omega_A'\right)^{a_i-k-1}\
   \Delta \tilde{T}_i^{A}(p_A,q')
\\
&= 
    \frac{4\pi\zeta_i^A}{2\pi i}\int_1^\infty d \omega_A'\ 
    \left(\omega_A'\right)^{a_i-k-1}\
    \tilde{W}_i^{A}(p_A,q')
    \nonumber\\
    &+
    \frac{4\pi\zeta_i^A}{2\pi i}\int_{-\infty}^{-1} 
    d \omega_A'\ 
    \left(\omega_A'\right)^{a_i-k-1}\
    (-1)^{b_i}\
    {\tilde{W}}_i^{A}(p_A,-q')
\\
&= 
    (-2i)\zeta_i^A\ \int_0^1 d x_A'\ \left(x_A'\right)^{k-1-a_i}\
    \tilde{W}_i^{A}(x_A',Q^2)
    \nonumber\\
    &+
    (-2i)\zeta_i^A\ \int_{-1}^{0} d x_A'\ \left(x_A'\right)^{k-1-a_i}\
    (-1)^{b_i}\
    {\tilde{W}}_i^{A}(-x_A',Q^2)\ .
\end{align}
The last line is obtained by changing to the variable $x'_A = (\omega_A')^{-1}$. Changing the integration variable of the second term to $z_A = -x_A'$, collecting factors of $(-1)$, 
using $(-1)^{a_i}=(-1)^{-a_i}$,
and relabeling gives
\begin{align}
t_{i(k)}^{A}(Q^2) &= 
    (-2i)\ \int_0^1 d x_A'\ \left(x_A'\right)^{k-1}
    \nonumber\\
     &\quad \times 
\left[
\zeta_i\ \left(x_A'\right)^{-a_i}\
    \tilde{W}_i^{A}(x_A',Q^2)
    + \left(-1\right)^{k-1+a_i+b_i} 
 \zeta_i\ \left(x_A'\right)^{-a_i}\   
    {\tilde{W}}_i^{A}(x_A',Q^2)
\right]    
    \\
    &=
    (-2i)\ \left[1    + (-1)^{k-1+a_i+b_i}    \right]\ 
    \int_0^1 d x_A'\ \left(x_A'\right)^{k-1}\ 
    \tilde{F}_i^{A}(x_A', Q^2).
\end{align}

We now denote the $N^{th}$ Mellin moment of the function $M(z)$ by $M^N$ and fix normalizations such that a Mellin transformation and its inverse (over a path $c$) are:
\begin{equation}
 M^N = \int_0^1 dz ~ z^{N-1} ~ M(z) \quad\text{with}\quad
 M(z) = \frac{1}{2\pi i}\int^{c+i\infty}_{c-i\infty} dN ~ z^{-N} ~ M^N.
\label{eq:mellinDef}
\end{equation}
Under this normalization, the moments of $\tilde{W}^A_i(x_A)$ are related to those of $\tilde{F}^A_i(x_A)$ by 
\\
\begin{subequations}
\label{eq:app_FW_moments_norm}
\begin{minipage}{.55\textwidth}
\begin{align}
    \tilde{F}_1^{AN} &= \tilde{W}_1^{AN}\ ,
    \\
\tilde{F}_i^{AN} &= 
\left(\frac{Q^2}{2 M_A^2}\right)\ \tilde{W}_i^{A(N-1)}\quad \text{for}\quad i=2,5,6\ ,
\end{align}
\end{minipage}
\begin{minipage}{.4\textwidth}
\begin{align}
    \tilde{F}_3^{AN} &= \left(\frac{Q^2}{M_A^2}\right)\ \tilde{W}_3^{A(N-1)}\ ,
    \\
     \tilde{F}_4^{AN} &= \left(\frac{Q^2}{2 M_A^2}\right)\ \tilde{W}_4^{AN}\ .
\end{align}
\end{minipage}
\end{subequations}
\\
Both are related to the $N^{th}$ coefficient function $t_{i(k=N)}^{A}(Q^2)$ by 
\begin{subequations}
\label{eq:dispersionTermByTerm}
\begin{align}
t_{i(N)}^{A}(Q^2) &= -2i\ \left[1    + (-1)^{N-1+a_i+b_i}    \right]\ \tilde{F}_i^{AN}(Q^2)
\\
&=
\left\{\begin{matrix}
0, & N=\text{even}\\
-4i\ \tilde{F}^{AN}_i(Q^2), & N=\text{odd}
\end{matrix}\right. , \quad\text{for}\quad i = 1,4,5
\ ,
\\
&=
\left\{\begin{matrix}
-4i\ \tilde{F}^{AN}_i(Q^2), & N=\text{even} \\ 
0, & N=\text{odd} 
\end{matrix}\right. , \quad\text{for}\quad i = 2,3,6
\ .
\end{align}
\end{subequations}
This allows us to rewrite the expansion in Eq.~\eqref{eq:opeDTtaylor2} as
\begin{align}
\label{eq:dispersionStrFn}
    \zeta_i^A\ \times\ \omega_A^{a_i}\ \times\
     \Delta\tilde{T}_i^{A}(Q^2, \omega_A)\ 
    =\ -4i\ 
    \sum_{N}^\infty\  
    \tilde{F}_i^{AN}(Q^2)\
    \omega_A^{N}\ ,
\end{align}
where it is implied that the index $N$ runs only over odd or even integers.

We now make a few brief comments.
First, in order for Eq.~\eqref{eq:dispersionStrFn} to hold, $\vert\omega_A\vert<1$ must be satisfied, i.e., one must be in the DIS limit. 
Second, as $N$ increases, the dominant contribution to $\tilde{F}_i^{AN}$ is when the argument of $\tilde{F}_i^{A}(z)$ approaches unity since (according to the definition of Eq.~\eqref{eq:mellinDef})   
$\tilde{F}_i^{AN}$ would otherwise be suppressed by a small number.  
Finally, since $\omega_A=1$ corresponds to the elastic limit (see Eq.~\eqref{eq:app_kinDef_hadronicMass}), 
$Q^2$ must be made increasingly large for large-$N$ moments to be well-defined and for the invariant mass of the hadronic system to remain in the perturbative regime.

\subsection*{Expanding $T_{\mu\nu}^A$ with the OPE}

 Using the OPE, the leading behavior  of $T_{\mu\nu}^A$ in powers of $(1/Q^p)$
 can be decomposed into a sum of operators $\mathcal{O}^{\mu_1,\dots,\mu_k}$ and Wilson coefficients $c_{\mu\nu\mu_1...\mu_k}^{\tau,\iota}$. This expansion is given by~\cite{Christ:1972ms,Georgi:1976ve,Georgi:1976vf}
 \begin{align}
 \label{eq:app_ope_compton}
\lim_{z\to0}\ 
{T}^A_{\mu\nu}(p_A,q) \quad \overset{\rm OPE}{=} \quad
& 
-2i\
\sum_{k,\iota} ~ 
c_{\mu\nu\mu_1...\mu_k}^{\tau=2,\iota}(q)\
\langle A(p_A) \vert \mathcal{O}_{\iota,\tau=2}^{\mu_{1}...\mu_{k}}\vert A(p_A)\rangle \ 
+ \mathcal{O}(\tau>2)
\ .
\end{align}
At leading power, the composite operators $\mathcal{O}_{\iota,\tau}^{\mu_1\dots\mu_k}$ are symmetric, quark billinears (or pairs of gluon field strengths) that sandwich uncontracted covariant derivatives. 
Tallying up the number of uncontracted Lorentz indices implies that $\mathcal{O}_{\iota,\tau}^{\mu_1\dots\mu_k}$ carries a spin of $k$.
The operators are organized according to their twist $\tau\equiv d-n$, where $d$ is the dimensionality of $\mathcal{O}_{\iota,\tau}^{\mu_1\dots\mu_k}$ in the standard sense of dimension power counting in an effective field theory. 
For a fixed spin and twist, the index $\iota$ catalogs all the Lorentz structures that can possibly contract with $\mathcal{O}_{\iota,\tau}^{\mu_1\dots\mu_k}$.
Coupling and renormalization factors are sequestered into the effective Wilson coefficient $c_{\mu\nu\mu_1...\mu_k}^{\tau,\iota}(q)$.
For fixed spin $k$, a larger twist $\tau$ corresponds to a larger $(1/Q)$ suppression in the Wilson coefficient.
 Schematically, Eq.~\eqref{eq:app_ope_compton} stipulates that in the short-distance limit, the time-ordered matrix element ${T}_{\mu\nu}^A$, which is a function of $p_A$ and $q$, can be expressed in terms of Wilson coefficients, which are only functions of $q$, and hadronic matrix elements, which are only functions of $p_A$.

The operators $\mathcal{O}^{\mu_1,\dots,\mu_k}$ can be decomposed into symmetric (traceless) and non-symmetric parts,
with  $\mathcal{O}^{\mu_1,\dots,\mu_k}\sim   \hat{P}_A^{\mu_1}\dots \hat{P}_A^{\mu_{k}} + {\rm Tr}$. 
The trace term captures all contributions proportional to the spacetime metric and derivatives, which after contractions or applications of equations of motion give rise to powers of quark masses~\cite{Georgi:1976ve,Georgi:1976vf}. 
Neglecting quark masses\footnote{In practice, finite quark masses can be incorporated in TMCs through a rescaling of the Nachtmann scaling variable. See Appendix A.2 of Ref.~\cite{Schienbein:2007gr} and references therein.}, the matrix elements that follow from acting on $\mathcal{O}^{\mu_1,\dots,\mu_{2k}}$ are given by
\begin{subequations}
\label{eq:app_had_matrix}
\begin{align}
\langle A \vert \mathcal{O}_{\iota,\tau=2}^{\mu_{1}...\mu_{2k}}\vert A\rangle  
&=
A_{\tau=2}^{2k} \ \times \ \tilde{\Pi}^{\mu_1\dots\mu_{2k} }, \quad\text{where}
\\
\tilde{\Pi}^{\mu_1\dots\mu_{2k} }
& =
\sum_{j=0}^{k}(-1)^{j}\ \frac{(2k-j)!}{2^{j}(2k)!} \ 
\eta(j,2k-2j) \
{\underbrace{\{ g...g\}}_{\scriptstyle j\ g^{\mu_{n}\mu_{m}}{}'s}}
\quad
{\underbrace{\{ p_A...p_A\}}_{\scriptstyle (2k-2j)\ p_A^{\mu_{n}}{}'s}}
(p_A^{2})^{j}\ .
\end{align}
\end{subequations}
Note the index change from $k$ to $2k$. The factor $A_{\tau=2}^{2k}$ is the scalar-valued ``reduced'' hadronic matrix element and describes long-distance hadronic dynamics.
(The $2k$ in $A_{\tau=2}^{2k}$ is an index, not an exponent.) 
The index $j$ sums over all permutations of $p_A^{\mu_n}$ and $g^{\mu_n \mu_m}$,
with $\mu_n,\mu_m \in\{\mu_1,\dots,\mu_{2k}\}$,
that contract with a particular Wilson coefficient.
For a given $k$ and $j$, the two 
$\{\dots\}$ brackets denote 
$j$ spacetime metrics $g^{\mu_n \mu_m}$
and $(2k-2j)$ factors of momentum $p_A^{\mu_m}$. The permutation multiplicity $\eta$ is
\begin{equation}
  \eta\left(j~\text{metrics},\ 2k-2j~\text{factors~of~}p_A\right) = \frac{1}{2^j} \frac{(2k)!}{j! (2k-2j)!} \ .
  \label{eq:permutationFactor}
\end{equation}
The numerator of $\eta$ is calculated from $(2k)! = [(2k-2j) + (2j)]!$  The denominator factor of $2^j$ accounts for the two-fold symmetry of $j$ symmetric metric tensors, i.e., $g^{\mu_n \mu_m}=g^{\mu_m \mu_n}$.
If a metric or momentum factor is pulled from either $\{\dots\}$ bracket, $\eta$ is updated accordingly.

In the massless target limit, i.e., when $(M_A^2/Q^2)\to 0$, one neglects $j>0$ terms since they generate powers of $(p_A^2)^j = M_A^{2j}$.
For this reason, the $j>0$ terms are sometimes called ``kinematical power corrections''~\cite{Ellis:1982cd,Ellis:1982wd}.
In the $j=0$ limit, Eq.~\eqref{eq:app_had_matrix} reduces to  $2k$ factors of $p_A^{\mu_m}$:
\begin{equation}
   \langle A \vert 
 \mathcal{O}_{\iota,\tau=2}^{\mu_1,\dots,\mu_{2k}}
 \vert A\rangle \Big\vert_{(M_A/Q)^2\to0}  = 
  A_{\tau=2}^{2k}(p_A^2) \times \tilde{\Pi}_{\mu_1\dots\mu_{2k} } \Big\vert_{j=0} 
 =  
 A_{\tau=2}^{2k}(p_A^2) \times
 (p_A^{\mu_1}\dots p_A^{\mu_{2k}})
 \ .
 \label{eq:massME}
\end{equation}

At leading power of $\tau$ and for a fixed $2k>2$, the $c$ functions in Eq.~\eqref{eq:app_ope_compton} can be decomposed as
\begin{align}
\label{eq:app_wilson_def}
c_{\mu\nu\mu_1,\dots,\mu_{2k}}^{\tau=2,\iota}(q)
~=~ &
\Bigl[
-2g_{\mu\nu} q_{\mu_1}q_{\mu_2} C_1^{2k}
+ g_{\mu\mu_1}g_{\nu\mu_2} Q^2 C_2^{2k} 
- i\epsilon_{\mu\nu\alpha\beta}
g^{\alpha}_{\mu_1}q^\beta q_{\mu_2}C_3^{2k}
 \nonumber\\
+
4\frac{q_\mu q_\nu}{Q^2}q_{\mu_1}q_{\mu_2} C_4^{2k}
&+2( g_{\mu\mu_1}q_\nu q_{\mu_2} \pm  g_{\nu\mu_1}q_\mu q_{\mu_2} )C_{5,6}^{2k}
\Bigr]
\times \frac{2^{2k}}{(Q^2)^{2k}} 
\times \left(\prod_{m=3}^{2k} q_{\mu_m}\right).
\end{align}
Here, the $C_{\iota=1,\dots,6}^{2k}$ are scalar-valued coefficients that parameterize the normalization of each $c_{\mu\nu\mu_1,\dots,\mu_{2k}}^{\tau=2,\iota}$. The $C_{\iota}$ are defined to all orders in QCD but can be identified and matched to quantities in fixed-order perturbation theory. %
The tensor part,
i.e., the part carrying Lorentz indices, can then be organized according to Lorentz structures as in Eq.~\eqref{eq:forwardFull}.
Conventional factors of $2$ in Eq.~\eqref{eq:app_wilson_def} are pulled from the $C_{\iota}^{2k}$ in order to simplify later expressions.

Assembling these ingredients and contracting over all $2k>k_{\min}$ indices, the OPE gives
\begin{align} 
\lim_{z\to0}{T}^A_{\mu\nu}(q,p_A)\   \overset{\rm OPE}{=}\ 
& -2i\ 
\sum_{k=k_{\min}}^\infty\
\Bigl[ 
-2g_{\mu\nu} q_{\mu_1}q_{\mu_2} C_1^{2k}
+ g_{\mu\mu_1}g_{\nu\mu_2} Q^2 C_2^{2k} 
- i\epsilon_{\mu\nu\alpha\beta}
g^{\alpha}_{\mu_1}q^\beta q_{\mu_2}C_3^{2k}
\label{eq:tmunuFull}
\nonumber\\
& +
4 \frac{q_\mu q_\nu}{Q^2}q_{\mu_1}q_{\mu_2} C_4^{2k}
+2( g_{\mu\mu_1}q_\nu q_{\mu_2} \pm  g_{\nu\mu_1}q_\mu q_{\mu_2} )C_{5,6}^{2k}
\Bigr]
\nonumber\\
&
\times \frac{2^{2k}}{(Q^2)^{2k}} 
\times \left(\prod_{m=3}^{2k} q_{\mu_m}\right)\times
A_{\tau=2}^{2k}(p_A^2) \times \tilde{\Pi}^{\mu_1\dots\mu_{2k} }
+ \mathcal{O}(\tau>2)
\\
\equiv ~ & ~
\Delta \tilde{T}^{A}_{1\mu\nu} +
\Delta \tilde{T}^{A}_{2\mu\nu} +
\Delta \tilde{T}^{A}_{3\mu\nu} +
\Delta \tilde{T}^{A}_{4\mu\nu} +
\Delta \tilde{T}^{A}_{5\mu\nu} +
\Delta \tilde{T}^{A}_{6\mu\nu} + 
\mathcal{O}(\tau>2)
\label{eq:tmunuDecomp}   
 \\
= ~ & ~
-g_{\mu\nu}                         
\Delta \tilde{T}^{A}_{1} +
\frac{p_{A\mu}p_{A\nu}}{M_A^2}      
\Delta \tilde{T}^{A}_{2} -
i\epsilon_{\mu\nu\alpha\beta} \frac{ p^\alpha_A q^\beta}{M_A^2} 
\Delta \tilde{T}^{A}_{3} +
\frac{q_\mu q_\nu}{M_A^2}           
\Delta \tilde{T}^{A}_{4}      
\nonumber\\
& ~ +
\frac{(p_{A\mu}q_\nu  +  p_{A\nu}q_\mu)}{M_A^2} \Delta \tilde{T}^{A}_{5}
+
\frac{(p_{A\mu}q_\nu  -  p_{A\nu}q_\mu)}{M_A^2} \Delta \tilde{T}^{A}_{6}
+ \mathcal{O}(\tau>2).
\label{eq:forwardScattDecomp}   
\end{align}
The starting point for the summation over $k$ depends on the particular Wilson coefficient;
specifically, 
$k_{\min}=2$ for $C_2$
while 
$k_{\min}=1$ for the other  $C_i$ coefficients.  
In Eq.~\eqref{eq:forwardScattDecomp}, the scalar-valued coefficients $\Delta \tilde{T}_{i=1,\dots,6}^{A}$, i.e., the quantities \textit{without} external Lorentz indices,  are related to the hadronic structure functions $W_i^{A}$ through the dispersion relationships of Eqs.~\eqref{eq:app_WFdispersion}
and \eqref{eq:app_StrFndispersion}, 
up to $(1/Q)$ corrections.
The $\Delta \tilde{T}_i$  are given explicitly in terms of summations over $k$ and $j$ below in Eqs.~\eqref{eq:strFunT1FullExpression}-\eqref{eq:strFunT6FullExpression}.

In the intermediate step Eq.~\eqref{eq:tmunuDecomp}, each $\Delta \tilde{T}_{\iota=1,\dots,6\ \mu\nu}^{A}$ denotes the collection of contractions that are respectively proportional to the coefficient $C_{\iota}$. 
We introduce this step because
$\Delta \tilde{T}_{\iota\ \mu\nu}^{A}$ and $\Delta \tilde{T}_{i}^{A}$ 
do not have a 
one-to-one correspondence when $M_A\neq0$. 
As a consequence of the normalizations adopted for $\tilde{W}_i^A$ in Eq.~\eqref{eq:strFnExpandW}, 
and subsequently those for $\Delta\tilde{T}_i^A$ in Eq.~\eqref{eq:forwardFull}, 
some $\Delta \tilde{T}_{i}^{A}$ are
sourced by multiple
OPE operators when $M_A \neq 0$.
This phenomenon of $\tilde{W}_i^A$ (or $\Delta\tilde{T}_i^A$) being sourced by two or more $C_{\iota}$ is sometimes called ``structure function mixing,''
and is discussed in App.~\ref{app:opeMixing}.

We now derive all six $\Delta \tilde{T}_{\iota\ \mu\nu}^{A}$ from Eq.~\eqref{eq:tmunuFull} by contracting all $\mu_m$ indices. 
After, we reorganize $\Delta \tilde{T}_{\iota\ \mu\nu}^{A}$ and group terms according to their Lorentz structure, e.g., collect all terms proportional to $g_{\mu\nu}$ or all terms proportional to $\epsilon_{\mu\nu\alpha\beta} p^\alpha_A q^\beta$. 
We then identify the $\Delta \tilde{T}_{i}^{A}$,
which are in terms of hadronic matrix elements $A_{\tau=2}^{2k}$, Wilson coefficients $C_{\iota}^{2k}$, and nested summations.
In App.~\ref{app:opeMassive}, the summations are first evaluated for the case of a massless target.
The results for the $(M_A^2/Q^2)\neq0$ case
are then expressed in terms of structure functions for massless targets.

\subsection*{Organization of Lorentz contractions in the OPE}

Explicit evaluation of all contractions of Lorentz indices in Eq.~\eqref{eq:tmunuFull} is an exacting task. Therefore, as a first step, we organize everything according to
$\Delta \tilde{T}_{\iota\ \mu\nu}^{A}$, and hence $C_{\iota}$. As a consequence, patterns in  $\Delta \tilde{T}_{\iota\ \mu\nu}^{A}$ appear and simplify the work.
The outcome is summarized in App.~\ref{app:opeMixing}.

\begin{description}
\item[\textbf{(i) $\Delta \tilde{T}^{A}_{1\mu\nu}$:}]
We start with $\Delta \tilde{T}^{A}_{1\mu\nu}$ for $M_A\neq0$ because, like $\Delta \tilde{T}^{A}_{4\mu\nu}$, the covariant tensor structure for $\mu_{m=1}\dots\mu_{m=2k}$ consists entirely of momentum factors $q$. That is to say, all lowered internal indices $\mu_{m}$ are carried by $q_{\mu_m}$. This results in the simplest possible permutations of metrics and momenta in $\tilde{\Pi}^{\mu_{1}\dots\mu_{2k}}$. Combining terms, one gets
\begin{align}
    \Delta \tilde{T}^{A}_{1\mu\nu}  
    & = -4i\  \sum_{k=1}^\infty
\left[-g_{\mu\nu} C_1^{2k}A_{\tau=2}^{2k}\right]
\times \frac{2^{2k}}{(Q^2)^{2k}} 
\times \left(\prod_{m=1}^{2k} q_{\mu_m}\right)\times
\tilde{\Pi}^{\mu_1\dots\mu_{2k} }
+ \mathcal{O}(\tau>2),
\end{align}
where the production factor $(\prod_{m=1}^{2k}q_{\mu_m})$ has (re)absorbed $q_{\mu_1}$ and $q_{\mu_2}$. 

The contraction of $\prod_{m=1}^{2k} q_{\mu_m}$ and the momentum factor $\tilde{\Pi}$ combines all possible permutations of $2k$ instances of $q_{\mu_m}$ with $j$ metrics $g^{\mu_m \mu_n}$ in  $\{g\dots g\}$ and $(2k-2j)$ momenta $p_A^{\mu_m}$ in $\{p_A\dots p_A\}$. This results in $(2k-2j)$ products of $(q\cdot p_A)$. The remaining $(2k-0)-(2k-2j)=2j$ instances of $q_{\mu_m}$ then contract with the $2j$ indices from $j$ metrics. As no instance of $g^{\mu_m \mu_n}$ or $p_A^{\mu_m}$ have been pulled from $\tilde{\Pi}$, its permutation factor $\eta$ is the same as in Eq.~\eqref{eq:permutationFactor}. The outcome is
\begin{align}
\left(\prod_{m=1}^{2k} q_{\mu_m}\right)
\tilde{\Pi}^{\mu_1\dots\mu_{2k} }
&
=
\sum_{j=0}^{k}(-1)^{j}\ \frac{(2k-j)!}{2^{j}(2k)!}\ 
\frac{1}{2^j} \frac{(2k)!}{j! (2k-2j)!}\
\nonumber \\
& \qquad \times \qquad 
(q^2)^j
(q\cdot p_A)^{(2k-2j)}
(p_A^{2})^{j}
\\
& =
\frac{(Q^2)^{2k}}{2^{2k}}\ 
\sum_{j=0}^{k}\ 
\frac{(2k-j)!}{j! (2k-2j)!}\ 
\left(\frac{M_A^2}{Q^2}\right)^j\ 
x_A^{-(2k-2j)}
\ ,
\end{align}
where $x_A = Q^2/(2p_A\cdot q)$ is the Bjorken scaling variable and $p_A^2= M_A^2$. 
Altogether, 
\begin{align}
 \Delta \tilde{T}^{A}_{1\mu\nu}  
     \quad & = \quad   
    -g_{\mu\nu}\ 
    \times\
    (-4i)\
    \sum_{k=1}^\infty
\left[C_1^{2k}A_{\tau=2}^{2k}\right]\
\sum_{j=0}^{k}\ 
\frac{(2k-j)!}{j! (2k-2j)!}\ 
\left(\frac{M_A^2}{Q^2}\right)^j\ 
x_A^{-(2k-2j)}\ 
\nonumber\\
&    + \quad \mathcal{O}(\tau>2).
\label{eq:opeOrgDT1}
\end{align}
We stress that that the quantity to the right of $g_{\mu\nu}$ here is not $\Delta \tilde{T}^{A}_{1}$ in Eq.~\eqref{eq:forwardScattDecomp}.
In general, the coefficients $\Delta \tilde{T}^{A}_{i}$ receive contributions from more than one $\Delta \tilde{T}^{A}_{\iota\ \mu\nu}$.

\item[\textbf{(ii) $\Delta \tilde{T}^{A}_{4\mu\nu}$:}]
The next term we construct is $\Delta \tilde{T}^{A}_{4\mu\nu}$ for $M_A\neq0$, which has an identical tensor structure of contracted covariant indices as $\Delta \tilde{T}^{A}_{1\mu\nu}$.
In other words, like $\Delta \tilde{T}^{A}_{1\mu\nu}$, all the lowered internal indices $\mu_{m}$ are carried exclusively by factors of $q_{\mu_m}$. The result is then, 
\begin{align}
 \Delta \tilde{T}^{A}_{4\mu\nu} \quad 
    & = \quad  
    \frac{q_\mu q_\nu}{Q^2}\ 
    \times\
    (-8i)\
    \sum_{k=1}^\infty
\left[C_4^{2k}A_{\tau=2}^{2k}\right]\
\sum_{j=0}^{k}\ 
\frac{(2k-j)!}{j! (2k-2j)!}\ 
\left(\frac{M_A^2}{Q^2}\right)^j\ 
x_A^{-(2k-2j)}\ 
\nonumber\\
&    + \quad \mathcal{O}(\tau>2).
\label{eq:opeOrgDT4}
\end{align}

\item[\textbf{(iii) $\Delta \tilde{T}^{A}_{3\mu\nu}$:}]
Moving onto $\Delta \tilde{T}^{A}_{3\mu\nu}$ for $M_A\neq0$, exactly one covariant index from the collection $\mu_{m=1}\dots\mu_{m=2k}$ is carried by the metric $g_{\mu_1}^\alpha$. All other covariant instances of $\mu_m$ are carried by $(2k-1)$ instances of $q_{\mu_m}$. This difference among the lowered indices introduces a complication that was absent in the previous cases but nevertheless allows us to write
\begin{align}
    \Delta \tilde{T}^{A}_{3\mu\nu}  
    & =
    -2i\
    \sum_{k=1}^\infty
\left[
- i\epsilon_{\mu\nu\alpha\beta}\ q^\beta\ C_3^{2k}A_{\tau=2}^{2k}
\right]\
\times
\frac{2^{2k}}{(Q^2)^{2k}} 
\times
g^{\alpha}_{\mu_1}
\times
\left(\prod_{m=2}^{2k} q_{\mu_m}\right)
 \times
 \tilde{\Pi}^{\mu_1\dots\mu_{2k} }\
 \nonumber \\
& 
+ \quad \mathcal{O}(\tau>2)\ .
\end{align}

In principle, $g_{\mu_1}^\alpha$ can contract with a metric from the collection $\{g\dots g\}$ when it contains the index $\mu_{m=1}$. 
This extracts a factor of $q^\alpha$ out of the 
production factor $(\prod_{m=2}^{2k} q_{\mu_m})$, and hence potential permutations, 
through the chain $q^\alpha = g_{\mu_1}^\alpha q_{\mu_n} g^{\mu_n \mu_1}$.
However, this generates a term proportional to $ \epsilon_{\mu\nu\alpha\beta}q^\beta q^\alpha$, which is zero due to the antisymmetric tensor. 

It is also possible for the metric $g_{\mu_1}^\alpha$ to contract with a momentum factor from the collection $\{p_A\dots p_A\}$. This extracts a factor of $p_A^\alpha = g_{\mu_1}^\alpha p_A^{\mu_1}$, and reduces the possible number of permutations in the momentum factor $\tilde{\Pi}$. The contractions within $\Delta \tilde{T}^{A}_{3\mu\nu}$ then involve $(2k-1)$ instances of $q_{\mu_m}$ contracting with  $(2k-2j-1)$ instances of $p_A^{\mu_m}$, to make $(2k-2j-1)$ powers of $(q\cdot p_A)$. The remaining $(2k-1)-(2k-2j-1)=2j$ instances of $q_{\mu_m}$ then contract with the remaining $j$ metrics. 
Finally, since the number of available $p_A^{\mu_m}$ has shifted (by one unit), 
the multiplicity factor in Eq.~\eqref{eq:permutationFactor} must be updated accordingly.
With the updated $\eta$ factor, the outcome is
\begin{align}
g_{\mu_1}^\alpha 
\left(\prod_{m=2}^{2k} q_{\mu_m}\right)
\tilde{\Pi}^{\mu_1\dots\mu_{2k} }
\quad = \quad 
& p_A^\alpha\ 
\sum_{j=0}^{k-1}(-1)^{j}\ \frac{(2k-j)!}{2^{j}(2k)!}\ 
\frac{1}{2^j} \frac{(2k-1)!}{j! (2k-2j-1)!}\
\nonumber\\
& \qquad \times \
(q^2)^j\
(q\cdot p_A)^{(2k-2j-1)}\
(p_A^{2})^{j}
\\
 =
\frac{2p_A^\alpha}{Q^2}\
\frac{(Q^2)^{2k}}{2^{2k}}\ 
&\sum_{j=0}^{k-1}\ 
\frac{(2k-j)!(2k-1)!}{(2k)!j!(2k-2j-1)!}\ 
\left(\frac{M_A^2}{Q^2}\right)^j\ 
x_A^{-(2k-2j-1)}
\ .
\label{eq:opeOrgDT3Pre}
\end{align}
Note that the summation over $j$ extends only to $j=k-1$ since there are fewer allowed permutations; this can be surmised from the denominator factor $1/(2k-2j-1)!$, 
which would reduce to $1/(-1)!$ were $j=k$.
After assembling the different terms, we obtain
\begin{align}
    \Delta \tilde{T}^{A}_{3\mu\nu}  \quad 
    & = \quad    
    - i\epsilon_{\mu\nu\alpha\beta}  \frac{p_A^\alpha q^\beta}{Q^2}\
    \times\ (-4i)\
    \sum_{k=1}^\infty\
\left[ C_3^{2k}A_{\tau=2}^{2k}\right]\
\nonumber
\\
&  \times\
\sum_{j=0}^{k-1}\ 
\frac{(2k-j)!(2k-1)!}{(2k)!j!(2k-2j-1)!}\ 
\left(\frac{M_A^2}{Q^2}\right)^j\ 
x_A^{-(2k-2j-1)}
\quad + \quad \mathcal{O}(\tau>2)\ .
\label{eq:opeOrgDT3}
\end{align}
Unlike the other $\Delta \tilde{T}^{A}_{i}$ in Eq.~\eqref{eq:forwardScattDecomp},
we can identify the quantity to the right of the antisymmetric tensor $\epsilon_{\mu\nu\alpha\beta}$ as $\Delta \tilde{T}^{A}_{3}$ since there is a one-to-one correspondence.
\item[\textbf{(iv) $\Delta \tilde{T}^{A}_{5\mu\nu}$:}]
The steps to determine $\Delta \tilde{T}^{A}_{5\mu\nu}$ (and $\Delta \tilde{T}^{A}_{6\mu\nu}$)
for $M_A\neq0$ closely follow those of $\Delta \tilde{T}^{A}_{3\mu\nu}$ due to the presence of only one metric carrying a lowered index from the collection $\mu_{m=1}\dots\mu_{m=2k}$. The difference, however, is the absence of an antisymmetric tensor. The contractions of $g_{\mu\mu_1}$ and $g_{\nu\mu_1}$ with the collections $\{g\dots g\}$ and  $\{p_A\dots p_A\}$ will thus generate two sets of terms each.
To obtain both sets of terms, we first write 
\begin{align}
    \Delta \tilde{T}^{A}_{5\mu\nu}   
     &=  -4i\   
     \sum_{k=1}^\infty\
\left[ ( g_{\mu\mu_1}q_\nu  +  g_{\nu\mu_1}q_\mu  )C_{5}^{2k}A_{\tau=2}^{2k}\right]\
\times \frac{2^{2k}}{(Q^2)^{2k}} 
\times \left(\prod_{m=2}^{2k} q_{\mu_m}\right)
 \times \tilde{\Pi}^{\mu_1\dots\mu_{2k} }\
 \nonumber \\
& 
+ \quad \mathcal{O}(\tau>2)\ .
 \label{eq:opeOrgDT5Pre}
\end{align}

For the case of the metric $g_{\alpha\mu_1}$ contracting with $\{p_A\dots p_A\}$, we have from Eq.~\eqref{eq:opeOrgDT3Pre}
\begin{align}
g_{\alpha\mu_1}
\left(\prod_{m=2}^{2k} q_{\mu_m}\right)
\tilde{\Pi}^{\mu_1\dots\mu_{2k} }
&\Bigg\vert_{\mu_1\in\{p_A\dots p_A\}}
 =
\nonumber\\
\frac{2p_{A\alpha}}{Q^2}\
\frac{(Q^2)^{2k}}{2^{2k}}\ 
& \sum_{j=0}^{k-1}\ 
\frac{(2k-j)!(2k-1)!}{(2k)!j!(2k-2j-1)!}\ 
\left(\frac{M_A^2}{Q^2}\right)^j\ 
x_A^{-(2k-2j-1)}
\ .
\end{align}

For the second case, $g_{\alpha\mu_1}$ contracts with a metric in $\{g\dots g\}$ and extracts a factor of $q_\alpha = g_{\alpha\mu_1} q_{\mu_n} g^{\mu_n \mu_1}$ out of the production factor $(\prod_{m=2}^{2k}q_{\mu_m})$. 
The precise factor of $q_{\mu_n}$ that is extracted can be any one of the $(2k-1)$ elements in the product factor, implying an additional multiplicity factor of this size~\cite{Kretzer:2003iu}.
The remaining $(2k-2)$ instances of $q_{\mu_m}$ then contract with $(2k-2j)$ instances of $p_A^{\mu_m}$ to make $(2k-2j)$ powers of $(q\cdot p_A)$. The remaining $(2k-2)-(2k-2j)=2j-2$ factors of $q_{\mu_m}$ then contract with the $2(j-1)$ indices of the remaining $(j-1)$ metrics. After updating  $\eta$ and noting that the sum now starts at $j=1$, the outcome is
\begin{align}
g_{\alpha\mu_1}
\left(\prod_{m=2}^{2k} q_{\mu_m}\right)
\tilde{\Pi}^{\mu_1\dots\mu_{2k} }
& \Bigg\vert_{\mu_1\in\{g\dots g\}}
 =  
 q_{\alpha}\ 
\sum_{j=1}^{k}(-1)^{j}\ \frac{(2k-j)!}{2^{j}(2k)!}\ 
\frac{1}{2^{j-1}} \frac{(2k-2)!}{(j-1)! (2k-2j)!}\
\times(2k-1)
\nonumber\\
& \qquad \times \
(q^2)^{(j-1)}\
(q\cdot p_A)^{(2k-2j)}\
(p_A^{2})^{j}
\\
 =
-\frac{2q_{\alpha}}{Q^2}\
\frac{(Q^2)^{2k}}{2^{2k}}\ 
\sum_{j=1}^{k}\ 
& \frac{(2k-j)! (2k-1)!} {(2k)!(j-1)!(2k-2j)!}\ 
\left(\frac{M_A^2}{Q^2}\right)^j\ 
x_A^{-(2k-2j)}
\ . 
\end{align}

Assembling all terms for both $g_{\mu\mu_1}$ and $g_{\nu\mu_1}$, we obtain the expression
\begin{align}
\Delta \tilde{T}^{A}_{5\mu\nu}  
&=  
\frac{(p_{A\mu}q_\nu+p_{A\nu}q_\mu)}{Q^2}\ 
\nonumber\\
& \quad \times\
(-8i)\
\sum_{k=1}^\infty\
\left[C_{5}^{2k}A_{\tau=2}^{2k}\right]\
 \sum_{j=0}^{k-1}\ 
\frac{(2k-j)!(2k-1)!}{(2k)!j!(2k-2j-1)!}\ 
\left(\frac{M_A^2}{Q^2}\right)^j\ 
x_A^{-(2k-2j-1)}
\nonumber
\\
&
-\frac{(q_{\mu}q_\nu+q_{\nu}q_\mu)}{Q^2}\
\nonumber\\
& \quad \times\
(-8i)\
\sum_{k=1}^\infty\
\left[C_{5}^{2k}A_{\tau=2}^{2k}\right]\
\sum_{j=1}^{k}\ 
 \frac{(2k-j)! (2k-1)!} {(2k)!(j-1)!(2k-2j)!}\ 
\left(\frac{M_A^2}{Q^2}\right)^j\ 
x_A^{-(2k-2j)}\ 
 \nonumber \\
& 
+ \quad \mathcal{O}(\tau>2)\ .
\label{eq:opeOrgDT5}
\end{align}
Importantly, there are two tensor structures:
The first is $(p_{A\mu}q_\nu+p_{A\nu}q_\mu)$, whose coefficient contributes to $\Delta \tilde{T}^A_5$ in Eq.~\eqref{eq:forwardScattDecomp}.
The second is $q_{\mu}q_\nu$, whose coefficient contributes to $\Delta \tilde{T}^A_4$  in Eq.~\eqref{eq:forwardScattDecomp}. 
Note that the second term
can equally be written as $2(q_{\mu}q_\nu)$; we do not (yet) condense the result in order to emphasize that $g_{\mu\mu_1}$ and $g_{\nu\mu_1}$ each generates two terms.

\item[\textbf{(v) $\Delta \tilde{T}^{A}_{6\mu\nu}$:}]
The next term, $\Delta \tilde{T}^{A}_{6\mu\nu}$ for $M_A\neq0$, can be obtained directly from $\Delta \tilde{T}^{A}_{5\mu\nu}$ by making the replacement $q_\mu\to-q_\mu$ and $C_5\to C_6$. When making this replacement, the second part of Eq.~\eqref{eq:opeOrgDT5} vanishes since $(q_{\mu}q_\nu+q_{\nu}q_\mu)\to(q_{\mu}q_\nu-q_{\nu}q_\mu)=0$.
The final result for $\Delta \tilde{T}^{A}_{6\mu\nu}$ is  
\begin{align}
     \Delta \tilde{T}^{A}_{6\mu\nu}  
      =  &  
\frac{(p_{A\mu}q_\nu-p_{A\nu}q_\mu)}{Q^2}\
\times
(-8i)\times
\sum_{k=1}^\infty\
\left[C_{6}^{2k}A_{\tau=2}^{2k}\right]\
\nonumber
\\
& \times 
 \sum_{j=0}^{k-1}\ 
\frac{(2k-j)!(2k-1)!}{(2k)!j!(2k-2j-1)!}\ 
\left(\frac{M_A^2}{Q^2}\right)^j\ 
x_A^{-(2k-2j-1)}
\quad + \quad \mathcal{O}(\tau>2)\ .
\label{eq:opeOrgDT6}
\end{align}

\item[\textbf{(vi) $\Delta \tilde{T}^{A}_{2\mu\nu}$:}]
The final term $\Delta \tilde{T}^{A}_{2\mu\nu}$ for $M_A\neq0$, which is given by 
\begin{align}
    \Delta \tilde{T}^{A}_{2\mu\nu} 
          &=   
          (-2i)\ 
     \sum_{k=2}^\infty
\left[Q^2 C_2^{2k}A_{\tau=2}^{2k}\right]\
\times \frac{2^{2k}}{(Q^2)^{2k}} 
\times g_{\mu\mu_1}g_{\nu\mu_2}
\times \left(\prod_{m=3}^{2k} q_{\mu_m}\right)
 \times \tilde{\Pi}^{\mu_1\dots\mu_{2k} }\ 
  \nonumber \\
& 
+ \quad \mathcal{O}(\tau>2)\ .
\end{align}
is the most nuanced because the covariant tensor structure involves two metrics. That is to say, two lowered indices from the collection $\mu_{m=1}\dots\mu_{m=2k}$ are carried by $g_{\mu\mu_1}$ and $g_{\nu\mu_2}$. The contraction of these metrics with the collections $\{g\dots g\}$ and $\{p_A\dots p_A\}$, in turn,  generates five terms with four distinct Lorentz structures. The five possible ways are:
\begin{enumerate}[(a)]
    \item both $g_{\mu\mu_1}$ and $g_{\nu\mu_2}$ contract with momenta from $\{p_A\dots p_A\}$ and extract $p_{A\mu}p_{A\nu}$;
    \item both $g_{\mu\mu_1}$ and $g_{\nu\mu_2}$ contract with \textit{different} metrics from $\{g\dots g\}$ and extract the quantity $q_{\mu}q_{\nu}$  from the production factor $(\prod_{m=3}^{2k}q_{\mu_m})$;
    \item both $g_{\mu\mu_1}$ and $g_{\nu\mu_2}$ contract with \textit{the same} metric from $\{g\dots g\}$ and extract $g_{\mu\nu}$;
    \item $g_{\mu\mu_1}$ contracts with $\{p_A\dots p_A\}$ and 
$g_{\nu\mu_2}$ with $\{g\dots g\}$ to extract $(p_{A\mu}q_\nu)$;
    \item same as (d) but with the index exchange $\mu\leftrightarrow\nu$ to extract the quantity $(p_{A\nu}q_\mu)$.
\end{enumerate}

In case (a), contracting the metrics with the collection $\{p_A\dots p_A\}$ leaves $(2k-2j-2)$ instances of $p_A^{\mu_m}$ to contract with $(2k-2)$ instances of $q_{\mu_m}$. The remaining $(2k-2)-(2k-2j-2)=2j$ instances of $q_{\mu_m}$ then contracts with the $2j$ indices of the $j$ metrics in the collection $\{g\dots g\}$. Updating the permutation factor, we get
\begin{align}
g_{\mu\mu_1}g_{\nu\mu_2}
\left(\prod_{m=3}^{2k} q_{\mu_m}\right)
\tilde{\Pi}^{\mu_1\dots\mu_{2k} }
& \Bigg\vert_{\mu_1,\mu_2\in\{p_A\dots p_A\}}
 =  
p_{A\mu}p_{A\nu}\
\sum_{j=0}^{k-1}\ \frac{(2k-j)!}{2^{j}(2k)!}\ 
\frac{1}{2^{j}} \frac{(2k-2)!}{j! (2k-2j-2)!}\
\nonumber\\
& \qquad \times \
(-1)^{j}(q^2)^{j}\
(q\cdot p_A)^{(2k-2j-2)}\
(p_A^{2})^{j}
\\
 =
\frac{4p_{A\mu}p_{A\nu}}{Q^4}\
\frac{(Q^2)^{2k}}{2^{2k}}\ 
\sum_{j=0}^{k-1}\ 
& \frac{(2k-j)!(2k-2)!}{(2k)!j!(2k-2j-2)!}\ 
\left(\frac{M_A^2}{Q^2}\right)^j\ 
x_A^{-(2k-2j-2)}
\ .
\end{align}
Note that the summation over $j$ extends only to $j=k-1$ since there are fewer allowed permutations; this can be inferred from the denominator factor $1/(2k-2j-2)!$.

In case (b), contracting $g_{\mu\mu_1}$ and $g_{\nu\mu_2}$ with different metrics from the collection $\{g\dots g\}$ leaves $(2k-4)$ instances of $q_{\mu_m}$ to contract with $(2k-2j)$ instances of $p_A^{\mu_m}$. 
The precise $q_{\mu_m}$ that are extracted from the product factor $\left(\prod_{m=3}^{2k} q_{\mu_m}\right)$ can be any two of its $(2k-2)$ elements, implying the additional multiplicity factor $(2k-2)\times(2k-3)$.
The remaining $(2k-4)-(2k-2j)=(2j-4)$ factors of $q_{\mu_m}$ then contract with the $(j-2)$ metrics  remaining in $\{g\dots g\}$. Updating the $\eta$ factor and noting the summation over $j$ starts at $j=2$, we get
\begin{align}
g_{\mu\mu_1}g_{\nu\mu_2}&
\left(\prod_{m=3}^{2k} q_{\mu_m}\right)
\tilde{\Pi}^{\mu_1\dots\mu_{2k} }
 \Bigg\vert_{\mu_1,\mu_2\in\{g\dots g\}}
\nonumber\\   
& =  
 q_{\mu}q_{\nu}\ 
\sum_{j=2}^{k}\ \frac{(2k-j)!}{2^{j}(2k)!}\ 
\frac{1}{2^{j-2}} 
\frac{(2k-4)!}{(j-2)! (2k-2j)!}\
\times\ (2k-3)(2k-2)
\
\nonumber\\
& \qquad \times \
(-1)^{j}(q^2)^{(j-2)}\
(q\cdot p_A)^{(2k-2j)}\
(p_A^{2})^{j}
\\
& =
\frac{4q_{\mu}q_{\nu}}{Q^4}\
\frac{(Q^2)^{2k}}{2^{2k}}\ 
\sum_{j=2}^{k}\ 
 \frac{(2k-j)! (2k-2)!} {(2k)!(j-2)!(2k-2j)!}\ 
\left(\frac{M_A^2}{Q^2}\right)^j\ 
x_A^{-(2k-2j)}
\ .
\end{align}

In case (c), contracting $g_{\mu\mu_1}$ and $g_{\nu\mu_2}$ with the same metric from the collection $\{g\dots g\}$ leaves $(2k-2)$ instances of $q_{\mu_m}$ to contract with all $(2k-2j)$ instances of $p_A^{\mu_m}$. The $(2k-2)-(2k-2j)=(2j-2)$ uncontracted instances of $q_{\mu_m}$ are then matched with the $(j-1)$ metrics that are left in $\{g\dots g\}$. Updating $\eta$ and the summation over $j$, which now starts from $j=1$, we get
\begin{align}
g_{\mu\mu_1}g_{\nu\mu_2}\
&
\left(\prod_{m=3}^{2k} q_{\mu_m}\right)
\tilde{\Pi}^{\mu_1\dots\mu_{2k} }
\Bigg\vert_{g^{\mu_1\mu_2}\in\{g\dots g\}}\
 =\  
g_{\mu\nu}\
\sum_{j=1}^{k}\ \frac{(2k-j)!}{2^{j}(2k)!}\ 
\nonumber\\
& \qquad \times \
\frac{1}{2^{j-1}} 
\frac{(2k-2)!}{(j-1)! (2k-2j)!}\
(-1)^{j}(q^2)^{(j-1)}\
(q\cdot p_A)^{(2k-2j)}\
(p_A^{2})^{j}
\\
& =\
-\frac{2g_{\mu\nu}}{Q^2}\
\frac{(Q^2)^{2k}}{2^{2k}}\ 
\sum_{j=1}^{k}\ 
 \frac{(2k-j)!(2k-2)!}{(2k)!(j-1)!(2k-2j)!}\ 
\left(\frac{M_A^2}{Q^2}\right)^j\ 
x_A^{-(2k-2j)}
\ .
\end{align}

In case (d), $g_{\mu\mu_1}$ contracts with $\{p_A\dots p_A\}$, leaving $(2k-2j-1)$ instances of $p_A^{\mu_m}$, while $g_{\nu\mu_2}$ contracts with $\{g\dots g\}$, and eventually $(\prod_{\mu=3}^{2k}q_{\mu_n})$, leaving a total of $(2k-3)$  instances of $q_{\mu_m}$. The result is $(2k-2j-1)$ powers of $(q\cdot p_A)$ and $(2k-3)-(2k-2j-1)=(2j-2)$ factors of $q_{\mu_m}$ to contract with $(j-1)$ metrics. 
The $q_{\mu_m}$ that is extracted from the product factor can be any one of its initial $(2k-2)$ elements, implying an additional  multiplicity factor of this size.
Updating $\eta$ and noting the summation over $j$ runs from $j=1$ to $j=k-1$, we obtain
\begin{align}
& g_{\mu\mu_1}g_{\nu\mu_2}
\left(\prod_{m=3}^{2k} q_{\mu_m}\right)
\tilde{\Pi}^{\mu_1\dots\mu_{2k} }
 \Bigg\vert^{\mu_1\in\{p_A\dots p_A\}}_{\mu_2\in\{g\dots g\}}
 =  
 p_{A\mu}q_{\nu}\
\sum_{j=1}^{k-1}\ \frac{(2k-j)!}{2^{j}(2k)!}\ 
\nonumber\\
& \qquad \times \
\frac{1}{2^{j-1}} \frac{(2k-3)!}{(j-1)! (2k-2j-1)!}\
\times(2k-2)\ 
(-1)^{j}(q^2)^{(j-1)}\
(q\cdot p_A)^{(2k-2j-1)}\
(p_A^{2})^{j}
\\
& =
-\frac{4p_{A\mu}q_{\nu}}{Q^4}\
\frac{(Q^2)^{2k}}{2^{2k}}\ 
\sum_{j=1}^{k-1}\ 
 \frac{(2k-j)! (2k-2)!} {(2k)!(j-1)!(2k-2j-1)!}\ 
\left(\frac{M_A^2}{Q^2}\right)^j\ 
x_A^{-(2k-2j-1)}
\ .
\end{align}
Case (e) follows from case (d) but with $\mu\leftrightarrow\nu$ exchanged.

After assembling all five cases and regrouping, we find
\begin{align}
    \Delta \tilde{T}^{A}_{2\mu\nu} 
    &   =  
    \frac{2p_{A\mu}p_{A\nu}}{Q^2}\ \times (-4i)\
\sum_{k=1}^\infty\left[C_2^{2k}A_{\tau=2}^{2k}\right]\
\nonumber
\\
& \quad ~  \times
\sum_{j=0}^{k-1}\ 
 \frac{(2k-j)!(2k-2)!}{(2k)!j!(2k-2j-2)!}\ 
\left(\frac{M_A^2}{Q^2}\right)^j\ 
x_A^{-(2k-2j-2)}
\nonumber
\\
 -g_{\mu\nu}\ & \times (-4i)\
\sum_{k=1}^\infty\left[C_2^{2k}A_{\tau=2}^{2k}\right]\
\sum_{j=1}^{k}\ 
 \frac{(2k-j)!(2k-2)!}{(2k)!(j-1)!(2k-2j)!}\ 
\left(\frac{M_A^2}{Q^2}\right)^j\ 
x_A^{-(2k-2j)}
\nonumber
\\
+ 
\frac{2q_{\mu}q_{\nu}}{Q^2}\ & \times (-4i)\
\sum_{k=2}^\infty\left[C_2^{2k}A_{\tau=2}^{2k}\right]\
\sum_{j=2}^{k}\ 
 \frac{(2k-j)! (2k-2)!} {(2k)!(j-2)!(2k-2j)!}\ 
\left(\frac{M_A^2}{Q^2}\right)^j\ 
x_A^{-(2k-2j)}
\nonumber
\\
& -\frac{2(p_{A\mu}q_{\nu}+p_{A\nu}q_{\mu})}{Q^2}\  \times (-4i)\
\sum_{k=2}^\infty\left[C_2^{2k}A_{\tau=2}^{2k}\right]\
\nonumber
\\
& \quad ~  \times
\sum_{j=1}^{k-1}\ 
 \frac{(2k-j)!  (2k-2)!} {(2k)!(j-1)!(2k-2j-1)!}\ 
\left(\frac{M_A^2}{Q^2}\right)^j\ 
x_A^{-(2k-2j-1)}\
 +\  \mathcal{O}(\tau>2)\  .
\label{eq:opeOrgDT2}
\end{align}
The result implies that $\Delta \tilde{T}_{2\mu\nu}^A$, and hence the Wilson coefficient $C_2^{2k}$, contributes to $\Delta T_1^A$, $\Delta T_2^A$, $\Delta T_4^A$, and $\Delta T_5^A$ in Eq.~\eqref{eq:forwardScattDecomp}. The absence of contributions to $\Delta T_3^A$ and $\Delta T_6^A$, which are asymmetric under $\mu\leftrightarrow\nu$ exchange, can be traced to the tensor coefficient multiplying $C_2^{2k}$ in $c_{\mu\nu\mu_1,\dots,\mu_{2k}}^{\tau=2,\iota}$, which is symmetric under $\mu\leftrightarrow\nu$ exchange.

\end{description}

\subsection{Nuclear structure functions from the OPE II: 
structure-function mixing
}\label{app:opeMixing}

Using the expressions for $\Delta \tilde{T}_{\iota\ \mu\nu}^{A}$, we finally obtain the quantities $\Delta \tilde{T}_{i}^{A}$ in Eq.~\eqref{eq:forwardScattDecomp}. For $\Delta \tilde{T}^{A}_{1}$ and $M_A\neq0$, we collect like-terms from $\Delta \tilde{T}^{A}_{1\mu\nu}$ and $\Delta \tilde{T}^{A}_{2\mu\nu}$ to obtain
\begin{align}
\Delta \tilde{T}^{A}_{1}  = 
(-4i)\ &\ \sum_{k=1}^\infty\
\left[C_1^{2k}A_{\tau=2}^{2k}\right]\
\sum_{j=0}^{k}\ 
\frac{(2k-j)! (2k)!}{(2k)! j! (2k-2j)!}\ 
\left(\frac{M_A^2}{Q^2}\right)^j\ 
x_A^{-(2k-2j)}\ 
\nonumber
\\
 +  (-4i)\ &\ 
 \sum_{k=1}^\infty\left[C_2^{2k}A_{\tau=2}^{2k}\right]\
\sum_{j=1}^{k}\ 
 \frac{(2k-j)!(2k-2)!}{(2k)!(j-1)!(2k-2j)!}\ 
\left(\frac{M_A^2}{Q^2}\right)^j\ 
x_A^{-(2k-2j)}\
+\ \mathcal{O}(\tau>2)\ 
\\ 
= 
(-4i)\ &\
\sum_{l=0}^\infty\
x_A^{-(2l+1)}\
\sum_{j=0}^\infty\
\frac{(2l+j+1)!}{j!(2l+1)!}\ 
\left(\frac{M_A^2}{Q^2}\right)^j\ 
\nonumber\\
&\ \quad\ \times\
\left[
\left(C_1^{2l+2j+1}A_{\tau=2}^{2l+2j+1}\right)
+
\frac{j\left(C_2^{2l+2j+1}A_{\tau=2}^{2l+2j+1}\right)}{(2l+2j+1)(2l+2j)}
\right]\
 +\ \mathcal{O}(\tau>2)\ .
\label{eq:strFunT1FullExpression}
\end{align}
To reach Eq.~\eqref{eq:strFunT1FullExpression}, a factor of $1=(2k)(2k-1)j/(2k)(2k-1)j$ is introduced into the $C_2$ term to complete the factorials.
The summation in the $C_2$ term can then be extended to include $j=0$ since the contribution is zero.  
We also re-index the summation over $k$ into a summation over {$2l=2k-2j-1$}. After this shift  it becomes manifest that the contribution from the $C_2$ Wilson coefficient vanishes when the summation over $j$ is truncated at $j=0$. It also becomes clear that structure-function mixing is due to TMCs since neglecting terms where $j>0$ is equivalent to taking the $(M_A^2/Q^2)\to0$ limit. 

The coefficient $\Delta \tilde{T}^{A}_{2}$ is sourced entirely from $\Delta \tilde{T}^{A}_{\mu\nu2}$, and is given by
\begin{align}
\Delta \tilde{T}^{A}_{2}  &=  
(-4i)\
\frac{2M_A^2}{Q^2}\
\sum_{k=1}^\infty\left[C_2^{2k}A_{\tau=2}^{2k}\right]\
\sum_{j=0}^{k-1}\ 
 \frac{(2k-j)!(2k-2)!}{(2k)!j!(2k-2j-2)!}\ 
\left(\frac{M_A^2}{Q^2}\right)^j\ 
x_A^{-(2k-2j-2)}\
\nonumber\\
&
+ \quad \mathcal{O}(\tau>2)\ 
\\ 
&=  
(-4i)\ 
\frac{2M_A^2}{Q^2}\
\sum_{l=0}^\infty\
x_A^{-(2l-1)}\
\sum_{j=0}^\infty\
 \frac{(2l+j+1)!}{j!(2l+1)!}\ 
 \left(\frac{M_A^2}{Q^2}\right)^j 
 (2l+1)(2l)\ 
\frac{\left(C_2^{(2l+2j+1)}A_{\tau=2}^{(2l+2j+1)}\right)}{(2l+2j+1)(2l+2j)}\
\nonumber\\
& 
+ \quad \mathcal{O}(\tau>2)\ 
.
\label{eq:strFunT2FullExpression}
\end{align}
In reaching Eq.~\eqref{eq:strFunT2FullExpression} we again completed the factorials and set $2l=2k-2j-1$. 

Similarly, $\Delta \tilde{T}^{A}_{3}$ for $M_A\neq0$ is sourced entirely from $\Delta \tilde{T}^{A}_{\mu\nu3}$. It is given by
\begin{align}
\Delta \tilde{T}^{A}_{3}   = &
(-4i)\ \frac{M_A^2}{Q^2}\
\sum_{k=1}^\infty\
\left[C_3^{2k}A_{\tau=2}^{2k}\right]\
\sum_{j=0}^{k-1}\ 
\frac{(2k-j)!(2k-1)!}{(2k)!j!(2k-2j-1)!}\ 
\left(\frac{M_A^2}{Q^2}\right)^j\ 
x_A^{-(2k-2j-1)}\
\nonumber\\
& + \quad \mathcal{O}(\tau>2)\ 
\\
   = &
(-4i)\ \frac{M_A^2}{Q^2}\
\sum_{l=0}^\infty\
x_A^{-(2l-1)}\
\sum_{j=0}^\infty\
\frac{(2l+j)!}{j!(2l)!}\ 
\left(\frac{M_A^2}{Q^2}\right)^j\ 
(2l)\
\frac{\left(C_3^{(2l+2j)}A_{\tau=2}^{(2l+2j)}\right)}{(2l+2j)}\
\nonumber\\
& + \quad \mathcal{O}(\tau>2)\ 
,
\label{eq:strFunT3FullExpression}
\end{align}
where the index $l=k-j$ was used after completing the factorials.

For $\Delta \tilde{T}^{A}_{4}$ and $M_A\neq0$, we collect terms from $\Delta \tilde{T}^{A}_{\mu\nu4}$, $\Delta \tilde{T}^{A}_{\mu\nu2}$, and $\Delta \tilde{T}^{A}_{\mu\nu6}$. The result is
\begin{align}
& \Delta \tilde{T}^{A}_{4}  \quad =  
(-4i)\
\frac{2M_A^2}{Q^2}\ 
\sum_{k=1}^\infty\
\left[C_4^{2k}A_{\tau=2}^{2k}\right]\
\sum_{j=0}^{k}\ 
\frac{(2k-j)!(2k)!}{(2k)! j! (2k-2j)!}\ 
\left(\frac{M_A^2}{Q^2}\right)^j\ 
x_A^{-(2k-2j)}\ .
\nonumber
\\
+&(-4i)\
\frac{2M_A^2}{Q^2}\
\sum_{k=2}^\infty\left[C_2^{2k}A_{\tau=2}^{2k}\right]\
\sum_{j=2}^{k}\ 
 \frac{(2k-j)!{(2k-2)!}}{(2k)!(j-2)!(2k-2j)!}\ 
\left(\frac{M_A^2}{Q^2}\right)^j\ 
x_A^{-(2k-2j)}\
\nonumber
\\
-&(-4i)\
\frac{4M_A^2}{Q^2}\
\sum_{k=1}^\infty\
\left[C_{5}^{2k}A_{\tau=2}^{2k}\right]\
\sum_{j=1}^{k}\ 
 \frac{(2k-j)!{(2k-1)!}}{(2k)!(j-1)!(2k-2j)!}\ 
\left(\frac{M_A^2}{Q^2}\right)^j\ 
x_A^{-(2k-2j)}\ 
\nonumber\\
& +  \mathcal{O}(\tau>2)\ 
\\  
=& (-4i)\
\frac{2M_A^2}{Q^2}\
\sum_{l=0}^\infty x_A^{-(2l+1)}\ 
\sum_{j=0}^\infty \frac{(2l+j+1)!}{j!(2l+1)!}\ 
\left(\frac{M_A^2}{Q^2}\right)^j\ 
\nonumber\\
& \times 
\left[ 
\left(C_{4}^{(2l+2j+1)}A_{\tau=2}^{(2l+2j+1)}\right)
+\
\frac{j\ (j-1)\ \left(C_{2}^{(2l+2j+1)}A_{\tau=2}^{(2l+2j+1)}\right)}{
(2l+2j+1)(2l+2j)
}
-\
\frac{2j\ \left(C_{5}^{(2l+2j+1)}A_{\tau=2}^{(2l+2j+1)}\right)}{(2l+2j+1)}\ 
\right]\ 
\nonumber\\
&  + \mathcal{O}(\tau>2)\ .
\label{eq:strFunT4FullExpression}
\end{align}
We complete the factorials using factors of $1=(2k)j/(2j)j$, etc., and use 
{$2l=2k-2j-1$} to re-index the summation over $k$. The summations over $j$ can also be extended to $j=0$ since the additional terms are proportional to $j$ or $j(j-1)$, i.e., are zero at $j=0$ and/or $j=1$, and hence do not contribute.

The coefficient $\Delta \tilde{T}^{A}_{5}$ for $M_A\neq0$
is generated by $\Delta \tilde{T}^{A}_{2\mu\nu}$
and $\Delta \tilde{T}^{A}_{5\mu\nu}$. The result is
\begin{align}
 \Delta \tilde{T}^{A}_{5}  &=  
(-4i)\ \frac{2M_A^2}{Q^2}\
\sum_{k=1}^\infty\
\left[C_{5}^{2k}A_{\tau=2}^{2k}\right]\
 \sum_{j=0}^{k-1}\ 
\frac{(2k-j)!(2k-1)!}{(2k)!j!(2k-2j-1)!}\ 
\left(\frac{M_A^2}{Q^2}\right)^j\ 
x_A^{-(2k-2j-1)}
\nonumber
\\
&-(-4i)\
\frac{2M_A^2}{Q^2}\
\sum_{k=2}^\infty\left[C_2^{2k}A_{\tau=2}^{2k}\right]\
\sum_{j=1}^{k-1}\ 
 \frac{(2k-j)!{(2k-2)!}}{(2k)!(j-1)!(2k-2j-1)!}\ 
\left(\frac{M_A^2}{Q^2}\right)^j\ 
x_A^{-(2k-2j-1)}\
\nonumber\\
& + \quad \mathcal{O}(\tau>2)\ 
\\
& =(-4i)\
\frac{2M_A^2}{Q^2}\
\sum_{l=0}^\infty x_A^{-2l}\ 
\sum_{j=0}^\infty \frac{(2l+j+1)!}{j!(2l+1)!}\ 
\left(\frac{M_A^2}{Q^2}\right)^j\ 
\nonumber\\
& \times    
\left[
\frac{(2l+1)\ \left(C_{5}^{(2l+2j+1)}A_{\tau=2}^{(2l+2j+1)}\right)}{(2l+2j+1)}\
-
\frac{j\ (2l+1)\ \left(C_{2}^{(2l+2j+1)}A_{\tau=2}^{(2l+2j+1)}\right)}{
(2l+2j+1)(2l+2j)
}\
\right]
\nonumber\\
&\
 + \mathcal{O}(\tau>2)\ 
,
\label{eq:strFunT5FullExpression}
\end{align}
where the index reassignment is {$2l=2k-2j-1$}.  

Finally, the coefficient $\Delta \tilde{T}^{A}_{6}$
is generated solely by $\Delta \tilde{T}^{A}_{6\mu\nu}$. It is given by
\begin{align}
\Delta \tilde{T}^{A}_{6}  = & 
(-4i)\
\frac{2M_A^2}{Q^2}\
\sum_{k=1}^\infty\
\left[C_{6}^{2k}A_{\tau=2}^{2k}\right]\
\sum_{j=0}^{k-1}\ 
\frac{(2k-j)!(2k-1)!}{(2k)!j!(2k-2j-1)!}\ 
\left(\frac{M_A^2}{Q^2}\right)^j\ 
x_A^{-(2k-2j-1)}\
\nonumber\\
& 
+ \quad \mathcal{O}(\tau>2)\ 
\\
=  & 
(-4i)\
\frac{2M_A^2}{Q^2}\
\sum_{l=0}^\infty\
x_A^{-(2l-1)}\
\sum_{j=0}^\infty\
\frac{(2l+j)!}{j!(2l)!}\ 
\left(\frac{M_A^2}{Q^2}\right)^j\ 
(2l)\
\frac{\left(C_{6}^{(2l+2j)}A_{\tau=2}^{(2l+2j)}\right)}{(2l+2j)}\
\nonumber\\
& 
+ \quad \mathcal{O}(\tau>2)\ 
,
\label{eq:strFunT6FullExpression}
\end{align}
where the index $l=k-j$ was used after completing the factorials.

We briefly stress that the $(M_A^2/Q^2)$ prefactor in all $\Delta \tilde{T}^{A}_{i}$ is conventional. They exist to match the normalization of $\tilde{T}^{A}_{i}$ in Eq.~\eqref{eq:forwardScattDecomp} and structure functions $\tilde{W}^{A}_{i}$ in Eq.~\eqref{eq:strFnExpandW}. The massless target limit should be understood as truncating the summation over $j$ at $j=0$. In addition, the re-indexing from $k$ to $l$ in the above expressions is not random. Different choices of relabeling can be found in the literature and depend on the precise interpretation of the quantities to the right of the $x_A^{-N}$ factor.
Our choices are motivated by the attempt to align the ratio of factorials into binomial distributions, e.g., $\binom{k+j}{j} = (k+j)!/j!k!$.

\subsection{Nuclear structure fns. from the OPE III:  massless and massive targets}\label{app:opeMassive}

\subsection*{Massless nuclear targets}

As a final preliminary step to obtaining TMCs at leading twist, we
establish the connection between the effective Wilson coefficients $C_{\iota}$, the reduced matrix element $A_{\tau=2}$, and structure functions in the limit 
that $(M_A^2/Q^2) \to 0$; in this limit, the TMCs will not be present, and we label this result as ``No~TMC'' in Eq.~\eqref{eq:app_ope_matching}).
For each $\Delta \tilde{T}^{A}_{i}$ above, truncating terms with $j>0$ and using
the Taylor expansion of $\Delta \tilde{T}^{A}_{i}$ from Eq.~\eqref{eq:dispersionStrFn} gives the following:
\begin{subequations}
\label{eq:app_ope_matching}
\begin{align}
\Delta \tilde{T}^{A}_{1} \Bigg\vert_{j=0}  &=
-4i
\sum_{l=0}^\infty\
x_A^{-(2l+1)}
\left(C_1^{2l+1}A_{\tau=2}^{2l+1}\right)
+ \mathcal{O}(\tau>2)
=
\sum_{l=0}^\infty\
\omega_A^{2l+1}
\tilde{F}_1^{A(2l+1)}\Bigg\vert_{\rm No~TMC}
,
\\
\left(\frac{Q^2}{2x_A M_A^2}\right)\
\Delta \tilde{T}^{A}_{2} \Bigg\vert_{j=0} &=  
-4i
\sum_{l=0}^\infty\
x_A^{-2l}\
\left(C_2^{2l+1}A_{\tau=2}^{2l+1}\right)\
+ \mathcal{O}(\tau>2)
=
\sum_{l=0}^\infty\
\omega_A^{2l}\
\tilde{F}_2^{A(2l)}\Bigg\vert_{\rm No~TMC}
,
\\
\left(\frac{Q^2}{x_A M_A^2}\right)\
\Delta \tilde{T}^{A}_{3} \Bigg\vert_{j=0} &=  
-4i 
\sum_{l=0}^\infty\
x_A^{-2l}\
\left(C_3^{2l}A_{\tau=2}^{2l}\right)
+ \mathcal{O}(\tau>2)
=
\sum_{l=0}^\infty\
\omega_A^{2l}\
\tilde{F}_3^{A(2l)}\Bigg\vert_{\rm No~TMC}
,
\\
\left(\frac{Q^2}{2 M_A^2}\right)
\Delta \tilde{T}^{A}_{4} \Bigg\vert_{j=0} & =  
-4i
\sum_{l=0}^\infty\
x_A^{-(2l+1)}
\left(C_4^{2l+1}A_{\tau=2}^{2l+1}\right)
+ \mathcal{O}(\tau>2)
=  
\sum_{l=0}^\infty\
\omega_A^{2l+1}
\tilde{F}_4^{A(2l+1)}\Bigg\vert_{\rm No~TMC}
,
\\
\left(\frac{Q^2}{2x_A M_A^2}\right)\
\Delta \tilde{T}^{A}_{5} \Bigg\vert_{j=0} &=
-4i
\sum_{l=0}^\infty\
x_A^{-(2l+1)}
\left(C_5^{2l+1}A_{\tau=2}^{2l+1}\right)
+ \mathcal{O}(\tau>2)
= 
\sum_{l=0}^\infty\
\omega_A^{2l+1}
\tilde{F}_5^{A(2l+1)}\Bigg\vert_{\rm No~TMC}
,
\\
\left(\frac{Q^2}{2x_A M_A^2}\right)\
\Delta \tilde{T}^{A}_{6} \Bigg\vert_{j=0} &= 
-4i
\sum_{l=0}^\infty\
x_A^{-2l}\
\left(C_{6}^{2l}A_{\tau=2}^{2l}\right)\
+ \mathcal{O}(\tau>2)
=
\sum_{l=0}^\infty\
\omega_A^{2l}\
\tilde{F}_6^{A(2l)}\Bigg\vert_{\rm No~TMC}
\ 
.
\end{align}
\end{subequations}

The conclusion is that, in the massless target limit, one can identify the leading power of the product of the $N^{\rm th}$ 
Wilson coefficient $C_{i}^N$ (note the index shift from $\iota$ to $i$) and hadronic matrix element $A_{\tau=2}^N$ 
as a Mellin moment of structure functions themselves.
This is summarized by
\begin{subequations}
\label{eq:masterMomentAppendix}
\begin{align}
\tilde{F}_i^{AN}\Bigg\vert_{\rm No~TMC} &= C_i^N A_{\tau=2}^N\ +\  \mathcal{O}(\tau>2) \quad\text{for}\quad i=1,3-6\ ,
\\
\tilde{F}_2^{A(N-1)}\Bigg\vert_{\rm No~TMC} &= C_2^N A_{\tau=2}^N\ +\  \mathcal{O}(\tau>2)\ .
\end{align}
\end{subequations}
As a technical remark, the normalization of Eq.~\eqref{eq:masterMomentAppendix} fixes the normalizations of $C_i$ in Eqs.~\eqref{eq:app_wilson_def} and \eqref{eq:tmunuFull}.

\subsection*{Massive nuclear targets}

Given the relationship between the moments of $\tilde{F}_i^{A}\big\vert_{\rm No~TMC}$ and the product $(C_i A_{\tau=2})$ in the massless target limit, we are now in position to bootstrap expressions for $\tilde{F}_i^{A}$ with target mass corrections. 
The key to this is comparing the expansion of $\Delta \tilde{T}^{A}_{i}$ in Eq.~\eqref{eq:dispersionStrFn} with the full  expressions 
for $\Delta \tilde{T}^{A}_{i}$ in
Eqs.~\eqref{eq:strFunT1FullExpression}-\eqref{eq:strFunT6FullExpression}.
One can identify the moments of structure functions with TMCs as everything encapsulated by the summation over $j$,
i.e., the coefficients multiplying the  $x_A^{-2l(\pm1)}$ factor.
Using Eq.~\eqref{eq:masterMomentAppendix}, the unknown/non-perturbative $(C_i A_{\tau=2})$ can then be replaced by the measurable $\tilde{F}_i^{A}\big\vert_{\rm No~TMC}$.
Expressions in $x_A$-space are obtained by inverse Mellin transformations. 
The transformations and summations over $j$ are solvable using various identities.

\begin{description}
\item[$\tilde{F}_2^A$:] To build $\tilde{F}_2^A$ for a non-zero target mass, we start from $\Delta \tilde{T}^{A}_{2}$ in Eq.~\eqref{eq:strFunT2FullExpression}.
Using Eq.~\eqref{eq:dispersionStrFn} (or Eq.~\eqref{eq:app_ope_matching}), 
we identify everything to the right of $x_A^{-(2l-1)}$ in Eq.~\eqref{eq:strFunT2FullExpression} as the $(2l)^{th}$ moment of $\tilde{F}_2^A(x_A,Q^2)$ with TMCs. This allows us to write
\begin{align}
\tilde{F}_2^{A(2l)} &= 
\sum_{j=0}^\infty\
 \frac{(2l+j+1)!}{j!(2l+1)!}\ 
 \left(\frac{M_A^2}{Q^2}\right)^j\ 
 (2l+1)(2l)\ 
\frac{\left(C_2^{(2l+2j+1)}A_{\tau=2}^{(2l+2j+1)}\right)}{(2l+2j+1)(2l+2j)}\
.
\end{align}
For an arbitrary integrable function $B(y)$ over $y\in[0,1]$ and $m\geq0$, we have the identities
\begin{subequations}
\begin{align}
    \frac{1}{(m+1)}\ \int_0^1 dy\ y^{m+1}\ B(y) &= \int_0^1 dy\ y^m\ H(y),\ 
    \label{eq:integralDefPower1}
    \\
    \frac{1}{(m+2)(m+1)}\ \int_0^1 dy\ y^{(m+2)}\ B(y) &= \int_0^1 dy\ y^m\ G(y),\
    \label{eq:integralDefPower2}
    \\
     \ \text{where} \qquad  \qquad 
    H(y) &= \int_y^1 dy'\ B(y'),\ 
    \\
    \quad\text{and}\qquad 
    G(y) = \int_y^1 dy'\ \int_{y'}^1 dy'' \ B(y'') &= 
    \int_y^1 dy'\  (y'-y) \ B(y')\
\end{align}
\end{subequations}
These identities will be used repeatedly for the different structure functions.
Schematically, the identities show that each factor of ``$1/m$,'' which originates from completing the (numerator) factorials in $\Delta \tilde{T}^{A}_{i}$, corresponds to an additional integral over a structure function.
Using the second identity 
and Eq.~\eqref{eq:masterMomentAppendix}, we have
\begin{align}
\frac{C_2^{(2l+2j+1)}A_{\tau=2}^{(2l+2j+1)}}{(2l+2j+1)(2l+2j)}
&= 
\frac{1}{(2l+2j+1)(2l+2j)}\ 
\int_0^1 dy\ y^{2l+2j+1}\cdot  y^{-2} \tilde{F}_2^A(y)\Bigg\vert_{\rm No~TMCs} 
\label{eq:appDefg2_pre}
\\
& = \int_0^1 dy\  y^{2l+2j-1}\  \tilde{g}_2(y), 
\quad\text{where}
\\ 
\tilde{g}_2(y) &\equiv \int_y^1 dy' \int_{y'}^1 dy''\ (y'')^{-2}\ \tilde{F}_2^A(y'')\Bigg\vert_{\rm No~TMCs}\ 
.
\label{eq:appDefg2}
\end{align}

This allows allows us to evaluate the summation over $j$:
\begin{align}
\tilde{F}_2^{A(2l)} &= 
\int_0^1 dy\  y^{2l-1}\  \tilde{g}_2(y)\
 (2l+1)(2l)\ 
\sum_{j=0}^\infty\
 \frac{(2l+j+1)!}{j!(2l+1)!}\ 
 \left(\frac{y^2 M_A^2}{Q^2}\right)^j\ 
\\
 &= 
\int_0^1 dy\  y^{-1}\  \tilde{g}_2(y)\
 \frac{(2l+1)(2l)\ y^{2l}}{(1-y^2 M_A^2 / Q^2)^{2l+2}}\ 
 ,
\end{align}
where we used the first of the following  identities, which also hold for $z>1$:
\begin{subequations}
\begin{align}
\sum_{j=0}^\infty\  \frac{(n+j)!}{j!\ n!}\ z^j &= \frac{1}{(1-z)^{(n+1)}}\ ,
    \\
\sum_{j=1}^\infty\  j\ \frac{(n+j)!}{j!\ n!}\ z^j &= \frac{(n+1)\ z}{(1-z)^{(n+2)}}\ ,
    \\
\sum_{j=2}^\infty\  j\ (j-1)\ \frac{(n+j)!}{j!\ n!}\ z^j &= \frac{(n+1)\ (n+2)\ z^2}{(1-z)^{(n+3)}}\
.    
\end{align}
\label{eq:summationClosedForm}
\end{subequations}
The first identity is derived by Taylor expanding the right-hand side and completing the factorial. 
The second follows from taking the derivative of the first identity and multiplying by $z$; similarly, the third by taking two derivatives and scaling by $z^2$.

After relabeling $N=2l$, and using the second of the following identities
\begin{subequations}
    \label{eq:derivativeIdentity}
\begin{align}
    (n + a) z^{-n} &= -z^{(1+a)}\ \frac{d}{dz}\ z^{-(n+a)}\ ,
\\
    (n+a)(n+a+1) z^{-n} &=\ z^{(2+a)} \frac{d^2}{dz^2}\ z^{-(n+a)}\ ,
\end{align}
\end{subequations}
we can take the inverse Mellin transformation of $\tilde{F}_2^{AN}$.
We remark that these derivative identities will be used repeatedly for the different structure functions. Schematically, they show that each factor of ``$n$'' or ``$(n+1)$,'' which  originates from completing the factorials in $\Delta \tilde{T}^{A}_{i}$, corresponds to a derivative over the generating functions $G(y)$ and $H(y)$.  
 
Inserting the derivative identity and taking the inverse Mellin transformation gives 
\begin{align}
\tilde{F}_2^{A}(x_A) &= \frac{1}{2\pi i}\int_{c-i\infty}^{c+i\infty} dN\ x_A^{-N}\ \tilde{F}_2^{AN} 
\\
& = 
\frac{x_A^2}{2\pi i}  \frac{d^2}{dx_A^2}
\int_0^1 dy\  
\frac{\tilde{g}_2(y)}{y(1-y^2 M_A^2 / Q^2)^2}\
\int_{c-i\infty}^{c+i\infty} dN\
 \frac{(y/x_A)^N}{(1-y^2 M_A^2 / Q^2)^{N}}\ 
 \\
 &=
 x_A^2\  \frac{d^2}{dx_A^2}
\int_0^1 dy\  
\frac{\tilde{g}_2(y)}{y(1-y^2 M_A^2 / Q^2)^2}\
 \delta\left[
\log\left(\frac{(y/x_A)}{(1-y^2 M_A^2 / Q^2)}\right)
\right]\
,
\end{align}
where we employed the contour integral
\begin{align}
    \int_{c-i\infty}^{c+i\infty} dN\ z^N = (2\pi i)\ \delta\left[\log(z)\right]\ 
    .
    \label{eq:contour}
\end{align}

To evaluate the $\delta$ function, we use the decomposition formula
\begin{align}
    \delta\left[g(z)\right] = \sum_{z_k}\ \frac{1}{\vert g'(z)\vert}\ \delta(z-z_k), \quad \text{such~that}\quad g(z=z_k) = 0.
    \label{eq:deltaIdentity}
\end{align}
This allows us to express the $\delta$ function in $\tilde{F}_2^A(x_A)$ in terms of the zeros of the logarithm, i.e., when $(y/x) = (1-y^2 M_A^2 / Q^2)$. While the logarithm vanishes for two values of $y$, only one is physical. The unphysical solution $y_2$ lies outside the domain $y\in[0,1]$, meaning the contribution from $\delta(y-y_2)$ is always zero.
The physical solution is the Nachtmann scaling variable~\cite{Nachtmann:1973mr}:
\begin{align}
    \xi_A = \frac{2x_A}{1+r_A}\ , \quad\text{where}\quad 
    r_A \equiv \sqrt{1+4 x_A^2 M_A^2/Q^2} 
    = \frac{(1+\xi_A^2 M_A^2/Q^2)}{(1-\xi_A^2 M_A^2/Q^2)}
    \
    ,
\end{align}
and corresponds to the following solution for the $\delta$ function:
\begin{align}
 \delta\left[
\log\left(\frac{(y/x_A)}{(1-y^2 M_A^2 / Q^2)}\right)
\right]\
 =\ 
    \left[\frac{y(1-y^2 M_A^2 / Q^2)}{(1+ y^2 M_A^2 / Q^2)}\right]\ \delta(\xi_A-y)\ .
\label{eq:deltaDecom}
\end{align}

Some identities that are useful for mapping between results in the literature include
\begin{align}
    \frac{x_A}{r_A \xi_A} = \frac{1}{(1+ \xi_A^2 M_A^2 / Q^2)} \quad\text{and}\quad
    \frac{x_A}{\xi_A} = \frac{1}{(1- \xi_A^2 M_A^2 / Q^2)}\ .
\label{eq:deltaDecom2}
\end{align}
Replacing the $\delta$ function with Eq.~\eqref{eq:deltaDecom} allows us to finally write
\begin{align}
\tilde{F}_2^{A}(x_A) &= 
 x_A^2\  \frac{d^2}{dx_A^2}
\int_0^1 dy\  
\frac{\tilde{g}_2(y)}{y(1-y^2 M_A^2 / Q^2)^2}\
\left[\frac{y(1-y^2 M_A^2 / Q^2)}{(1+ y^2 M_A^2 / Q^2)}\right]\ \delta(\xi-y)
\
\\
\Aboxed{ 
\tilde{F}_2^{A}(x_A) 
&=\   
x_A^2\  \frac{d^2}{dx_A^2}\ \left[\frac{(1+r_A)^2}{4r_A}\ \tilde{g}_2(\xi_A)\right]\  \ .
} %
\end{align}
Allowing the derivative to act on the quantities inside the bracket recovers Eq.~\eqref{eq:master2}. The final expression is summarized in Eq.~\eqref{eq:app_summary}.
We omit intermediate steps, which is a tedious but straightforward 
exercise in (one dimensional) differentiation 
when $r_A$ and $\xi_A$ are treated as functions of $x_A$.
After successive applications of the product rule, i.e.,
$(u\cdot v)'=u'\cdot v + u\cdot v'$,
and the chain rule, i.e.,
$(g\circ \xi\circ x)' = (g'\circ\xi)\cdot\xi'(x)$,
the $\mathcal{O}(\tilde{g}_2)$ term in 
Eqs.~\eqref{eq:master2} and \eqref{eq:app_summary}
is generated when both derivatives act on the $(1+r_A)^2/4r_A$ factor.
Likewise,
the $\mathcal{O}(\tilde{h}_2)$ term is generated when only one derivative acts on $\tilde{g}_2(\xi_A)$,
and 
the $\mathcal{O}(\tilde{F}^A_2)$ term is generated when both derivatives act on $\tilde{g}_2(\xi_A)$.

Some identities that are helpful to obtain the final result include:
\\
\begin{subequations}
\label{eq:app_chainrule_f2}
\begin{minipage}{.43\textwidth}
\begin{align}
\frac{d}{dx_A}\ r_A &= \frac{(r_A^2-1)}{x_A r_A}\ ,
    \\
\frac{d}{dx_A}\ \frac{1}{r_A^n} &= 
\frac{(-n)}{x_A}
\frac{(r_A^2-1)}{r_A^{n+2}}\ ,
\\
\frac{d}{dx_A}\ \left(\frac{(1+r_A)^2}{4r_A}\right) &=
\frac{(r_A^2-1)^2}{4x_A r_A^2}\ ,
\\
r_A^2-1 &=  \frac{4x_A^2M_A^2}{Q^2}\ ,
\end{align}
\end{minipage}
\begin{minipage}{.52\textwidth}
\begin{align}
\frac{d}{dx_A}\ \xi_A &= \frac{2}{r_A(1+r_A)}\ ,
\\
\frac{d}{dx_A}\ \tilde{g}_2(\xi_A) &= 
\left(\frac{d\xi_A}{dx_A}\right)\
\frac{d}{d\xi_A} \tilde{g}_2(\xi_A)
\nonumber\\
&=
-\left(\frac{d\xi_A}{dx_A}\right)\ \tilde{h}_2(\xi_A)\ ,
\\
\frac{d}{dx_A}\ \tilde{h}_2(\xi_A) &= 
\left(\frac{d\xi_A}{dx_A}\right)\
\frac{d}{d\xi_A} \tilde{h}_2(\xi_A)
\nonumber\\
&=
-\left(\frac{d\xi_A}{dx_A}\right)\ \xi_A^{-2}\ \tilde{F}^A_2(\xi_A)\Big\vert_{\rm No~TMCs}\ .
\end{align}
\end{minipage}
\end{subequations}

\item[$\tilde{F}_3^A$:] To build $\tilde{F}_3^A$, we similarly use Eq.~\eqref{eq:dispersionStrFn} and identify everything to the right of $x_A^{-(2l-1)}$
in Eq.~\eqref{eq:strFunT3FullExpression} as the $2l^{th}$ moment of $\tilde{F}_3^A(x_A,Q^2)$ with TMCs.
We then write
\begin{align}
\tilde{F}_3^{A(2l)} &= 
\sum_{j=0}^\infty\
\frac{(2l+j)!}{j!(2l)!}\ 
\left(\frac{M_A^2}{Q^2}\right)^j\ 
(2l)\
\frac{\left(C_3^{(2l+2j)}A_{\tau=2}^{(2l+2j)}\right)}{(2l+2j)}\
.
\end{align}
Using the integral identity of Eq.~\eqref{eq:integralDefPower1} and the relationship in Eq.~\eqref{eq:masterMomentAppendix} between the moment of $F_3^A$ and the product $(C_3A_{\tau=2})$ when $(M_A^2/Q^2)\to0$, we have 
\begin{align}
\frac{C_3^{(2l+2j)}A_{\tau=2}^{(2l+2j)}}{(2l+2j)}
&= 
\frac{1}{(2l+2j)}\ 
\int_0^1 dy\ y^{2l+2j}\cdot  y^{-1} \tilde{F}_3^A(y)\Bigg\vert_{\rm No~TMCs}\ 
\\
& 
= 
\int_0^1 dy\  y^{2l+2j-1}\  \tilde{h}_3(y), 
\quad\text{where}
\\ 
\tilde{h}_3(y) &\equiv \int_y^1 dy'\ (y')^{-1}\ \tilde{F}_3^A(y')\Bigg\vert_{\rm No~TMCs}\ 
.
\end{align}
Using this, the summation over $j$ is then given by
\begin{align}
\tilde{F}_3^{A(2l)} &= 
\int_0^1 dy\
 y^{2l-1}\  \tilde{h}_3(y)\ (2l)\
\sum_{j=0}^\infty\
\frac{(2l+j)!}{j!(2l)!}\ 
\left(\frac{y^2 M_A^2}{Q^2}\right)^j\ 
\\
& = 
\int_0^1 dy\
  y^{-1}\ \tilde{h}_3(y)\ 
  \frac{(2l)\ y^{2l}}{(1-y^2 M_A^2 / Q^2)^{2l+1}}\
  .
\end{align}

After relabeling $N=2l$, using the first derivative identity in Eq.~\eqref{eq:derivativeIdentity}, 
the contour integral of Eq.~\eqref{eq:contour},
the $\delta$ function decomposition of Eq.~\eqref{eq:deltaDecom},
and the identities for $\xi_A$,
we get as the inverse Mellin transformation of $\tilde{F}_3^A$
\begin{align}
\tilde{F}_3^{A}(x_A) &= \frac{1}{2\pi i}\int_{c-i\infty}^{c+i\infty} dN\ x_A^{-N}\ \tilde{F}_3^{AN} 
\\
& = 
\frac{-x_A}{2\pi i}
\frac{d}{dx_A}\
\int_0^1 dy\
\frac{\tilde{h}_3(y)}{y(1-y^2 M_A^2 / Q^2)}\
\int_{c-i\infty}^{c+i\infty} dN\ 
  \frac{(y/x_A)^{N}}{(1-y^2 M_A^2 / Q^2)^{N}}\
\\
& = 
-x_A\frac{d}{dx_A}\
\int_0^1 dy\
\frac{\tilde{h}_3(y)}{y(1-y^2 M_A^2 / Q^2)}\
\left[\frac{y(1-y^2 M_A^2 / Q^2)}{(1+ y^2 M_A^2 / Q^2)}\right]\ \delta(\xi_A-y)\ ,
\\
\Aboxed{
\tilde{F}_3^{A}(x_A)
& =\   
-x_A\frac{d}{dx_A}\ 
\left[\frac{(1+r_A)}{2r_A}\ \tilde{h}_3(\xi_A)\right]\ \ .
} %
\end{align}
Allowing the derivative to act on the quantities inside the bracket recovers Eq.~\eqref{eq:master3}. The final expression is summarized in Eq.~\eqref{eq:app_summary}.
In addition to Eq.~\eqref{eq:app_chainrule_f2}, helpful identities 
include:
\begin{align}
\label{eq:app_chainrule_f3}
\frac{d}{dx_A}\ \left(\frac{(1+r_A)}{2r_A}\right) &=
-\frac{(r_A^2-1)}{2x_A r_A^3}\ 
\quad\text{and}\quad
\frac{d}{dx_A}\ \tilde{h}_3(\xi_A) = 
-\left(\frac{d\xi_A}{dx_A}\right)\ \xi_A^{-1}\ \tilde{F}^A_3(\xi_A)\Big\vert_{\rm No~TMCs}\ .
\end{align}

\item[$\tilde{F}_6^A$:] 
Building $\tilde{F}_6^A$ with TMCs proceeds identically to $\tilde{F}_3^A$.
We first identify everything to the right of $x_A^{-(2l-1)}$ in Eq.~\eqref{eq:strFunT6FullExpression} as the $2l^{th}$ moment of $\tilde{F}_6^A(x_A,Q^2)$. Explicitly,
\begin{align}
\tilde{F}_3^{A(2l)} &= 
\sum_{j=0}^\infty\
\frac{(2l+j)!}{j!(2l)!}\ 
\left(\frac{M_A^2}{Q^2}\right)^j\ 
(2l)\
\frac{\left(C_{6}^{(2l+2j)}A_{\tau=2}^{(2l+2j)}\right)}{(2l+2j)}\
.
\end{align}
The procedure for this case is similar to $\tilde{F}_3^A$. Using the relationship in Eq.~\eqref{eq:masterMomentAppendix}, we obtain
\begin{empheq}[box=\fbox]{align}
\tilde{F}_6^{A}(x_A) &= 
-x_A\frac{d}{dx_A}\ 
\left[\frac{(1+r_A)}{2r_A}\ \tilde{h}_6(y)\right]\
, 
\quad\text{where}\quad
\tilde{h}_6(y) \equiv \int_y^1 dy' y'^{-1}\ \tilde{F}_6^A(y')\Bigg\vert_{\rm No~TMCs}\ 
.
\end{empheq}
The final expression is summarized in Eq.~\eqref{eq:app_summary}
and can be obtained by using identities analogous to Eq.~\eqref{eq:app_chainrule_f3}.
To our knowledge, the TMCs to the $\tilde{F_6}$ structure function have not been previously reported.

\item[$\tilde{F}_1^A$:] 

In principle, building $\tilde{F}_1^A$ follows the same procedure as the previous cases. However, the nuance here is that there are contributions from $(C_1 A_{\tau=2})$ and $(C_2 A_{\tau=2})$ when $j>0$. The solution is to decompose these  into the moments of $\tilde{F}_1^A$ and $\tilde{F}_2^A$, respectively. 
Using Eqs.~\eqref{eq:dispersionStrFn} and \eqref{eq:app_ope_matching}, we identify
everything to the right of $x_A^{-(2l+1)}$
in Eq.~\eqref{eq:strFunT1FullExpression} as the $(2l+1)^{st}$ moment of $\tilde{F}_1^A(x_A,Q^2)$ with TMCs.
Symbolically, this is given by
\begin{align}
\tilde{F}_1^{A(2l+1)} &= 
\sum_{j=0}^\infty\
\frac{(2l+j+1)!}{j!\ (2l+1)!}\ 
\left(\frac{M_A^2}{Q^2}\right)^j\ 
\left[
\left(C_1^{2l+2j+1}A_{\tau=2}^{2l+2j+1}\right)
+
\frac{j\left(C_2^{2l+2j+1}A_{\tau=2}^{2l+2j+1}\right)}{(2l+2j+1)(2l+2j)}
\right]\    
\\
=& \int_0^1 dy\ 
\sum_{j=0}^\infty
\frac{(2l+j+1)!}{j!\ (2l+1)!}\ 
\left(\frac{M_A^2}{Q^2}\right)^j
\left[
y^{2l+2j} \tilde{F}_1^{A}(y)\Big\vert_{\rm No~TMCs}
+
j\ y^{2l+2j-1} \tilde{g}_2(y)
\right]
,
\end{align}
where we expressed $(C_2 A_{\tau=2})$ as an integral over $\tilde{g}_2(y)$ according to Eq.~\eqref{eq:appDefg2_pre}. Using  Eq.~\eqref{eq:summationClosedForm}, we evaluate the summation of each term separately, giving
\begin{align}
\tilde{F}_1^{A(2l+1)}  = & 
\int_0^1 dy\ 
\frac{\tilde{F}_1^{A}(y)\vert_{\rm No~TMCs}\ }{y(1-y^2 M_A^2/Q^2)}\
\left[
\frac{y^{2l+1}}{(1-y^2 M_A^2/Q^2)^{2l+1}}
\right]
\nonumber\\
&+
\left(\frac{M_A^2}{Q^2}\right)\
\int_0^1 dy\ 
\frac{ \tilde{g}_2(y)}{(1-y^2 M_A^2/Q^2)^{2}}\
\left[\frac{(2l+2) y^{2l+1} }{(1-y^2 M_A^2/Q^2)^{2l+1}}\right]
.
\end{align}
We now set $N=(2l+1)$, take the inverse Mellin transformation with respect to $N$, and use the first derivative identity in Eq.~\eqref{eq:derivativeIdentity} to remove the $(N+1)$. The result is
\begin{align}
\tilde{F}_1^{A}(x_A)  = & \frac{1}{2\pi i}\int_{c-i\infty}^{c+i\infty} dN\ x_A^{-N}\ \tilde{F}_1^{AN} 
\\
 = &
\frac{1}{2\pi i}
\int_0^1 dy\ 
\frac{\tilde{F}_1^{A}(y)\vert_{\rm No~TMCs}\ }{y(1-y^2 M_A^2/Q^2)}\
\int_{c-i\infty}^{c+i\infty} dN\
\left[
\frac{(y/x_A)^{N}}{(1-y^2 M_A^2/Q^2)^{N}}
\right]
\nonumber\\
-
\left(\frac{M_A^2 x_A^2}{2\pi i Q^2}\right)\
&
\frac{d}{dx_A}
\int_0^1 dy\ 
\frac{ \tilde{g}_2(y)}{(1-y^2 M_A^2/Q^2)^{2}}\
\int_{c-i\infty}^{c+i\infty} dN\ 
\left[\frac{(y/x_A)^{N}\ x_A^{-1} }{(1-y^2 M_A^2/Q^2)^{N}}\right]
\\
  = &
\left[\frac{\tilde{F}_1^{A}(y)\vert_{\rm No~TMCs}\ }{(1+ y^2 M_A^2 / Q^2)}\right]_{y=\xi_A}\
\nonumber\\
& -
\left(\frac{M_A^2 x_A^2}{Q^2}\right)\
\frac{d}{dx_A}\
\left[\frac{\tilde{g}_2(y)\ x_A^{-1}\ y }{(1-y^2 M_A^2/Q^2)(1+ y^2 M_A^2 / Q^2)}\right]_{y=\xi_A}\
.
\end{align}
To reach the last line, we used the contour integral of Eq.~\eqref{eq:contour} and the $\delta$ function decomposition of Eq.~\eqref{eq:deltaDecom}.
Using the $\xi_A$ identities, the final expressions is 
\begin{empheq}[box=\fbox]{align}
\tilde{F}_1^{A}(x_A) = &
\frac{x_A}{r_A \xi_A}\ \tilde{F}_1^{A}(\xi_A)\Bigg\vert_{\rm No~TMCs}\ 
 -
\left(\frac{M_A^2 x_A^2}{Q^2}\right)\
\frac{d}{dx_A}\
\left[\frac{(1+r_A)}{2 r_A}\ \tilde{g}_2(\xi_A)\right]
\quad .
\label{eq:appF1Atilde}
\end{empheq}
Applying the derivative recovers Eq.~\eqref{eq:master1}.
Helpful identities are given in Eqs.~\eqref{eq:app_chainrule_f2} and \eqref{eq:app_chainrule_f3}.
Note that the factor of $1/2$ in the square brackets is absent in Eq.~(14) of Ref.~\cite{Schienbein:2007gr}. 
This appears to be a typo in Ref.~\cite{Schienbein:2007gr} as the final result 
matches  Eq.~\eqref{eq:master1}.
Furthermore, after  employing $\xi_A$ identities, we find agreement with Eq.~(3.1) of Ref.~\cite{Kretzer:2003iu}.
The final expression is summarized in Eq.~\eqref{eq:app_summary}.

\item[$\tilde{F}_5^A$:] 
Building $\tilde{F}_5^A$ with TMCs, which involves mixing with $\tilde{F}_2^A$, is similar to $\tilde{F}_1^A$. Subsequently, we identify everything to the right of $x_A^{-(2l)}$ in Eq.~\eqref{eq:strFunT5FullExpression} as the $(2l+1)^{th}$ moment of $\tilde{F}_5^A(x_A,Q^2)$:
\begin{align}
&\tilde{F}_5^{A(2l+1)} = 
\sum_{j=0}^\infty \frac{(2l+j+1)!}{j!\ (2l+1)!}\ 
\left(\frac{M_A^2}{Q^2}\right)^j\ 
\nonumber\\
\times &
\left[
\frac{(2l+1)\ \left(C_{5}^{(2l+2j+1)}A_{\tau=2}^{(2l+2j+1)}\right)}{(2l+2j+1)}
-
\frac{j\ (2l+1)\ \left(C_{2}^{(2l+2j+1)}A_{\tau=2}^{(2l+2j+1)}\right)}{
(2l+2j+1)(2l+2j)
}\
\right]
 + \mathcal{O}(\tau>2)\ 
 \\
&
=  
\int_0^1 dy\
\sum_{j=0}^\infty \frac{(2l+j+1)!}{j!\ (2l+1)!}\ 
\left(\frac{M_A^2}{Q^2}\right)^j 
\left[
\frac{(2l+1)}{2}\
 y^{2l+2j}\ \tilde{h}_5(y)
-
j (2l+1)\ y^{2l+2j-1}\ \tilde{g}_2(y)
\right]\ 
\nonumber\\
 & \qquad  +\  \mathcal{O}(\tau>2),\
\end{align}
where we used Eqs.~\eqref{eq:masterMomentAppendix} and \eqref{eq:appDefg2_pre} to decompose both   $(C_i A_{\tau=2})$ products as integrals over 
$\tilde{F}_i^A(y)\vert_{\rm No~TMCs}$,
and have defined via the identity  Eq.~\eqref{eq:integralDefPower1} 
the generating integral
\begin{align}
\label{eq:app_gen_fun_h5}
    \tilde{h}_5(y)\  &\equiv\  \int_y^1 dy'\ (y')^{-1}\ 2\tilde{F}_5^A(y')\Bigg\vert_{\rm No~TMCs}\ .
\end{align} 
Note the factor of $2$ in the definition of $\tilde{h}_5$, which follows the convention of Eq.~(3.15) in Ref.~\cite{Kretzer:2003iu}.

Using Eq.~\eqref{eq:summationClosedForm}, we evaluate the summation of each term separately. This gives
\begin{align}
\tilde{F}_5^{A(2l+1)}   = 
&\int_0^1 dy\
\frac{\tilde{h}_5(y)}{2y(1-y^2 M_A^2/Q^2)}\
\left[\frac{(2l+1)\ y^{2l+1} }{(1-y^2 M_A^2/Q^2)^{(2l+1)}}\right]
\nonumber\\
-
\left(\frac{M_A^2}{Q^2}\right)\ 
&\int_0^1 dy\
\frac{\tilde{g}_2(y)}{(1-y^2 M_A^2/Q^2)^2}\
\left[\frac{(2l+2)\ (2l+1)\ y^{2l+1}}{(1-y^2 M_A^2/Q^2)^{(2l+1)}}\right]
\
  + \mathcal{O}(\tau>2).\
\end{align}

Setting $N=(2l+1)$, taking the inverse Mellin transform with respect to $x_A$, and using the derivative identities of Eq.~\eqref{eq:derivativeIdentity} allows us to obtain $\tilde{F}_5^A(x_A)$. It is given by
\begin{align}
\tilde{F}_5^A(x_A) &= \frac{1}{2\pi i}\ \int_{c-i\infty}^{c+i\infty}dN\ x_A^{-N}\ \tilde{F}_5^{A(N)}
\\
& = 
\left(\frac{-x_A}{4\pi i}\right)\
\frac{d}{dx_A}\
\int_0^1 dy\
\frac{\tilde{h}_5(y)}{y\ (1-y^2 M_A^2/Q^2)}\
\int_{c-i\infty}^{c+i\infty}dN\
\left[\frac{(y/x_A)^{N} }{(1-y^2 M_A^2/Q^2)^{N}}\right]
\nonumber\\
&-
\left(\frac{x_A^2 M_A^2}{2\pi i Q^2}\right)\ 
\frac{d^2}{dx_A^2}
\int_0^1 dy\
\frac{\tilde{g}_2(y)}{(1-y^2 M_A^2/Q^2)^2}\
\int_{c-i\infty}^{c+i\infty}dN\
\left[\frac{(y/x_A)^{N}}{(1-y^2 M_A^2/Q^2)^{N}}\right]
\nonumber\\
& 
  + \mathcal{O}(\tau>2).\
\end{align}
Applying the contour integral of Eq.~\eqref{eq:contour}
and the identity of Eq.~\eqref{eq:deltaDecom}, we get
\begin{empheq}[box=\fbox]{align}
\tilde{F}_5^A(x_A) 
&  = 
\left(\frac{-x_A}{2}\right)\
\frac{d}{dx_A}\
\left[\frac{(1+r_A)}{2r_A}\ \tilde{h}_5(\xi_A)\right]\
-
\left(\frac{x_A^2 M_A^2}{Q^2}\right)\ 
\frac{d^2}{dx_A^2}\left[\frac{(1+r_A)^2}{4r_A}\ 
\xi_A\  
\tilde{g}_2(\xi_A)\right]\
\nonumber\\
& 
\quad  + \mathcal{O}(\tau>2)    \  .
\end{empheq}
After accounting for the difference between $\tilde{W}_5^A$ and $\tilde{F}_5^A$, this agrees with Eq.~(3.5) of Ref.~\cite{Kretzer:2003iu}.
The final expression is summarized in Eq.~\eqref{eq:app_summary}.
In addition to Eqs.~\eqref{eq:app_chainrule_f2} and \eqref{eq:app_chainrule_f3}, a helpful identity is
\begin{align}
\label{eq:app_chainrule_f5}
    \frac{d}{dx_A}\ \tilde{h}_5(\xi_A) = 
-2\left(\frac{d\xi_A}{dx_A}\right)\ \xi_A^{-1}\ \tilde{F}^A_5(\xi_A)\Big\vert_{\rm No~TMCs}\ .
\end{align}

\item[$\tilde{F}_4^A$:]
Building $\tilde{F}_4^A$ follows the same procedure as above with the complication that the  result is sourced by three contributions for $j>0$.   
Using Eqs.~\eqref{eq:dispersionStrFn} and \eqref{eq:app_ope_matching}, we identify 
everything to the right of of $x_A^{-(2l+1)}$ in Eq.~\eqref{eq:strFunT4FullExpression} as the $(2l+1)^{st}$ moment of $\tilde{F}_4^A(x_A,Q^2)$. This is given by
\begin{align}
\tilde{F}^{A(2l+1)}_{4} &= 
\sum_{j=0}^\infty \frac{(2l+j+1)!}{j!\ (2l+1)!}\ 
\left(\frac{M_A^2}{Q^2}\right)^j\ 
\nonumber\\
&\times 
\left[ 
\left(C_{4}^{(2l+2j+1)}A_{\tau=2}^{(2l+2j+1)}\right)
+
\frac{j(j-1) \left(C_{2}^{(2l+2j+!)}A_{\tau=2}^{(2l+2j+1)}\right)}{
(2l+2j+1)\ (2l+2j)
}
-
\frac{2j \left(C_{5}^{(2l+2j+1)}A_{\tau=2}^{(2l+2j+1)}\right)}{(2l+2j+1)} 
\right]\ 
\nonumber\\
& \qquad  +\  \mathcal{O}(\tau>2),\
\\  
& = \int_0^1 dy\  
\sum_{j=0}^\infty \frac{(2l+j+1)!}{j!\ (2l+1)!}\ 
\left(\frac{M_A^2}{Q^2}\right)^j\ 
\nonumber\\
  \times &
\left[ 
 y^{2l+2j} \tilde{F}_4^{A}(y)\vert_{\rm No~TMC}\ 
+\ 
j (j-1) y^{2l+2j-1} \tilde{g}_2(y)\
-\
j  y^{2l+2j} \tilde{h}_5(y)
\right]\ 
 +\  \mathcal{O}(\tau>2)\ .
\end{align}
Here, we again used  the Eqs.~\eqref{eq:masterMomentAppendix}, \eqref{eq:appDefg2_pre}, 
and 
\eqref{eq:app_gen_fun_h5}
to rewrite $(C_i A_{\tau=2})$.

Distributing the summations and using the   identities listed in Eq.~\eqref{eq:summationClosedForm}, we obtain
\end{description}
\begin{align}
\tilde{F}^{A(2l+1)}_{4} =   
\int_0^1 dy\  
\Bigg [y^{2l}\
\tilde{F}_4^{A}(y)\vert_{\rm No~TMC}   
&\sum_{j=0}^\infty \frac{(2l+j+1)!}{j!\ (2l+1)!} 
\left(\frac{M_A^2 y^2}{Q^2}\right)^j\ 
\nonumber\\
+\ 
y^{2l-1}\ \tilde{g}_2(y) 
&\sum_{j=0}^\infty j (j-1) \frac{(2l+j+1)!}{j!\ (2l+1)!} 
\left(\frac{M_A^2 y^2}{Q^2}\right)^j 
 \nonumber\\
-\
y^{2l}\
\tilde{h}_5(y)\ 
&\sum_{j=0}^\infty j\frac{(2l+j+1)!}{j!\ (2l+1)!}\ 
\left(\frac{M_A^2 y^2}{Q^2}\right)^j\ 
 \Bigg ]
  +\  \mathcal{O}(\tau>2)\ 
\\
= 
\int_0^1 dy\  
\Bigg [
\frac{\tilde{F}_4^{A}(y)\vert_{\rm No~TMC}}{y(1-y^2 M_A^2 / Q^2)}\
&  
\left[\frac{y^{2l+1} }{(1-y^2 M_A^2 / Q^2)^{2l+1}}\right]\ 
\nonumber\\
+\ 
\left(\frac{M_A^2}{Q^2}\right)^2\
\frac{y^2\ \tilde{g}_2(y)}{(1-y^2 M_A^2 / Q^2)^{3}}\
&\left[
\frac{(2l+2)(2l+3)y^{2l+1}}{(1-y^2 M_A^2 / Q^2)^{2l+1}}\right]\ 
\nonumber\\ 
-\
\left(\frac{M_A^2}{Q^2}\right)\ 
\frac{y\ \tilde{h}_5(y)}{(1-y^2 M_A^2 / Q^2)}
&\left[\frac{(2l+2)y^{2l+1}}{(1-y^2 M_A^2 / Q^2)^{2l+1}}\right]\
 \Bigg ]\ 
  +\  \mathcal{O}(\tau>2)\ .
\end{align}

Relabeling $N=(2l+1)$ and using the derivative identities in Eq.~\eqref{eq:derivativeIdentity} gives
\begin{align}
\tilde{F}_4^{A}(x_A) = \frac{1}{2\pi i}
&\int_{c-i\infty}^{c+i\infty} dN\ x_A^{-N}\ \tilde{F}_4^{AN} 
\\
 =  
\frac{1}{2\pi i}\ 
&
\int_0^1 dy\  
\frac{\tilde{F}_4^{A}(y)\vert_{\rm No~TMC}}{y(1-y^2 M_A^2 / Q^2)}\  
\int_{c-i\infty}^{c+i\infty} dN\
\left[
 \frac{(y/x_A)^{N} }{(1-y^2 M_A^2 / Q^2)^{N}}\ 
 \right]
 \nonumber\\   
 +\ 
\left(\frac{M_A^2}{Q^2}\right)^2\    
\left(\frac{x_A^3}{2\pi i}\right)\ 
\frac{d^2}{dx_A^2}\
&
\int_0^1 dy\  
\frac{y^2\ x_A^{-1}\ \tilde{g}_2(y)}{(1-y^2 M_A^2 / Q^2)^3}\ 
\int_{c-i\infty}^{c+i\infty} dN\
\left[
\frac{(y/x_A)^{N}}{(1-y^2 M_A^2 / Q^2)^{N}}\
\right]
 \nonumber\\ 
+(-1)^2\
\left(\frac{M_A^2}{Q^2}\right)\ 
\left(\frac{x_A^2}{2\pi i}\right)\ 
\frac{d}{dx_A}\
& 
\int_0^1 dy\  
\frac{y\ x_A^{-1}\ \tilde{h}_5(y)}{(1-y^2 M_A^2 / Q^2)^2}\ 
\int_{c-i\infty}^{c+i\infty} dN\
\left[
\frac{(y/x_A)^{N}}{(1-y^2 M_A^2 / Q^2)^{N}}\
\right]
\nonumber
\\
& +\  \mathcal{O}(\tau>2)\ .
\end{align}
Finally, using contour integral of Eq.~\eqref{eq:contour} and the $\xi_A$ identities, we get
\begin{empheq}[box=\fbox]{align}
\tilde{F}_4^{A}(x_A) = 
&
\frac{(1+r_A)}{2r_A}\
\tilde{F}_4^{A}(\xi_A)\Bigg\vert_{\rm No~TMC}
 +\ 
\left(\frac{M_A^2}{Q^2}\right)^2\    
x_A^3\ 
\frac{d^2}{dx_A^2}\
\left[
\left(\frac{x_A^2}{r_A}\right)\
\tilde{g}_2(\xi_A)
\right]
 \nonumber\\ 
& +
\left(\frac{M_A^2}{Q^2}\right)\ 
x_A^2\ 
\frac{d}{dx_A}\
\left[
\left(\frac{x_A}{r_A}\right)\ \tilde{h}_5(\xi_A)
\right]\
 +\  
 \mathcal{O}(\tau>2)\ .
\end{empheq}
After accounting for differences between $\tilde{W}_4^A$ and $\tilde{F}_4^A$, this agrees with Eq.~(3.4) of Ref.~\cite{Kretzer:2003iu}.
The final expression is summarized in Eq.~\eqref{eq:app_summary}.
In addition to Eqs.~\eqref{eq:app_chainrule_f2},
\eqref{eq:app_chainrule_f3},
and \eqref{eq:app_chainrule_f5}, a helpful identity is
\begin{align}
\label{eq:app_chainrule_f4}
    \frac{d}{dx_A}\ \frac{x_A^n}{r_A} = 
\frac{(r_A^2+n-1)x_A^{n-1}}{r_A}\ .
\end{align}

\subsection*{Summary}

We now give the full expressions for all six structure functions with TMCs. These are obtained by successive applications of the product and chain rules to the several expressions above. The final results are:
\begin{subequations}
\label{eq:app_summary}
\begin{align}
\tilde{F}_1^{A}(x_A) &= 
\frac{x_A}{\xi_A r_A}\ \tilde{F}_1^{A}(\xi_A)\Bigg\vert_{\rm No~TMCs}\ 
+\
\left(\frac{M_A^2 x_A^2}{Q^2 r_A^2}\right)\ \tilde{h}_2(\xi_A)\
+\
\left(\frac{2M_A^4 x_A^3}{Q^4 r_A^3}\right)\ \tilde{g}_2(\xi_A)\ 
 +\  
 \mathcal{O}(\tau>2)\ ,
\label{eq:app_summary_f1}
\\
\tilde{F}_2^{A}(x_A) &= 
\frac{x_A^2}{\xi_A^2 r_A^3}\ \tilde{F}_2^{A}(\xi_A)\Bigg\vert_{\rm No~TMCs}\ 
+\
\left(\frac{6 M_A^2 x_A^3}{Q^2 r_A^4}\right)\ \tilde{h}_2(\xi_A)\
+\
\left(\frac{12M_A^4 x_A^4}{Q^4 r_A^5}\right)\ \tilde{g}_2(\xi_A)\ 
\nonumber \\ 
& \qquad\ +\ \mathcal{O}(\tau>2)\ ,
\label{eq:app_summary_f2}
\\  
\tilde{F}_3^{A}(x_A) &= 
\frac{x_A}{\xi_A r_A^2}\
\tilde{F}_3^{A}(\xi_A) \Bigg\vert_{\rm No~TMCs}\ 
+\
\left(\frac{2M_A^2 x_A^2}{Q^2 r_A^3}\right)\ \tilde{h}_3(\xi_A)\ 
 +\  
 \mathcal{O}(\tau>2)\ ,
\label{eq:app_summary_f3}
\\
\tilde{F}_4^{A}(x_A) &= 
\frac{x_A}{\xi_A r_A}\ \tilde{F}_4^{A}(\xi_A)\Bigg\vert_{\rm No~TMCs}\ 
-\ 
\left(\frac{2M_A^2 x_A^2}{Q^2 r_A^2}\right)\ 
\tilde{F}_5^{A}(\xi_A)\Bigg\vert_{\rm No~TMCs}\ 
+\ 
\left(\frac{M_A^4 x_A^3}{Q^4 r_A^3}\right)\ 
\tilde{F}_2^{A}(\xi_A)\Bigg\vert_{\rm No~TMCs}\ 
\nonumber \\ 
& 
\qquad\ +\   
\left(\frac{M_A^2 x_A^2}{Q^2 r_A^3}\right)\ 
\tilde{h}_5^{A}(\xi_A)\
-\ 
\left(\frac{2M_A^4 x_A^4}{Q^4 r_A^4}\right)\ 
\left(2-\xi_A^2 M_A^2/Q^2\right)\ 
\tilde{h}_2^{A}(\xi_A)\ 
\nonumber \\ 
& 
\qquad\ +\
\left(\frac{2M_A^4 x_A^3}{Q^4 r_A^5}\right)\ 
\left(1-2 x_A^2 M_A^2/Q^2\right)\ 
\tilde{g}_2^{A}(\xi_A)\ 
+\   
 \mathcal{O}(\tau>2)\ ,
\label{eq:app_summary_f4}
\\
\tilde{F}_5^{A}(x_A) &= 
\frac{x_A}{\xi_A r_A^2}\ \tilde{F}_5^{A}(\xi_A)\Bigg\vert_{\rm No~TMCs}\ 
-\ 
\left(\frac{M_A^2 x_A^2}{Q^2 r_A^3 \xi_A}\right)\ 
\tilde{F}_2^{A}(\xi_A)\Bigg\vert_{\rm No~TMCs}\ 
\nonumber \\ 
& 
\qquad\ +\   
\left(\frac{M_A^2 x_A^2}{Q^2 r_A^3}\right)\ 
\tilde{h}_5^{A}(\xi_A)\
-\ 
\left(\frac{2M_A^2 x_A^2}{Q^2 r_A^4}\right)\ 
\left(1-x_A \xi_A  M_A^2/Q^2\right)\ 
\tilde{h}_2^{A}(\xi_A)\ 
\nonumber \\ 
& 
\qquad\ +\
\left(\frac{6 M_A^4 x_A^3}{Q^4 r_A^5}\right)\ 
\tilde{g}_2^{A}(\xi_A)\ 
+\   
 \mathcal{O}(\tau>2)\ ,
\label{eq:app_summary_f5}
\\
\tilde{F}_6^{A}(x_A) &= 
\frac{x_A}{\xi_A r_A^2}\
\tilde{F}_6^{A}(\xi_A) \Bigg\vert_{\rm No~TMCs}\ 
+\
\left(\frac{2M_A^2 x_A^2}{Q^2 r_A^3}\right)\ \tilde{h}_6(\xi_A)\ 
 +\  
 \mathcal{O}(\tau>2)\ .
\label{eq:app_summary_f6}
\end{align}
\end{subequations}
Importantly, the above expressions for \textit{nuclear} structure functions $\tilde{F}_1,\dots, \tilde{F}_5$ agree with those for \textit{nucleon} structure functions ${F}_1,\dots, {F}_5$ in Eq.~(3.17) of Ref.~\cite{Schienbein:2007gr}. This is a main conclusion of our work: TMCs for unpolarized protons and neutrons are the same for unpolarized nuclei.

Using Table~\ref{tab:helicity_expansion}, 
the TMCs to the $\tilde{F}_+^A$, $\tilde{F}_-^A$, and $\tilde{F}_0^A$
structure functions in the helicity basis are explicitly 
\begin{align}
\label{eq:app_summary_helicity}
\tilde{F}_\pm^A &= \tilde{F}_1^A\ \mp\ r_A  \tilde{F}_3^A
\\
&=
\frac{x_A}{\xi_A r_A}\ 
\left[\tilde{F}_1^{A}(\xi_A)\ \mp\ \tilde{F}_3^{A}(\xi_A)
\right]_{\rm No~TMCs}\
+\
\left(\frac{M_A^2 x_A^2}{Q^2 r_A^2}\right)\ 
\left[\tilde{h}_2(\xi_A)\ \mp\ 2\tilde{h}_3(\xi_A)\right]\ 
\nonumber\\
&\qquad +\
\left(\frac{2M_A^4 x_A^3}{Q^4 r_A^3}\right)\ \tilde{g}_2(\xi_A)\ 
 +\  
 \mathcal{O}(\tau>2)\ ,
\\
\tilde{F}_0^A &= \left(\frac{r_A^2}{2x_A}\right)\ \tilde{F}_2^A - \tilde{F}_1^A
\\
&=
\frac{x_A}{2\xi_A^2 r_A}\
\left[
 \tilde{F}_2^{A}(\xi_A)
-
2\xi \tilde{F}_1^{A}(\xi_A)
\right]_{\rm No~TMCs}\ 
+\
\left(\frac{2 M_A^2 x_A^2}{Q^2 r_A^2}\right)\ \tilde{h}_2(\xi_A)\
+\
\left(\frac{4M_A^4 x_A^3}{Q^4 r_A^3}\right)\ \tilde{g}_2(\xi_A)\ 
\nonumber \\ 
& \qquad\ +\ \mathcal{O}(\tau>2)\ .
\end{align}
As noted in Eq.\,(\ref{eq:FLF0}), there is a relative factor of 
$2x_A$ between the helicity $\tilde{F}_0^A$ and the longitudinal $\tilde{F}_L^A$ 
structure functions such that $\tilde{F}_L^A = 2x_A \tilde{F}_0^A$.

\section{Derivation of the Full/Leading TMC Parameterization}
\label{app:par}

In this appendix, we derive our parameterizations  of Eq.~\eqref{eqs:approx} by approximating the general form of the structure functions in the TMC region. 
First,  we assume the structure functions $F^{(0)}_a(x)$, $a=1,2,3$, vanish at $x=1$.  
This follows from our working assumption that  the nPDFs vanish at $x\geq1$ 
{c.f.}, Sec.\,\ref{sec:largex}.

Next, we use  a finite difference formula to   approximate  derivatives of $F^{j, (0)}_{a}(x=\xi)$ as
\begin{align}
 F_a^{'(0)}(u=\xi) &\approx  - \frac{ \gamma_a F_a^{(0)}(\xi)}{1-\xi} \\
 F_a^{j, (0)}(u=\xi) & \approx \frac{0- F_a^{j-1, (0)}(\xi)}{(1-\xi)}= (-1)^{j} \frac{ \gamma_a F_a^{(0)}(\xi)}{(1-\xi)^j} \quad . 
\end{align}
Here,  $\gamma_a$ is a universal correction factor for the first derivative which we assume to 
be independent of $x$ and the underlying PDFs, and  to have a mild $Q$ dependence.
We can then expand $F_a^{(0)}(u)$ about $u=\xi$ as 
\begin{align}
    F_a^{(0)}(u)& = 
    \sum_{j=0}^\infty \frac{1}{j!}  F_a^{j, (0)}(\xi) (u-\xi)^j \\
    &\approx  F_a^{(0)}(\xi) \ 
    \left(1+\sum_{j=1}^\infty \frac{1}{j!} (-1)^{j} \frac{ \gamma_a }{(1-\xi)^j} (u-\xi)^j  \right) \\
   & \equiv F_a^{(0)}(\xi)  \, K_a(u, \xi, \gamma_a) \quad . 
\end{align}
Here, 
\begin{equation}
K_a(u, \xi, \gamma_a) = 1+\sum_{j=1}^\infty \frac{1}{j!} (-1)^{j} \frac{ \gamma_a }{(1-\xi)^j} (u-\xi)^j\,.
\end{equation}
As $K_a(u, \xi, \gamma)$ is independent of the PDFs,  this implies that the ratios 
\begin{align}
   \frac{h_a(\xi)}{ F_a^{(0)}(\xi)}  = \int_\xi^1 L_a(u) K_a(\xi, x, \gamma) du\\
   \frac{g_2(\xi)}{ F_2^{(0)}(\xi)} =  \int_\xi^1 \frac{u-\xi}{u^2} K_a(\xi, x, \gamma) du
\end{align}
are also independent of the underlying PDFs. 
Here we have defined $L_1(u) = 2/u$, $L_2(u) = 1/u^2$, and $L_3(u) = 1/u $.

Evaluating the explicit expression for $K_a(u,\xi, \gamma_a)$ in the above relations, we obtain (\ref{h2of2}), (\ref{h3of3}), and (\ref{g2of2}). To calculate $h_2(\xi)/F_1^{(0)}(\xi)$ and $g_2(\xi)/F_1^{(0)}(\xi)$, which are needed to calculate the ratio $F_1^{TMC}(x) / F_1^{Leading-TMC}$, one can assume an approximate Callan-Gross relation. Deviations from the Callan-Gross relation can be absorbed into the fitted $\gamma_1(Q)$. We note that in practice, we calculate the massless structure function using 
Eq.~\eqref{eq:master_acot}, namely : 
\begin{subequations}
\begin{align}
 F_{1}^{(0)}(\xi) & = F_{1}^{TMC-ACOT}(x), \\
 F_{2}^{(0)}(\xi) & =\frac{\xi \, r^{2}}{x} F_{2}^{TMC-ACOT}(x), \\
 F_{3}^{(0)}(\xi) & = r F_{3}^{TMC-ACOT}(x)  ,
\end{align}
\end{subequations}
Assuming Callan-Gross relation at the level of the TMC-ACOT structure functions, implies 
\begin{equation}\label{Eqs:F2F1}
    F_{2}^{(0)}(\xi) =2\xi r^2  F_{1}^{(0)}(\xi) 
\end{equation}
Using Eq.~\eqref{Eqs:F2F1}, one can easily obtain $h_2(\xi) / F_1^{(0)}(\xi)$ and $g_2(\xi) / F_1^{(0)}(\xi)$ from $h_2(\xi) / F_2^{(0)}(\xi)$ and $g_2(\xi) / F_2^{(0)}(\xi)$, as shown in Eqs.~\eqref{h2of1} and \eqref{g2of1}.

If we perform a single parameter fit and take $\gamma_a$ as a constant, 
we find reasonable agreement with the exact TMC result at the level of $\lesssim 0.75\%$ for $Q=1.3$~GeV and $Q=2$~GeV. We can do even better if we perform a 2-parameter fit and replace $\gamma_a$ with: 
\begin{equation}
     \gamma_a \rightarrow \gamma_a(Q) = \lambda_a\ln(Q)^{\delta_a}
\end{equation}
Thus, we give a mild $Q$ dependence to the parameter $\gamma_a$. This  2-parameter fit is superior to the one parameter fit as 
the agreement is $\lesssim 0.6\%$ for $Q=1.3$~GeV, and 
   $\lesssim 0.2\%$ for $Q=2$~GeV which is a typical kinematic cut for the global PDF fits.

\clearpage

\bibliography{nTMC_refs,nTMC_refs_extra}

\end{document}